\documentclass[12pt]{article}

  \usepackage{graphicx}
  \usepackage{epsfig}
  \usepackage{amssymb}
  \usepackage{subfigure}
    \usepackage{feynmf}%%%{feynmp}
\evensidemargin - 1.truecm
\begin{document}
\begin{fmffile}{NONabFF}

\newcommand{\be}{\begin{equation}}
\newcommand{\ee}{\end{equation}}
\newcommand{\nn}{\nonumber}
\newcommand{\bea}{\begin{eqnarray}}
\newcommand{\eea}{\end{eqnarray}}
\newcommand{\bfig}{\begin{figure}}
\newcommand{\efig}{\end{figure}}
\newcommand{\bc}{\begin{center}}
\newcommand{\ec}{\end{center}}
\newcommand{\bd}{\begin{displaymath}}
\newcommand{\ed}{\end{displaymath}}

\begin{titlepage}
\nopagebreak
{\flushright{
        \begin{minipage}{5cm}
        PITHA 04/09\\
        Freiburg-THEP 04/08\\
        ZU-TH 09/04\\
        NSF-KITP-04-59\\
        UCLA/04/TEP/22\\
%        CERN-TH/2004-XXX\\
        {\tt hep-ph/0406046}\\
        \end{minipage}        }

}
\vspace*{-1.5cm}                        
\vskip 2.5cm
\begin{center}
\boldmath
{\Large \bf Two-loop QCD Corrections to the Heavy Quark \\[2mm]
Form Factors: the Vector Contributions}\unboldmath
\vskip 1.cm
{\large  W.~Bernreuther$\rm \, ^{a, \,}$\footnote{Email: 
{\tt breuther@physik.rwth-aachen.de}}}
{\large  R.~Bonciani$\rm \, ^{b, \,}$\footnote{Email: 
{\tt Roberto.Bonciani@physik.uni-freiburg.de}}}
{\large T.~Gehrmann$\rm \, ^{c, \, a, \,}$\footnote{Email: 
{\tt gehrt@physik.unizh.ch}}},
{\large R.~Heinesch$\rm \, ^{a, \,}$\footnote{Email: 
{\tt heinesch@physik.rwth-aachen.de}}}, \\[2mm] 
{\large T.~Leineweber$\rm \, ^{a, \,}$\footnote{Email: 
{\tt leineweber@physik.rwth-aachen.de}}}, 
{\large P.~Mastrolia$\rm \, ^{d, \,}$\footnote{Email: 
{\tt mastrolia@physics.ucla.edu}}}, and
{\large E.~Remiddi$\rm \, ^{e, \, f, \,}$\footnote{Email: 
{\tt Ettore.Remiddi@bo.infn.it}}}
\vskip .7cm
{\it $\rm ^a$ Institut f\"ur Theoretische Physik, RWTH Aachen,
D-52056 Aachen, Germany} 
\vskip .3cm
{\it $\rm ^b$ Fakult\"at f\"ur Mathematik und Physik, Albert-Ludwigs-Universit\"at
Freiburg, \\ D-79104 Freiburg, Germany} 
\vskip .3cm
{\it $\rm ^c$ Institut f\"ur Theoretische Physik, 
Universit\"at Z\"urich, CH-8057 Z\"urich, Switzerland}
\vskip .3cm
{\it $\rm ^d$ Department of Physics and Astronomy, UCLA,
Los Angeles, CA 90095-1547} 
\vskip .3cm
{\it $\rm ^e$ Theory Division, CERN, CH-1211 Geneva 23, Switzerland} 
\vskip .3cm
{\it $\rm ^f$ Dipartimento di Fisica dell'Universit\`a di Bologna, and
INFN, Sezione di Bologna, I-40126 Bologna, Italy} 
\end{center}
\vskip .4cm

\begin{abstract} 
We present closed analytic expressions of the electromagnetic vertex form 
factors for heavy quarks at the two-loop level in QCD for arbitrary momentum 
transfer. The calculation is carried out in dimensional regularization. 
The electric and magnetic form factors are expressed in terms of 
1-dimensional harmonic polylogarithms of maximum weight 4.

\flushright{
        \begin{minipage}{12.3cm}
{\it Key words}:  Feynman diagrams, Multi-loop calculations,  Vertex diagrams,
\hspace*{18.5mm} Heavy quarks.\\
{\it PACS}: 11.15.Bt, 12.38.Bx, 14.65.Fy, 14.65.Ha
        \end{minipage}        }
\end{abstract}
\vfill
\end{titlepage}

\section{Introduction \label{Intro}}

The forward-backward asymmetry $A_{fb}^b$ in the production of bottom quarks 
at electron-positron colliders displays presently a substantial discrepancy 
between experimental measurement and theoretical expectation~\cite{lepewwg}. 
This theoretical expectation is determined using the electroweak parameters
obtained from a global fit to a number of different electroweak precision 
observables, including this forward-backward asymmetry itself. In turn, 
$A_{fb}^b$ exercises a strong pull on the global fit, in particular towards
larger masses of the Higgs boson.  In particular, among the electroweak 
precision observables, $A_{fb}^b$ is more sensitive on the Higgs mass than 
most other quantities. 

The present theoretical description of  $A_{fb}^b$ includes the 
fully massive next-to-leading order (NLO)  electroweak~\cite{onel1ew}
and fully massive NLO QCD~\cite{Arbuzov,Djouadi} corrections as well as 
the leading terms from the next-to-next-to-leading order (NNLO)  
QCD corrections~\cite{seymour} (see also \cite{Altarelli,vanNeerven}), which 
were obtained based on the massless approximation plus leading logarithmic 
mass terms. Given the substantial discrepancy between the experimental result
and the theoretical expectation and the high impact on the Higgs mass 
determination,  a more precise theoretical understanding of $A_{fb}^b$
is clearly desired. 

At a future linear collider~\cite{tesla}, 
precision determinations of electroweak parameters will again involve 
the forward-backward asymmetries. In this setting, the 
top quark asymmetry $A_{fb}^t$, is experimentally accessible, and of high 
interest in understanding the interplay of quark mass generation and 
electroweak symmetry breaking. For a precise theoretical description of this 
asymmetry, inclusion of mass corrections is clearly mandatory.

The NNLO QCD corrections to $A_{fb}^Q$ for massive quarks $Q$
involve three classes of contributions: (1) the tree level matrix elements for 
the decay of a vector boson into four partons, (at least) two of which being 
the heavy quark-antiquark pair; (2) the one-loop corrected matrix elements for 
the decay of a vector boson into a heavy quark-antiquark pair plus a gluon;
(3) the two-loop corrections to the decay of a vector boson into a 
heavy quark-antiquark pair. While the former two contributions can be obtained
\cite{Bernreuther:2000zx}  along the lines of the calculations
of three jet production involving heavy quarks
\cite{Brandenburg:1997pu,OleNas,Rodrigo:1999qg}, the latter remain to
be calculated. 

It is the aim of this paper (and of two companion papers)
to contribute to the NNLO QCD corrections to 
$A^Q_{fb}$ for massive quarks by 
computing the virtual two-loop QCD corrections 
to the form factors of a massive quark. These form factors describe the full 
structure of the $(Z^*,\gamma^*)\to 
Q\bar Q$ vertex function, involving the vector and 
axial vector couplings of the vector boson.  In the present paper, we derive 
the two-loop corrections to the vector form factors, 
while two following papers 
will discuss the parity-violating form factors at two loops. The two-loop 
corrections to the vector form factors were considered previously only  
in~\cite{teubner}, where the contribution from 
closed fermion loops was calculated.

This paper is organized as follows.

In Section \ref{FFact} we give the notations followed throughout all the paper,
defining the form factors and their expansions in terms of the coupling
constant.
In Section \ref{unrenorm} we give the virtual contributions to the one- and 
two-loop form factors before the subtraction of the UV divergences.
In Section \ref{renorm} we discuss in detail the renormalization and
in Section \ref{FFrenorm} we give the UV-renormalized form factors at the one- and
two-loop level. These results are presented for the case of choosing the 
renormalization scale $\mu$ equal to the mass $m$ of the heavy quark. 
In the Subsection \ref{munotm} the logarithms of the ratio 
$m/\mu$, that are present if $\mu \neq m$, are explicitly given.
Finally, Section \ref{analytical} deals with the analytical continuation of 
our formulas above the threshold.

%%%%%%%%%%%%%%%%%%%% one-loop Vertex %%%%%%%%%%%%%%%%%%%%%%%%%%%%%%%%%%%%%
\bfig
\bc
\subfigure[]{
\begin{fmfgraph*}(30,30)
\fmfleft{i}
\fmfright{o1,o2}
\fmfforce{0.8w,0.93h}{v2}
\fmfforce{0.8w,0.07h}{v1}
\fmfforce{0.2w,0.5h}{v5}
%\fmfforce{0.66w,0.5h}{v55}
\fmf{double}{v1,o1}
\fmf{double}{v2,o2}
\fmf{photon}{i,v5}
\fmflabel{$p_2$}{o1}
\fmflabel{$p_1$}{o2}
\fmflabel{$Q$}{i}
\fmf{double,tension=.3}{v2,v5}
\fmf{double,tension=.3}{v1,v5}
\end{fmfgraph*}}
%
%%%%%%%%%%%%%%%%%%%%%%%
%
\hspace{20mm}
\subfigure[]{
\begin{fmfgraph*}(30,30)
\fmfleft{i}
\fmfright{o1,o2}
\fmfforce{0.8w,0.93h}{v2}
\fmfforce{0.8w,0.07h}{v1}
\fmfforce{0.2w,0.5h}{v5}
\fmf{double}{v1,o1}
\fmf{double}{v2,o2}
\fmf{photon}{i,v5}
\fmflabel{$p_2$}{o1}
\fmflabel{$p_1$}{o2}
\fmflabel{$Q$}{i}
\fmf{double,tension=.3}{v2,v5}
\fmf{double,tension=.3}{v1,v5}
\fmf{gluon,tension=0}{v1,v2}
\end{fmfgraph*}}
%
%%%%%%%%%%%%%%%%%%%%%%%
%
\vspace*{8mm}
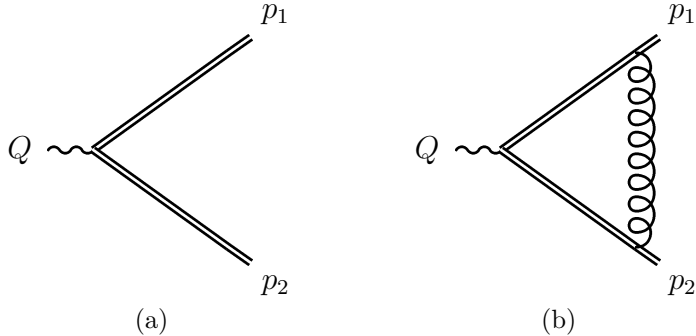
\caption{\label{fig1} Tree-level and one-loop diagrams, involved in the
calculation of the heavy-quark vertex form factors. The curly 
line represents a gluon; the double straight lines, quarks of 
mass $m$. The external fermion lines are on the mass-shell: 
$p_1^2 = p_2^2 = m^{2}$. The wavy line on the l.h.s. carries  
momentum $Q=p_{1}+p_{2}$, with the metrical convention $Q^{2}<0$ 
when $Q$ is space-like.}
\ec
\efig
%%%%%%%%%%%%%%%%%%%%%%%%%%%%%%%%%%%%%%%%%%%%%%%%%%%%%%%%%%%%%%%%%%%%%%%%

%%%%%%%%%%%%%%%%%%%% two-loop Vertex ABELIAN %%%%%%%%%%%%%%%%%%%%%%%%%%%%%%%
\bfig
\bc
\subfigure[]{
\begin{fmfgraph*}(30,30)
\fmfleft{i}
\fmfright{o1,o2}
\fmf{double}{v1,o1}
\fmf{double}{v2,o2}
\fmf{photon}{i,v5}
\fmflabel{$p_{2}$}{o1}
\fmflabel{$p_{1}$}{o2}
\fmflabel{$Q$}{i}
\fmf{double,tension=.4}{v2,v3}
\fmf{double,tension=.3}{v3,v5}
\fmf{double,tension=.4}{v1,v4}
\fmf{double,tension=.3}{v4,v5}
\fmf{gluon,tension=0}{v1,v2}
\fmf{gluon,tension=0}{v4,v3}
\end{fmfgraph*} }
%
%%%%%%%%%%%%%%%%%%%%%%%
%
%
\hspace{8mm}
\subfigure[]{
\begin{fmfgraph*}(30,30)
\fmfleft{i}
\fmfright{o1,o2}
\fmf{double}{v1,o1}
\fmf{double}{v2,o2}
\fmf{photon}{i,v5}
\fmflabel{$p_{2}$}{o1}
\fmflabel{$p_{1}$}{o2}
\fmflabel{$Q$}{i}
\fmf{double,tension=.3}{v2,v3}
\fmf{double,tension=.3}{v3,v5}
\fmf{double,tension=.3}{v1,v4}
\fmf{double,tension=.3}{v4,v5}
\fmf{gluon,tension=0}{v4,v2}
\fmf{gluon,tension=0}{v1,v3}
\end{fmfgraph*} }
%
%%%%%%%%%%%%%%%%%%%%%%%
%
\hspace{8mm}
\subfigure[]{
\begin{fmfgraph*}(30,30)
\fmfleft{i}
\fmfright{o1,o2}
\fmfforce{0.8w,0.93h}{v2}
\fmfforce{0.8w,0.07h}{v1}
\fmfforce{0.8w,0.5h}{v3}
\fmfforce{0.2w,0.5h}{v5}
\fmfforce{0.8w,0.3h}{v10}
\fmf{double}{v1,o1}
\fmf{double}{v2,o2}
\fmf{photon}{i,v5}
\fmflabel{$p_{2}$}{o1}
\fmflabel{$p_{1}$}{o2}
\fmflabel{$Q$}{i}
\fmf{double,tension=0}{v2,v5}
\fmf{double,tension=0}{v3,v4}
\fmf{gluon,tension=.4}{v4,v1}
\fmf{double,tension=.4}{v4,v5}
\fmf{double,tension=0}{v1,v3}
\fmf{gluon,tension=0}{v3,v2}
\end{fmfgraph*} } \\
%
%%%%%%%%%%%%%%%%%%%%%%%
%
\subfigure[]{
\begin{fmfgraph*}(30,30)
\fmfleft{i}
\fmfright{o1,o2}
\fmfforce{0.8w,0.93h}{v2}
\fmfforce{0.8w,0.07h}{v1}
\fmfforce{0.2w,0.5h}{v5}
\fmfforce{0.8w,0.4h}{v9}
\fmfforce{0.5w,0.45h}{v10}
\fmfforce{0.8w,0.5h}{v11}
\fmf{double}{v1,o1}
\fmf{double}{v2,o2}
\fmf{photon}{i,v5}
\fmflabel{$p_{2}$}{o1}
\fmflabel{$p_{1}$}{o2}
\fmflabel{$Q$}{i}
\fmf{double}{v2,v3}
\fmf{gluon,tension=.25,right}{v3,v4}
\fmf{double,tension=.25}{v3,v4}
\fmf{double}{v4,v5}
\fmf{double}{v1,v5}
\fmf{gluon}{v1,v2}
\end{fmfgraph*} }
%
%%%%%%%%%%%%%%%%%%%%%%%
%
\hspace{8mm}
\subfigure[]{
\begin{fmfgraph*}(30,30)
\fmfleft{i}
\fmfright{o1,o2}
\fmfforce{0.8w,0.93h}{v2}
\fmfforce{0.8w,0.07h}{v1}
\fmfforce{0.8w,0.3h}{v3}
\fmfforce{0.8w,0.7h}{v4}
\fmfforce{0.2w,0.5h}{v5}
\fmf{double}{v1,o1}
\fmf{double}{v2,o2}
\fmf{photon}{i,v5}
\fmflabel{$p_{2}$}{o1}
\fmflabel{$p_{1}$}{o2}
\fmflabel{$Q$}{i}
\fmf{double}{v2,v5}
\fmf{gluon}{v1,v3}
\fmf{gluon}{v4,v2}
\fmf{double}{v1,v5}
\fmf{double,right}{v4,v3}
\fmf{double,right}{v3,v4}
\end{fmfgraph*} }
%
%%%%%%%%%%%%%%%%%%%%%%%
%
\hspace{8mm}
\subfigure[]{
\begin{fmfgraph*}(30,30)
\fmfleft{i}
\fmfright{o1,o2}
\fmfforce{0.8w,0.93h}{v2}
\fmfforce{0.8w,0.07h}{v1}
\fmfforce{0.8w,0.3h}{v3}
\fmfforce{0.8w,0.7h}{v4}
\fmfforce{0.2w,0.5h}{v5}
\fmf{double}{v1,o1}
\fmf{double}{v2,o2}
\fmf{photon}{i,v5}
\fmflabel{$p_{2}$}{o1}
\fmflabel{$p_{1}$}{o2}
\fmflabel{$Q$}{i}
\fmf{double}{v2,v5}
\fmf{gluon}{v1,v3}
\fmf{gluon}{v4,v2}
\fmf{double}{v1,v5}
\fmf{plain,right}{v4,v3}
\fmf{plain,right}{v3,v4}
\end{fmfgraph*} } \\
%
%%%%%%%%%%%%%%%%%%%%%%%
%
\subfigure[]{
\begin{fmfgraph*}(30,30)
\fmfleft{i}
\fmfright{o1,o2}
\fmfforce{0.8w,0.93h}{v2}
\fmfforce{0.8w,0.07h}{v1}
\fmfforce{0.8w,0.3h}{v3}
\fmfforce{0.8w,0.7h}{v4}
\fmfforce{0.2w,0.5h}{v5}
\fmf{double}{v1,o1}
\fmf{double}{v2,o2}
\fmf{photon}{i,v5}
\fmflabel{$p_{2}$}{o1}
\fmflabel{$p_{1}$}{o2}
\fmflabel{$Q$}{i}
\fmf{double}{v2,v5}
\fmf{gluon}{v1,v3}
\fmf{gluon}{v4,v2}
\fmf{double}{v1,v5}
\fmf{dashes,right}{v4,v3}
\fmf{dashes,right}{v3,v4}
\end{fmfgraph*} }
%
%%%%%%%%%%%%%%%%%%%%%%%
%
\hspace{8mm}
\subfigure[]{
\begin{fmfgraph*}(30,30)
\fmfleft{i}
\fmfright{o1,o2}
\fmfforce{0.8w,0.93h}{v2}
\fmfforce{0.8w,0.07h}{v1}
\fmfforce{0.8w,0.3h}{v3}
\fmfforce{0.8w,0.7h}{v4}
\fmfforce{0.2w,0.5h}{v5}
\fmf{double}{v1,o1}
\fmf{double}{v2,o2}
\fmf{photon}{i,v5}
\fmflabel{$p_{2}$}{o1}
\fmflabel{$p_{1}$}{o2}
\fmflabel{$Q$}{i}
\fmf{double}{v2,v5}
\fmf{gluon}{v1,v3}
\fmf{gluon}{v4,v2}
\fmf{double}{v1,v5}
\fmf{gluon,left}{v3,v4}
\fmf{gluon,left}{v4,v3}
\end{fmfgraph*} } 
%
%%%%%%%%%%%%%%%%%%%%
%
\hspace{8mm}
\subfigure[]{
\begin{fmfgraph*}(30,30)
\fmfleft{i}
\fmfright{o1,o2}
\fmfforce{0.8w,0.93h}{v2}
\fmfforce{0.8w,0.07h}{v1}
\fmfforce{0.8w,0.5h}{v3}
\fmfforce{0.2w,0.5h}{v5}
\fmf{double}{v1,o1}
\fmf{double}{v2,o2}
\fmf{photon}{i,v5}
\fmflabel{$p_{2}$}{o1}
\fmflabel{$p_{1}$}{o2}
\fmflabel{$Q$}{i}
\fmf{double,tension=0}{v1,v5}
\fmf{gluon,tension=0}{v4,v3}
\fmf{double,tension=.4}{v2,v4}
\fmf{double,tension=.4}{v4,v5}
\fmf{gluon,tension=0}{v1,v3}
\fmf{gluon,tension=0}{v3,v2}
\end{fmfgraph*}}
%
%%%%%%%%%%%%%%%%%%%%
%
\vspace*{5mm}
\caption{\label{fig2} 
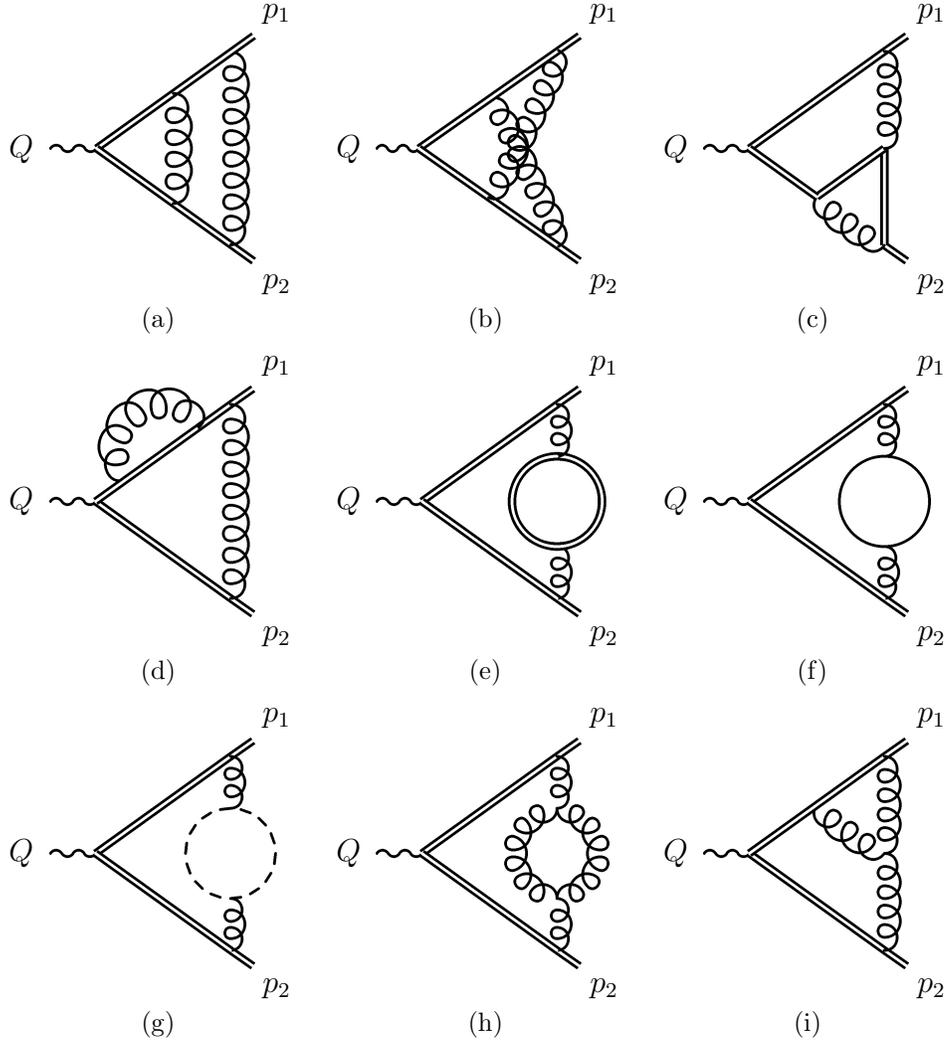
The two-loop vertex diagrams involved in the calculation of the 
form factors at order ${\mathcal O}(\alpha_{S}^{2})$. 
The curly lines are gluons; the double straight lines, quarks of 
mass $m$; the single straight lines, massless quarks and the dashed
lines ghosts. The external fermion lines are on the mass-shell: 
$p_1^2 = p_2^2 = m^{2}$. The wavy line on the l.h.s. carries  
momentum $Q=p_{1}+p_{2}$, with the metrical convention $Q^{2}<0$ 
when $Q$ is space-like. } 
\ec
\efig
%%%%%%%%%%%%%%%%%%%%%%%%%%%%%%%%%%%%%%%%%%%%%%%%%%%%%%%%%%%%%%%%%%%%%%%%

\section{The QCD Form Factors \label{FFact}}

We call $V^{\mu}_{c_{1}c_{2}}(p_1,p_2)$ the QCD vertex amplitude,
corresponding to the decay of a virtual photon of momentum $Q=p_{1}+p_{2}$
into a quark and an anti-quark, of momenta $p_1$, $p_2$ and colors $c_{1}$ 
and $c_{2}$ respectively. The two fermions are on the mass-shell, 
$p_{1}^{2}=p_{2}^{2}=m^2$, where $m$ denotes the mass of the heavy quark in the
on-shell scheme. Let us define the following two vectors:
\be
Q^{\mu} = p_{1}^{\mu}+p_{2}^{\mu} \, , \qquad 
t^{\mu} = p_{2}^{\mu}-p_{1}^{\mu} \, , 
\label{b0001}
\ee
such that $Q^{2}=S$, where $S$ is the c.m. energy squared, 
and the dimensionless variable
\be
s = \frac{S}{m^2} = \frac{Q^2}{m^2} \ . 
\label{defq2ands} 
\ee 

Within QCD the $V^{\mu}_{c_{1}c_{2}}(p_1,p_2)$ can be expressed in terms 
of two dimensionless scalar form factors $F_i(s), i=1,2$, 
depending only on the dimensionless variable $s$ of Eq.~(\ref{defq2ands}), 
as follows: 
\bea
\hspace{-5mm}
 V^{\mu}_{c_{1}c_{2}}(p_1,p_2) & = & \bar{u}_{c_{1}}(p_1) 
 \Gamma^{\mu}_{c_{1}c_{2}}(p_1,p_2) v_{c_{2}}(p_2) \, , \\ 
\hspace{-5mm}
\Gamma^{\mu}_{c_{1}c_{2}}(p_1,p_2) & = & - i \,
v_Q \, \delta_{c_{1}c_{2}} 
\Biggl[ F_{1}(s) \ \gamma^{\mu} 
  + \frac{1}{2m} F_{2}(s) \, i \, \sigma^{\mu \nu} Q_{\nu} \Biggr] , 
\label{b0002} 
\eea
where $\bar{u}_{c_{1}}(p_1), v_{c_{2}}(p_2)$ are the spinor wave functions 
of the quark and the anti-quark, 
$\sigma^{\mu \nu} = \frac{i}{2}[ \gamma^{\mu},\gamma^{\nu}]$, and where:
\be
v_Q = e \, Q_Q  \, ,
\ee
$Q_Q$ being the charge of the heavy quark in units of the positron charge $e > 0$.

The indices $c_{1}, c_{2}$, in Eq.~(\ref{b0002}), are the color indices: 
$c_{1},c_{2}=1, \cdots, N_c$, where $N_c$ is the number of colors.

The extraction of each form factor $F_i(s)$ from Eq.~(\ref{b0002}) 
can be carried out by the following general projector operators 
$P_{\mu}^{(i)}$:
\be
\! \! \! \! P_{\mu}^{(i)}(m,p_1,p_2) = 
         \frac{ \not{\! p_{2}} \! - \! m }{m} 
\biggl[  i \, g_{1}^{(i)} \gamma_{\mu} 
        + \frac{i}{2m} g_{2}^{(i)} t_{\mu}  \biggr] 
         \frac{ \not{\! p_{1}} \! + \! m }{m} ,
\label{b0003} 
\ee 
performing a trace over the spinor and the color indices. 
The constants $g_{j}^{(i)}, j=1,2$, are properly chosen such that:
\be
{\rm Tr}\left( P_{\mu}^{(i)}(m,p_1,p_2) \Gamma^{\mu}(p_{1},p_{2}) \right) 
  = F_{i}(s) .
\label{b0004}
\ee
Let us observe that since we work in a $D$-dimensional space (to 
regularize the divergences arising in the computation) the {\it trace} 
over the spinor indices is consistently  performed in $D$ dimensions 
as well. We use the convention of keeping the
trace of the unit matrix equal to four also in D dimensions.

The explicit values of the constants are:
\bea
g_{1}^{(1)} & = & - \frac{1}{v_Q N_{c}} \, \frac{1}{4(1- \epsilon)} 
\frac{1}{(s-4)} \, ,
\label{b0005} \\
g_{2}^{(1)} & = & \frac{1}{v_Q N_{c}} \, \frac{(3-2 \epsilon)}{(1- \epsilon)} 
\frac{1}{(s-4)^2} \, ,
\label{b0006} \\
g_{1}^{(2)} & = & \frac{1}{v_Q N_{c}} \, \frac{1}{(1- \epsilon)}
\frac{1}{s(s-4)} \, ,
\label{b0008} \\
g_{2}^{(2)} & = & - \frac{1}{v_Q N_{c}} \, \frac{1}{(1- \epsilon)}
\frac{1}{(s-4)^2} \left[ \frac{4}{s} - 2 + 2 \epsilon \right]  \, .
\label{b0009} 
\eea

The form factors are given as an expansion in powers of 
$\alpha_S/(2\pi)$, where $\alpha_S$ is defined as the standard
$\overline{\mathrm{MS}}$ coupling in QCD (with $N_f$ massless and one 
massive quark) at the renormalization scale $\mu$:
\bea
\hspace{-5mm}
F_{1} \Bigl( \epsilon,s,\frac{\mu^2}{m^2} \Bigr) & = & 
   1 + \left( \frac{\alpha_S}{2 \pi} \right) 
        F^{(1l)}_{1,R} \Bigl( \epsilon,s,\frac{\mu^2}{m^2} \Bigr)
     + \left( \frac{\alpha_S}{2 \pi} \right)^{2} 
        F^{(2l)}_{1,R} \Bigl( \epsilon,s,\frac{\mu^2}{m^2} \Bigr) \nn\\
& &  + {\mathcal O} 
     \Bigg( \! \left( \frac{\alpha_S}{2 \pi} \right) ^{3} \! \Bigg)
     , 
\label{b00020} 
\\
\hspace{-5mm}
F_{2} \Bigl(\epsilon,s,\frac{\mu^2}{m^2} \Bigr) & = & 
       \left( \frac{\alpha_S}{2 \pi} \right) 
        F^{(1l)}_{2,R} \Bigl( \epsilon,s,\frac{\mu^2}{m^2} \Bigr)
     + \left( \frac{\alpha_S}{2 \pi} \right) ^{2}
       F^{(2l)}_{2,R} \Bigl( \epsilon,s,\frac{\mu^2}{m^2} \Bigr) \nn\\
& &  + {\mathcal O} 
     \Bigg( \! \left( \frac{\alpha_S}{2 \pi} \right) ^{3} \! \Bigg) 
     \, .
\label{b00021} 
\eea
The superscripts ``$1l$'' and ``$2l$'' stand for one- and two-loop
contributions and the first term $1$ in $F_{1}(\epsilon,s,\mu^2/m^2)$ 
is the tree-level approximation. The subscript ``$R$'' stands for 
``renormalized'', meaning that 
%the functions 
$F^{(1l)}_{i,R}(\epsilon,s,\mu^2/m^2)$ 
and $F^{(2l)}_{i,R}(\epsilon,s,\mu^2/m^2)$ come from the sum of the 
contributions of the virtual diagrams of Figs.~\ref{fig1} and \ref{fig2} 
and the relative counterterms, shown in Fig.~\ref{fig3}, for the 
subtraction of the UV divergences. In the following the expressions of 
the unsubtracted as well as the UV-renormalized form factors will be 
given at the one- and two-loop level.

\section{Unsubtracted Contributions \label{unrenorm}}

In this Section we consider the contribution of the diagrams shown in
Figs.~\ref{fig1} and \ref{fig2} to the expansions in 
Eqs.~(\ref{b00020},\ref{b00021}). As explained in 
\cite{RoPieRem1,RoPieRem2,RoPieRem3}, 
the form factors coming from the 
computation of the trace operation introduced in the previous Section 
are expressed in terms of several hundreds of scalar integrals. It is 
possible to express all these integrals as a combination of only 17 
independent scalar integrals, called the Master Integrals (MIs) of 
the problem, by means of the so-called Laporta algorithm \cite{Lap},
using integration-by-parts identities \cite{Chet}, 
Lorentz invariance \cite{Rem3} and general symmetry relations. This 
{\it reduction} algorithm is performed exactly in $D=(4-2 \epsilon)$ 
dimensions \cite{DimReg}. Once the expression in terms of the MIs is 
found, we expand the result in powers of $\epsilon$ around $\epsilon=0$ 
($D=4$), using the expansions of the MIs given in \cite{RoPieRem1,RoPieRem3} 
(the MIs are evaluated with the differential equations method 
\cite{Kot,Rem1,Rem2}).
The result will be therefore given as a Laurent expansion in $\epsilon$ 
where both UV- and IR-poles are regularized with the same parameter 
$\epsilon$.

The form factors that we are going to present in this Section are 
given for space-like $Q$ ($S=Q^2<0$) and they are expressed in terms 
of 1-dimensional harmonic polylogarithms (HPLs) \cite{Polylog,Polylog3} of 
the variable:
\be
x = \frac{\sqrt{-s+4} - \sqrt{-s} }{\sqrt{-s+4} + \sqrt{-s} } 
  = \frac{\sqrt{-S+4m^2} - \sqrt{-S} }{\sqrt{-S+4m^2} + \sqrt{-S}} ,
\qquad ( 0 \leq x \leq 1 )
\label{xvar}
\ .
\ee

$C_{F}$ and $C_{A}$ are the Casimir operators for the fundamental and
adjoint representation of the color gauge group $SU(N_c)$ respectively:
\be
C_{F} = \frac{N_c^2-1}{2N_c} \, , \quad C_{A} = N_c \, ,
\ee
where $N_c$ is the number of colors. $T_R$ is the normalization factor
of the generators of the fundamental representation:
\be
tr \left( t^a t^b \right) = T_R \delta^{ab} = \frac{1}{2} \delta^{ab}
\, .
\ee

Because all the calculations are done in $D$ dimensions, we have to 
take into account for each loop a factor
\be
C(\epsilon) \, \left( \frac{\mu^{2}}{m^2} \right)^{\epsilon} =
(4 \pi)^{\epsilon} \, \Gamma \left( 1+\epsilon \right) 
\, \left( \frac{\mu^{2}}{m^2} \right)^{\epsilon} \, ,
\ee
where $\mu$ is the mass scale of dimensional regularization, that we
choose equal to the renormalization scale.

\subsection{One-Loop Unsubtracted Form Factors}

At the one-loop level only the diagram shown in Fig.~\ref{fig1} (b) 
contributes to the form factors $F_1(\epsilon,s,\mu^2/m^2)$ and 
$F_2(\epsilon,s,\mu^2/m^2)$. 
We will write:
\be
F_{i,{\mathrm{Fig.1(b)}}}^{(1l)} \Bigl(\epsilon,s,\frac{\mu^2}{m^2} \Bigr)
= C(\epsilon) \, \left( \frac{\mu^{2}}{m^2} \right)^{\epsilon} 
{\mathcal F}_{i}^{(1l)}(\epsilon,s) \, , \qquad \mbox{with} \; \, i=1,2
\, ,
\ee
where:
\bea
\hspace*{-5mm} {\mathcal F}_1^{(1l)}(\epsilon,s) & = &  
\frac{1}{\epsilon} \, \Biggl\{ C_{F} \,\Biggl[ 
           \frac{1}{2}
       + \biggl(
            1
          - \frac{1}{1+x}
          - \frac{1}{1-x}
          \biggr) H(0;x) 
\Biggr] \Biggr\}  \nn\\
\hspace*{-5mm} & &  
  + \, C_{F} \Biggl[
         \frac{1}{2} \biggl(
            1
          - \frac{2}{1-x} \biggr) H(0;x)
       - \biggl(
            1
          - \frac{1}{1+x}
          - \frac{1}{1-x}
          \biggr) \bigl( \zeta(2) \nn\\
\hspace*{-5mm} & &  \hspace*{12mm} 
   - H(0;x) 
   - H(0,0;x)
   + 2 H(-1,0;x) \bigr) \Biggr] \nn\\
\hspace*{-5mm} & &  
- \, \epsilon \, \Biggl\{ C_{F} \Biggl[ \frac{1}{2} \biggl(
            1
          - \frac{2}{1-x} \biggr) \bigl( \zeta(2)
   - H(0,0;x) 
   + 2 H(-1,0;x) \bigr) \nn\\
\hspace*{-5mm} & & \hspace*{20mm} 
        + \biggl(
            1
          - \frac{1}{1+x}
          - \frac{1}{1-x}
          \biggr) \bigl( \zeta(2)
   + 2 \zeta(3) \nn\\
\hspace*{-5mm} & & \hspace*{20mm} 
          - (4 \! -  \! \zeta(2)) H(0;x)  \! 
   - \!  2 \zeta(2) H(-1;x) \! 
   - \!  H(0,0;x) \nn\\
\hspace*{-5mm} & & \hspace*{20mm} 
   + 2 H(-1,0;x)
   - H(0,0,0;x)
   + 2 H(-1,0,0;x) \nn\\
\hspace*{-5mm} & & \hspace*{20mm} 
   + 2 H(0, \! -1,0;x) \! 
   - \! 4 H(-1, \! -1,0;x) \bigr) \Biggr] \! \Biggr\} 
+ {\mathcal O} \left( \epsilon^2 \right) \! , 
\label{1loopF1} \\
\hspace*{-5mm} {\mathcal F}_2^{(1l)}(\epsilon,s) & = & 
   - \, C_{F}  \,\biggl(
     \frac{1}{1-x}
   - \frac{1}{1+x} \biggr) H(0;x) \nn\\
\hspace*{-5mm} & &  
+ \, \epsilon \, \Biggl\{ C_{F} \Biggl[ 
          \biggl(
     \frac{1}{1-x}
   - \frac{1}{1+x} \biggr) \bigl( \zeta(2)
   - 4 H(0;x)
          - H(0,0;x) \nn\\
\hspace*{-5mm} & & \hspace*{20mm} 
   + 2 H(-1,0;x) \bigr) \Biggr] \Biggr\} 
+ \, {\mathcal O} \left( \epsilon^2 \right) \, .
\label{1loopF2}
\eea

\subsection{Two-Loop Unsubtracted Form Factors}

At the two-loop level, all the 9 Feynman diagrams of Fig.~\ref{fig2}
contribute to the form factors $F_1(\epsilon,s,\mu^2/m^2)$ and 
$F_2(\epsilon,s,\mu^2/m^2)$. 
The total contribution comes from the sum of the diagrams involving 
only the heavy quark, Fig.~\ref{fig2}~(a)--(e) and (g)--(i), and the 
diagrams in which a light quark runs in the internal loop, 
Fig.~\ref{fig2}~(f). If we consider $N_f$ light quarks, the latter 
contribution is simply the contribution of diagram (f), in which the 
quark in the internal loop is considered as massless, multiplied by $N_f$.

We will write:
\be
F_{i,{\mathrm{Fig.2}}}^{(2l)} \Bigl(\epsilon,s,\frac{\mu^2}{m^2} \Bigr) = 
C^{2}(\epsilon) \, \left( \frac{\mu^{2}}{m^2} \right)^{2 \epsilon} 
{\mathcal F}_{i}^{(2l)}(\epsilon,s) \, , \qquad \mbox{with} \; \, i=1,2 \, ,
\ee
where:
\bea
{\mathcal F}_1^{(2l)}(\epsilon,s) &=& 
     \frac{1}{\epsilon^2} \Bigg\{
         C_F C_A \Bigg[
         \frac{11}{24}
       +   \Biggl(
            \frac{11}{12}
          - \frac{11}{12 (1-x)}
          - \frac{11}{12 (1+x)}
          \Biggr) H(0;x)
        \Bigg] \nn\\
& & \hspace*{8mm}
       + \, C_F T_R N_f \Bigg[
       - \frac{1}{6}
       -   \Biggl(
            \frac{1}{3}
          - \frac{1}{3 (1-x)}
          - \frac{1}{3 (1+x)}
          \Biggr) H(0;x)
        \Bigg] \nn\\
& & \hspace*{8mm}
       + \, C_F T_R \Bigg[
       - \frac{1}{3}
       -   \Biggl(
            \frac{2}{3}
          - \frac{2}{3 (1-x)}
          - \frac{2}{3 (1+x)}
          \Biggr) H(0;x)
        \Bigg] \nn\\
& & \hspace*{8mm}
       + \, C_F^2 \Bigg[
            \frac{25}{8}
          + \frac{6}{(1+x)^2}
          - \frac{6}{(1+x)}
       +   \Biggl(
           \frac{1}{2}
          - \frac{7}{2 (1-x)}
          + \frac{6}{(1+x)^3} \nn\\
& & \hspace*{20mm}
          - \frac{9}{(1+x)^2}
          + \frac{11}{2 (1+x)}
          \Biggr) H(0;x)
       +   \Biggl(
           1
          + \frac{1}{(1-x)^2} \nn\\
& & \hspace*{20mm}
          - \frac{1}{(1-x)}
          + \frac{1}{(1+x)^2}
          - \frac{1}{(1+x)}
          \Biggr) H(0,0;x)
        \Bigg]
     \Bigg\}
     \nn\\
& & 
    + \frac{1}{\epsilon} \Bigg\{
         C_F C_A \Bigg[
           \frac{185}{144}
          + \frac{4 \zeta(2)}{3 (1-x)}
          + \frac{4 \zeta(2)}{3 (1+x)}
          - \frac{ \zeta(2)}{3}
          - \frac{ \zeta(3)}{2 (1-x)^2} \nn\\
& & \hspace*{20mm}
          + \frac{ \zeta(3)}{2 (1-x)}
          - \frac{ \zeta(3)}{2 (1+x)^2}
          + \frac{ \zeta(3)}{2 (1+x)}
          - \frac{ \zeta(3)}{2} \nn\\
& & \hspace*{20mm}
       -   \Biggl(
            \frac{14}{3}
          - \frac{14}{3 (1-x)}
          - \frac{14}{3 (1+x)}
          \Biggr) H(-1,0;x) \nn\\
& & \hspace*{20mm}
       +   \Biggl(
            \frac{83}{18}
          - \frac{\zeta(2)}{2 (1-x)^2}
          + \frac{3 \zeta(2)}{2 (1-x)}
          - \frac{\zeta(2)}{2 (1+x)^2}
          + \frac{3 \zeta(2)}{2 (1+x)} \nn\\
& & \hspace*{20mm}
          - \frac{3 \zeta(2)}{2} \! 
          - \! \frac{199}{36 (1-x)} \! 
          - \! \frac{133}{36 (1\!+\!x)}
          \Biggr) H(0;x)
       + \Biggl(
            1 \! 
          + \! \frac{1}{(1-x)^2} \nn\\
& & \hspace*{20mm}
          - \frac{1}{(1-x)}
          + \frac{1}{(1+x)^2}
          - \frac{1}{(1+x)}
          \Biggr) H(0,-1,0;x) \nn\\
& & \hspace*{20mm}
       +  \Biggl(
            \frac{23}{6} \! 
          - \! \frac{17}{6 (1-x)}\! 
          - \! \frac{17}{6 (1+x)}
          \Biggr) H(0,0;x)
       -  \!  \Biggl(
            2\! 
          + \! \frac{1}{(1-x)^2} \nn\\
& & \hspace*{20mm}
          - \frac{2}{ (1-x)}
          + \frac{1}{(1+x)^2}
          - \frac{2}{(1+x)}
          \Biggr) H(0,0,0;x) \nn\\
& & \hspace*{20mm}
       -   \Biggl(
            1 \! 
          +  \! \frac{1}{(1-x)^2} \! 
          -  \! \frac{1}{(1-x)} \! 
          +  \! \frac{1}{(1 \! + \! x)^2} \! 
          -  \! \frac{1}{(1 \! + \! x)}
          \Biggr) H(0, \! 1, \! 0;x) \nn\\
& & \hspace*{20mm}
       +  \Biggl(
            1
          - \frac{1}{(1-x)}
          - \frac{1}{(1+x)}
          \Biggr) H(1,0;x) 
        \Bigg] \nn\\
& & \hspace*{8mm}
       + \, C_F T_R N_f \Bigg[
          - \frac{13}{36} \! 
          -  \! \frac{2 \zeta(2)}{3 (1-x)} \! 
          -  \! \frac{2 \zeta(2)}{3 (1\!+\!x)} \! 
          + \! \frac{2 \zeta(2)}{3}
       +  \Biggl(
            \frac{4}{3} \! 
          - \! \frac{4}{3 (1-x)} \nn\\
& & \hspace*{20mm}
          - \frac{4}{3 (1 \! + \! x)}
          \Biggr) H(-1,0;x) \! 
       -   \!  \Biggl(
            \frac{14}{9} \! 
          -  \! \frac{17}{9 (1-x)}  \nn\\
& & \hspace*{20mm}
          -  \frac{11}{9 (1 \! + \! x)}
          \Biggr) H(0;x)\!
       -  \! \Biggl(
            \frac{2}{3} \!
          - \!\frac{2}{3 (1\!-\!x)} \!
          - \!\frac{2}{3 (1\!+\!x)}
          \Biggr) H(0,0;x)
        \Bigg] \nn\\
& & \hspace*{8mm}
       + \, C_F T_R \Bigg[
          - \frac{1}{4}
          - \frac{2 \zeta(2)}{3 (1-x)}
          - \frac{2 \zeta(2)}{3 (1+x)}
          + \frac{2 \zeta(2)}{3}
       +   \Biggl(
            \frac{4}{3}
          - \frac{4}{3 (1-x)} \nn\\
& & \hspace*{20mm}
          -  \frac{4}{3 (1 \!+ \!x)}
          \Biggr) H(-1,0;x) \!
       -   \! \Biggl(
            1 \!
          -  \!\frac{4}{3 (1\!-\!x)} \!
          -  \!\frac{2}{3 (1 \!+ \!x)}
          \Biggr) H(0;x) \nn\\
& & \hspace*{20mm}
       -   \Biggl(
            \frac{2}{3}
          - \frac{2}{3 (1-x)}
          - \frac{2}{3 (1+x)}
          \Biggr) H(0,0;x)
        \Bigg] \nn\\
& & \hspace*{8mm}
       + \, C_F^2 \Bigg[
          - \frac{35}{16}
          + \frac{7 \zeta(2)}{2 (1-x)}
          - \frac{6 \zeta(2)}{ (1+x)^3}
          + \frac{9 \zeta(2)}{ (1+x)^2}
          - \frac{11 \zeta(2)}{2 (1+x)} \nn\\
& & \hspace*{20mm}
          - \frac{\zeta(2)}{2}
          + \frac{8}{ (1+x)^2}
          - \frac{8}{ (1+x)}
       -   \Biggl(
            1
          - \frac{7}{ (1-x)}
          + \frac{12}{ (1+x)^3} \nn\\
& & \hspace*{20mm}
          - \frac{18}{ (1+x)^2}
          + \frac{11}{ (1+x)}
          \Biggr) H(-1,0;x)
       -   \Biggl(
            4
          + \frac{4}{ (1-x)^2} \nn\\
& & \hspace*{20mm}
          - \frac{4}{ (1-x)}
          + \frac{4}{ (1+x)^2}
          - \frac{4}{ (1+x)}
          \Biggr) H(-1,0,0;x) \nn\\
& & \hspace*{20mm}
       +   \Biggl(
            \frac{3}{4}
          - \frac{\zeta(2)}{ (1-x)^2}
          + \frac{\zeta(2)}{ (1-x)}
          - \frac{\zeta(2)}{ (1+x)^2}
          + \frac{\zeta(2)}{ (1+x)}
          - \zeta(2) \nn\\
& & \hspace*{20mm}
          - \frac{8}{ (1-x)}
          + \frac{20}{ (1+x)^3}
          - \frac{30}{ (1+x)^2}
          + \frac{33}{2 (1+x)}
          \Biggr) H(0;x) \nn\\
& & \hspace*{20mm}
       -  \Biggl( \! 
            2 \! 
          +  \! \frac{2}{ (1-x)^2} \! 
          -  \! \frac{2}{ (1-x)} \! 
          +  \! \frac{2}{ (1 \! + \! x)^2} \! 
          -  \! \frac{2}{ (1 \! + \! x)} \! 
          \Biggr) H(0, \! -1, \! 0;x) \nn\\
& & \hspace*{20mm}
       +   \Biggl(
            \frac{7}{2}
          + \frac{4}{ (1-x)^2}
          - \frac{15}{2 (1-x)}
          + \frac{6}{ (1+x)^3}
          - \frac{7}{ (1+x)^2} \nn\\
& & \hspace*{20mm}
          + \frac{7}{2 (1+x)}
          \Biggr) H(0,0;x)
       +   \Biggl(
            3
          + \frac{3}{ (1-x)^2}
          - \frac{3}{ (1-x)} \nn\\
& & \hspace*{20mm}
          + \frac{3}{ (1+x)^2}
          - \frac{3}{ (1+x)}
          \Biggr) H(0,0,0;x)
        \Bigg]
     \Bigg\}
     \nn\\
& & 
       + \, C_F C_A \Bigg[
            \frac{2585}{864}
          + \frac{36 \zeta(2) \log(2)}{ (1+x)^2}
          - \frac{36 \zeta(2) \log(2)}{ (1+x)}
          + 18 \zeta(2) \log(2) \nn\\
& & \hspace*{20mm}
          + \frac{145 \zeta(2)}{18 (1-x)}
          - \frac{168 \zeta(2)}{ (1+x)^4}
          + \frac{363 \zeta(2)}{ (1+x)^3}
          - \frac{543 \zeta(2)}{2 (1+x)^2} \nn\\
& & \hspace*{20mm}
          + \frac{1501 \zeta(2)}{18 (1+x)}
          - \frac{179 \zeta(2)}{9}
          - \frac{37 \zeta^2(2)}{20 (1-x)^2}
          + \frac{679 \zeta^2(2)}{160 (1-x)} \nn\\
& & \hspace*{20mm}
          - \frac{879 \zeta^2(2)}{10 (1+x)^5}
          + \frac{879 \zeta^2(2)}{4 (1+x)^4}
          - \frac{7481 \zeta^2(2)}{40 (1+x)^3}
          + \frac{943 \zeta^2(2)}{16 (1+x)^2} \nn\\
& & \hspace*{20mm}
          - \frac{441 \zeta^2(2)}{160 (1+x)}
          - \frac{71 \zeta^2(2)}{20}
          + \frac{53 \zeta(3)}{6 (1-x)}
          + \frac{324 \zeta(3)}{ (1+x)^4} \nn\\
& & \hspace*{20mm}
          - \frac{648 \zeta(3)}{ (1+x)^3}
          + \frac{374 \zeta(3)}{ (1+x)^2}
          - \frac{247 \zeta(3)}{6 (1+x)}
          - \frac{35 \zeta(3)}{6}
          + \frac{3}{ (1+x)^2} \nn\\
& & \hspace*{20mm}
          - \frac{3}{ (1+x)}
       -   \Biggl(
            \frac{19 \zeta(2)}{3 (1-x)}
          + \frac{90 \zeta(2)}{ (1+x)^4}
          - \frac{180 \zeta(2)}{ (1+x)^3}
          + \frac{135 \zeta(2)}{ (1+x)^2} \nn\\
& & \hspace*{20mm}
          - \frac{116 \zeta(2)}{3 (1+x)}
          - \frac{\zeta(2)}{3}
          - \frac{2 \zeta(3)}{ (1-x)^2}
          + \frac{2 \zeta(3)}{ (1-x)}
          - \frac{2 \zeta(3)}{ (1+x)^2} \nn\\
& & \hspace*{20mm}
          + \frac{2 \zeta(3)}{ (1+x)}
          - 2 \zeta(3)
          \Biggr) H(-1;x)
       +   \Biggl(
            \frac{74}{3}
          - \frac{74}{3 (1-x)} \nn\\
& & \hspace*{20mm}
          - \frac{74}{3 (1+x)}
          \Biggr) H(-1,-1,0;x)
       -   \Biggl(
            \frac{161}{18}
          - \frac{2 \zeta(2)}{ (1-x)^2} \nn\\
& & \hspace*{20mm}
          - \frac{2 \zeta(2)}{ (1+x)^2}
          - \frac{109}{9 (1-x)}
          - \frac{6}{ (1+x)^3}
          + \frac{9}{ (1+x)^2} \nn\\
& & \hspace*{20mm}
          - \frac{79}{9 (1+x)}
          \Biggr) H(-1,0;x)
       -   \Biggl(
            4
          + \frac{4}{ (1-x)^2}
          - \frac{4}{ (1-x)} \nn\\
& & \hspace*{20mm}
          + \frac{4}{ (1\! +\! x)^2}\! 
          - \! \frac{4}{ (1\! +\! x)}\! 
          \Biggr) H(-1,0,\! -1,0;x)
       -   \Biggl(
            \frac{64}{3}\! 
          - \! \frac{46}{3 (1\!-\!x)} \nn\\
& & \hspace*{20mm}
          + \frac{282}{ (1\! +\! x)^4}
          - \frac{564}{ (1\! +\! x)^3}\! 
          + \! \frac{339}{ (1\! +\! x)^2}
          - \frac{217}{3 (1\! +\! x)}
          \Biggr) H(-1,\!0,\!0;x) \nn\\
& & \hspace*{20mm}
       +   \Biggl(
            2
          + \frac{4}{ (1-x)^2}
          - \frac{2}{ (1-x)}
          + \frac{4}{ (1+x)^2} \nn\\
& & \hspace*{20mm}
          - \frac{2}{ (1+x)}
          \Biggr) H(-1,0,0,0;x)
       +   \Biggl(
            4
          + \frac{4}{ (1-x)^2}
          - \frac{4}{ (1-x)} \nn\\
& & \hspace*{20mm}
          + \frac{4}{ (1+x)^2}
          - \frac{4}{ (1+x)}
          \Biggr) H(-1,0,1,0;x)
       -   \Biggl(
            6
          - \frac{6}{ (1-x)} \nn\\
& & \hspace*{20mm}
          - \frac{6}{ (1+x)}
          \Biggr) H(-1,1,0;x)
       +   \Biggl(
            \frac{4129}{216}
          + \frac{11 \zeta(2)}{2 (1-x)} \nn\\
& & \hspace*{20mm}
          - \frac{258 \zeta(2)}{ (1+x)^5}
          + \frac{744 \zeta(2)}{ (1+x)^4}
          - \frac{779 \zeta(2)}{ (1+x)^3}
          + \frac{357 \zeta(2)}{ (1+x)^2}
          - \frac{145 \zeta(2)}{2 (1+x)} \nn\\
& & \hspace*{20mm}
          + 3 \zeta(2)
          - \frac{\zeta(3)}{2 (1-x)^2}
          + \frac{2 \zeta(3)}{ (1-x)}
          + \frac{324 \zeta(3)}{ (1+x)^5}
          - \frac{810 \zeta(3)}{ (1+x)^4} \nn\\
& & \hspace*{20mm}
          + \frac{662 \zeta(3)}{ (1+x)^3}
          - \frac{367 \zeta(3)}{2 (1+x)^2}
          + \frac{20 \zeta(3)}{ (1+x)}
          - \frac{15 \zeta(3)}{2}
          - \frac{563}{27 (1-x)} \nn\\
& & \hspace*{20mm}
          + \frac{9}{ (1+x)^3}
          - \frac{27}{2 (1+x)^2}
          - \frac{1391}{108 (1+x)}
          \Biggr) H(0;x) \nn\\
& & \hspace*{20mm}
       +  \Biggl(
            \frac{\zeta(2)}{ (1-x)^2} \! 
          - \! \frac{35 \zeta(2)}{8 (1-x)}\! 
          - \! \frac{90 \zeta(2)}{ (1+x)^5}\! 
          + \! \frac{225 \zeta(2)}{ (1\! +\! x)^4}\! 
          - \frac{363 \zeta(2)}{2 (1\!+\!x)^3} \nn\\
& & \hspace*{20mm}
          +  \frac{193 \zeta(2)}{4 (1 \! + \! x)^2} \! 
          -  \! \frac{83 \zeta(2)}{8 (1 \! + \! x)} \! 
          +  \! 7 \zeta(2)
          \Biggr) H(0, \! -1;x)\!
       -   \!  \Biggl(
            10 \! 
          +  \! \frac{10}{ (1 \! - \! x)^2} \nn\\
& & \hspace*{20mm}
          - \frac{10}{ (1-x)}
          + \frac{10}{ (1+x)^2}
          - \frac{10}{ (1+x)}
          \Biggr) H(0,-1,-1,0;x) \nn\\
& & \hspace*{20mm}
       -   \Biggl(
            \frac{79}{3}
          - \frac{46}{3 (1-x)}
          - \frac{12 }{(1+x)^4}
          + \frac{24}{ (1+x)^3}
          - \frac{14 }{(1+x)^2} \nn\\
& & \hspace*{20mm}
          - \frac{40}{3 (1+x)}
          \Biggr) H(0,-1,0;x)
       +   \Biggl(
            14
          + \frac{8}{ (1-x)^2}
          - \frac{83}{8 (1-x)} \nn\\
& & \hspace*{20mm}
          - \frac{282}{ (1+x)^5}
          + \frac{705}{ (1+x)^4}
          - \frac{1171}{2 (1+x)^3}
          + \frac{725}{4 (1+x)^2} \nn\\
& & \hspace*{20mm}
          - \frac{227}{8 (1+x)}
          \Biggr) H(0,-1,0,0;x)
       +   \Biggl(
            6
          + \frac{6}{ (1-x)^2} \nn\\
& & \hspace*{20mm}
          - \frac{6}{ (1-x)}
          + \frac{6}{ (1+x)^2}
          - \frac{6}{ (1+x)}
          \Biggr) H(0,-1,1,0;x) \nn\\
& & \hspace*{20mm}
       -   \Biggl(
            \frac{59}{18}
          - \frac{\zeta(2)}{ (1-x)^2}
          - \frac{5 \zeta(2)}{16 (1-x)}
          + \frac{69 \zeta(2)}{ (1+x)^5}
          - \frac{345 \zeta(2)}{2 (1+x)^4} \nn\\
& & \hspace*{20mm}
          + \frac{563 \zeta(2)}{4 (1+x)^3}
          - \frac{317 \zeta(2)}{8 (1+x)^2}
          + \frac{59 \zeta(2)}{16 (1+x)}
          - \zeta(2)
          + \frac{91}{18 (1-x)} \nn\\
& & \hspace*{20mm}
          - \frac{24}{ (1+x)^4}\ ! 
          + \! \frac{39}{ (1+x)^3}\! 
          - \! \frac{3}{2 (1+x)^2}\! 
          - \! \frac{227}{18 (1+x)}
          \Biggr) H(0,0;x) \nn\\
& & \hspace*{20mm}
       +   \Biggl(
            22
          + \frac{12}{ (1-x)^2}
          - \frac{75}{4 (1-x)}
          + \frac{12}{ (1+x)^5}
          - \frac{30}{ (1+x)^4} \nn\\
& & \hspace*{20mm}
          + \frac{49}{ (1+x)^3}
          - \frac{63}{2 (1+x)^2}
          - \frac{51}{4 (1+x)}
          \Biggr) H(0,0,-1,0;x) \nn\\
& & \hspace*{20mm}
       +   \Biggl(
            \frac{50}{3}
          - \frac{19}{6 (1-x)}
          - \frac{258}{ (1+x)^5}
          + \frac{816}{ (1+x)^4}
          - \frac{923}{ (1+x)^3} \nn\\
& & \hspace*{20mm}
          + \frac{436}{ (1+x)^2}
          - \frac{529}{6 (1+x)}
          \Biggr) H(0,0,0;x)
       -   \Biggl(
            12
          + \frac{6}{ (1-x)^2} \nn\\
& & \hspace*{20mm}
          - \frac{193}{16 (1-x)}\! 
          - \! \frac{3}{ (1\!+\!x)^5}\! 
          + \! \frac{15}{2 (1\!+\!x)^4}\! 
          - \! \frac{17}{4 (1+x)^3}\! 
          + \! \frac{39}{8 (1\! +\! x)^2} \nn\\
& & \hspace*{20mm}
          - \frac{177}{16 (1\!+\!x)}
          \Biggr) H(0,0,0,0;x)
       -   \Biggl(
            14
          + \frac{8}{ (1-x)^2}
          - \frac{12}{ (1-x)} \nn\\
& & \hspace*{20mm}
          - \frac{48}{ (1+x)^5}
          + \frac{120}{ (1+x)^4}
          - \frac{88}{ (1+x)^3}
          + \frac{20}{ (1+x)^2} \nn\\
& & \hspace*{20mm}
          - \frac{12}{ (1+x)}
          \Biggr) H(0,0,1,0;x)
       +   \Biggl(
            \frac{\zeta(2)}{ (1-x)^2}
          - \frac{\zeta(2)}{ (1-x)} \nn\\
& & \hspace*{20mm}
          + \frac{\zeta(2)}{ (1+x)^2}
          - \frac{\zeta(2)}{ (1+x)}
          + \zeta(2)
          \Biggr) H(0,1;x)
       +   \Biggl(
            6
          + \frac{6}{ (1-x)^2} \nn\\
& & \hspace*{20mm}
          - \frac{6}{ (1-x)}
          + \frac{6}{ (1+x)^2}
          - \frac{6}{ (1+x)}
          \Biggr) H(0,1,-1,0;x) \nn\\
& & \hspace*{20mm}
       +   \Biggl(
            8
          - \frac{4}{ (1-x)}
          + \frac{48}{ (1+x)^4}
          - \frac{96}{ (1+x)^3}
          + \frac{56}{ (1+x)^2} \nn\\
& & \hspace*{20mm}
          - \frac{12}{ (1+x)}
          \Biggr) H(0,1,0;x)
       -   \Biggl(
            2
          + \frac{4}{ (1-x)^2}
          - \frac{5}{ (1-x)} \nn\\
& & \hspace*{20mm}
          - \frac{24}{ (1+x)^5}
          + \frac{60}{ (1+x)^4}
          - \frac{72}{ (1+x)^3}
          + \frac{52}{ (1+x)^2} \nn\\
& & \hspace*{20mm}
          - \frac{11}{ (1+x)}
          \Biggr) H(0,1,0,0;x)
       -   \Biggl(
            2
          + \frac{2}{ (1-x)^2}
          - \frac{2}{ (1-x)} \nn\\
& & \hspace*{20mm}
          + \frac{2}{ (1\!+\!x)^2}\! 
          - \! \frac{2}{ (1\!+\!x)}
          \Biggr) H(0,1,1,0;x)\! 
       +  \!  \Biggl(
            \frac{\zeta(2)}{ (1-x)}\! 
          + \! \frac{\zeta(2)}{ (1\!+\!x)} \nn\\
& & \hspace*{20mm}
          - \zeta(2)
          - \frac{2 \zeta(3)}{ (1-x)^2}
          + \frac{2 \zeta(3)}{ (1-x)}
          - \frac{2 \zeta(3)}{ (1+x)^2}
          + \frac{2 \zeta(3)}{ (1+x)} \nn\\
& & \hspace*{20mm}
          - 2 \zeta(3)
          \Biggr) H(1;x)\! 
       -  \!  \Biggl(
            6\! 
          - \! \frac{6}{ (1-x)}\! 
          - \! \frac{6}{ (1\! +\! x)}
          \Biggr) H(1,\! -1,\! 0;x) \nn\\
& & \hspace*{20mm}
       -   \Biggl(
            3
          + \frac{2 \zeta(2)}{ (1-x)^2}
          - \frac{2 \zeta(2)}{ (1-x)}
          + \frac{336 \zeta(2)}{ (1+x)^5}
          - \frac{840 \zeta(2)}{ (1+x)^4} \nn\\
& & \hspace*{20mm}
          + \frac{688 \zeta(2)}{ (1+x)^3}
          - \frac{190 \zeta(2)}{ (1+x)^2}
          + \frac{18 \zeta(2)}{ (1+x)}
          - 4 \zeta(2)
          - \frac{2}{ (1-x)} \nn\\
& & \hspace*{20mm}
          - \frac{24}{ (1\!+\!x)^3}\! 
          + \! \frac{36}{ (1\!+\!x)^2}\! 
          - \! \frac{16}{ (1\! +\! x)}
          \Biggr) H(1,0;x)\! 
       +  \!  \Biggl(
            4\! 
          + \! \frac{4}{ (1-x)^2} \nn\\
& & \hspace*{20mm}
          - \frac{4}{ (1-x)}
          + \frac{4}{ (1+x)^2}
          - \frac{4}{ (1+x)}
          \Biggr) H(1,0,-1,0;x) \nn\\
& & \hspace*{20mm}
       -   \Biggl(
            4
          + \frac{4}{ (1-x)}
          - \frac{24}{ (1+x)^4}
          + \frac{48}{ (1+x)^3}
          - \frac{38}{ (1+x)^2} \nn\\
& & \hspace*{20mm}
          + \frac{18}{ (1\! +\! x)}
          \Biggr) H(1,\! 0,\! 0;x)\! 
       +  \!  \Biggl(
            2\! 
          - \! \frac{4}{ (1-x)^2}\! 
          + \! \frac{4}{ (1-x)}\! 
          - \! \frac{336}{ (1\! +\! x)^5} \nn\\
& & \hspace*{20mm}
          + \frac{840}{ (1+x)^4} \! 
          - \! \frac{688}{ (1+x)^3}\! 
          + \! \frac{188}{ (1\! +\! x)^2}\! 
          - \! \frac{16}{ (1\! +\! x)}
          \Biggr) H(1,0,0,0;x) \nn\\
& & \hspace*{20mm}
       -   \Biggl(
            4 \! 
          + \! \frac{4}{ (1\! -\! x)^2}\! 
          - \! \frac{4}{ (1\! -\! x)}\! 
          + \! \frac{4}{ (1\! +\! x)^2}\! 
          - \! \frac{4}{ (1\! +\! x)}
          \Biggr) H(1,\! 0,\! 1,\! 0;x) \nn\\
& & \hspace*{20mm}
       +   \Biggl(
            2
          - \frac{2}{ (1-x)}
          - \frac{2}{ (1+x)}
          \Biggr) H(1,1,0;x)
        \Bigg] \nn\\
& & 
       + \, C_F T_R N_f \Bigg[
          - \frac{169}{216}
          - \frac{34 \zeta(2)}{9 (1-x)}
          - \frac{22 \zeta(2)}{9 (1+x)}
          + \frac{22 \zeta(2)}{9}
          - \frac{8 \zeta(3)}{3 (1-x)} \nn\\
& & \hspace*{20mm}
          - \frac{8 \zeta(3)}{3 (1\! +\! x)}\! 
          + \! \frac{8 \zeta(3)}{3}\! 
       -  \! \frac{8 \zeta(2)}{3} \Biggl( \!
             1 \!
          - \! \frac{1}{(1\! -\! x)}\! 
          - \! \frac{1}{(1\! +\! x)}\! 
          \Biggr) H(-1;x) \nn\\
& & \hspace*{20mm}
       -   \Biggl(
            \frac{16}{3}
          - \frac{16}{3 (1-x)}
          - \frac{16}{3 (1+x)}
          \Biggr) H(-1,-1,0;x) \nn\\
& & \hspace*{20mm}
       +  \Biggl(
            \frac{56}{9} \! 
          - \! \frac{68}{9 (1\!-\!x)}\! 
          - \! \frac{44}{9 (1\! +\! x)}\! 
          \Biggr) H(-1,0;x) \! 
       +  \!  \Biggl(
            \frac{8}{3}\! 
          - \! \frac{8}{3 (1-x)} \nn\\
& & \hspace*{20mm}
          - \frac{8}{3 (1+x)}
          \Biggr) H(-1,0,0;x)
       -   \Biggl(
            \frac{353}{54}
          - \frac{178}{27 (1-x)} \nn\\
& & \hspace*{20mm}
          - \frac{175}{27 (1+x)}
          \Biggr) H(0;x)
       +   \Biggl(
            \frac{8}{3}
          - \frac{8}{3 (1-x)} \nn\\
& & \hspace*{20mm}
          - \frac{8}{3 (1+x)}
          \Biggr) H(0,-1,0;x)
       -   \Biggl(
            \frac{28}{9}
          - \frac{34}{9 (1-x)} \nn\\
& & \hspace*{20mm}
          - \frac{22}{9 (1+x)}
          \Biggr) H(0,0;x)
       -   \Biggl(
            \frac{4}{3}
          - \frac{4}{3 (1-x)} \nn\\
& & \hspace*{20mm}
          - \frac{4}{3 (1+x)}
          \Biggr) H(0,0,0;x)
        \Bigg] \nn\\
& & 
       + \, C_F T_R \Bigg[
            \frac{223}{216}
          - \frac{4 \zeta(2)}{3 (1-x)}
          + \frac{392 \zeta(2)}{3 (1+x)^4}
          - \frac{784 \zeta(2)}{3 (1+x)^3}
          + \frac{458 \zeta(2)}{3 (1+x)^2} \nn\\
& & \hspace*{20mm}
          - \frac{68 \zeta(2)}{3 (1+x)}
          + 8 \zeta(2)
          - \frac{4 \zeta(3)}{3 (1-x)}
          - \frac{4 \zeta(3)}{3 (1+x)}
          + \frac{4 \zeta(3)}{3} \nn\\
& & \hspace*{20mm}
          + \frac{196}{9 (1\! +\! x)^2}\! 
          - \frac{196}{9 (1\! +\! x)}\! 
       +  \!  \frac{4 \zeta(2)}{3} \! \Biggl(
            \frac{1}{(1\! -\! x)}\! 
          + \! \frac{1}{(1\! +\! x)} \nn\\
& & \hspace*{20mm}
          - 1
          \Biggr) H(-1;x)\! 
       -   \! \frac{8}{3} \Biggl(
            1 \! 
          - \! \frac{1}{(1\! -\! x)}\! 
          - \! \frac{1}{(1\! +\! x)}
          \Biggr) H(-1,\! -1,\! 0;x) \nn\\
& & \hspace*{20mm}
       +  \Biggl(
            2 \! 
          - \! \frac{8}{3 (1-x)}\! 
          - \! \frac{4}{3 (1\!+\!x)}
          \Biggr) H(-1,0;x)\! 
       +  \!  \Biggl(
            \frac{4}{3}\! 
          - \! \frac{4}{3 (1-x)} \nn\\
& & \hspace*{20mm}
          - \frac{4}{3 (1+x)}
          \Biggr) H(-1,0,0;x)
       -   \Biggl(
            \frac{409}{54}
          - \frac{\zeta(2)}{ (1-x)}
          + \frac{24 \zeta(2)}{ (1+x)^5} \nn\\
& & \hspace*{20mm}
          - \frac{60 \zeta(2)}{ (1+x)^4}
          + \frac{44 \zeta(2)}{ (1+x)^3}
          - \frac{6 \zeta(2)}{ (1+x)^2}
          - \frac{\zeta(2)}{ (1+x)}
          - \frac{308}{27 (1-x)} \nn\\
& & \hspace*{20mm}
          - \frac{356}{9 (1+x)^3}
          + \frac{178}{3 (1+x)^2}
          - \frac{635}{27 (1+x)}
          \Biggr) H(0;x) \nn\\
& & \hspace*{20mm}
       +   \Biggl(
            \frac{4}{3}
          - \frac{4}{3 (1-x)}
          - \frac{4}{3 (1+x)}
          \Biggr) H(0,-1,0;x) \nn\\
& & \hspace*{20mm}
       +   \Biggl(
            \frac{10}{9}
          + \frac{4}{3 (1-x)}
          + \frac{248}{9 (1+x)^4}
          - \frac{496}{9 (1+x)^3}
          + \frac{326}{9 (1+x)^2} \nn\\
& & \hspace*{20mm}
          - \frac{8}{ (1+x)}
          \Biggr) H(0,0;x)
       -   \Biggl(
            \frac{4}{3}
          - \frac{7}{3 (1-x)}
          + \frac{24}{ (1+x)^5} \nn\\
& & \hspace*{20mm}
          -  \frac{60 }{(1 \! + \! x)^4} \! 
          +  \! \frac{44}{ (1 \! + \! x)^3} \! 
          -  \! \frac{6}{ (1 \! + \! x)^2} \! 
          -  \! \frac{7}{3 (1 \! + \! x)}
          \Biggr) H(0, \! 0, \! 0;x)
        \Bigg] \Biggr\} \nn\\
& & 
       + \, C_F^2 \Bigg[
            \frac{383}{32} \! 
          -  \! \frac{72 \zeta(2) \log(2)}{ (1+x)^2} \! 
          +  \! \frac{72 \zeta(2) \log(2)}{ (1+x)} \! 
          -  \! 36 \zeta(2) \log(2) \! 
          +  \! \frac{14 \zeta(2)}{ (1-x)} \nn\\
& & \hspace*{10mm}
          - \frac{240 \zeta(2)}{ (1+x)^4}
          + \frac{508 \zeta(2)}{ (1+x)^3}
          - \frac{270 \zeta(2)}{ (1+x)^2}
          - \frac{8 \zeta(2)}{ (1+x)}
          + 29 \zeta(2) \nn\\
& & \hspace*{10mm}
          +  \frac{61 \zeta^2(2)}{10 (1-x)^2} \! 
          -  \! \frac{1219 \zeta^2(2)}{80 (1-x)} \! 
          -  \! \frac{171 \zeta^2(2)}{ (1 \! + \! x)^5} \! 
          +  \! \frac{855 \zeta^2(2)}{2 (1 \! + \! x)^4} \! 
          -  \! \frac{6867 \zeta^2(2)}{20 (1 \! + \! x)^3} \nn\\
& & \hspace*{10mm}
          + \frac{749 \zeta^2(2)}{8 (1+x)^2}
          - \frac{1731 \zeta^2(2)}{80 (1+x)}
          + \frac{181 \zeta^2(2)}{10}
          - \frac{4 \zeta(3)}{ (1-x)^2}
          + \frac{11 \zeta(3)}{ (1-x)} \nn\\
& & \hspace*{10mm}
          + \frac{168 \zeta(3)}{ (1+x)^4}
          - \frac{348 \zeta(3)}{ (1+x)^3}
          + \frac{208 \zeta(3)}{ (1+x)^2}
          - \frac{33 \zeta(3)}{ (1+x)}
          - 4 \zeta(3) \nn\\
& & \hspace*{10mm}
          + \frac{52}{ (1+x)^2}
          - \frac{52}{ (1+x)}
       -   \Biggl(
            \frac{7 \zeta(2)}{ (1-x)}
          - \frac{180 \zeta(2)}{ (1+x)^4}
          + \frac{348 \zeta(2)}{ (1+x)^3} \nn\\
& & \hspace*{10mm}
          - \frac{252 \zeta(2)}{ (1+x)^2}
          + \frac{79 \zeta(2)}{ (1+x)}
          - 13 \zeta(2)
          \Biggr) H(-1;x)
       +   \Biggl(
            2
          - \frac{14}{ (1-x)} \nn\\
& & \hspace*{10mm}
          + \frac{24}{ (1+x)^3}
          - \frac{36}{ (1+x)^2}
          + \frac{22}{ (1+x)}
          \Biggr) H(-1,-1,0;x)
       +   \Biggl(
            16 \nn\\
& & \hspace*{10mm}
          +  \frac{16}{ (1-x)^2} \! 
          -  \! \frac{16}{ (1-x)} \! 
          +  \! \frac{16}{ (1+x)^2} \! 
          -  \! \frac{16}{ (1+x)}
          \Biggr) H(-1, \! -1, \! 0, \! 0;x) \nn\\
& & \hspace*{10mm}
       -   \Biggl(
            36
          - \frac{4 \zeta(2)}{ (1-x)^2}
          + \frac{4 \zeta(2)}{ (1-x)}
          - \frac{4 \zeta(2)}{ (1+x)^2}
          + \frac{4 \zeta(2)}{ (1+x)}
          - 4 \zeta(2) \nn\\
& & \hspace*{10mm}
          - \frac{60}{ (1-x)}
          - \frac{152}{ (1+x)^3}
          + \frac{228}{ (1+x)^2}
          - \frac{88}{ (1+x)}
          \Biggr) H(-1,0;x) \nn\\
& & \hspace*{10mm}
       +   \Biggl(
            8 \! 
          +  \! \frac{8}{ (1-x)^2} \! 
          -  \! \frac{8}{ (1\!-\!x)} \! 
          +  \! \frac{8}{ (1 \! + \! x)^2} \! 
          -  \! \frac{8}{ (1 \! + \! x)}
          \Biggr) H(-1, \! 0, \! -1, \! 0;x) \nn\\
& & \hspace*{10mm}
       -   \Biggl(
            9
          + \frac{16}{ (1-x)^2}
          - \frac{23}{ (1-x)}
          + \frac{252}{ (1+x)^4}
          - \frac{492}{ (1+x)^3}
          + \frac{252}{ (1+x)^2} \nn\\
& & \hspace*{10mm}
          - \frac{7}{ (1+x)}
          \Biggr) H(-1,0,0;x)
       -   \Biggl(
            12
          + \frac{12}{ (1-x)^2}
          - \frac{12}{ (1-x)} \nn\\
& & \hspace*{10mm}
          + \frac{12}{ (1+x)^2}
          - \frac{12}{ (1+x)}
          \Biggr) H(-1,0,0,0;x)
       +   \Biggl(
            \frac{19}{8}
          - \frac{8 \zeta(2)}{ (1-x)^2} \nn\\
& & \hspace*{10mm}
          + \frac{6 \zeta(2)}{ (1-x)}
          - \frac{60 \zeta(2)}{ (1+x)^5}
          + \frac{156 \zeta(2)}{ (1+x)^4}
          - \frac{142 \zeta(2)}{ (1+x)^3}
          + \frac{15 \zeta(2)}{ (1+x)^2} \nn\\
& & \hspace*{10mm}
          +  \frac{52 \zeta(2)}{ (1+x)} \! 
          -  \! \frac{37 \zeta(2)}{2} \! 
          -  \! \frac{7 \zeta(3)}{ (1-x)} \! 
          +  \! \frac{168 \zeta(3)}{ (1+x)^5} \! 
          -  \! \frac{420 \zeta(3)}{ (1+x)^4} \! 
          +  \! \frac{364 \zeta(3)}{ (1+x)^3} \nn\\
& & \hspace*{10mm}
          -  \frac{126 \zeta(3)}{ (1+x)^2} \! 
          +  \! \frac{5 \zeta(3)}{ (1+x)} \! 
          +  \! 8 \zeta(3) \! 
          -  \! \frac{45}{2 (1-x)} \! 
          +  \! \frac{68}{ (1+x)^3} \! 
          -  \! \frac{102}{ (1+x)^2} \nn\\
& & \hspace*{10mm}
          + \frac{207}{4 (1+x)}
          \Biggr) H(0;x)
       +   \Biggl(
            \frac{2 \zeta(2)}{ (1-x)^2}
          + \frac{19 \zeta(2)}{4 (1-x)}
          + \frac{180 \zeta(2)}{ (1+x)^5} \nn\\
& & \hspace*{10mm}
          - \frac{450 \zeta(2)}{ (1+x)^4}
          + \frac{363 \zeta(2)}{ (1+x)^3}
          - \frac{185 \zeta(2)}{2 (1+x)^2}
          + \frac{67 \zeta(2)}{4 (1+x)} \nn\\
& & \hspace*{10mm}
          - 10 \zeta(2)
          \Biggr) H(0,-1;x)
       +   \Biggl(
            4
          + \frac{4}{ (1-x)^2}
          - \frac{4}{ (1-x)}
          + \frac{4}{ (1+x)^2} \nn\\
& & \hspace*{10mm}
          - \frac{4}{ (1+x)}
          \Biggr) H(0,-1,-1,0;x)
       +   \Biggl(
            13
          + \frac{7}{ (1-x)}
          + \frac{384}{ (1+x)^4} \nn\\
& & \hspace*{10mm}
          -  \frac{780}{ (1 \! + \! x)^3} \! 
          +  \! \frac{460}{ (1 \! + \! x)^2} \! 
          -  \! \frac{69}{ (1 \! + \! x)}
          \Biggr) H(0,-1,0;x) \! 
       -    \! \Biggl(
            14 \! 
          +  \! \frac{10}{ (1-x)^2} \nn\\
& & \hspace*{10mm}
          - \frac{67}{4 (1-x)}
          + \frac{252}{ (1+x)^5}
          - \frac{630}{ (1+x)^4}
          + \frac{517}{ (1+x)^3}
          - \frac{271}{2 (1+x)^2} \nn\\
& & \hspace*{10mm}
          - \frac{19}{4 (1+x)}
          \Biggr) H(0,-1,0,0;x)
       +   \Biggl(
            \frac{123}{2}
          + \frac{5 \zeta(2)}{ (1-x)^2}
          - \frac{105 \zeta(2)}{8 (1-x)} \nn\\
& & \hspace*{10mm}
          - \frac{66 \zeta(2)}{ (1+x)^5}
          + \frac{165 \zeta(2)}{ (1+x)^4}
          - \frac{281 \zeta(2)}{2 (1+x)^3}
          + \frac{203 \zeta(2)}{4 (1+x)^2}
          - \frac{137 \zeta(2)}{8 (1+x)} \nn\\
& & \hspace*{10mm}
          +  13 \zeta(2) \! 
          +  \! \frac{16}{ (1-x)^2} \! 
          -  \! \frac{54 }{(1-x)} \! 
          +  \! \frac{192}{ (1+x)^4} \! 
          -  \! \frac{484 }{(1 \! + \! x)^3} \! 
          +  \! \frac{436}{ (1 \! + \! x)^2} \nn\\
& & \hspace*{10mm}
          -  \frac{158}{ (1 \! + \! x)}
          \Biggr) H(0,0;x) \! 
       -   \!  \Biggl(
            22 \! 
          +  \! \frac{10}{ (1-x)^2} \! 
          -  \! \frac{21}{2 (1-x)} \! 
          -  \! \frac{384}{ (1+x)^5} \nn\\
& & \hspace*{10mm}
          + \frac{960}{ (1+x)^4} \! 
          -  \! \frac{746}{ (1+x)^3} \! 
          +  \! \frac{169}{ (1+x)^2} \! 
          -  \! \frac{45}{2 (1+x)}
          \Biggr) H(0,0,-1,0;x) \nn\\
& & \hspace*{10mm}
       -   \Biggl(
            \frac{17}{2}
          - \frac{4}{ (1-x)^2}
          + \frac{13 }{(1-x)}
          + \frac{60}{ (1+x)^5}
          - \frac{216}{ (1+x)^4}
          + \frac{250}{ (1+x)^3} \nn\\
& & \hspace*{10mm}
          - \frac{80}{ (1+x)^2}
          - \frac{40}{ (1+x)}
          \Biggr) H(0,0,0;x)
       +   \Biggl(
            27
          + \frac{17}{ (1-x)^2} \nn\\
& & \hspace*{10mm}
          - \frac{217}{8 (1-x)}
          - \frac{6}{ (1+x)^5}
          + \frac{15}{ (1+x)^4}
          - \frac{17}{2 (1+x)^3}
          + \frac{59}{4 (1+x)^2} \nn\\
& & \hspace*{10mm}
          - \frac{201}{8 (1+x)}
          \Biggr) H(0,0,0,0;x)
       +   \Biggl(
            12
          + \frac{4}{ (1-x)^2}
          - \frac{8}{ (1-x)} \nn\\
& & \hspace*{10mm}
          - \frac{96}{ (1+x)^5}
          + \frac{240}{ (1+x)^4}
          - \frac{176}{ (1+x)^3}
          + \frac{28}{ (1+x)^2} \nn\\
& & \hspace*{10mm}
          - \frac{8}{ (1+x)}
          \Biggr) H(0,0,1,0;x)
       -   \Biggl(
            10
          + \frac{8}{ (1-x)^2}
          - \frac{8}{ (1-x)} \nn\\
& & \hspace*{10mm}
          + \frac{96}{ (1+x)^4}
          - \frac{192}{ (1+x)^3}
          + \frac{116}{ (1+x)^2}
          - \frac{20}{ (1+x)}
          \Biggr) H(0,1,0;x) \nn\\
& & \hspace*{10mm}
       -   \Biggl(
            4
          + \frac{7}{ (1-x)}
          - \frac{360}{ (1+x)^5}
          + \frac{900}{ (1+x)^4}
          - \frac{700}{ (1+x)^3}
          + \frac{150}{ (1+x)^2} \nn\\
& & \hspace*{10mm}
          - \frac{5}{ (1+x)}
          \Biggr) H(0,1,0,0;x)
       +   \Biggl(
            \frac{4 \zeta(3)}{ (1-x)^2}
          - \frac{4 \zeta(3)}{ (1-x)}
          + \frac{4 \zeta(3)}{ (1+x)^2} \nn\\
& & \hspace*{10mm}
          - \frac{4 \zeta(3)}{ (1+x)}
          + 4 \zeta(3)
          \Biggr) H(1;x)
       +   \Biggl(
            16
          + \frac{4 \zeta(2)}{ (1-x)^2}
          - \frac{6 \zeta(2)}{ (1-x)} \nn\\
& & \hspace*{10mm}
          - \frac{144 \zeta(2)}{ (1+x)^5}
          + \frac{360 \zeta(2)}{ (1+x)^4}
          - \frac{312 \zeta(2)}{ (1+x)^3}
          + \frac{112 \zeta(2)}{ (1+x)^2}
          - \frac{14 \zeta(2)}{ (1+x)} \nn\\
& & \hspace*{10mm}
          +  4 \zeta(2) \! 
          -  \! \frac{16}{ (1-x)} \! 
          -  \! \frac{48}{ (1+x)^3} \! 
          +  \! \frac{72}{ (1+x)^2} \! 
          -  \! \frac{40}{ (1+x)}
          \Biggr) H(1,0;x) \nn\\
& & \hspace*{10mm}
       -   \Biggl(
            8 \! 
          +  \! \frac{8}{ (1-x)^2} \! 
          -  \! \frac{8}{ (1\!-\!x)} \! 
          +  \! \frac{8 }{(1 \! + \! x)^2} \! 
          -  \! \frac{8}{ (1 \! + \! x)}
          \Biggr) H(1, \! 0, \! -1, \! 0;x) \nn\\
& & \hspace*{10mm}
       +  \Biggl(
            16 \! 
          +  \! \frac{360}{ (1+x)^4} \! 
          -  \! \frac{720}{ (1+x)^3} \! 
          +  \! \frac{394}{ (1\!+\!x)^2} \! 
          -  \! \frac{34}{ (1 \! + \! x)}
          \Biggr) H(1,0,0;x) \nn\\
& & \hspace*{10mm}
       +   \Biggl(
            8
          + \frac{8}{ (1-x)^2}
          - \frac{10}{ (1-x)}
          - \frac{144}{ (1+x)^5}
          + \frac{360}{ (1+x)^4}
          - \frac{312}{ (1+x)^3} \nn\\
& & \hspace*{10mm}
          + \frac{116}{ (1+x)^2}
          - \frac{18}{ (1+x)}
          \Biggr) H(1,0,0,0;x)
       +   \Biggl(
            8
          + \frac{8}{ (1-x)^2} \nn\\
& & \hspace*{10mm}
          - \frac{8}{ (1-x)}
          + \frac{8}{ (1+x)^2}
          - \frac{8}{ (1+x)}
          \Biggr) H(1,0,1,0;x)
        \Bigg] 
\label{2loopF1}
\, , \\
%
%%%%%%%%%%%%%%%%%%%%%%%%%%%%%%%%%
%
{\mathcal F}_2^{(2l)}(\epsilon,s) &=& 
     \frac{1}{\epsilon^2} \Bigg\{
       C_F^2 \Bigg[
          - \frac{6}{ (1+x)^2}
          + \frac{6}{ (1+x)}
       +   \Biggl(
            \frac{3}{2 (1-x)}
          - \frac{6}{ (1+x)^3}
          + \frac{9}{ (1+x)^2} \nn\\
& & \hspace*{20mm}
          - \frac{9}{2 (1+x)}
          \Biggr) H(0;x)
        \Bigg]
     \Bigg\}
     \nn\\
& & 
    + \frac{1}{\epsilon} \Bigg\{
       C_F C_A \Bigg[
       -   \Biggl(
            \frac{11}{6 (1-x)}
          - \frac{11}{6 (1+x)}
          \Biggr) H(0;x)
        \Bigg] \nn\\
& & \hspace*{8mm}
       + \, C_F T_R N_f \Bigg[
       +   \Biggl(
            \frac{2}{3 (1-x)}
          - \frac{2}{3 (1+x)}
          \Biggr) H(0;x)
        \Bigg] \nn\\
& & \hspace*{8mm}
       + \, C_F T_R \Bigg[
       +   \Biggl(
            \frac{2}{3 (1-x)}
          - \frac{2}{3 (1+x)}
          \Biggr) H(0;x)
        \Bigg] \nn\\
& & \hspace*{8mm}
       + \, C_F^2 \Bigg[
          -  \! \frac{3 \zeta(2)}{2 (1-x)} \! 
          + \! \frac{6 \zeta(2)}{ (1+x)^3} \! 
          -  \! \frac{9 \zeta(2)}{ (1+x)^2} \! 
          +  \! \frac{9 \zeta(2)}{2 (1+x)}
          -  \! \frac{20}{ (1+x)^2} \nn\\
& & \hspace*{20mm}
          + \frac{20}{ (1+x)}
       -   \Biggl(
            \frac{3}{ (1-x)}
          - \frac{12}{ (1+x)^3}
          + \frac{18}{ (1+x)^2} \nn\\
& & \hspace*{20mm}
          -  \frac{9}{ (1+x)}
          \Biggr) H(-1,0;x) \! 
       + \Biggl(
            \frac{3}{2 (1-x)} \! 
          -  \! \frac{32}{ (1 \! + \! x)^3} \! 
          +  \! \frac{48}{ (1 \! + \! x)^2} \nn\\
& & \hspace*{20mm}
          - \frac{35}{2 (1+x)}
          \Biggr) H(0;x)
       +   \Biggl(
            \frac{2}{ (1-x)^2}
          - \frac{1}{2 (1-x)}
          - \frac{6}{ (1+x)^3} \nn\\
& & \hspace*{20mm}
          + \frac{7}{ (1+x)^2}
          - \frac{5}{2 (1+x)}
          \Biggr) H(0,0;x)
        \Bigg]
     \Bigg\}
     \nn\\
& & 
       + \, C_F C_A \Bigg[
          - \frac{24 \zeta(2) \log(2)}{ (1+x)^2}
          + \frac{24 \zeta(2) \log(2)}{ (1+x)}
          + \frac{3 \zeta(2)}{ (1-x)^2}
          - \frac{13 \zeta(2)}{12 (1-x)} \nn\\
& & \hspace*{10mm}
          + \frac{168 \zeta(2)}{ (1+x)^4}
          - \frac{363 \zeta(2)}{ (1+x)^3}
          + \frac{491 \zeta(2)}{2 (1+x)^2}
          - \frac{629 \zeta(2)}{12 (1+x)}
          + \frac{69 \zeta^2(2)}{40 (1-x)^3}\nn\\
& & \hspace*{10mm}
          -  \frac{207 \zeta^2(2)}{80 (1-x)^2} \! 
          +  \! \frac{45 \zeta^2(2)}{32 (1-x)} \! 
          +  \! \frac{879 \zeta^2(2)}{10 (1 \! + \! x)^5} \! 
          -  \! \frac{879 \zeta^2(2)}{4 (1 \! + \! x)^4} \! 
          +  \! \frac{1797 \zeta^2(2)}{10 (1 \! + \! x)^3} \nn\\
& & \hspace*{10mm}
          - \frac{249 \zeta^2(2)}{5 (1+x)^2}
          + \frac{45 \zeta^2(2)}{32 (1+x)}
          - \frac{324 \zeta(3)}{ (1+x)^4}
          + \frac{648 \zeta(3)}{ (1+x)^3}
          - \frac{344 \zeta(3)}{ (1+x)^2} \nn\\
& & \hspace*{10mm}
          + \frac{20 \zeta(3)}{ (1+x)}
          - \frac{3}{ (1+x)^2}
          + \frac{3}{ (1+x)}
       -   \Biggl(
            \frac{9 \zeta(2)}{2 (1-x)^2}
          - \frac{9 \zeta(2)}{2 (1-x)} \nn\\
& & \hspace*{10mm}
          - \frac{90 \zeta(2)}{ (1+x)^4}
          + \frac{180 \zeta(2)}{ (1+x)^3}
          - \frac{231 \zeta(2)}{2 (1+x)^2}
          + \frac{51 \zeta(2)}{2 (1+x)}
          \Biggr) H(-1;x) \nn\\
& & \hspace*{10mm}
       +   \Biggl(
            \frac{23}{6 (1-x)}
          - \frac{6}{ (1+x)^3}
          + \frac{9}{ (1+x)^2}
          - \frac{41}{6 (1+x)}
          \Biggr) H(-1,0;x) \nn\\
& & \hspace*{10mm}
       -   \Biggl(
            \frac{1}{2 (1-x)^2}
          - \frac{1}{2 (1-x)}
          - \frac{282}{ (1+x)^4}
          + \frac{564}{ (1+x)^3}
          - \frac{623}{2 (1+x)^2} \nn\\
& & \hspace*{10mm}
          +  \frac{59}{2 (1 \! + \! x)}
          \Biggr) H(-1,0,0;x) \! 
       +    \! \Biggl(
            \frac{3 \zeta(2)}{ (1-x)^3} \! 
          -  \! \frac{11 \zeta(2)}{4 (1-x)^2} \! 
          +  \! \frac{3 \zeta(2)}{8 (1-x)} \nn\\
& & \hspace*{10mm}
          + \frac{258 \zeta(2)}{ (1+x)^5}
          - \frac{744 \zeta(2)}{ (1+x)^4}
          + \frac{1515 \zeta(2)}{2 (1+x)^3}
          - \frac{621 \zeta(2)}{2 (1+x)^2}
          + \frac{307 \zeta(2)}{8 (1+x)} \nn\\
& & \hspace*{10mm}
          + \frac{13 \zeta(3)}{4 (1-x)}
          - \frac{324 \zeta(3)}{ (1+x)^5}\!
          + \!\frac{810 \zeta(3)}{ (1+x)^4}
          - \frac{635 \zeta(3)}{ (1+x)^3}\!
          + \!\frac{285 \zeta(3)}{2 (1+x)^2} \nn\\
& & \hspace*{10mm}
          + \frac{13 \zeta(3)}{4 (1\! +\! x)}\! 
          - \! \frac{259}{18 (1\!-\!x)}\! 
          - \! \frac{9}{(1\! +\! x)^3}\! 
          + \! \frac{27}{2 (1\! +\! x)^2}\! 
          + \! \frac{89}{9 (1\! +\! x)}
          \Biggr) H(0;x) \nn\\
& & \hspace*{10mm}
       -   \Biggl(
            \frac{9 \zeta(2)}{2 (1-x)^3}
          - \frac{27 \zeta(2)}{4 (1-x)^2}
          + \frac{21 \zeta(2)}{8 (1-x)}
          - \frac{90 \zeta(2)}{ (1+x)^5}
          + \frac{225 \zeta(2)}{ (1+x)^4} \nn\\
& & \hspace*{10mm}
          - \frac{174 \zeta(2)}{ (1+x)^3}
          + \frac{36 \zeta(2)}{ (1+x)^2}
          + \frac{21 \zeta(2)}{8 (1+x)}
          \Biggr) H(0,-1;x)
       -   \Biggl(
            \frac{1}{(1-x)^2} \nn\\
& & \hspace*{10mm}
          - \frac{1}{(1\!-\!x)}\! 
          + \! \frac{12}{ (1\! +\! x)^4}\! 
          - \! \frac{24}{ (1\! +\! x)^3}\! 
          + \! \frac{21}{ (1\! +\! x)^2}\! 
          - \! \frac{9}{ (1\! +\! x)}
          \Biggr) H(0,\! -1,\! 0;x) \nn\\
& & \hspace*{10mm}
       -   \Biggl(
            \frac{1}{2 (1-x)^3}
          - \frac{3}{4 (1-x)^2}
          + \frac{5}{8 (1-x)}
          - \frac{282}{ (1+x)^5}
          + \frac{705}{ (1+x)^4} \nn\\
& & \hspace*{10mm}
          - \frac{562}{ (1+x)^3}
          + \frac{138 }{(1+x)^2}
          + \frac{5}{8 (1+x)}
          \Biggr) H(0,-1,0,0;x) \nn\\
& & \hspace*{10mm}
       +   \Biggl(
            \frac{7 \zeta(2)}{4 (1-x)^3}
          - \frac{21 \zeta(2)}{8 (1-x)^2}
          - \frac{5 \zeta(2)}{16 (1-x)}
          + \frac{69 \zeta(2)}{ (1+x)^5}
          - \frac{345 \zeta(2)}{2 (1+x)^4} \nn\\
& & \hspace*{10mm}
          +  \frac{135 \zeta(2)}{ (1+x)^3} \! 
          -  \! \frac{30 \zeta(2)}{ (1+x)^2} \! 
          -  \! \frac{5 \zeta(2)}{16 (1+x)} \! 
          +  \! \frac{3}{ (1-x)^2} \! 
          -  \! \frac{59}{12 (1-x)} \nn\\
& & \hspace*{10mm}
          - \frac{24}{ (1+x)^4}
          + \frac{39}{ (1+x)^3}
          - \frac{7}{2 (1+x)^2}
          - \frac{115}{12 (1+x)}
          \Biggr) H(0,0;x) \nn\\
& & \hspace*{10mm}
       -   \Biggl(
            \frac{1}{(1-x)^3}
          - \frac{3}{2 (1-x)^2}
          + \frac{25}{4 (1-x)}
          + \frac{12}{ (1+x)^5}
          - \frac{30}{ (1+x)^4} \nn\\
& & \hspace*{10mm}
          + \frac{48}{ (1+x)^3}
          - \frac{42}{ (1+x)^2}
          + \frac{25}{4 (1+x)}
          \Biggr) H(0,0,-1,0;x) \nn\\
& & \hspace*{10mm}
       +   \Biggl(
            \frac{3}{ (1-x)^3}
          - \frac{15}{4 (1-x)^2}
          + \frac{11}{8 (1-x)}
          + \frac{258}{ (1+x)^5}
          - \frac{816}{ (1+x)^4} \nn\\
& & \hspace*{10mm}
          + \frac{1803}{2 (1+x)^3}
          - \frac{767}{2 (1+x)^2}
          + \frac{315}{8 (1+x)}
          \Biggr) H(0,0,0;x) \nn\\
& & \hspace*{10mm}
       +   \Biggl(
            \frac{3}{4 (1-x)^3}
          - \frac{9}{8 (1-x)^2}
          + \frac{11}{16 (1-x)}
          - \frac{3}{ (1+x)^5}
          + \frac{15}{2 (1+x)^4} \nn\\
& & \hspace*{10mm}
          - \frac{4}{ (1+x)^3}
          - \frac{3}{2 (1+x)^2}
          + \frac{11}{16 (1+x)}
          \Biggr) H(0,0,0,0;x) \nn\\
& & \hspace*{10mm}
       +   \Biggl(
            \frac{3}{ (1-x)}
          - \frac{48}{ (1+x)^5}
          + \frac{120}{ (1+x)^4}
          - \frac{84}{ (1+x)^3}
          + \frac{6}{ (1+x)^2} \nn\\
& & \hspace*{10mm}
          + \frac{3}{ (1+x)}
          \Biggr) H(0,0,1,0;x)
       -   \Biggl(
            \frac{48}{ (1+x)^4}
          - \frac{96}{ (1+x)^3} \nn\\
& & \hspace*{10mm}
          + \frac{48}{ (1+x)^2}
          \Biggr) H(0,1,0;x)
       -   \Biggl(
            \frac{11}{2 (1-x)}
          + \frac{24}{ (1+x)^5}
          - \frac{60}{ (1+x)^4} \nn\\
& & \hspace*{10mm}
          + \frac{70}{ (1+x)^3}
          - \frac{45}{ (1+x)^2}
          + \frac{11}{2 (1+x)}
          \Biggr) H(0,1,0,0;x) \nn\\
& & \hspace*{10mm}
       -   \Biggl(
            \frac{3 \zeta(2)}{ (1-x)}
          - \frac{336 \zeta(2)}{ (1+x)^5}
          + \frac{840 \zeta(2)}{ (1+x)^4}
          - \frac{660 \zeta(2)}{ (1+x)^3}
          + \frac{150 \zeta(2)}{ (1+x)^2} \nn\\
& & \hspace*{10mm}
          + \frac{3 \zeta(2)}{ (1+x)}
          + \frac{24}{ (1+x)^3}
          - \frac{36}{ (1+x)^2}
          + \frac{12}{ (1+x)}
          \Biggr) H(1,0;x) \nn\\
& & \hspace*{10mm}
       -   \Biggl(
            \frac{24}{ (1+x)^4}
          - \frac{48}{ (1+x)^3}
          + \frac{40}{ (1+x)^2}
          - \frac{16}{ (1+x)}
          \Biggr) H(1,0,0;x) \nn\\
& & \hspace*{10mm}
       -   \Biggl(
            \frac{3}{ (1-x)}
          - \frac{336}{ (1+x)^5}
          + \frac{840}{ (1+x)^4}
          - \frac{660}{ (1+x)^3}
          + \frac{150}{ (1+x)^2} \nn\\
& & \hspace*{10mm}
          + \frac{3}{ (1+x)}
          \Biggr) H(1,0,0,0;x)
        \Bigg] \nn\\
& & 
       + \, C_F T_R N_f \Bigg[
          - \frac{4 \zeta(2)}{3 (1-x)}
          + \frac{4 \zeta(2)}{3 (1+x)}
       -  \Biggl(
            \frac{8}{3 (1-x)} \nn\\
& & \hspace*{10mm}
          - \frac{8}{3 (1+x)}
          \Biggr) H(-1,0;x) 
       +   \Biggl(
            \frac{49}{9 (1-x)}
          - \frac{49}{9 (1+x)}
          \Biggr) H(0;x) \nn\\
& & \hspace*{10mm}
       +   \Biggl(
            \frac{4}{3 (1-x)}
          - \frac{4}{3 (1+x)}
          \Biggr) H(0,0;x)
        \Bigg] \nn\\
& & 
       + \, C_F T_R \Bigg[
          -  \frac{2 \zeta(2)}{3 (1-x)} \! 
          -  \! \frac{136 \zeta(2)}{ (1+x)^4} \! 
          +  \! \frac{272 \zeta(2)}{ (1+x)^3} \! 
          -  \! \frac{132 \zeta(2)}{ (1+x)^2} \! 
          -  \! \frac{10 \zeta(2)}{3 (1\!+\!x)} \nn\\
& & \hspace*{10mm}
          - \frac{68}{3 (1+x)^2} \! 
          +  \! \frac{68}{3 (1+x)} \! 
       +   \!  \Biggl(
          - \frac{4}{3 (1-x)} \! 
          +  \! \frac{4}{3 (1+x)}
          \Biggr) H(-1,0;x) \nn\\
& & \hspace*{10mm}
       +   \Biggl(
          - \frac{3 \zeta(2)}{2 (1-x)}
          + \frac{24 \zeta(2)}{ (1+x)^5}
          - \frac{60 \zeta(2)}{ (1+x)^4}
          + \frac{42 \zeta(2)}{ (1+x)^3}
          - \frac{3 \zeta(2)}{ (1+x)^2} \nn\\
& & \hspace*{10mm}
          -   \frac{3 \zeta(2)}{2 (1  +  x)}  
          -   \frac{44}{9 (1-x)}  
          -   \frac{124}{3 (1  +  x)^3}  
          +   \frac{62}{ (1  +  x)^2}   \nn\\
& & \hspace*{10mm}
          -   \frac{142}{9 (1  +  x)}
          \Biggr) H(0;x)
       +     \Biggl(
            \frac{2}{3 (1-x)}  
          -   \frac{88}{3 (1  +  x)^4}  
          +   \frac{176}{3 (1  +  x)^3}  \nn\\
& & \hspace*{10mm} 
          -   \frac{92}{3 (1  +  x)^2}  
          +   \frac{2}{3 (1  +  x)}
          \Biggr) H(0,0;x)
       -   \Biggl(
            \frac{3}{2 (1-x)}
          - \frac{24}{ (1+x)^5} \nn\\
& & \hspace*{10mm}
          + \frac{60}{ (1+x)^4}
          - \frac{42}{ (1+x)^3}
          + \frac{3}{ (1+x)^2}
          + \frac{3}{2 (1+x)}
          \Biggr) H(0,0,0;x)
        \Bigg] \nn\\
& & 
       + \, C_F^2 \Bigg[
            \frac{48 \zeta(2) \log(2)}{ (1 \! + \! x)^2} \! 
          -  \! \frac{48 \zeta(2) \log(2)}{ (1 \! + \! x)} \! 
          -  \! \frac{9 \zeta(2)}{ (1-x)} \! 
          +  \! \frac{240 \zeta(2)}{ (1 \! + \! x)^4} \! 
          -  \! \frac{496 \zeta(2)}{ (1 \! + \! x)^3} \nn\\
& & \hspace*{10mm}
          + \frac{256 \zeta(2)}{ (1+x)^2}
          + \frac{9 \zeta(2)}{ (1+x)}
          - \frac{69 \zeta^2(2)}{20 (1-x)^3}
          + \frac{207 \zeta^2(2)}{40 (1-x)^2}
          - \frac{327 \zeta^2(2)}{80 (1-x)} \nn\\
& & \hspace*{10mm}
          +  \frac{171 \zeta^2(2)}{ (1+x)^5} \! 
          -  \! \frac{855 \zeta^2(2)}{2 (1+x)^4} \! 
          + \!  \frac{3291 \zeta^2(2)}{10 (1+x)^3} \! 
          -  \! \frac{1323 \zeta^2(2)}{20 (1+x)^2} \! 
          -  \! \frac{327 \zeta^2(2)}{80 (1 \! + \! x)} \nn\\
& & \hspace*{10mm}
          -  \frac{10 \zeta(3)}{ (1-x)^2} \! 
          +  \! \frac{7 \zeta(3)}{ (1-x)} \! 
          -  \! \frac{168 \zeta(3)}{ (1+x)^4} \! 
          +  \! \frac{348 \zeta(3)}{ (1+x)^3} \! 
          -  \! \frac{200 \zeta(3)}{ (1+x)^2} \! 
          +  \! \frac{23 \zeta(3)}{ (1+x)} \nn\\
& & \hspace*{10mm}
          - \frac{44}{ (1+x)^2}
          + \frac{44}{ (1+x)}
       +   \Biggl(
            \frac{9 \zeta(2)}{ (1-x)^2}
          - \frac{6 \zeta(2)}{ (1-x)}
          - \frac{180 \zeta(2)}{ (1+x)^4} \nn\\
& & \hspace*{10mm}
          + \frac{348 \zeta(2)}{ (1+x)^3}
          - \frac{213 \zeta(2)}{ (1+x)^2}
          + \frac{42 \zeta(2)}{ (1+x)}
          \Biggr) H(-1;x)
       +   \Biggl(
            \frac{6}{ (1-x)} \nn\\
& & \hspace*{10mm}
          - \frac{24}{ (1+x)^3}
          + \frac{36}{ (1+x)^2}
          - \frac{18}{ (1+x)}
          \Biggr) H(-1,-1,0;x) \nn\\
& & \hspace*{10mm}
       -  \Biggl(
            \frac{10}{ (1-x)}
          + \frac{128}{ (1+x)^3}
          - \frac{192}{ (1+x)^2}
          + \frac{54}{ (1+x)}
          \Biggr) H(-1,0;x)  \nn\\
& & \hspace*{10mm}
       +   \Biggl(
            \frac{1}{(1-x)^2}
          - \frac{4}{ (1-x)}
          + \frac{252}{ (1+x)^4}
          - \frac{492}{ (1+x)^3}
          + \frac{239}{ (1+x)^2} \nn\\
& & \hspace*{10mm}
          + \frac{4}{ (1+x)}
          \Biggr) H(-1,0,0;x)
       -   \Biggl(
            \frac{19 \zeta(2)}{2 (1-x)^2}
          - \frac{75 \zeta(2)}{4 (1-x)}
          - \frac{60 \zeta(2)}{ (1+x)^5} \nn\\
& & \hspace*{10mm}
          + \frac{156 \zeta(2)}{ (1+x)^4}
          - \frac{137 \zeta(2)}{ (1+x)^3}
          + \frac{19 \zeta(2)}{ (1+x)^2}
          + \frac{125 \zeta(2)}{4 (1+x)}
          + \frac{7 \zeta(3)}{2 (1-x)} \nn\\
& & \hspace*{10mm}
          + \frac{168 \zeta(3)}{ (1+x)^5}
          - \frac{420 \zeta(3)}{ (1+x)^4}
          + \frac{350 \zeta(3)}{ (1+x)^3}
          - \frac{105 \zeta(3)}{ (1+x)^2}
          + \frac{7 \zeta(3)}{2 (1+x)} \nn\\
& & \hspace*{10mm}
          - \frac{15}{4 (1-x)}
          + \frac{84}{ (1+x)^3}
          - \frac{126}{ (1+x)^2}
          + \frac{183}{4 (1+x)}
          \Biggr) H(0;x) \nn\\
& & \hspace*{10mm}
       +   \Biggl(
            \frac{9 \zeta(2)}{ (1-x)^3}
          - \frac{27 \zeta(2)}{2 (1-x)^2}
          + \frac{21 \zeta(2)}{4 (1-x)}
          - \frac{180 \zeta(2)}{ (1+x)^5}
          + \frac{450 \zeta(2)}{ (1+x)^4} \nn\\
& & \hspace*{10mm}
          - \frac{348 \zeta(2)}{ (1+x)^3}
          + \frac{72 \zeta(2)}{ (1+x)^2}
          + \frac{21 \zeta(2)}{4 (1+x)}
          \Biggr) H(0,-1;x) \nn\\
& & \hspace*{10mm}
       -   \Biggl(
            \frac{2}{ (1-x)^2}
          + \frac{1}{(1-x)}
          + \frac{384}{ (1+x)^4}
          - \frac{780}{ (1+x)^3}
          + \frac{412}{ (1+x)^2} \nn\\
& & \hspace*{10mm}
          - \frac{19}{ (1+x)}
          \Biggr) H(0,-1,0;x)
       +   \Biggl(
            \frac{1}{(1-x)^3}
          - \frac{3}{2 (1-x)^2}
          - \frac{7}{4 (1-x)} \nn\\
& & \hspace*{10mm}
          + \frac{252}{ (1+x)^5}
          - \frac{630}{ (1+x)^4}
          + \frac{496}{ (1+x)^3}
          - \frac{114}{ (1+x)^2} \nn\\
& & \hspace*{10mm}
          - \frac{7}{4 (1+x)}
          \Biggr) H(0,-1,0,0;x)
       -   \Biggl(
            \frac{7 \zeta(2)}{2 (1-x)^3}
          - \frac{21 \zeta(2)}{4 (1-x)^2} \nn\\
& & \hspace*{10mm}
          + \frac{\zeta(2)}{8 (1-x)}
          - \frac{66 \zeta(2)}{ (1+x)^5}
          + \frac{165 \zeta(2)}{ (1+x)^4}
          - \frac{135 \zeta(2)}{ (1+x)^3}
          + \frac{75 \zeta(2)}{2 (1+x)^2} \nn\\
& & \hspace*{10mm}
          + \frac{\zeta(2)}{8 (1+x)}
          - \frac{2}{ (1-x)^2}
          - \frac{1}{(1-x)}
          + \frac{192}{ (1+x)^4}
          - \frac{472}{ (1+x)^3} \nn\\
& & \hspace*{10mm}
          + \frac{394}{ (1+x)^2}
          - \frac{111}{ (1+x)}
          \Biggr) H(0,0;x)
       +   \Biggl(
            \frac{2}{ (1-x)^3}
          - \frac{3}{ (1-x)^2} \nn\\
& & \hspace*{10mm}
          + \frac{14}{ (1-x)}
          - \frac{384}{ (1+x)^5}
          + \frac{960}{ (1+x)^4}
          - \frac{714}{ (1+x)^3}
          + \frac{111}{ (1+x)^2} \nn\\
& & \hspace*{10mm}
          + \frac{14}{ (1+x)}
          \Biggr) H(0,0,-1,0;x)
       -   \Biggl(
            \frac{7}{2 (1-x)^2}
          - \frac{63}{4 (1-x)} \nn\\
& & \hspace*{10mm}
          -  \frac{60}{ (1 \! + \! x)^5} \! 
          +  \! \frac{216}{ (1 \! + \! x)^4} \! 
          -  \! \frac{245}{ (1 \! + \! x)^3} \! 
          +  \! \frac{79}{ (1 \! + \! x)^2} \! 
          +  \! \frac{89}{4 (1 \! + \! x)}
          \Biggr) H(0,\!0,\!0;x) \nn\\
& & \hspace*{10mm}
       -   \Biggl(
            \frac{3}{2 (1-x)^3}
          - \frac{9}{4 (1-x)^2}
          + \frac{11}{8 (1-x)}
          - \frac{6}{ (1+x)^5}
          + \frac{15}{ (1+x)^4} \nn\\
& & \hspace*{10mm}
          - \frac{8}{ (1+x)^3}
          - \frac{3}{ (1+x)^2}
          + \frac{11}{8 (1+x)}
          \Biggr) H(0,0,0,0;x) \nn\\
& & \hspace*{10mm}
       -   \Biggl(
            \frac{6}{ (1-x)}
          - \frac{96}{ (1+x)^5}
          + \frac{240}{ (1+x)^4}
          - \frac{168}{ (1+x)^3}
          + \frac{12}{ (1+x)^2} \nn\\
& & \hspace*{10mm}
          + \frac{6}{ (1+x)}
          \Biggr) H(0,0,1,0;x)
       -   \Biggl(
            \frac{4}{ (1-x)^2}
          - \frac{4}{ (1-x)} \nn\\
& & \hspace*{10mm}
          - \frac{96}{ (1+x)^4}
          + \frac{192}{ (1+x)^3}
          - \frac{100}{ (1+x)^2}
          + \frac{4}{ (1+x)}
          \Biggr) H(0,1,0;x) \nn\\
& & \hspace*{10mm}
       +   \Biggl(
            \frac{25}{2 (1-x)}
          - \frac{360}{ (1+x)^5}
          + \frac{900}{ (1+x)^4}
          - \frac{670}{ (1+x)^3}
          + \frac{105}{ (1+x)^2} \nn\\
& & \hspace*{10mm}
          + \frac{25}{2 (1+x)}
          \Biggr) H(0,1,0,0;x)
       +   \Biggl(
            \frac{3 \zeta(2)}{ (1-x)}
          + \frac{144 \zeta(2)}{ (1+x)^5} 
          - \frac{360 \zeta(2)}{ (1+x)^4}\nn\\
& & \hspace*{10mm}
          + \frac{300 \zeta(2)}{ (1+x)^3}
          - \frac{90 \zeta(2)}{ (1+x)^2}
          + \frac{3 \zeta(2)}{ (1+x)}
          - \frac{4}{ (1-x)}
          + \frac{48}{ (1+x)^3} \nn\\
& & \hspace*{10mm}
          - \frac{72}{ (1+x)^2}
          + \frac{28}{ (1+x)}
          \Biggr) H(1,0;x)
       -   \Biggl(
            \frac{4}{ (1-x)^2}
          - \frac{4}{ (1-x)} \nn\\
& & \hspace*{10mm}
          + \frac{360}{ (1+x)^4}
          - \frac{720}{ (1+x)^3}
          + \frac{356}{ (1+x)^2}
          + \frac{4}{ (1+x)}
          \Biggr) H(1,0,0;x) \nn\\
& & \hspace*{10mm}
       +   \Biggl(
            \frac{3}{ (1-x)}
          + \frac{144}{ (1+x)^5}
          - \frac{360}{ (1+x)^4}
          + \frac{300}{ (1+x)^3}
          - \frac{90}{ (1+x)^2} \nn\\
& & \hspace*{10mm}
          + \frac{3}{ (1+x)}
          \Biggr) H(1,0,0,0;x)
        \Bigg] 
\, .
\label{2loopF2}
\eea

\section{Renormalization \label{renorm}}

%%%%%%%%%%%%%%%%%%%% one-loop Vertex %%%%%%%%%%%%%%%%%%%%%%%%%%%%%%%%%%%
\bfig
\bc
\subfigure[]{\begin{fmfgraph*}(25,25)
\fmfleft{i}
\fmfright{o1,o2}
\fmfforce{0.8w,0.93h}{v2}
\fmfforce{0.8w,0.07h}{v1}
\fmfforce{0.5w,0.7h}{v3}
\fmfforce{0.5w,0.58h}{v33}
\fmfforce{0.2w,0.5h}{v5}
\fmfforce{0.9w,0.5h}{v55}
\fmf{double}{v1,o1}
\fmf{double}{v2,o2}
%\fmfv{label=$m$}{v55}
\fmfv{label=$\otimes$}{v33}
\fmfv{l=$\delta m^{(1l)}$,l.a=120,l.d=.15w}{v3}
\fmf{photon}{i,v5}
\fmf{double}{v2,v5}
\fmf{gluon}{v1,v2}
\fmf{double}{v1,v5}
\end{fmfgraph*} } 
%
%%%%%%%%%%%%%%%%%%%%%%%
%
\hspace{6mm}
\subfigure[]{\begin{fmfgraph*}(25,25)
\fmfleft{i}
\fmfright{o1,o2}
\fmfforce{0.8w,0.93h}{v2}
\fmfforce{0.8w,0.07h}{v1}
\fmfforce{0.5w,0.3h}{v3}
\fmfforce{0.5w,0.42h}{v33}
\fmfforce{0.2w,0.5h}{v5}
\fmfforce{0.9w,0.5h}{v55}
\fmf{double}{v1,o1}
\fmf{double}{v2,o2}
%\fmfv{label=$m$}{v55}
\fmfv{label=$\otimes$}{v33}
\fmfv{l=$\delta m^{(1l)}$,l.a=-120,l.d=.15w}{v3}
\fmf{photon}{i,v5}
\fmf{double}{v2,v5}
\fmf{gluon}{v1,v2}
\fmf{double}{v1,v5}
\end{fmfgraph*} }  
%
%%%%%%%%%%%%%%%%%%%%%%%
%
\hspace{6mm}
\subfigure[]{\begin{fmfgraph*}(25,25)
\fmfleft{i}
\fmfright{o1,o2}
\fmfforce{0.8w,0.93h}{v2}
\fmfforce{0.8w,0.07h}{v1}
\fmfforce{0.5w,0.7h}{v3}
\fmfforce{0.5w,0.58h}{v33}
\fmfforce{0.2w,0.5h}{v5}
\fmfforce{0.9w,0.5h}{v55}
\fmf{double}{v1,o1}
\fmf{double}{v2,o2}
%\fmfv{label=$m$}{v55}
\fmfv{label=$\otimes$}{v33}
\fmfv{l=$Z_2^{(1l)}$,l.a=120,l.d=.15w}{v3}
\fmf{photon}{i,v5}
\fmf{double}{v2,v5}
\fmf{gluon}{v1,v2}
\fmf{double}{v1,v5}
\end{fmfgraph*} } 
%
%%%%%%%%%%%%%%%%%%%%%%%
%
\hspace{6mm}
\subfigure[]{\begin{fmfgraph*}(25,25)
\fmfleft{i}
\fmfright{o1,o2}
\fmfforce{0.8w,0.93h}{v2}
\fmfforce{0.8w,0.07h}{v1}
\fmfforce{0.5w,0.3h}{v3}
\fmfforce{0.5w,0.42h}{v33}
\fmfforce{0.2w,0.5h}{v5}
\fmfforce{0.9w,0.5h}{v55}
\fmf{double}{v1,o1}
\fmf{double}{v2,o2}
%\fmfv{label=$m$}{v55}
\fmfv{label=$\otimes$}{v33}
\fmfv{l=$Z_2^{(1l)}$,l.a=-120,l.d=.15w}{v3}
\fmf{photon}{i,v5}
\fmf{double}{v2,v5}
\fmf{gluon}{v1,v2}
\fmf{double}{v1,v5}
\end{fmfgraph*} }  \\
%
%%%%%%%%%%%%%%%%%%%%%%%
%
\subfigure[]{\begin{fmfgraph*}(25,25)
\fmfleft{i}
\fmfright{o1,o2}
\fmfforce{0.8w,0.93h}{v2}
\fmfforce{0.8w,0.07h}{v1}
\fmfforce{0.8w,0.5h}{v3}
\fmfforce{0.7w,0.5h}{v33}
\fmfforce{0.2w,0.5h}{v5}
\fmfforce{0.9w,0.5h}{v55}
\fmf{double}{v1,o1}
\fmf{double}{v2,o2}
\fmfv{label=$\otimes$}{v33}
\fmfv{l=$Z_3^{(1l)}$,l.a=0,l.d=.15w}{v3}
\fmf{photon}{i,v5}
\fmf{double}{v2,v5}
\fmf{gluon}{v1,v2}
\fmf{double}{v1,v5}
\end{fmfgraph*} } 
%
%%%%%%%%%%%%%%%%%%%%%%%
%
\hspace{6mm}
\subfigure[]{\begin{fmfgraph*}(25,25)
\fmfleft{i}
\fmfright{o1,o2}
\fmfforce{0.8w,0.93h}{v2}
\fmfforce{0.8w,0.07h}{v1}
\fmfforce{0.5w,0.7h}{v3}
\fmfforce{0.7w,0.2h}{v33}
\fmfforce{0.2w,0.5h}{v5}
\fmfforce{0.9w,0.5h}{v55}
\fmf{double}{v1,o1}
\fmf{double}{v2,o2}
\fmfv{label=$\otimes$}{v33}
\fmfv{l=$Z_{1F}^{(1l)}$,l.a=-140,l.d=.05w}{v1}
\fmf{photon}{i,v5}
\fmf{double}{v2,v5}
\fmf{gluon}{v1,v2}
\fmf{double}{v1,v5}
\end{fmfgraph*} }
%
%%%%%%%%%%%%%%%%%%%%%%%
%
\hspace{6mm}
\subfigure[]{\begin{fmfgraph*}(25,25)
\fmfleft{i}
\fmfright{o1,o2}
\fmfforce{0.8w,0.93h}{v2}
\fmfforce{0.8w,0.07h}{v1}
\fmfforce{0.5w,0.7h}{v3}
\fmfforce{0.7w,0.8h}{v33}
\fmfforce{0.2w,0.5h}{v5}
\fmfforce{0.9w,0.5h}{v55}
\fmf{double}{v1,o1}
\fmf{double}{v2,o2}
\fmfv{label=$\otimes$}{v33}
\fmfv{l=$Z_{1F}^{(1l)}$,l.a=140,l.d=.05w}{v2}
\fmf{photon}{i,v5}
\fmf{double}{v2,v5}
\fmf{gluon}{v1,v2}
\fmf{double}{v1,v5}
\end{fmfgraph*} }
%
%%%%%%%%%%%%%%%%%%%%%%%
%
\hspace{6mm}
\subfigure[]{\begin{fmfgraph*}(25,25)
\fmfleft{i}
\fmfright{o1,o2}
\fmfforce{0.8w,0.93h}{v2}
\fmfforce{0.8w,0.07h}{v1}
\fmfforce{0.5w,0.7h}{v3}
\fmfforce{0.34w,0.5h}{v33}
\fmfforce{0.2w,0.5h}{v5}
\fmfforce{0.9w,0.5h}{v55}
\fmf{double}{v1,o1}
\fmf{double}{v2,o2}
\fmfv{label=$\otimes$}{v33}
\fmfv{l=$Z_2^{(1l)}$,l.a=110,l.d=.15w}{v5}
\fmf{photon}{i,v5}
\fmf{double}{v2,v5}
\fmf{gluon}{v1,v2}
\fmf{double}{v1,v5}
\end{fmfgraph*} } \\
%
%%%%%%%%%%%%%%%%%%%%%%%
%
\subfigure[]{
\begin{fmfgraph*}(25,25)
\fmfleft{i}
\fmfright{o1,o2}
\fmfforce{0.8w,0.93h}{v2}
\fmfforce{0.8w,0.07h}{v1}
\fmfforce{0.2w,0.5h}{v5}
\fmfforce{0.34w,0.5h}{v55}
\fmf{double}{v1,o1}
\fmf{double}{v2,o2}
\fmf{photon}{i,v5}
\fmfv{label=$\otimes$}{v55}
\fmfv{l=$Z_{2}^{(2l)}$,l.a=110,l.d=.12w}{v5}
\fmf{double,tension=.3}{v2,v5}
\fmf{double,tension=.3}{v1,v5}
\end{fmfgraph*}}
%
%%%%%%%%
%
\hspace{6mm}
\subfigure[]{
\begin{fmfgraph*}(25,25)
\fmfleft{i}
\fmfright{o1,o2}
\fmfforce{0.8w,0.93h}{v2}
\fmfforce{0.8w,0.07h}{v1}
\fmfforce{0.2w,0.5h}{v5}
\fmfforce{0.34w,0.5h}{v55}
\fmf{double}{v1,o1}
\fmf{double}{v2,o2}
\fmf{photon}{i,v5}
\fmfv{label=$\otimes$}{v55}
\fmfv{l=$Z_{2}^{(1l)}$,l.a=110,l.d=.12w}{v5}
\fmf{double,tension=.3}{v2,v5}
\fmf{double,tension=.3}{v1,v5}
\end{fmfgraph*}}
%
%%%%%%%%
%
\vspace*{5mm}
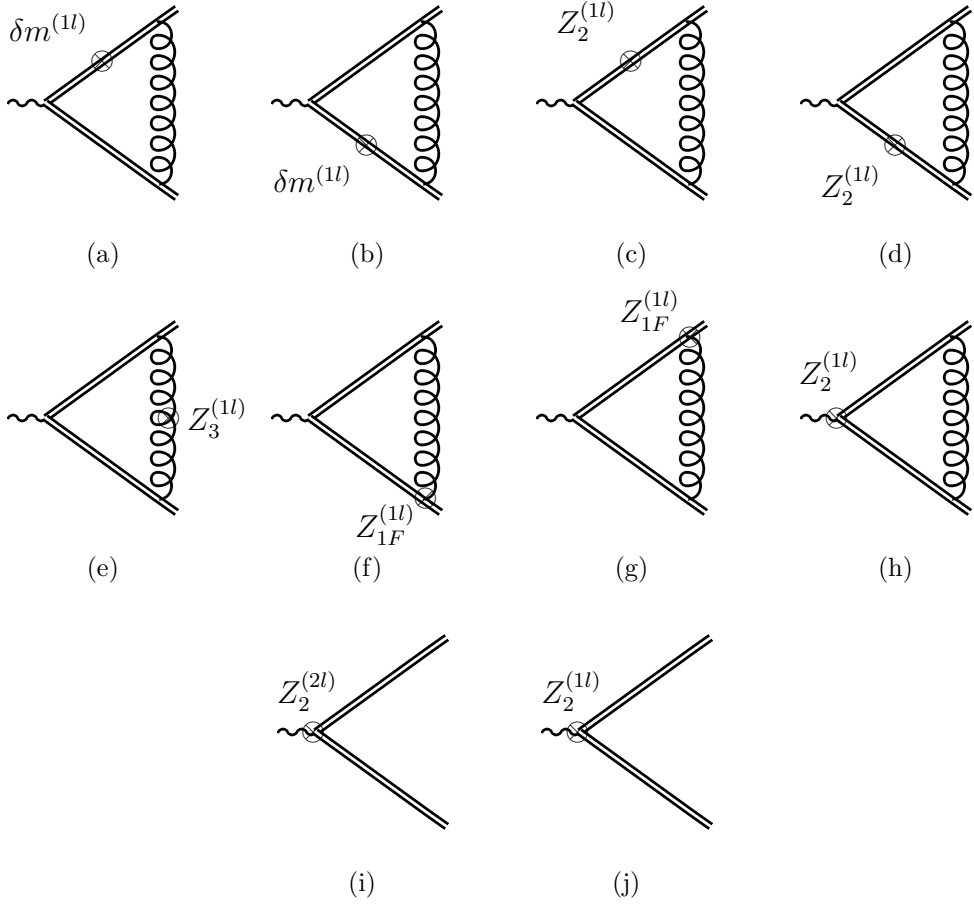
\caption{\label{fig3} Counterterm diagrams. For the renormalization of the
two-loop form factors we use diagrams (a)--(i). Diagram (j) is employed in the
renormalization of the one-loop form factors.}
\ec
\efig
%%%%%%%%%%%%%%%%%%%%%%%%%%%%%%%%%%%%%%%%%%%%%%%%%%%%%%%%%%%%%%%%%%%%%%
The subtraction of the one-loop subdivergences and the two-loop overall 
divergence in the two-loop graphs of Fig.~\ref{fig2} is performed in 
a hybrid scheme: we renormalize the heavy-quark wave function and 
mass in the {\it on-shell} (OS) renormalization scheme, while the 
coupling $\alpha_S$ and the gluon wave function are renormalized 
in the modified minimal subtraction ($\overline{\mathrm{MS}}$) scheme. 
The counterterm diagrams to add to the unsubtracted form factors 
given in  Eqs.~(\ref{1loopF1},\ref{1loopF2},\ref{2loopF1},\ref{2loopF2}), 
are shown in Fig.~\ref{fig3}.

\subsection{One-Loop Counterterms}

The renormalization of the one-loop form factors and the subtraction of the
one-loop subdivergences from the two-loop graphs shown in Fig.~\ref{fig2} require
the renormalization constants $Z_{g,\overline{\mathrm{MS}}}(\epsilon)$,
$Z_{3,\overline{\mathrm{MS}}}(\epsilon)$, 
$\delta m_{{\mathrm OS}}(\epsilon,m,\mu^2/m^2)$ and
$Z_{2, {\mathrm OS}}(\epsilon,\mu^2/m^2)$, at the one-loop level. Once 
$Z_{g,\overline{\mathrm{MS}}}(\epsilon)$,
$Z_{3,\overline{\mathrm{MS}}}(\epsilon)$ and 
$Z_{2, {\mathrm OS}}(\epsilon,\mu^2/m^2)$ are defined, the expression of
$Z_{1F}(\epsilon,\mu^2/m^2)$ is obtained using the Slavnov-Taylor identities.

The expressions for the coupling and gluon wave function renormalization
constants (the latter one in the Feynman gauge) are in the 
$\overline{\mathrm{MS}}$-scheme:
\bea
Z_{g,\overline{\mathrm{MS}}}^{(1l)}(\epsilon) & = & 
- \frac{\alpha_S}{2 \pi}  \, C(\epsilon) \,
  \frac{1}{4 \epsilon} \left( \frac{11}{3} C_{A} 
- \frac{4}{3} T_{R} (N_f +1) \right)
\label{c001} 
\ , \\
Z_{3,\overline{\mathrm{MS}}}^{(1l)}(\epsilon) & = & \hspace{3mm}
\frac{\alpha_S}{2 \pi}   \, C(\epsilon) \,
  \frac{1}{2 \epsilon} \left( \frac{5}{3} C_{A} 
- \frac{4}{3} T_{R} (N_f +1) \right)
\label{c002} 
\ .
\eea

For what concerns the renormalization of the heavy quark mass and wave function,
we need the following constants at the one-loop level, defined in the  
OS-scheme:
\bea
\delta m_{{\mathrm OS}}^{(1l)} \Bigl( \epsilon,m,\frac{\mu^2}{m^2} \Bigr) 
& = & - \, m \ \frac{\alpha_S}{2 \pi}  \, C(\epsilon) \, 
\left( \frac{\mu^{2}}{m^2} \right)^{\epsilon} \
\frac{C_{F}}{2} \frac{(3-2 \epsilon)}{\epsilon \, (1-2 \epsilon)} \ ,  
\label{c003} \\
Z_{2, {\mathrm OS}}^{(1l)} \Bigl(\epsilon,\frac{\mu^2}{m^2} \Bigr) 
& = & - \frac{\alpha_S}{2 \pi}  \, C(\epsilon) \, 
\left( \frac{\mu^{2}}{m^2} \right)^{\epsilon} \
\frac{C_{F}}{2} \frac{(3-2 \epsilon)}{\epsilon \, (1-2 \epsilon)}
\label{c004} 
 \ .
\eea

The constant $Z_{1F}^{(1l)}(\epsilon,\mu^2/m^2)$, needed for the 
renormalization of the $Q \bar{Q}$-gluon vertex, is obtained from 
a Slavnov-Taylor identity as follows:
\bea
Z_{1F}^{(1l)} \Bigl(\epsilon,\frac{\mu^2}{m^2} \Bigr) & = & 
Z_{g,\overline{\mathrm{MS}}}^{(1l)}(\epsilon) 
+ Z_{2,{\mathrm OS}}^{(1l)} \Bigl(\epsilon,\frac{\mu^2}{m^2} \Bigr) 
+ \frac{1}{2} Z_{3,\overline{\mathrm{MS}}}^{(1l)}(\epsilon) \\
& = & - \frac{\alpha_S}{2 \pi}  \, C(\epsilon) \, 
\frac{1}{2 \epsilon} \Biggl[ C_{A} +
\left( \frac{\mu^{2}}{m^2} \right)^{\epsilon} 
C_{F} \frac{(3-2 \epsilon)}{(1-2 \epsilon)} 
\Biggr]
\, .
\eea

The renormalization of the UV divergences of the one-loop form factors,
Eqs.~(\ref{1loopF1},\ref{1loopF2}), is straightforward. It is sufficient to 
add the counterterm of Fig.~\ref{fig3}~(j), defined as:
\be
\parbox{20mm}{\begin{fmfgraph*}(15,15)
\fmfleft{i}
\fmfright{o1,o2}
\fmfforce{0.8w,0.93h}{v2}
\fmfforce{0.8w,0.07h}{v1}
\fmfforce{0.2w,0.5h}{v5}
\fmfforce{0.45w,0.5h}{v55}
\fmf{double}{v1,o1}
\fmf{double}{v2,o2}
\fmf{photon}{i,v5}
\fmfv{label=$\otimes$}{v55}
\fmfv{l=$Z_{2}^{(1l)}$,l.a=110,l.d=.12w}{v5}
\fmf{double,tension=.3}{v2,v5}
\fmf{double,tension=.3}{v1,v5}
\end{fmfgraph*} }   \stackrel{def}{=}  
 \ Z_{2, {\mathrm OS}}^{(1l)} \Bigl(\epsilon,\frac{\mu^2}{m^2} \Bigr) \ 
 \times 
\parbox{20mm}{\begin{fmfgraph*}(15,15) 
\fmfleft{i}
\fmfright{o1,o2}
\fmfforce{0.8w,0.93h}{v2}
\fmfforce{0.8w,0.07h}{v1}
\fmfforce{0.2w,0.5h}{v5}
\fmf{double}{v1,o1}
\fmf{double}{v2,o2}
\fmf{photon}{i,v5}
\fmf{double,tension=.3}{v2,v5}
\fmf{double,tension=.3}{v1,v5}
\end{fmfgraph*} }  \, ,
\label{CTtreeL}
\ee
where the constant $Z_{2, {\mathrm OS}}^{(1l)}(\epsilon,\mu^2/m^2)$ is given by 
Eq.~(\ref{c004}), to the diagram of Fig.~\ref{fig1}~(b). The corresponding 
UV-renormalized form factors are given in Section \ref{FFrenorm}. Let us note
that ${\mathcal F}_{2}^{(1l)}(\epsilon,s)$ in Eq.~(\ref{1loopF2}) is IR and UV 
finite. The counterterm diagram defined in Eq.~(\ref{CTtreeL}) is, in fact,
proportional to $\gamma^{\mu}$ and affects only 
${\mathcal F}_{1}^{(1l)}(\epsilon,s)$.

For the subtraction of the one-loop subdivergences from the two-loop diagrams
of Fig.~\ref{fig2}, we have to define as well the counterterm diagrams 
shown in Fig.~\ref{fig3}~(a)--(h).

The first two involve the constant 
$\delta m_{{\mathrm OS}}^{(1l)}(\epsilon,m,\mu^2/m^2)$:
\\

$\bullet$ graph (a) in Fig.~(\ref{fig3}):

\be
\parbox{20mm}{\begin{fmfgraph*}(15,15)
\fmfleft{i}
\fmfright{o1,o2}
\fmfforce{0.8w,0.93h}{v2}
\fmfforce{0.8w,0.07h}{v1}
\fmfforce{0.5w,0.7h}{v3}
\fmfforce{0.5w,0.53h}{v33}
\fmfforce{0.2w,0.5h}{v5}
\fmfforce{0.9w,0.5h}{v55}
\fmf{double}{v1,o1}
\fmf{double}{v2,o2}
%\fmfv{label=$m$}{v55}
\fmfv{label=$\otimes$}{v33}
\fmfv{l=$\delta m^{(1l)}$,l.a=120,l.d=.15w}{v3}
\fmf{photon}{i,v5}
\fmf{double}{v2,v5}
\fmf{gluon}{v1,v2}
\fmf{double}{v1,v5}
\end{fmfgraph*} }   \stackrel{def}{=} -
\ \frac{1}{m} \ \delta m_{{\mathrm OS}}^{(1l)} 
\Bigl(\epsilon,m,\frac{\mu^2}{m^2} \Bigr) \ \times \left( m \, \, 
\parbox{20mm}{\begin{fmfgraph*}(15,15)
\fmfleft{i}
\fmfright{o1,o2}
\fmfforce{0.8w,0.93h}{v2}
\fmfforce{0.8w,0.07h}{v1}
\fmfforce{0.5w,0.7h}{v3}
\fmfforce{0.5w,0.53h}{v33}
\fmfforce{0.2w,0.5h}{v5}
\fmfforce{0.9w,0.5h}{v55}
\fmf{double}{v1,o1}
\fmf{double}{v2,o2}
%\fmfv{label=$m$}{v55}
\fmfv{label=$\otimes$}{v33}
%\fmfv{l=$\delta m^{(1l)}$,l.a=120,l.d=.15w}{v3}
\fmf{photon}{i,v5}
\fmf{double}{v2,v5}
\fmf{gluon}{v1,v2}
\fmf{double}{v1,v5}
\end{fmfgraph*} } \right) ;
\label{c1} 
\ee

$\bullet$ graph (b) in Fig.~(\ref{fig3}):

\be
\parbox{20mm}{\begin{fmfgraph*}(15,15)
\fmfleft{i}
\fmfright{o1,o2}
\fmfforce{0.8w,0.93h}{v2}
\fmfforce{0.8w,0.07h}{v1}
\fmfforce{0.5w,0.3h}{v3}
\fmfforce{0.5w,0.47h}{v33}
\fmfforce{0.2w,0.5h}{v5}
\fmfforce{0.9w,0.5h}{v55}
\fmf{double}{v1,o1}
\fmf{double}{v2,o2}
%\fmfv{label=$m$}{v55}
\fmfv{label=$\otimes$}{v33}
\fmfv{l=$\delta m^{(1l)}$,l.a=-120,l.d=.15w}{v3}
\fmf{photon}{i,v5}
\fmf{double}{v2,v5}
\fmf{gluon}{v1,v2}
\fmf{double}{v1,v5}
\end{fmfgraph*} }   \stackrel{def}{=} -
\ \frac{1}{m} \ \delta m_{{\mathrm OS}}^{(1l)} 
\Bigl(\epsilon,m,\frac{\mu^2}{m^2} \Bigr) \ \times \left( m \, \,
\parbox{20mm}{\begin{fmfgraph*}(15,15)
\fmfleft{i}
\fmfright{o1,o2}
\fmfforce{0.8w,0.93h}{v2}
\fmfforce{0.8w,0.07h}{v1}
\fmfforce{0.5w,0.3h}{v3}
\fmfforce{0.5w,0.47h}{v33}
\fmfforce{0.2w,0.5h}{v5}
\fmfforce{0.9w,0.5h}{v55}
\fmf{double}{v1,o1}
\fmf{double}{v2,o2}
%\fmfv{label=$m$}{v55}
\fmfv{label=$\otimes$}{v33}
%\fmfv{l=$\delta m^{(1l)}$,l.a=-120,l.d=.15w}{v3}
\fmf{photon}{i,v5}
\fmf{double}{v2,v5}
\fmf{gluon}{v1,v2}
\fmf{double}{v1,v5}
\end{fmfgraph*} } \right) ;
\label{c2} 
\ee

\vspace*{3mm}

The one-loop diagram multiplying 
$\delta m_{\mathrm OS}^{(1l)}(\epsilon,m,\mu^2/m^2)/m$ 
in Eq.~(\ref{c1}) is defined as follows:
\bea
m \, \, 
\parbox{20mm}{\begin{fmfgraph*}(15,15)
\fmfleft{i}
\fmfright{o1,o2}
\fmfforce{0.8w,0.93h}{v2}
\fmfforce{0.8w,0.07h}{v1}
\fmfforce{0.5w,0.7h}{v3}
\fmfforce{0.5w,0.53h}{v33}
\fmfforce{0.2w,0.5h}{v5}
\fmfforce{0.9w,0.5h}{v55}
\fmf{double}{v1,o1}
\fmf{double}{v2,o2}
%\fmfv{label=$m$}{v55}
\fmfv{label=$\otimes$}{v33}
%\fmfv{l=$\delta m^{(1l)}$,l.a=120,l.d=.15w}{v3}
\fmf{photon}{i,v5}
\fmf{double}{v2,v5}
\fmf{gluon}{v1,v2}
\fmf{double}{v1,v5}
\end{fmfgraph*} }  \! \! \! \! \! \! & = & 
m \, C_{F} \, \frac{\alpha_S}{2 \pi} 
C(\epsilon) \, \left( \frac{\mu^{2}}{m^2} \right)^{\epsilon} \nn\\
\! \! \! \! \!&  & \hspace*{5mm} \times
\int {\mathfrak{D}}^Dk \ 
\frac{{\mathcal U}^{\mu}}
{\bigl[(p_1+k)^2-m^2 \bigr]^2 \bigl[(p_2-k)^2-m^2 \bigr] k^2  } , 
\eea
where the measure ${\mathfrak{D}}^Dk$ is such that:
\be
\int{\mathfrak D}^Dk = \frac{1}{C(\epsilon)} 
\left( \frac{m^2}{\mu^{2}} \right)^{\epsilon} 
\int \frac{d^D k}{(2 \pi)^{2(1-\epsilon)}} \, ,
\ee
and where:
\be
\! \! \! \! \! \! \! \! \! \! {\mathcal U}^{\mu} =  v_Q
\, \gamma_{\sigma} 
[\not{\! p_{1}} + \! \! \not{\! k} \! +\! m] 
[\not{\! p_{1}} + \! \! \not{\! k} \! +\! m] 
\gamma^{\mu} 
[\not{\! k} - \! \! \not{\! p_{2}} \! +\! m] 
\gamma_{\sigma} \ .
\ee

The corresponding form factors are:
\be
F_{i,(\otimes)}^{(1l)} \Bigl( \epsilon,s, \frac{\mu^2}{m^2} \Bigr) = 
C(\epsilon) \, \left( \frac{\mu^{2}}{m^2} \right)^{\epsilon} 
{\mathcal F}_{i}^{(\otimes)}(\epsilon,s) 
\qquad \mbox{with} \; \, i=1,2 \, ,
\ee
where:
\bea
\hspace*{-5mm} {\mathcal F}_1^{(\otimes)}(\epsilon,s) & = &  
- \frac{1}{\epsilon} 
\Biggl\{ C_F \Biggl[
            1
          - \frac{2}{1+x}
          - \frac{2}{(1+x)^2}
 - \biggl( 
            \frac{1}{1-x}
          - \frac{2}{1+x} \nn\\
\hspace*{-5mm} & & \hspace*{18mm} 
          + \frac{3}{(1+x)^2}
          - \frac{2}{(1+x)^3}
          \biggr) H(0;x)
\Biggr] \Biggr\}   \nn\\
\hspace*{-5mm} & & 
+ \, C_F \Biggl[
     2 
   - \biggl( \frac{1}{1+x}
   - \frac{3}{(1+x)^2}
   + \frac{2}{(1+x)^3} \biggr) H(0;x) \nn\\
\hspace*{-5mm} & &  \hspace*{8mm}
      - \biggl( 
            \frac{1}{1-x} \! 
          -  \! \frac{2}{1 \! + \! x}  \! 
          +  \! \frac{3}{(1 \! + \! x)^2} \! 
          -  \! \frac{2}{(1 \! + \! x)^3} \biggr)  \! 
   \bigl[ \zeta(2) \! 
   - H(0;x)  \nn\\
\hspace*{-5mm} & &  \hspace*{8mm}
   - H(0,0;x)
   + 2 H(-1,0;x) \bigr] \Biggr] \nn\\
\hspace*{-5mm} & & - 
\epsilon \, \Biggl\{ C_F \Biggl[
     4
          - \frac{12}{1+x}
          + \frac{12}{(1+x)^2}   \nn\\
\hspace*{-5mm} & & \hspace*{12mm} 
       + \!  \biggl[ \biggl(
            \frac{1}{1-x} \! 
          -  \! \frac{2}{1 \! + \! x}  \! 
          +  \! \frac{3}{(1 \! + \! x)^2} \! 
          -  \! \frac{2}{(1 \! + \! x)^3} \biggr) \zeta(2)   \nn\\
\hspace*{-5mm} & & \hspace*{18mm}  
       -  \frac{2}{1-x}
          + \frac{12}{1+x} 
          - \frac{18}{(1+x)^2}  
          + \frac{4}{(1+x)^3} \biggr] H(0;x)  \nn\\
\hspace*{-5mm} & & \hspace*{12mm}  
       + \!  \biggl[ 
            \frac{2}{1-x} \! 
          -  \! \frac{4}{1 \! + \! x}  \! 
          +  \! \frac{6}{(1 \! + \! x)^2} \! 
          -  \! \frac{4}{(1 \! + \! x)^3} \biggr] \bigl[ \zeta(3)  \nn\\
\hspace*{-5mm} & & \hspace*{18mm}  
          - H(-1;x) \bigr] \nn\\
\hspace*{-5mm} & & \hspace*{12mm}  
       +  \biggl[ 
            \frac{1}{1-x} \! 
          -  \! \frac{3}{1 \! + \! x}  \! 
          +  \! \frac{6}{(1 \! + \! x)^2} \! 
          -  \! \frac{4}{(1 \! + \! x)^3} \biggr]  \bigl[ \zeta(2) \nn\\
\hspace*{-5mm} & & \hspace*{18mm}  
          - H(0,0;x)
          + 2 H(-1,0;x) \bigr] \nn\\
\hspace*{-5mm} & & \hspace*{12mm}  
       -  \biggl[ 
            \frac{1}{1-x} \! 
          -  \! \frac{2}{1+x}  \! 
          +  \! \frac{3}{(1+x)^2}  \! 
          -  \! \frac{2}{(1+x)^3} \biggr] \times  \nn\\
\hspace*{-5mm} & & \hspace*{18mm} 
        \times \bigl[ 
     H(0,0,0;x)
   + 4 H(-1,-1,0;x) \nn\\
\hspace*{-5mm} & & \hspace*{18mm}  
   - 2 H(-1,0,0;x)
   - 2 H(0,-1,0;x) \bigr] \Biggr]
\Biggr\} + {\mathcal O} \left( \epsilon^2 \right) \, , 
\label{FFDenQuadrato1} \\
\hspace*{-5mm} {\mathcal F}_2^{(\otimes)}(\epsilon,s) & = &  
- \frac{1}{\epsilon} 
\Biggl\{ C_F \Biggl[
            \frac{2}{(1+x)} \biggl(
     1
          - \frac{1}{1+x} \biggr)
 + \frac{1}{2} \biggl( 
            \frac{1}{1-x}
          - \frac{3}{1+x} \nn\\
\hspace*{-5mm} & & \hspace*{18mm} 
          + \frac{6}{(1+x)^2}
          - \frac{4}{(1+x)^3}
          \biggr) H(0;x)
\Biggr] \Biggr\}   \nn\\
\hspace*{-5mm} & & 
   + C_F \Biggl[ 
      - \frac{4}{(1+x)} \biggl(
     1
          - \frac{1}{1+x} \biggr)\nn\\
\hspace*{-5mm} & &  
       + \frac{4}{1+x} \biggl( 
            1
   - \frac{3}{1+x}
   + \frac{2}{(1+x)^2} \biggr) H(0;x) \nn\\
\hspace*{-5mm} & &  
      + \frac{1}{2} \biggl( 
            \frac{1}{1-x}
          - \frac{3}{1+x} 
          + \frac{6}{(1+x)^2}
          - \frac{4}{(1+x)^3} \biggr)
   \bigl[ \zeta(2) \nn\\
\hspace*{-5mm} & & 
   - H(0,0;x)
   + 2 H(-1,0;x) \bigr] \Biggr] \nn\\
\hspace*{-5mm} &  & -
\epsilon \, \Biggl\{ C_F \Biggl[ \frac{4}{(1+x)} \biggl( 1 - \frac{1}{1+x} 
\biggr) \nn\\
\hspace*{-5mm} & & \hspace*{12mm}
       + \frac{2}{(1+x)} \biggl(
             1 
   - \frac{3}{1+x}
   + \frac{2}{1+x} \biggr)
   \bigl[ 2 \zeta(2) \nn\\
\hspace*{-5mm} & & \hspace*{18mm}
          - 3 H(0;x)
   - 2 H(0,0;x) 
   + 4 H(-1,0;x) \bigr] \nn\\
\hspace*{-5mm} & & \hspace*{12mm}
       - \frac{1}{2} \biggl( 
            \frac{1}{1 \! - \! x} \! 
          -  \! \frac{3}{1 \! + \! x}  \! 
          +  \! \frac{6}{(1 \! + \! x)^2} \! 
          -  \! \frac{4}{(1 \! + \! x)^3} \biggr)
   \bigl[ 2 \zeta(3) \nn\\
\hspace*{-5mm} & & \hspace*{18mm}
          + \zeta(2) ( H(0;x)
   - 2 H(-1;x) ) 
   - H(0,0,0;x)  \nn\\
\hspace*{-5mm} & & \hspace*{18mm}
         - 4 H(-1,-1,0;x)
  + 2 H(-1,0,0;x) \nn\\
\hspace*{-5mm} & & \hspace*{18mm}
  + 2 H(0,-1,0;x)\bigr]  \Biggr]
\Biggr\} + {\mathcal O} \left( \epsilon^2 \right) \, . 
\label{FFDenQuadrato2} 
\eea

The one-loop diagram appearing in Eq.~(\ref{c2}) has exactly the same form
factors given in Eqs.~(\ref{FFDenQuadrato1},\ref{FFDenQuadrato2}). Therefore, 
in the following we will refer to both diagrams using the picture of the first 
one only.

The counterterm diagrams which involve 
$Z_{2, {\mathrm OS}}(\epsilon,\mu^2/m^2)$, $Z_{1F}(\epsilon,\mu^2/m^2)$, and
$Z_{3,\overline{\mathrm{MS}}}(\epsilon)$  are defined 
as the product of the renormalization constants times the one-loop vertex
diagram: \\

\vspace*{2mm}

$\bullet$ graph (c) in Fig.~(\ref{fig3}):

\be
\parbox{20mm}{\begin{fmfgraph*}(15,15)
\fmfleft{i}
\fmfright{o1,o2}
\fmfforce{0.8w,0.93h}{v2}
\fmfforce{0.8w,0.07h}{v1}
\fmfforce{0.5w,0.7h}{v3}
\fmfforce{0.5w,0.53h}{v33}
\fmfforce{0.2w,0.5h}{v5}
\fmfforce{0.9w,0.5h}{v55}
\fmf{double}{v1,o1}
\fmf{double}{v2,o2}
\fmfv{label=$\otimes$}{v33}
\fmfv{l=$Z_2^{(1l)}$,l.a=120,l.d=.15w}{v3}
\fmf{photon}{i,v5}
\fmf{double}{v2,v5}
\fmf{gluon}{v1,v2}
\fmf{double}{v1,v5}
\end{fmfgraph*} }  \stackrel{def}{=} -
\ Z_{2, {\mathrm OS}}^{(1l)} \Bigl( \epsilon, \frac{\mu^2}{m^2} \Bigr) 
\ \times 
\parbox{20mm}{\begin{fmfgraph*}(15,15) 
\fmfleft{i}
\fmfright{o1,o2}
\fmfforce{0.8w,0.93h}{v2}
\fmfforce{0.8w,0.07h}{v1}
\fmfforce{0.5w,0.7h}{v3}
\fmfforce{0.5w,0.53h}{v33}
\fmfforce{0.2w,0.5h}{v5}
\fmfforce{0.9w,0.5h}{v55}
\fmf{double}{v1,o1}
\fmf{double}{v2,o2}
\fmf{photon}{i,v5}
\fmf{double}{v2,v5}
\fmf{gluon}{v1,v2}
\fmf{double}{v1,v5}
\end{fmfgraph*} }  \, ;
\label{c3} 
\ee

$\bullet$ graph (d) in Fig.~(\ref{fig3}):

\be
\parbox{20mm}{\begin{fmfgraph*}(15,15)
\fmfleft{i}
\fmfright{o1,o2}
\fmfforce{0.8w,0.93h}{v2}
\fmfforce{0.8w,0.07h}{v1}
\fmfforce{0.5w,0.3h}{v3}
\fmfforce{0.5w,0.47h}{v33}
\fmfforce{0.2w,0.5h}{v5}
\fmfforce{0.9w,0.5h}{v55}
\fmf{double}{v1,o1}
\fmf{double}{v2,o2}
\fmfv{label=$\otimes$}{v33}
\fmfv{l=$Z_2^{(1l)}$,l.a=-120,l.d=.15w}{v3}
\fmf{photon}{i,v5}
\fmf{double}{v2,v5}
\fmf{gluon}{v1,v2}
\fmf{double}{v1,v5}
\end{fmfgraph*} }  \stackrel{def}{=} -
\ Z_{2, {\mathrm OS}}^{(1l)} \Bigl( \epsilon, \frac{\mu^2}{m^2} \Bigr) 
\ \times 
\parbox{20mm}{\begin{fmfgraph*}(15,15) 
\fmfleft{i}
\fmfright{o1,o2}
\fmfforce{0.8w,0.93h}{v2}
\fmfforce{0.8w,0.07h}{v1}
\fmfforce{0.5w,0.3h}{v3}
\fmfforce{0.5w,0.47h}{v33}
\fmfforce{0.2w,0.5h}{v5}
\fmfforce{0.9w,0.5h}{v55}
\fmf{double}{v1,o1}
\fmf{double}{v2,o2}
\fmf{photon}{i,v5}
\fmf{double}{v2,v5}
\fmf{gluon}{v1,v2}
\fmf{double}{v1,v5}
\end{fmfgraph*} }  \, ;
\label{c4} 
\ee

$\bullet$ graph (e) in Fig.~(\ref{fig3}):

\be
\parbox{20mm}{\begin{fmfgraph*}(15,15)
\fmfleft{i}
\fmfright{o1,o2}
\fmfforce{0.8w,0.93h}{v2}
\fmfforce{0.8w,0.07h}{v1}
\fmfforce{0.9w,0.5h}{v3}
\fmfforce{0.7w,0.5h}{v33}
\fmfforce{0.2w,0.5h}{v5}
\fmfforce{0.9w,0.5h}{v55}
\fmf{double}{v1,o1}
\fmf{double}{v2,o2}
\fmfv{label=$\otimes$}{v33}
\fmfv{l=$Z_3^{(1l)}$,l.a=0,l.d=.15w}{v3}
\fmf{photon}{i,v5}
\fmf{double}{v2,v5}
\fmf{gluon}{v1,v2}
\fmf{double}{v1,v5}
\end{fmfgraph*} }  \quad \stackrel{def}{=} -
 \ Z_{3,\overline{\mathrm{MS}}}^{(1l)}(\epsilon) \ \times 
\parbox{20mm}{\begin{fmfgraph*}(15,15) 
\fmfleft{i}
\fmfright{o1,o2}
\fmfforce{0.8w,0.93h}{v2}
\fmfforce{0.8w,0.07h}{v1}
\fmfforce{0.8w,0.5h}{v3}
\fmfforce{0.7w,0.5h}{v33}
\fmfforce{0.2w,0.5h}{v5}
\fmfforce{0.9w,0.5h}{v55}
\fmf{double}{v1,o1}
\fmf{double}{v2,o2}
\fmf{photon}{i,v5}
\fmf{double}{v2,v5}
\fmf{gluon}{v1,v2}
\fmf{double}{v1,v5}
\end{fmfgraph*} }  \, ;
\label{c5} 
\ee

$\bullet$ graph (f) in Fig.~(\ref{fig3}):

\be
\parbox{20mm}{\begin{fmfgraph*}(15,15)
\fmfleft{i}
\fmfright{o1,o2}
\fmfforce{0.8w,0.93h}{v2}
\fmfforce{0.8w,0.07h}{v1}
\fmfforce{0.5w,0.7h}{v3}
\fmfforce{0.64w,0.27h}{v33}
\fmfforce{0.2w,0.5h}{v5}
\fmfforce{0.9w,0.5h}{v55}
\fmf{double}{v1,o1}
\fmf{double}{v2,o2}
\fmfv{label=$\otimes$}{v33}
\fmfv{l=$Z_{1F}^{(1l)}$,l.a=-140,l.d=.05w}{v1}
\fmf{photon}{i,v5}
\fmf{double}{v2,v5}
\fmf{gluon}{v1,v2}
\fmf{double}{v1,v5}
\end{fmfgraph*} }  \stackrel{def}{=} 
 \ Z_{1F}^{(1l)} \Bigl( \epsilon, \frac{\mu^2}{m^2} \Bigr) \ \times 
\parbox{20mm}{\begin{fmfgraph*}(15,15) 
\fmfleft{i}
\fmfright{o1,o2}
\fmfforce{0.8w,0.93h}{v2}
\fmfforce{0.8w,0.07h}{v1}
\fmfforce{0.5w,0.7h}{v3}
\fmfforce{0.7w,0.2h}{v33}
\fmfforce{0.2w,0.5h}{v5}
\fmfforce{0.9w,0.5h}{v55}
\fmf{double}{v1,o1}
\fmf{double}{v2,o2}
\fmf{photon}{i,v5}
\fmf{double}{v2,v5}
\fmf{gluon}{v1,v2}
\fmf{double}{v1,v5}
\end{fmfgraph*} }  \, ;
\label{c6} 
\ee

\vspace*{5mm}

$\bullet$ graph (g) in Fig.~(\ref{fig3}):

\vspace*{3mm}

\be
\parbox{20mm}{\begin{fmfgraph*}(15,15)
\fmfleft{i}
\fmfright{o1,o2}
\fmfforce{0.8w,0.93h}{v2}
\fmfforce{0.8w,0.07h}{v1}
\fmfforce{0.5w,0.7h}{v3}
\fmfforce{0.64w,0.73h}{v33}
\fmfforce{0.2w,0.5h}{v5}
\fmfforce{0.9w,0.5h}{v55}
\fmf{double}{v1,o1}
\fmf{double}{v2,o2}
\fmfv{label=$\otimes$}{v33}
\fmfv{l=$Z_{1F}^{(1l)}$,l.a=140,l.d=.05w}{v2}
\fmf{photon}{i,v5}
\fmf{double}{v2,v5}
\fmf{gluon}{v1,v2}
\fmf{double}{v1,v5}
\end{fmfgraph*} }  \stackrel{def}{=} 
 \ Z_{1F}^{(1l)} \Bigl( \epsilon, \frac{\mu^2}{m^2} \Bigr) \ \times 
\parbox{20mm}{\begin{fmfgraph*}(15,15) 
\fmfleft{i}
\fmfright{o1,o2}
\fmfforce{0.8w,0.93h}{v2}
\fmfforce{0.8w,0.07h}{v1}
\fmfforce{0.5w,0.7h}{v3}
\fmfforce{0.7w,0.8h}{v33}
\fmfforce{0.2w,0.5h}{v5}
\fmfforce{0.9w,0.5h}{v55}
\fmf{double}{v1,o1}
\fmf{double}{v2,o2}
\fmf{photon}{i,v5}
\fmf{double}{v2,v5}
\fmf{gluon}{v1,v2}
\fmf{double}{v1,v5}
\end{fmfgraph*} }  \, ;
\label{c7} 
\ee

$\bullet$ graph (h) in Fig.~(\ref{fig3}):

\be
\parbox{20mm}{\begin{fmfgraph*}(15,15)
\fmfleft{i}
\fmfright{o1,o2}
\fmfforce{0.8w,0.93h}{v2}
\fmfforce{0.8w,0.07h}{v1}
\fmfforce{0.5w,0.7h}{v3}
\fmfforce{0.45w,0.5h}{v33}
\fmfforce{0.2w,0.5h}{v5}
\fmfforce{0.9w,0.5h}{v55}
\fmf{double}{v1,o1}
\fmf{double}{v2,o2}
\fmfv{label=$\otimes$}{v33}
\fmfv{l=$Z_2^{(1l)}$,l.a=110,l.d=.15w}{v5}
\fmf{photon}{i,v5}
\fmf{double}{v2,v5}
\fmf{gluon}{v1,v2}
\fmf{double}{v1,v5}
\end{fmfgraph*} }  \stackrel{def}{=} 
\ Z_{2, {\mathrm OS}}^{(1l)} \Bigl( \epsilon, \frac{\mu^2}{m^2} \Bigr) 
\ \times 
\parbox{20mm}{\begin{fmfgraph*}(15,15) 
\fmfleft{i}
\fmfright{o1,o2}
\fmfforce{0.8w,0.93h}{v2}
\fmfforce{0.8w,0.07h}{v1}
\fmfforce{0.5w,0.7h}{v3}
\fmfforce{0.34w,0.5h}{v33}
\fmfforce{0.2w,0.5h}{v5}
\fmfforce{0.9w,0.5h}{v55}
\fmf{double}{v1,o1}
\fmf{double}{v2,o2}
\fmf{photon}{i,v5}
\fmf{double}{v2,v5}
\fmf{gluon}{v1,v2}
\fmf{double}{v1,v5}
\end{fmfgraph*} }  \, ;
\label{c8} 
\ee

The one-loop vertex diagram, appearing in Eqs.~(\ref{c3}--\ref{c8}) 
is defined as:
\be
\parbox{20mm}{\begin{fmfgraph*}(15,15)
\fmfleft{i}
\fmfright{o1,o2}
\fmfforce{0.8w,0.93h}{v2}
\fmfforce{0.8w,0.07h}{v1}
\fmfforce{0.5w,0.7h}{v3}
\fmfforce{0.5w,0.53h}{v33}
\fmfforce{0.2w,0.5h}{v5}
\fmfforce{0.9w,0.5h}{v55}
\fmf{double}{v1,o1}
\fmf{double}{v2,o2}
\fmf{photon}{i,v5}
\fmf{double}{v2,v5}
\fmf{gluon}{v1,v2}
\fmf{double}{v1,v5}
\end{fmfgraph*} }  \! \! \! \! \! \! = C_{F} \, \frac{\alpha_S}{2 \pi} 
C(\epsilon) \, \left( \frac{\mu^{2}}{m^2} \right)^{\epsilon} 
\int {\mathfrak{D}}^Dk \ 
\frac{{\mathcal V}^{\mu}}
{\bigl[(p_1+k)^2-m^2 \bigr] \bigl[(p_2-k)^2-m^2 \bigr] k^2  } , 
\ee
where
\be
\! \! \! \! \! \! \! \! \! \! {\mathcal V}^{\mu} =  v_Q
\, \gamma_{\sigma} 
[\not{\! p_{1}} + \! \! \not{\! k} \! +\! m] 
\gamma^{\mu} 
[\not{\! k} - \! \! \not{\! p_{2}} \! +\! m] 
\gamma_{\sigma} \ ,
\ee

and the corresponding form factors are given in 
Eqs.~(\ref{1loopF1},\ref{1loopF2}).

\subsection{Two-Loop Counterterm}

In order to complete the UV-renormalization of the form factor $F_1$, 
we have also to subtract its value at $s = 0$, or, which is the same, 
to add the counterterm diagram shown in Fig.~\ref{fig3}~(i). We need,
therefore, the constant $Z_{2, {\mathrm OS}}(\epsilon,\mu^2/m^2)$ at 
the two-loop level, that was computed in \cite{Broad,Mel}. We use the 
result of \cite{Mel} and we express it in terms of the renormalized 
$\overline{\mathrm{MS}}$ coupling $\alpha_S$, finding:
\bea
Z_{2, {\mathrm OS}}^{(2l)} \Bigl( \epsilon, \frac{\mu^2}{m^2} \Bigr) 
\! \! \! \! & = & \! \! \! \! \left( \frac{\alpha_S}{2 \pi}
\right)^2 C^2(\epsilon) \, \Biggl\{
\left( \frac{\mu^2}{m^2} \right)^{2 \epsilon} \Biggl[
C_{F}^2 \Biggl( 
     \frac{9}{8 \epsilon^2} \!
   + \!\frac{51}{16 \epsilon} \!
   + \!\frac{433}{32} 
   - 6 \zeta(3) \!
   + \!24 \zeta(2) \log{2}  \nn\\
\! \! \! \! & & \! \! \! \!
   - \frac{39}{2} \zeta(2) \Biggr)
   + C_{F} C_{A} \Biggl( 
   - \frac{11}{8 \epsilon^2} 
   - \frac{101}{16 \epsilon} 
   - \frac{803}{32} 
   + 3 \zeta(3) 
   - 12 \zeta(2) \log{2}  \nn\\
\! \! \! \! & & \! \! \! \!
   + \frac{15}{2} \zeta(2) \Biggr)
   + C_{F} T_{R} N_f \left( \frac{1}{2 \epsilon^2} \!
   + \!\frac{9}{4 \epsilon} \!
   + \!\frac{59}{8} \!
   + \!2 \zeta(2) \right)\!
   + C_{F} T_{R} \Biggl( 
     \frac{1}{\epsilon^2} \!
   + \!\frac{19}{12 \epsilon}  \nn\\
\! \! \! \! & & \! \! \! \!
   + \frac{1139}{72} 
   - 8 \zeta(2) \Biggr) \Biggr]
   + \left( \frac{\mu^2}{m^2} \right)^{\epsilon} \Biggl[
     C_{F} C_{A} \Biggl( 
     \frac{33}{12 \epsilon^2} 
   + \frac{11}{3 \epsilon}
   + \frac{22}{3} \Biggr)  \nn\\
\! \! \! \! & & \! \! \! \!
   - C_{F} T_{R} (N_f+1) \Biggl( 
     \frac{1}{\epsilon^2} 
   + \frac{4}{3 \epsilon}
   + \frac{8}{3} \Biggr)
\Biggr] \Biggr\}
\, .
\label{c006} 
\eea

The counterterm diagram is defined, therefore, as follows:
\be
\parbox{20mm}{\begin{fmfgraph*}(15,15)
\fmfleft{i}
\fmfright{o1,o2}
\fmfforce{0.8w,0.93h}{v2}
\fmfforce{0.8w,0.07h}{v1}
\fmfforce{0.2w,0.5h}{v5}
\fmfforce{0.45w,0.5h}{v55}
\fmf{double}{v1,o1}
\fmf{double}{v2,o2}
\fmf{photon}{i,v5}
\fmfv{label=$\otimes$}{v55}
\fmfv{l=$Z_{2}^{(2l)}$,l.a=110,l.d=.12w}{v5}
\fmf{double,tension=.3}{v2,v5}
\fmf{double,tension=.3}{v1,v5}
\end{fmfgraph*} }   \stackrel{def}{=} 
\ Z_{2, {\mathrm OS}}^{(2l)} \Bigl( \epsilon, \frac{\mu^2}{m^2} \Bigr)  
\ \times 
\parbox{20mm}{\begin{fmfgraph*}(15,15) 
\fmfleft{i}
\fmfright{o1,o2}
\fmfforce{0.8w,0.93h}{v2}
\fmfforce{0.8w,0.07h}{v1}
\fmfforce{0.2w,0.5h}{v5}
\fmf{double}{v1,o1}
\fmf{double}{v2,o2}
\fmf{photon}{i,v5}
\fmf{double,tension=.3}{v2,v5}
\fmf{double,tension=.3}{v1,v5}
\end{fmfgraph*} }  \, .
\label{CTtreeL2l}
\ee

\section{Renormalized Form Factors \label{FFrenorm}}

We report now the analytic expression of the UV-renormalized form 
factors at the one- and two-loop level, 
$F^{(1l)}_{i,R}(\epsilon,s,\mu^2/m^2)$ and
$F^{(2l)}_{i,R}(\epsilon,s,\mu^2/m^2)$, in the space-like region 
$s = q^2< 0$, in terms of HPLs of the variable $x$ already introduced in 
Eq.~(\ref{xvar}), 
$$ x=\frac{\sqrt{-s+4}-\sqrt{-s}}{\sqrt{-s+4}+\sqrt{-s}} \ . $$ 
The one-loop form factors are given up to the first term in the 
expansion in $\epsilon$, while the two-loop ones up to the finite part.

The renormalization of the UV divergences is carried out in the hybrid scheme
explained in the previous Section. Note that adding the diagrams of
Figs.~\ref{fig1} and \ref{fig2} and the corresponding counterterms we will
have a non-trivial dependence on the renormalization scale $\mu$, due to the
fact that the virtual contributions coming from the one- and two-loop diagrams
reported in Section \ref{unrenorm} depend on the ratio 
$(\mu^2/m^2)^{2 \epsilon}$, while in the counterterms we have a dependence
on $(\mu^2/m^2)^{2 \epsilon}$ (those calculated in the OS-scheme) as well 
as on $(\mu^2/m^2)^{\epsilon}$ (for the counterterms in the
$\overline{\mathrm{MS}}$-scheme). The expansion of these ratios in powers 
of $\epsilon$ generates terms proportional 
to $\log{(\mu^2/m^2)}$ and $\log^2{(\mu^2/m^2)}$. In this Section we give
the one- and two-loop UV-renormalized form factors for $\mu=m$. For the terms
proportional to $\log{(\mu^2/m^2)}$ and $\log^2{(\mu^2/m^2)}$ we refer the
reader to Section~\ref{munotm}.

\subsection{One-Loop UV-Renormalized Form Factors}

The one-loop UV-renormalized form factors are recovered adding the diagram of
Fig.~\ref{fig1}~(b) and the corresponding counterterm:
\be
{\mathrm CT}^{(1l)}(\epsilon,s) =  
\ Z_{2, {\mathrm OS}}^{(1l)}(\epsilon) \ \times 
\parbox{20mm}{\begin{fmfgraph*}(15,15) 
\fmfleft{i}
\fmfright{o1,o2}
\fmfforce{0.8w,0.93h}{v2}
\fmfforce{0.8w,0.07h}{v1}
\fmfforce{0.2w,0.5h}{v5}
\fmf{double}{v1,o1}
\fmf{double}{v2,o2}
\fmf{photon}{i,v5}
\fmf{double,tension=.3}{v2,v5}
\fmf{double,tension=.3}{v1,v5}
\end{fmfgraph*} } .
\ee

We put $\mu=m$ and we define:
\be
F_{i,R}^{(1l)}(\epsilon,s) = 
C(\epsilon) \, 
{\mathcal F}_{i,R}^{(1l)}(\epsilon,s) \, , \qquad \mbox{with} \; \, i=1,2
\, ,
\ee
finding:
\bea
{\mathcal F}^{(1l)}_{1,R}(\epsilon,s) & = &  \frac{1}{\epsilon} 
 \Biggl\{ C_{F} \Biggl[ - 1 + \biggl( 1 - \frac{1}{1-x} - \frac{1}{1+x} \biggr) 
      H(0;x)
\Biggr] \Biggl\}
\nn\\
& & 
- C_{F} \Biggl[
            2
          + \biggl( 
            \frac{1}{2} \! 
          - \! \frac{1}{1\! +\! x}
          \biggr) H(0;x) 
  +  \biggl( 1 \! - \! 
   \frac{1}{1\! -\! x} \! - \! \frac{1}{1\! +\! x} \biggr) \bigl[ 
   \zeta(2) \nn\\
& & \hspace{12mm} 
   - 2 H(0;x) - H(0,0;x) 
   + 2 H(-1,0;x) \bigr] \Biggr] \nn\\
& & 
  - \, \epsilon \Biggl\{ C_{F} \Biggl[ 4 
       - \frac{1}{2} \biggl( 
            1 \! 
          -  \! \frac{2}{1 \! + \! x} \biggr)
         \bigl[ \zeta(2)  \! 
          -  \! H(0,0;x)  \! 
          +  \! 2 H( \! -1,0;x) \bigr] \nn\\
& & \hspace*{11mm}
       + \! \biggl( 
            1 \! 
          - \! \frac{1}{1\! -\! x} \! 
          - \! \frac{1}{1\! +\! x} \biggr) 
         \bigl[
            2( \zeta(2) \! + \! \zeta(3) ) \!
          + \! ( \zeta(2)  \! -  \! 4 ) H(0;x)  \nn\\
& & \hspace*{16mm} 
          -  2 \zeta(2) H(-1;x) \! 
          -  2 H(0, \! 0;x)  \!
          + \!  4 H( \! -1,0;x)   \nn\\
& & \hspace*{16mm}
          -  \! H(0,0,0;x)
          -  \! 4 H( \! -1, \! -1,0;x)  \!
          +  \! 2 H(-1,0,0;x)  \nn\\
& & \hspace*{16mm}
          +  2 H(0,-1,0;x) \bigr] \Biggr]  \Biggr\} 
+ \, {\mathcal O} \left( \epsilon^2 \right)
\label{1lrenF1}
\, , \\
{\mathcal F}^{(1l)}_{2,R}(\epsilon,s) & = &  
- \, C_{F} \left( \frac{1}{1-x} -
\frac{1}{1+x} \right) H(0;x) \nn\\
&  &  
+ \, \epsilon \Biggl\{ C_{F} \Biggl[ \left( \frac{1}{1-x} -
\frac{1}{1+x} \right) \bigl[ \zeta(2) 
           - 4 H(0;x)   \nn\\
& & \hspace*{11mm}
           - H(0,0;x) 
           + 2 H(-1,0;x)
\bigr] \Biggl] \Biggr\} 
+ \, {\mathcal O} \left( \epsilon^2 \right)
\label{1lrenF2}  \, .
\eea

\subsection{Two-Loop UV-Renormalized Form Factors}

Adding the contributions of Eqs.~(\ref{c1},\ref{c2}) and (\ref{c3}--\ref{c8})
and (\ref{CTtreeL2l}), we find, diagrammatically, the following counterterm:
\bea
\! \! \! \! {\mathrm CT}^{(2l)}(\epsilon,s) \! \! & = &  \! \! 
- \, 2 \, \frac{\delta m_{{\mathrm OS}}^{(1l)}(\epsilon,m)}{m}
\times \left( m \, \, 
\parbox{20mm}{\begin{fmfgraph*}(15,15)
\fmfleft{i}
\fmfright{o1,o2}
\fmfforce{0.8w,0.93h}{v2}
\fmfforce{0.8w,0.07h}{v1}
\fmfforce{0.5w,0.7h}{v3}
\fmfforce{0.5w,0.53h}{v33}
\fmfforce{0.2w,0.5h}{v5}
\fmfforce{0.9w,0.5h}{v55}
\fmf{double}{v1,o1}
\fmf{double}{v2,o2}
\fmfv{label=$\otimes$}{v33}
\fmf{photon}{i,v5}
\fmf{double}{v2,v5}
\fmf{gluon}{v1,v2}
\fmf{double}{v1,v5}
\end{fmfgraph*} } \! \! \! \right)   \nn\\
\! \! \! \! \! \! & &  \! \! 
+ \Bigl[ 2 Z_{g,\overline{\mathrm{MS}}}^{(1l)}(\epsilon) +
Z_{2, {\mathrm OS}}^{(1l)}(\epsilon) \Bigr] \times \
\parbox{20mm}{\begin{fmfgraph*}(15,15) 
\fmfleft{i}
\fmfright{o1,o2}
\fmfforce{0.8w,0.93h}{v2}
\fmfforce{0.8w,0.07h}{v1}
\fmfforce{0.5w,0.7h}{v3}
\fmfforce{0.34w,0.5h}{v33}
\fmfforce{0.2w,0.5h}{v5}
\fmfforce{0.9w,0.5h}{v55}
\fmf{double}{v1,o1}
\fmf{double}{v2,o2}
\fmf{photon}{i,v5}
\fmf{double}{v2,v5}
\fmf{gluon}{v1,v2}
\fmf{double}{v1,v5}
\end{fmfgraph*} }   \hspace{-4mm} +
Z_{2, {\mathrm OS}}^{(2l)}(\epsilon) \times 
\parbox{20mm}{\begin{fmfgraph*}(15,15) 
\fmfleft{i}
\fmfright{o1,o2}
\fmfforce{0.8w,0.93h}{v2}
\fmfforce{0.8w,0.07h}{v1}
\fmfforce{0.2w,0.5h}{v5}
\fmf{double}{v1,o1}
\fmf{double}{v2,o2}
\fmf{photon}{i,v5}
\fmf{double,tension=.3}{v2,v5}
\fmf{double,tension=.3}{v1,v5}
\end{fmfgraph*} } \! \! \hspace{-4mm} .
\label{CTdiagr}
\eea

The two-loop UV-renormalized form factors are recovered adding the diagrams in
Fig.~\ref{fig2} and the counterterm in Eq.~(\ref{CTdiagr}).

We put $\mu=m$ and we define:
\be
F_{i,R}^{(2l)}(\epsilon,s) = 
C^2(\epsilon) \, 
{\mathcal F}_{i,R}^{(2l)}(\epsilon,s) \, , \qquad \mbox{with} \; \, i=1,2
\, ,
\ee
finding:
\bea
\hspace*{-5mm} {\mathcal F}^{(2l)}_{1,R}(\epsilon,s) & = & \frac{1}{\epsilon^2}
    \Biggl\{
         \frac{11}{12} C_{F} C_{A}  \Biggl[ 
            1
          - \Biggl(
            1
          - \frac{1}{1-x}
          - \frac{1}{1+x} \Biggl)  H(0;x) \Biggr] \nn\\
& &  \hspace{6mm}
       - \frac{1}{3} C_{F} T_{R} N_{f}   \Biggl[ 
            1
          - \Biggl(
            1
          - \frac{1}{1-x}
          - \frac{1}{1+x} \Biggl)  H(0;x) \Biggr]  \nn\\
& &  \hspace{6mm}
       + \, C_{F}^2   \Biggl[
            \frac{1}{2}
         - \Biggl(
            1
          - \frac{1}{1-x}
          - \frac{1}{1+x} \Biggl)  H(0;x) \nn\\
& &  \hspace{12mm}
        + \Biggl(
            1 \! 
          - \! \frac{1}{1-x} \! 
          + \! \frac{1}{(1-x)^2}\! 
          - \! \frac{1}{1\! +\! x} \! 
          + \! \frac{1}{(1\! +\! x)^2} \Biggl) H(0,0;x)
          \Biggr]
     \Biggr\} \nn\\
& & + \frac{1}{\epsilon}  \Biggl\{
         C_{F} C_{A} \Biggl[
          - \frac{49}{36}
          + \zeta(2)
          + H(0,0;x) \nn\\
& &  \hspace{12mm}
       + \Biggl(
            1
          - \frac{1}{1-x}
          - \frac{1}{1+x} \Biggl) \Biggl[
            \frac{\zeta(2)}{2}
          + \Biggl( \frac{67}{36} - \zeta(2) \Biggr) H(0;x) \nn\\
& &  \hspace{12mm}
          + H(0,0;x)
          - H(-1,0;x)
          + H(1,0;x)
          - H(0,0,0;x) \Biggr] \nn\\
& &  \hspace{12mm}
        - \Biggl(
            1
          - \frac{1}{1-x}
          + \frac{1}{(1-x)^2}
          - \frac{1}{1+x} 
          + \frac{1}{(1+x)^2} \Biggl) \Biggl[
            \frac{\zeta(3)}{2} \nn\\
& &  \hspace{12mm}
          + \frac{1}{2} \zeta(2) H(0;x)
          - H(0,-1,0;x)
          + H(0,0,0;x) \nn\\
& &  \hspace{12mm}
          + H(0,1,0;x) \Biggr] \nn\\
& &  \hspace{6mm}
       + \frac{5}{9} C_{F} T_{R} N_{f}   \Biggl[
            1
          - \Biggl(
            1
          - \frac{1}{1-x}
          - \frac{1}{1+x} \Biggl)  H(0;x) 
          \Biggr] \nn\\
& &  \hspace{6mm}
       + C_{F}^2   \Biggl[
            2
          - \Biggl( \frac{1}{2}
           - \frac{1}{1-x} \Biggr)  H(0;x) \nn\\
& &  \hspace{12mm}
          + \Biggl( \frac{1}{1+x}
          - \frac{1}{1-x}
          - \frac{2}{(1+x)^2} \Biggr) H(0,0;x)  \nn\\
& &  \hspace{12mm}
       + \Biggl(
            1
          - \frac{1}{1-x}
          - \frac{1}{1+x} \Biggl) \Bigl(
            \zeta(2)
          - 3 H(0;x)
          - 2 H(0,0;x) \nn\\
& &  \hspace{12mm}
          + 2 H(-1,0;x)  \Bigr)
        - \Biggl(
            1
          - \frac{1}{1-x}
          + \frac{1}{(1-x)^2}
          - \frac{1}{1+x}  \nn\\
& &  \hspace{12mm}
          + \frac{1}{(1+x)^2} \Biggl) \Bigl(
            \zeta(2) H(0;x)
          - 4 H(0,0;x)
          - 3 H(0,0,0;x) \nn\\
& &  \hspace{12mm}
          + 2 H(0,-1,0;x)
          + 4 H(-1,0,0;x)  \Bigr) \Biggr]
\Biggr\} \nn\\
& & \hspace{6mm}
       + \, C_{F} C_{A}   \Biggl[
       - \frac{1595}{108}
          + \frac{3}{ (1+x)^2 }
          - \frac{3}{ (1+x)}
       + \zeta(2)   \Biggl(
          - \frac{347}{36} \nn\\
& &  \hspace{12mm}
          + \frac{36 \log{2}}{ (1+x)^2 }
          - \frac{36 \log{2}}{ (1+x)}
          + 6 \log{2}
          + \frac{79}{18 (1-x)}
          - \frac{168}{ (1+x)^4 } \nn\\
& &  \hspace{12mm}
          + \frac{363}{ (1+x)^3 }
          - \frac{543}{2 (1+x)^2 }
          + \frac{734}{9 (1+x)}
          \Biggr)
       + \zeta(2)^2    \Biggl(
          - \frac{71}{20} \nn\\
& &  \hspace{12mm}
          - \frac{37}{20 (1-x)^2 }
          + \frac{679}{160 (1-x)}
          - \frac{879}{10 (1+x)^5 }
          + \frac{879}{4 (1+x)^4 } \nn\\
& &  \hspace{12mm}
          - \frac{7481}{40 (1+x)^3 }
          + \frac{943}{16 (1+x)^2 }
          - \frac{441}{160 (1+x)}
          \Biggr)
       + \zeta(3)   \Biggl(
            \frac{5}{6}\nn\\
& &  \hspace{12mm}
          + \frac{31}{6 (1-x)} \! 
          + \! \frac{324}{ (1+x)^4 }\! 
          - \! \frac{648}{ (1+x)^3 }\! 
          + \! \frac{374}{ (1+x)^2 }\! 
          - \! \frac{269}{6 (1+x)}
          \Biggr) \nn\\
& &  \hspace{12mm}
       + \zeta(2)   \Biggl(
          - \frac{10}{3}\! 
          - \! \frac{8}{3 (1-x)}\! 
          - \! \frac{90}{ (1+x)^4 }\! 
          + \! \frac{180}{ (1+x)^3 }\! 
          - \! \frac{135}{ (1+x)^2 } \nn\\
& &  \hspace{12mm}
          + \frac{127}{3 (1+x)}
          \Biggr) H(-1;x)
       + \zeta(3)   \Biggl(
            2
          + \frac{2}{ (1-x)^2 }
          - \frac{2}{ (1-x)} \nn\\
& &  \hspace{12mm}
          + \frac{2}{ (1+x)^2 }
          - \frac{2}{ (1+x)}
          \Biggr) H(-1;x)
       +   \Biggl(
            \frac{52}{3}
          - \frac{52}{3 (1-x)} \nn\\
& &  \hspace{12mm}
          - \frac{52}{3 (1+x)}
          \Biggr) H(-1,-1,0;x) \nn\\
& &  \hspace{12mm}
       + \zeta(2)   \Biggl(
            \frac{2}{ (1-x)^2 }
          + \frac{2}{ (1+x)^2 }
          \Biggr) H(-1,0;x) \nn\\
& &  \hspace{12mm}
       +   \Biggl(
          - \frac{31}{9}
          + \frac{43}{9 (1-x)}
          + \frac{6}{ (1+x)^3 }
          - \frac{9}{ (1+x)^2 } \nn\\
& &  \hspace{12mm}
          + \frac{46}{9 (1+x)}
          \Biggr) H(-1,0;x)
       +   \Biggl(
          - 4
          - \frac{4}{ (1-x)^2 }
          + \frac{4}{ (1-x)} \nn\\
& &  \hspace{12mm}
          - \frac{4}{ (1+x)^2 }
          + \frac{4}{ (1+x)}
          \Biggr) H(-1,0,-1,0;x)
       +   \Biggl(
          - \frac{53}{3} \nn\\
& &  \hspace{12mm}
          + \frac{35}{3 (1-x)}
          - \frac{282}{ (1+x)^4 }
          + \frac{564}{ (1+x)^3 }
          - \frac{339}{ (1+x)^2 } \nn\\
& &  \hspace{12mm}
          + \frac{206}{3 (1+x)}
          \Biggr) H(-1,0,0;x)
       +   \Biggl(
            2
          + \frac{4}{ (1-x)^2 }
          - \frac{2}{ (1-x)} \nn\\
& &  \hspace{12mm}
          + \frac{4}{ (1+x)^2 } \! 
          - \! \frac{2}{ (1+x)}
          \Biggr) H(-1,0,0,0;x) \! 
       +   \!  \Biggl(\!
            4 \! 
          +  \! \frac{4}{ (1-x)^2 } \nn\\
& &  \hspace{12mm}
          - \frac{4}{ (1-x)}
          + \frac{4}{ (1+x)^2 }
          - \frac{4}{ (1+x)}
          \Biggr) H(-1,0,1,0;x) \nn\\
& &  \hspace{12mm}
       +  \Biggl(
          - 6
          + \frac{6}{ (1-x)}
          + \frac{6}{ (1+x)}
          \Biggr) H(-1,1,0;x)  \nn\\
& &  \hspace{12mm}
       + \zeta(2)   \Biggl(
            \frac{29}{6}\! 
          + \! \frac{11}{3 (1-x)}\! 
          - \frac{258}{ (1+x)^5 }\! 
          + \! \frac{744}{ (1+x)^4 }\! 
          - \frac{779}{ (1+x)^3 } \nn\\
& &  \hspace{12mm}
          + \frac{357}{ (1+x)^2 } \! 
          - \! \frac{223}{3 (1+x)}
          \Biggr) H(0;x)
       +  \! \zeta(3)   \Biggl(
          -  \! \frac{15}{2} \! 
          -  \! \frac{1}{2 (1-x)^2 } \nn\\
& &  \hspace{12mm}
          +  \frac{2}{ (1-x)} \! 
          +  \! \frac{324}{ (1+x)^5 } \! 
          -  \! \frac{810}{ (1+x)^4 } \! 
          +  \! \frac{662}{ (1+x)^3 } \! 
          -  \! \frac{367}{2 (1+x)^2 } \nn\\
& &  \hspace{12mm}
          + \frac{20}{ (1+x)}
          \Biggr) H(0;x)
       +   \Biggl(
            \frac{2545}{216}
          - \frac{365}{27 (1-x)}
          + \frac{9}{ (1+x)^3 } \nn\\
& &  \hspace{12mm}
          - \frac{27}{2 (1+x)^2 }\! 
          - \frac{599}{108 (1+x)}\! 
          \Biggr) H(0;x)
       + \! \zeta(2)   \Biggl(
            7\! 
          + \! \frac{1}{(1-x)^2 } \nn\\
& &  \hspace{12mm}
          - \frac{35}{8 (1-x)}\! 
          - \! \frac{90}{ (1+x)^5 }\! 
          + \! \frac{225}{ (1+x)^4 }\! 
          - \! \frac{363}{2 (1+x)^3 }\! 
          + \! \frac{193}{4 (1+x)^2 } \nn\\
& &  \hspace{12mm}
          - \frac{83}{8 (1+x)}
          \Biggr) H(0,-1;x)
       +   \Biggl(
          - 10
          - \frac{10}{ (1-x)^2 }
          + \frac{10}{ (1-x)} \nn\\
& &  \hspace{12mm}
          - \frac{10}{ (1+x)^2 }
          + \frac{10}{ (1+x)}
          \Biggr) H(0,-1,-1,0;x)
       +   \Biggl(
          - \frac{68}{3} \nn\\
& &  \hspace{12mm}
          + \frac{35}{3 (1-x)}
          + \frac{12}{ (1+x)^4 }
          - \frac{24}{ (1+x)^3 }
          + \frac{14}{ (1+x)^2 } \nn\\
& &  \hspace{12mm}
          + \frac{29}{3 (1+x)}
          \Biggr) H(0,-1,0;x)
       +   \Biggl(
            14
          + \frac{8}{ (1-x)^2 }
          - \frac{83}{8 (1-x)} \nn\\
& &  \hspace{12mm}
          - \frac{282}{ (1+x)^5 }
          + \frac{705}{ (1+x)^4 }
          - \frac{1171}{2 (1+x)^3 }
          + \frac{725}{4 (1+x)^2 } \nn\\
& &  \hspace{12mm}
          - \frac{227}{8 (1+x)}
          \Biggr) H(0,-1,0,0;x)\! 
       +  \!  \Biggl(
            6\! 
          + \! \frac{6}{ (1-x)^2 }
          - \frac{6}{ (1-x)} \nn\\
& &  \hspace{12mm}
          + \frac{6}{ (1+x)^2 }
          - \frac{6}{ (1+x)}
          \Biggr) H(0,-1,1,0;x) \nn\\
& &  \hspace{12mm}
       + \zeta(2)   \Biggl(
            1\! 
          + \! \frac{1}{(1-x)^2 } \! 
          + \! \frac{5}{16 (1-x)}\! 
          - \! \frac{69}{ (1+x)^5 }\! 
          + \! \frac{345}{2 (1+x)^4 } \nn\\
& &  \hspace{12mm}
          - \frac{563}{4 (1+x)^3 }
          + \frac{317}{8 (1+x)^2 }
          - \frac{59}{16 (1+x)}
          \Biggr) H(0,0;x)  \nn\\
& &  \hspace{12mm}
       -   \Biggl(
            \frac{217}{36} \! 
          +  \! \frac{25}{18 (1-x)} \! 
          - \frac{24}{ (1+x)^4 } \! 
          +  \! \frac{39}{ (1+x)^3 } \! 
          - \frac{3}{2 (1+x)^2 } \nn\\
& &  \hspace{12mm}
          - \frac{130}{9 (1+x)}
          \Biggr) H(0,0;x)
       +   \Biggl(
            22
          + \frac{12}{ (1-x)^2 }
          - \frac{75}{4 (1-x)} \nn\\
& &  \hspace{12mm}
          + \frac{12}{ (1+x)^5 }
          - \frac{30}{ (1+x)^4 }
          + \frac{49}{ (1+x)^3 }
          - \frac{63}{2 (1+x)^2 } \nn\\
& &  \hspace{12mm}
          - \frac{51}{4 (1+x)} \! 
          \Biggr) H(0,0,-1,0;x) \! 
       +  \!   \Biggl(
            \frac{89}{6} \! 
          -  \! \frac{4}{3 (1-x)} \! 
          -  \! \frac{258}{ (1+x)^5 } \nn\\
& &  \hspace{12mm}
          +  \frac{816}{ (1+x)^4 } \! 
          -  \! \frac{923}{ (1+x)^3 } \! 
          +  \! \frac{436}{ (1+x)^2 } \! 
          -  \! \frac{259}{3 (1+x)}
          \Biggr) H(0,0,0;x) \nn\\
& &  \hspace{12mm}
       +   \Biggl(
          - 12
          - \frac{6}{ (1-x)^2 }
          + \frac{193}{16 (1-x)}
          + \frac{3}{ (1+x)^5 }
          - \frac{15}{2 (1+x)^4 } \nn\\
& &  \hspace{12mm}
          + \frac{17}{4 (1+x)^3 }
          - \frac{39}{8 (1+x)^2 }
          + \frac{177}{16 (1+x)}
          \Biggr) H(0,0,0,0;x) \nn\\
& &  \hspace{12mm}
       +   \Biggl(
          - 14
          - \frac{8}{ (1-x)^2 }
          + \frac{12}{ (1-x)}
          + \frac{48}{ (1+x)^5 }
          - \frac{120}{ (1+x)^4 } \nn\\
& &  \hspace{12mm}
          + \frac{88}{ (1+x)^3 }
          - \frac{20}{ (1+x)^2 }
          + \frac{12 }{(1+x)}
          \Biggr) H(0,0,1,0;x) \nn\\
& &  \hspace{12mm}
       + \zeta(2)   \Biggl(
            1
          + \frac{1}{(1-x)^2 }
          - \frac{1}{(1-x)}
          + \frac{1}{(1+x)^2 } \nn\\
& &  \hspace{12mm}
          - \frac{1}{(1+x)}
          \Biggr) H(0,1;x)
       +   \Biggl(
            6
          + \frac{6}{ (1-x)^2 }
          - \frac{6}{ (1-x)} \nn\\
& &  \hspace{12mm}
          + \frac{6}{ (1+x)^2 }
          - \frac{6}{ (1+x)}
          \Biggr) H(0,1,-1,0;x)
       +   \Biggl(
            8
          - \frac{4}{ (1-x)} \nn\\
& &  \hspace{12mm}
          +  \frac{48}{ (1+x)^4 } \! 
          -  \! \frac{96}{ (1+x)^3 } \! 
          +  \! \frac{56}{ (1+x)^2 } \! 
          -  \! \frac{12}{ (1+x)}
          \Biggr) H(0,1,0;x) \nn\\
& &  \hspace{12mm}
       +  \Biggl(
          - 2
          - \frac{4}{ (1-x)^2 }
          + \frac{5}{ (1-x)}
          + \frac{24}{ (1+x)^5 }
          - \frac{60}{ (1+x)^4 } \nn\\
& &  \hspace{12mm}
          + \frac{72}{ (1+x)^3 }
          - \frac{52}{ (1+x)^2 }
          + \frac{11}{ (1+x)}
          \Biggr) H(0,1,0,0;x)  \nn\\
& &  \hspace{12mm}
       +   \Biggl(
          - 2
          - \frac{2}{ (1-x)^2 }
          + \frac{2}{ (1-x)}
          - \frac{2}{ (1+x)^2 }  \nn\\
& &  \hspace{12mm}
          + \frac{2}{ (1+x)}
          \Biggr) H(0,1,1,0;x)
       + \zeta(2)   \Biggl(
          - 1
          + \frac{1}{(1-x)}  \nn\\
& &  \hspace{12mm}
          + \frac{1}{(1+x)}
          \Biggr) H(1;x)
       + \zeta(3)   \Biggl(
          - 2
          - \frac{2}{ (1-x)^2 }
          + \frac{2}{ (1-x)}  \nn\\
& &  \hspace{12mm}
          - \frac{2}{ (1+x)^2 }
          + \frac{2}{ (1+x)}
          \Biggr) H(1;x)
       +   \Biggl(
          - 6
          + \frac{6}{ (1-x)}  \nn\\
& &  \hspace{12mm}
          +  \frac{6}{ (1+x)}
          \Biggr) H(1,-1,0;x) \! 
       +  \! \zeta(2)   \Biggl(
            4 \! 
          -  \! \frac{2}{ (1-x)^2 } \! 
          +  \! \frac{2}{ (1-x)}  \nn\\
& &  \hspace{12mm}
          - \frac{336}{ (1+x)^5 }
          + \frac{840}{ (1+x)^4 }
          - \frac{688}{ (1+x)^3 }
          + \frac{190}{ (1+x)^2 }  \nn\\
& &  \hspace{12mm}
          - \frac{18}{ (1+x)}
          \Biggr) H(1,0;x)
       +   \Biggl(
          - 3
          + \frac{2}{ (1-x)}
          + \frac{24}{ (1+x)^3 }  \nn\\
& &  \hspace{12mm}
          - \frac{36}{ (1+x)^2 }
          + \frac{16}{ (1+x)}
          \Biggr) H(1,0;x)
       +   \Biggl(
            4
          + \frac{4}{ (1-x)^2 }  \nn\\
& &  \hspace{12mm}
          - \frac{4}{ (1-x)}
          + \frac{4}{ (1+x)^2 }
          - \frac{4}{ (1+x)}
          \Biggr) H(1,0,-1,0;x)  \nn\\
& &  \hspace{12mm}
       +   \Biggl(
          - 4
          - \frac{4}{ (1-x)}
          + \frac{24}{ (1+x)^4 }
          - \frac{48}{ (1+x)^3 }
          + \frac{38}{ (1+x)^2 }  \nn\\
& &  \hspace{12mm}
          - \frac{18}{ (1+x)}
          \Biggr) H(1,0,0;x)
       +   \Biggl(
            2
          - \frac{4}{ (1-x)^2 }
          + \frac{4}{ (1-x)}  \nn\\
& &  \hspace{12mm}
          - \frac{336}{ (1+x)^5 }
          + \frac{840}{ (1+x)^4 }
          - \frac{688}{ (1+x)^3 }
          + \frac{188}{ (1+x)^2 }  \nn\\
& &  \hspace{12mm}
          - \frac{16}{ (1+x)}
          \Biggr) H(1,0,0,0;x)
       +   \Biggl(
          - 4
          - \frac{4}{ (1-x)^2 }
          + \frac{4}{ (1-x)}  \nn\\
& &  \hspace{12mm}
          - \frac{4}{ (1+x)^2 }
          + \frac{4}{ (1+x)}
          \Biggr) H(1,0,1,0;x)
       +   \Biggl(
            2
          - \frac{2}{ (1-x)}  \nn\\
& &  \hspace{12mm}
          - \frac{2}{ (1+x)}
          \Biggr) H(1,1,0;x)
       \Biggr] \nn\\
& & \hspace{6mm}
       + C_{F} T_{R} N_{f}   \Biggl[
         \frac{106}{27}
       + \zeta(2)   \Biggl(
            \frac{31}{9}
          - \frac{22}{9 (1-x)}
          - \frac{16}{9 (1+x)}
          \Biggr)  \nn\\
& &  \hspace{12mm}
       + \zeta(3)   \Biggl(
            \frac{4}{3}
          - \frac{4}{3 (1-x)}
          - \frac{4}{3 (1+x)}
          \Biggr) \nn\\
& &  \hspace{12mm}
       + \zeta(2)   \Biggl(
          - \frac{4}{3}
          + \frac{4}{3 (1-x)} 
          + \frac{4}{3 (1+x)}
          \Biggr) H(-1;x)  \nn\\
& &  \hspace{12mm}
       +   \Biggl(
          - \frac{8}{3}
          + \frac{8}{3 (1-x)}
          + \frac{8}{3 (1+x)}
          \Biggr) H(-1,-1,0;x)  \nn\\
& &  \hspace{12mm}
       +   \Biggl(
            \frac{38}{9}
          - \frac{44}{9 (1-x)} 
          - \frac{32}{9 (1+x)}
          \Biggr) H(-1,0;x)  \nn\\
& &  \hspace{12mm}
       +   \Biggl(
            \frac{4}{3}
          - \frac{4}{3 (1-x)} 
          - \frac{4}{3 (1+x)}
          \Biggr) H(-1,0,0;x)  \nn\\
& &  \hspace{12mm}
       + \zeta(2)   \Biggl(
          - \frac{2}{3}
          + \frac{2}{3 (1-x)} 
          + \frac{2}{3 (1+x)}
          \Biggr) H(0;x)  \nn\\
& &  \hspace{12mm}
       +  \Biggl(
          - \frac{209}{54}
          + \frac{106}{27 (1-x)}
          + \frac{103}{27 (1+x)}
          \Biggr) H(0;x)   \nn\\
& &  \hspace{12mm}
       +   \Biggl(
            \frac{4}{3}
          - \frac{4}{3 (1-x)}
          - \frac{4}{3 (1+x)}
          \Biggr) H(0,-1,0;x)  \nn\\
& &  \hspace{12mm}
       +   \Biggl(
          - \frac{19}{9}
          + \frac{22}{9 (1-x)}
          + \frac{16}{9 (1+x)}
          \Biggr) H(0,0;x)  \nn\\
& &  \hspace{12mm}
       +   \Biggl(
          - \frac{2}{3}
          + \frac{2}{3 (1-x)}
          + \frac{2}{3 (1+x)}
          \Biggr) H(0,0,0;x)
       \Biggr] \nn\\
& & \hspace{6mm}
       + \, C_{F} T_{R}   \Biggl[
            \frac{383}{27}
          + \frac{196}{9 (1+x)^2 }
          - \frac{196}{9 (1+x)} \nn\\
& &  \hspace{12mm}
       - \zeta(2)   \Biggl(
            1 \! 
          -  \! \frac{392}{3 (1+x)^4 } \! 
          +  \! \frac{784}{3 (1+x)^3 }
          -  \! \frac{458}{3 (1+x)^2 } \! 
          +  \! \frac{22}{ (1+x)}
          \Biggr) \nn\\
& &  \hspace{12mm}
       +  \zeta(2)   \Biggl(
          -  \frac{2}{3} \! 
          +  \! \frac{5}{3 (1-x)} \! 
          -  \! \frac{24}{ (1+x)^5 } \! 
          +  \! \frac{60}{ (1+x)^4 } \! 
          -  \! \frac{44}{ (1+x)^3 }\nn\\
& &  \hspace{12mm}
          + \frac{6}{ (1+x)^2 }
          + \frac{5}{3 (1+x)}
          \Biggr) H(0;x)
       +   \Biggl(
          - \frac{265}{54}
          + \frac{236}{27 (1-x)}\nn\\
& &  \hspace{12mm}
          + \frac{356}{9 (1+x)^3 }
          - \frac{178}{3 (1+x)^2 }
          + \frac{563}{27 (1+x)}
          \Biggr) H(0;x)\nn\\
& &  \hspace{12mm}
       +   \Biggl(
            \frac{19}{9}
          + \frac{248}{9 (1+x)^4 }
          - \frac{496}{9 (1+x)^3 }
          + \frac{326}{9 (1+x)^2 }\nn\\
& &  \hspace{12mm}
          - \frac{26}{3 (1+x)}
          \Biggr) H(0,0;x)
       +   \Biggl(
          - \frac{2}{3}
          + \frac{5}{3 (1-x)}
          - \frac{24}{ (1+x)^5 } \nn\\
& &  \hspace{12mm}
          +  \frac{60}{ (1 \! + \! x)^4 } \! 
          -  \! \frac{44}{ (1 \! + \! x)^3 } \! 
          +  \! \frac{6}{ (1\!+\!x)^2 } \! 
          +  \! \frac{5}{3 (1+x)}
          \Biggr) H(0,0,0;x)
       \Biggr] \nn\\
& & \hspace{6mm}
       + \, C_{F}^2   \Biggl[
         \frac{23}{2} \! 
       +  \! \zeta(2)   \Biggl(
            \frac{55}{4} \! 
          - \frac{72 \log{2}}{ (1+x)^2 } \! 
          +  \! \frac{72 \log{2}}{ (1+x)} \! 
          - 12 \log{2} \! 
          +  \! \frac{2}{ (1-x)} \nn\\
& &  \hspace{12mm}
          - \frac{240}{ (1+x)^4 }
          + \frac{528}{ (1+x)^3 }
          - \frac{300}{ (1+x)^2 }
          + \frac{11}{2 (1+x)}
          \Biggr) \nn\\
& &  \hspace{12mm}
       + \zeta(2)^2    \Biggl(
            \frac{181}{10}
          + \frac{61}{10 (1-x)^2 }
          - \frac{1219}{80 (1-x)}
          - \frac{171}{ (1+x)^5 } \nn\\
& &  \hspace{12mm}
          + \frac{855}{2 (1+x)^4 }
          - \frac{6867}{20 (1+x)^3 }
          + \frac{749}{8 (1+x)^2 }
          - \frac{1731}{80 (1+x)}
          \Biggr) \nn\\
& &  \hspace{12mm}
       + \zeta(3)   \Biggl(
          - 7
          - \frac{4}{ (1-x)^2 }
          + \frac{2}{ (1-x)}
          + \frac{168}{ (1+x)^4 }
          - \frac{336}{ (1+x)^3 } \nn\\
& &  \hspace{12mm}
          + \frac{190}{ (1+x)^2 }
          - \frac{24}{ (1+x)}
          \Biggr)
       + \zeta(2)   \Biggl(
            10
          + \frac{2}{ (1-x)}
          + \frac{180}{ (1+x)^4 } \nn\\
& &  \hspace{12mm}
          -  \frac{360}{ (1+x)^3 } \! 
          +  \! \frac{270}{ (1+x)^2 } \! 
          -  \! \frac{88}{ (1+x)} \! 
          \Biggr) H(-1;x) \! 
       -   \!  \Biggl(
            4 \! 
          -  \! \frac{4}{ (1-x)} \nn\\
& &  \hspace{12mm}
          - \frac{4}{ (1+x)}
          \Biggr) H(-1,-1,0;x)
       +   \Biggl(
            16
          + \frac{16}{ (1-x)^2 }
          - \frac{16}{ (1-x)} \nn\\
& &  \hspace{12mm}
          + \frac{16}{ (1+x)^2 }
          - \frac{16}{ (1+x)}
          \Biggr) H(-1,-1,0,0;x)
       + \zeta(2)   \Biggl(
            4 \nn\\
& &  \hspace{12mm}
          + \frac{4}{ (1-x)^2 }
          - \frac{4}{ (1-x)}
          + \frac{4}{ (1+x)^2 }
          - \frac{4}{ (1+x)}
          \Biggr) H(-1,0;x) \nn\\
& &  \hspace{12mm}
       +   \Biggl(
          - \frac{55}{2}
          + \frac{36}{ (1-x)}
          + \frac{192}{ (1+x)^3 }
          - \frac{288}{ (1+x)^2 } \nn\\
& &  \hspace{12mm}
          + \frac{115}{ (1+x)}
          \Biggr) H(-1,0;x)
       +   \Biggl(
            8
          + \frac{8}{ (1-x)^2 }
          - \frac{8}{ (1-x)} \nn\\
& &  \hspace{12mm}
          + \frac{8}{ (1+x)^2 } \! 
          -  \! \frac{8}{ (1+x)} \! 
          \Biggr) H(-1,0,-1,0;x) \! 
       -   \!  \Biggl(
            6 \! 
          +  \! \frac{16}{ (1-x)^2 } \nn\\
& &  \hspace{12mm}
          - \frac{14}{ (1-x)}
          + \frac{252}{ (1+x)^4 }
          - \frac{504}{ (1+x)^3 }
          + \frac{270}{ (1+x)^2 } \nn\\
& &  \hspace{12mm}
          - \frac{16}{ (1+x)}
          \Biggr) H(-1,0,0;x)
       +   \Biggl(
          - 12
          - \frac{12}{ (1-x)^2 }
          + \frac{12}{ (1-x)} \nn\\
& &  \hspace{12mm}
          - \frac{12}{ (1+x)^2 }
          + \frac{12}{ (1+x)}
          \Biggr) H(-1,0,0,0;x)
       + \zeta(2)   \Biggl(
          - 17 \nn\\
& &  \hspace{12mm}
          -  \frac{8}{ (1-x)^2 } \! 
          +  \! \frac{3}{2 (1-x)} \! 
          -  \! \frac{60}{ (1+x)^5 } \! 
          +  \! \frac{156}{ (1+x)^4 } \! 
          -  \! \frac{136}{ (1+x)^3 } \nn\\
& &  \hspace{12mm}
          + \frac{6}{ (1+x)^2 }
          + \frac{113}{2 (1+x)}
          \Biggr) H(0;x)
       + \zeta(3)   \Biggl(
            8
          - \frac{7}{ (1-x)} \nn\\
& &  \hspace{12mm}
          + \frac{168}{ (1+x)^5 }
          - \frac{420}{ (1+x)^4 }
          + \frac{364}{ (1+x)^3 }
          - \frac{126}{ (1+x)^2 } \nn\\
& &  \hspace{12mm}
          +  \frac{5}{ (1+x)}
          \Biggr) H(0;x) \! 
       -   \!  \Biggl(
            \frac{85}{8} \! 
          - \frac{19}{2 (1-x)} \! 
          - \frac{47}{4 (1+x)} \! 
          \Biggr) H(0;x) \nn\\
& &  \hspace{12mm}
       + \zeta(2)   \Biggl(
          - 10
          + \frac{2}{ (1-x)^2 }
          + \frac{19}{4 (1-x)}
          + \frac{180}{ (1+x)^5 } \nn\\
& &  \hspace{12mm}
          -  \frac{450}{ (1 \! + \! x)^4 } \! 
          +  \! \frac{363}{ (1 \! + \! x)^3 } \! 
          -  \! \frac{185}{2 (1 \! + \! x)^2 } \! 
          +  \! \frac{67}{4 (1 \! + \! x)} \! 
          \Biggr) H(0,-1;x) \nn\\
& &  \hspace{12mm}
       +   \Biggl(
            4
          + \frac{4}{ (1-x)^2 }
          - \frac{4}{ (1-x)}
          + \frac{4}{ (1+x)^2 } \nn\\
& &  \hspace{12mm}
          - \frac{4}{ (1+x)}
          \Biggr) H(0,-1,-1,0;x) \! 
       +  \!   \Biggl(
            16 \! 
          -  \! \frac{2}{ (1-x)} \! 
          +  \! \frac{384}{ (1 \! + \! x)^4 } \nn\\
& &  \hspace{12mm}
          - \frac{768}{ (1+x)^3 }
          + \frac{442}{ (1+x)^2 }
          - \frac{60}{ (1+x)}
          \Biggr) H(0,-1,0;x) \nn\\
& &  \hspace{12mm}
       +   \Biggl(
          - 14
          - \frac{10}{ (1-x)^2 }
          + \frac{67}{4 (1-x)}
          - \frac{252}{ (1+x)^5 }
          + \frac{630}{ (1+x)^4 } \nn\\
& &  \hspace{12mm}
          - \frac{517}{ (1+x)^3 }
          + \frac{271}{2 (1+x)^2 }
          + \frac{19}{4 (1+x)}
          \Biggr) H(0,-1,0,0;x) \nn\\
& &  \hspace{12mm}
       + \zeta(2)   \Biggl(
            13
          + \frac{5}{ (1-x)^2 }
          - \frac{105}{8 (1-x)}
          - \frac{66}{ (1+x)^5 } \nn\\
& &  \hspace{12mm}
          +  \frac{165}{ (1+x)^4 } \! 
          -  \! \frac{281}{2 (1+x)^3 } \! 
          +  \! \frac{203}{4 (1+x)^2 } \! 
          -  \! \frac{137}{8 (1+x)}
          \Biggr) H(0,0;x) \nn\\
& &  \hspace{12mm}
       +   \Biggl(
            \frac{229}{4}
          + \frac{16}{ (1-x)^2 }
          - \frac{42}{ (1-x)}
          + \frac{192}{ (1+x)^4 }
          - \frac{504}{ (1+x)^3 } \nn\\
& &  \hspace{12mm}
          + \frac{466}{ (1+x)^2 }
          - \frac{343}{2 (1+x)}
          \Biggr) H(0,0;x)
       +   \Biggl(
          - 22
          - \frac{10}{ (1-x)^2 } \nn\\
& &  \hspace{12mm}
          +  \frac{21}{2 (1-x)} \! 
          +  \! \frac{384}{ (1+x)^5 } \! 
          -  \! \frac{960}{ (1+x)^4 } \! 
          +  \! \frac{746}{ (1+x)^3 } \! 
          -  \! \frac{169}{ (1+x)^2 } \nn\\
& &  \hspace{12mm}
          + \frac{45}{2 (1+x)}
          \Biggr) H(0,0,-1,0;x)
       +   \Biggl(
          - 10
          + \frac{4}{ (1-x)^2 } \nn\\
& &  \hspace{12mm}
          -  \frac{17}{2 (1-x)} \! 
          -  \! \frac{60}{ (1+x)^5 } \! 
          +  \! \frac{216}{ (1+x)^4 } \! 
          -  \! \frac{256}{ (1+x)^3 } \! 
          +  \! \frac{89}{ (1+x)^2 } \nn\\
& &  \hspace{12mm}
          + \frac{71}{2 (1+x)}
          \Biggr) H(0,0,0;x)
       +   \Biggl(
            27
          + \frac{17}{ (1-x)^2 }
          - \frac{217}{8 (1-x)} \nn\\
& &  \hspace{12mm}
          - \frac{6}{ (1+x)^5 }
          + \frac{15}{ (1+x)^4 }
          - \frac{17}{2 (1+x)^3 }
          + \frac{59}{4 (1+x)^2 } \nn\\
& &  \hspace{12mm}
          - \frac{201}{8 (1+x)}
          \Biggr) H(0,0,0,0;x)
       +   \Biggl(
            12
          + \frac{4}{ (1-x)^2 }
          - \frac{8}{ (1-x)} \nn\\
& &  \hspace{12mm}
          - \frac{96}{ (1+x)^5 }
          + \frac{240}{ (1+x)^4 }
          - \frac{176}{ (1+x)^3 }
          + \frac{28}{ (1+x)^2 } \nn\\
& &  \hspace{12mm}
          - \frac{8}{ (1+x)}
          \Biggr) H(0,0,1,0;x)
       +   \Biggl(
          - 10
          - \frac{8 }{(1-x)^2 }
          + \frac{8}{ (1-x)} \nn\\
& &  \hspace{12mm}
          -  \frac{96}{ (1+x)^4 } \! 
          +  \! \frac{192}{ (1+x)^3 } \! 
          -  \! \frac{116}{ (1+x)^2 } \! 
          +  \! \frac{20}{ (1+x)} \! 
          \Biggr) H(0,1,0;x) \nn\\
& &  \hspace{12mm}
       +   \Biggl(
          - 4
          - \frac{7}{ (1-x)}
          + \frac{360}{ (1+x)^5 }
          - \frac{900}{ (1+x)^4 }
          + \frac{700}{ (1+x)^3 } \nn\\
& &  \hspace{12mm}
          -  \frac{150}{ (1+x)^2 } \! 
          +  \! \frac{5}{ (1+x)} \! 
          \Biggr) H(0,1,0,0;x) \! 
       +  \! \zeta(3)   \Biggl(
            4 \! 
          +  \! \frac{4}{ (1-x)^2 } \nn\\
& &  \hspace{12mm}
          -  \frac{4}{ (1 \! - \! x)} \! 
          +  \! \frac{4}{ (1 \! + \! x)^2 } \! 
          -  \! \frac{4}{ (1 \! + \! x)}
          \Biggr) H(1;x) \! 
       +  \! \zeta(2)   \Biggl(
            4 \! 
          +  \! \frac{4}{ (1 \! - \! x)^2 } \nn\\
& &  \hspace{12mm}
          - \frac{6}{ (1-x)}
          - \frac{144}{ (1+x)^5 }
          + \frac{360}{ (1+x)^4 }
          - \frac{312}{ (1+x)^3 }
          + \frac{112}{ (1+x)^2 } \nn\\
& &  \hspace{12mm}
          - \frac{14}{ (1+x)}
          \Biggr) H(1,0;x)
       +   \Biggl(
            16
          - \frac{16}{ (1-x)}
          - \frac{48}{ (1+x)^3 } \nn\\
& &  \hspace{12mm}
          + \frac{72}{ (1+x)^2 }
          - \frac{40}{ (1+x)}
          \Biggr) H(1,0;x)
       +   \Biggl(
          - 8
          - \frac{8}{ (1-x)^2 } \nn\\
& &  \hspace{12mm}
          + \frac{8}{ (1-x)}
          - \frac{8}{ (1+x)^2 }
          + \frac{8}{ (1+x)}
          \Biggr) H(1,0,-1,0;x) \nn\\
& &  \hspace{12mm}
       +   \Biggl(
            16
          + \frac{360}{ (1+x)^4 }
          - \frac{720}{ (1+x)^3 }
          + \frac{394}{ (1+x)^2 } \nn\\
& &  \hspace{12mm}
          - \frac{34}{ (1+x)}
          \Biggr) H(1,0,0;x)
       +   \Biggl(
            8
          + \frac{8}{ (1-x)^2 }
          - \frac{10}{ (1-x)} \nn\\
& &  \hspace{12mm}
          - \frac{144}{ (1+x)^5 }
          + \frac{360}{ (1+x)^4 }
          - \frac{312}{ (1+x)^3 }
          + \frac{116}{ (1+x)^2 } \nn\\
& &  \hspace{12mm}
          - \frac{18}{ (1+x)}
          \Biggr) H(1,0,0,0;x)
       +   \Biggl(
            8
          + \frac{8}{ (1-x)^2 }
          - \frac{8}{ (1-x)} \nn\\
& &  \hspace{12mm}
          + \frac{8}{ (1+x)^2 }
          - \frac{8}{ (1+x)}
          \Biggr) H(1,0,1,0;x)
       \Biggr]
\label{FF1fr} \ , \\
\hspace*{-5mm} {\mathcal F}^{(2l)}_{2,R}(\epsilon,s) & = &  
\frac{1}{\epsilon}  \Biggl\{
         C_{F}^2 \Biggl[
        \biggl(
            \frac{1}{1-x}
          - \frac{1}{1+x} \Biggl) H(0;x)
        - 2 \biggl(
            \frac{1}{1-x} 
          - \frac{1}{(1-x)^2} \nn\\
& &  \hspace{12mm}
          - \frac{1}{1+x}
          + \frac{1}{(1+x)^2} \Biggl) H(0,0;x)
          \Biggr] \Biggr\} \nn\\
& & \hspace{6mm}
       + C_{F} C_{A}   \Biggl[
          - \frac{3}{ (1+x)^2 } \! 
          +  \! \frac{3}{ (1+x)} \! 
       +  \! \zeta(2)   \Biggl(
          - \frac{24 \log{2}}{ (1+x)^2 } \! 
          +  \! \frac{24 \log{2}}{ (1+x)} \nn\\
& &  \hspace{12mm}
          +  \frac{3}{ (1-x)^2 } \! 
          -  \! \frac{35}{12 (1-x)} \! 
          +  \! \frac{168}{ (1 \! + \! x)^4 } \! 
          -  \! \frac{363}{ (1 \! + \! x)^3 } \! 
          +  \! \frac{491}{2 (1 \! + \! x)^2 } \nn\\
& &  \hspace{12mm}
          - \frac{607}{12 (1+x)}
          \Biggr)
       + \zeta(2)^2    \Biggl(
            \frac{69}{40 (1-x)^3 }
          - \frac{207}{80 (1-x)^2 } \nn\\
& &  \hspace{12mm}
          + \frac{45}{32 (1-x)}
          + \frac{879}{10 (1+x)^5 }
          - \frac{879}{4 (1+x)^4 }
          + \frac{1797}{10 (1+x)^3 } \nn\\
& &  \hspace{12mm}
          -  \frac{249}{5 (1+x)^2 } \! 
          +  \! \frac{45}{32 (1+x)}
          \Biggr)
       +  \! \zeta(3)   \Biggl(
          -  \frac{324}{ (1+x)^4 } \! 
          +  \! \frac{648}{ (1+x)^3 } \nn\\
& &  \hspace{12mm}
          - \frac{344}{ (1+x)^2 } \! 
          +  \! \frac{20}{ (1+x)}
          \Biggr) \! 
       +  \! \zeta(2)   \Biggl(
          - \frac{9}{2 (1-x)^2 }
          + \frac{9}{2 (1-x)}\nn\\
& &  \hspace{12mm}
          +  \frac{90}{ (1 \! + \! x)^4 } \! 
          -  \! \frac{180}{ (1+x)^3 } \! 
          +  \! \frac{231}{2 (1+x)^2 }  \! 
          -  \! \frac{51}{2 (1+x)}
          \Biggr) H(-1;x) \nn\\
& &  \hspace{12mm}
       +   \!  \Biggl(
            \frac{1}{6 (1-x)} \! 
          -  \! \frac{6}{ (1 \! + \! x)^3 } \! 
          +  \! \frac{9}{ (1 \! + \! x)^2 } \! 
          -  \! \frac{19}{6 (1 \! + \! x)} \! 
          \Biggr) H(-1,0;x) \nn\\
& &  \hspace{12mm}
       +   \Biggl(
          - \frac{1}{2 (1-x)^2 }
          + \frac{1}{2 (1-x)}
          + \frac{282}{ (1+x)^4 }
          - \frac{564}{ (1+x)^3 } \nn\\
& &  \hspace{12mm}
          +  \frac{623}{2 (1+x)^2 } \! 
          -  \! \frac{59}{2 (1+x)}
          \Biggr) H(-1,0,0;x)
       +  \! \zeta(2)   \Biggl(
            \frac{3}{ (1-x)^3 } \nn\\
& &  \hspace{12mm}
          -  \frac{11}{4 (1-x)^2 } \! 
          +  \! \frac{3}{8 (1-x)} \! 
          +  \! \frac{258}{ (1+x)^5 } \! 
          -  \! \frac{744}{ (1+x)^4 } \! 
          +  \! \frac{1515}{2 (1+x)^3 } \nn\\
& &  \hspace{12mm}
          - \frac{621}{2 (1+x)^2 }
          + \frac{307}{8 (1+x)}
          \Biggr) H(0;x)
       + \zeta(3)   \Biggl(
            \frac{13}{4 (1-x)} \nn\\
& &  \hspace{12mm}
          - \frac{324}{ (1+x)^5 }
          + \frac{810}{ (1+x)^4 }
          - \frac{635}{ (1+x)^3 }
          + \frac{285}{2 (1+x)^2 } \nn\\
& &  \hspace{12mm}
          +  \frac{13}{4 (1 \! + \! x)}
          \Biggr) H(0;x) \! 
       +   \!  \Biggl(
          - \frac{127}{18 (1 \! - \! x)} \! 
          - \frac{9}{ (1 \! + \! x)^3 } \! 
          +  \! \frac{27}{2 (1 \! + \! x)^2 } \nn\\
& &  \hspace{12mm}
          + \frac{23}{9 (1+x)}
          \Biggr) H(0;x)
       + \zeta(2)   \Biggl(
          - \frac{9}{2 (1-x)^3 }
          + \frac{27}{4 (1-x)^2 } \nn\\
& &  \hspace{12mm}
          - \frac{21}{8 (1-x)} \! 
          +  \! \frac{90}{ (1+x)^5 }
          - \frac{225}{ (1+x)^4 } \! 
          + \!  \frac{174}{ (1+x)^3 }
          - \frac{36}{ (1+x)^2 } \nn\\
& &  \hspace{12mm}
          - \frac{21}{8 (1 \! + \! x)}
          \Biggr) H(0, \! -1;x) \! 
       +   \!  \Biggl(
          - \frac{1}{(1 \! - \! x)^2 } \! 
          +  \! \frac{1}{(1 \! - \! x)} \! 
          -  \! \frac{12}{ (1 \! + \! x)^4 } \nn\\
& &  \hspace{12mm}
          + \frac{24}{ (1+x)^3 }
          - \frac{21}{ (1+x)^2 }
          + \frac{9}{ (1+x)}
          \Biggr) H(0,-1,0;x) \nn\\
& &  \hspace{12mm}
       +   \Biggl(
          - \frac{1}{2 (1-x)^3 }
          + \frac{3}{4 (1-x)^2 }
          - \frac{5}{8 (1-x)}
          + \frac{282}{ (1+x)^5 } \nn\\
& &  \hspace{12mm}
          -  \frac{705}{ (1 \! + \! x)^4 } \! 
          +  \! \frac{562}{ (1 \! + \! x)^3 } \! 
          -  \! \frac{138}{ (1 \! + \! x)^2 } \! 
          -  \! \frac{5}{8 (1 \! + \! x)}
          \Biggr) H(0, \! -1,0,0;x) \nn\\
& &  \hspace{12mm}
       + \zeta(2)   \Biggl(
            \frac{7}{4 (1-x)^3 }
          - \frac{21}{8 (1-x)^2 }
          - \frac{5}{16 (1-x)}
          + \frac{69}{ (1+x)^5 } \nn\\
& &  \hspace{12mm}
          -  \frac{345}{2 (1+x)^4 } \! 
          +  \! \frac{135}{ (1+x)^3 } \! 
          -  \! \frac{30}{ (1+x)^2 } \! 
          -  \! \frac{5}{16 (1 \! + \! x)} \! 
          \Biggr) H(0,0;x) \nn\\
& &  \hspace{12mm}
       +   \Biggl(
            \frac{3}{ (1-x)^2 } \! 
          -  \! \frac{37}{12 (1-x)} \! 
          -  \! \frac{24}{ (1 \! + \! x)^4 } \! 
          +  \! \frac{39}{ (1 \! + \! x)^3 } \! 
          -  \! \frac{7}{2 (1 \! + \! x)^2 } \nn\\
& &  \hspace{12mm}
          - \frac{137}{12 (1+x)}
          \Biggr) H(0,0;x)
       +   \Biggl(
          - \frac{1}{(1-x)^3 }
          + \frac{3}{2 (1-x)^2 } \nn\\
& &  \hspace{12mm}
          - \frac{25}{4 (1-x)} \! 
          - \frac{12}{ (1+x)^5 } \! 
          +  \! \frac{30}{ (1+x)^4 } \! 
          - \frac{48}{ (1+x)^3 } \! 
          +  \! \frac{42}{ (1+x)^2 } \nn\\
& &  \hspace{12mm}
          - \frac{25}{4 (1+x)}
          \Biggr) H(0,0,-1,0;x)
       +   \Biggl(
            \frac{3}{ (1-x)^3 }
          - \frac{15}{4 (1-x)^2 } \nn\\
& &  \hspace{12mm}
          +  \frac{11}{8 (1-x)} \! 
          +  \! \frac{258}{ (1 \! + \! x)^5 } \! 
          -  \! \frac{816}{ (1+x)^4 } \! 
          +  \! \frac{1803}{2 (1+x)^3 } \! 
          -  \! \frac{767}{2 (1+x)^2 } \nn\\
& &  \hspace{12mm}
          + \frac{315}{8 (1+x)}
          \Biggr) H(0,0,0;x)
       +  \Biggl(
            \frac{3}{4 (1-x)^3 }
          - \frac{9}{8 (1-x)^2 } \nn\\
& &  \hspace{12mm}
          +  \frac{11}{16 (1-x)} \! 
          -  \! \frac{3}{ (1 \! + \! x)^5 } \! 
          +  \! \frac{15}{2 (1 \! + \! x)^4 } \! 
          -  \! \frac{4}{ (1 \! + \! x)^3 } \! 
          -  \! \frac{3}{2 (1 \! + \! x)^2 } \nn\\
& &  \hspace{12mm}
          + \frac{11}{16 (1+x)}
          \Biggr) H(0,0,0,0;x) 
       +   \Biggl(
            \frac{3}{ (1-x)}
          - \frac{48}{ (1+x)^5 } \nn\\
& &  \hspace{12mm}
          +  \frac{120}{ (1 \! + \! x)^4 } \! 
          -  \! \frac{84}{ (1 \! + \! x)^3 } \! 
          +  \! \frac{6}{ (1 \! + \! x)^2 } \! 
          +  \! \frac{3}{ (1 \! + \! x)}
          \Biggr) H(0,0,1,0;x) \nn\\
& &  \hspace{12mm}
       +   \Biggl(
          - \frac{48}{ (1+x)^4 }
          + \frac{96}{ (1+x)^3 }
          - \frac{48 }{(1+x)^2 }
          \Biggr) H(0,1,0;x) \nn\\
& &  \hspace{12mm}
       -   \Biggl(
            \frac{11}{2 (1-x)} \! 
          +  \! \frac{24}{ (1+x)^5 } \! 
          - \frac{60}{ (1+x)^4 } \! 
          +  \! \frac{70}{ (1+x)^3 } \! 
          - \frac{45}{ (1+x)^2 } \nn\\
& &  \hspace{12mm}
          + \frac{11}{2 (1+x)}
          \Biggr) H(0,1,0,0;x)
       + \zeta(2)   \Biggl(
          - \frac{3}{ (1-x)}
          + \frac{336}{ (1+x)^5 } \nn\\
& &  \hspace{12mm}
          - \frac{840}{ (1+x)^4 }
          + \frac{660}{ (1+x)^3 }
          - \frac{150}{ (1+x)^2 }
          - \frac{3}{ (1+x)}
          \Biggr) H(1,0;x) \nn\\
& &  \hspace{12mm}
       -   \Biggl(
            \frac{24}{ (1+x)^3 }
          - \frac{36}{ (1+x)^2 }
          + \frac{12}{ (1+x)}
          \Biggr) H(1,0;x) \nn\\
& &  \hspace{12mm}
       -   \Biggl(
            \frac{24}{ (1+x)^4 } \! 
          -  \! \frac{48}{ (1+x)^3 } \! 
          +  \! \frac{40}{ (1+x)^2 } \! 
          -  \! \frac{16}{ (1 \! + \! x)}
          \Biggr) H(1,0,0;x) \nn\\
& &  \hspace{12mm}
       -  \Biggl(
            \frac{3}{ (1-x)} \! 
          - \frac{336}{ (1+x)^5 } \! 
          +  \! \frac{840}{ (1+x)^4 } \! 
          - \frac{660}{ (1+x)^3 } \! 
          +  \! \frac{150}{ (1+x)^2 } \nn\\
& &  \hspace{12mm}
          + \frac{3}{ (1+x)}
          \Biggr) H(1,0,0,0;x) 
       \Biggr] \nn\\
& & \hspace{6mm}
       + \, C_{F} T_{R} N_{f}   \Biggl[
            \zeta(2)   \Biggl(
          - \frac{2}{3 (1-x)}
          + \frac{2}{3 (1+x)}
          \Biggr)
       +   \Biggl(
          - \frac{4}{3 (1-x)} \nn\\
& &  \hspace{12mm}
          +  \frac{4}{3 (1+x)}
          \Biggr) H(-1,0;x) \! 
       +   \!  \Biggl(
            \frac{25}{9 (1-x)} \! 
          -  \! \frac{25}{9 (1+x)}
          \Biggr) H(0;x) \nn\\
& &  \hspace{12mm}
       +   \Biggl(
            \frac{2}{3 (1-x)}
          - \frac{2}{3 (1+x)}
          \Biggr) H(0,0;x)
       \Biggr] \nn\\
& & \hspace{6mm}
       + \, C_{F} T_{R}   \Biggl[
          - \frac{68}{3 (1 \! + \! x)^2 } \! 
          +  \! \frac{68}{3 (1 \! + \! x)} \! 
       +  \! \zeta(2)   \Biggl(
          - \frac{136}{ (1 \! + \! x)^4 } \! 
          +  \! \frac{272}{ (1 \! + \! x)^3 } \nn\\
& &  \hspace{12mm}
          - \frac{132}{ (1+x)^2 }
          - \frac{4}{ (1+x)}
          \Biggr)
       + \zeta(2)   \Biggl(
          - \frac{3}{2 (1-x)}
          + \frac{24}{ (1+x)^5 } \nn\\
& &  \hspace{12mm}
          - \frac{60}{ (1+x)^4 }
          + \frac{42}{ (1+x)^3 }
          - \frac{3}{ (1+x)^2 }
          - \frac{3}{2 (1+x)}
          \Biggr) H(0;x) \nn\\
& &  \hspace{12mm}
       -   \Biggl(
            \frac{68}{9 (1-x)} \! 
          +  \! \frac{124}{3 (1+x)^3 }
          - \frac{62}{ (1+x)^2 } \! 
          +  \! \frac{118}{9 (1+x)}
          \Biggr) H(0;x) \nn\\
& &  \hspace{12mm}
       -  \Biggl(
            \frac{88}{3 (1 \! + \! x)^4 } \! 
          -  \! \frac{176}{3 (1 \! + \! x)^3 } \! 
          +  \! \frac{92}{3 (1 \! + \! x)^2 } \! 
          -  \! \frac{4}{3 (1 \! + \! x)}
          \Biggr) H(0,0;x) \nn\\
& &  \hspace{12mm}
       -   \Biggl(
            \frac{3}{2 (1-x)} \! 
          - \frac{24}{ (1+x)^5 } \! 
          +  \! \frac{60}{ (1+x)^4 } \! 
          - \frac{42}{ (1+x)^3 } \! 
          +  \! \frac{3}{ (1+x)^2 } \nn\\
& &  \hspace{12mm}
          + \frac{3}{2 (1+x)}
          \Biggr) H(0,0,0;x)
       \Biggr] \nn\\
& & \hspace{6mm}
       + \, C_{F}^2   \Biggl[
            \zeta(2)   \Biggl(
            \frac{48 \log{2}}{ (1+x)^2 }
          - \frac{48 \log{2}}{ (1+x)}
          - \frac{17}{2 (1-x)} \nn\\
& &  \hspace{12mm}
          + \frac{240}{ (1+x)^4 }
          - \frac{528}{ (1+x)^3 }
          + \frac{304}{ (1+x)^2 }
          - \frac{15}{2 (1+x)}
          \Biggr) \nn\\
& &  \hspace{12mm}
       - \zeta^2(2)   \Biggl(
            \frac{69}{20 (1-x)^3 } \! 
          -  \! \frac{207}{40 (1-x)^2 } \! 
          +  \! \frac{327}{80 (1-x)} \! 
          -  \! \frac{171}{ (1 \! + \! x)^5 } \nn\\
& &  \hspace{12mm}
          + \frac{855}{2 (1+x)^4 }
          - \frac{3291}{10 (1+x)^3 }
          + \frac{1323}{20 (1+x)^2 }
          + \frac{327}{80 (1+x)}
          \Biggr) \nn\\
& &  \hspace{12mm}
       + \zeta(3)   \Biggl(
          - \frac{10}{ (1-x)^2 }
          + \frac{10}{ (1-x)}
          - \frac{168}{ (1+x)^4 }
          + \frac{336}{ (1+x)^3 } \nn\\
& &  \hspace{12mm}
          - \frac{182}{ (1+x)^2 }
          + \frac{14}{ (1+x)}
          \Biggr)
       + \zeta(2)   \Biggl(
            \frac{9}{ (1-x)^2 }
          - \frac{9}{ (1-x)} \nn\\
& &  \hspace{12mm}
          - \frac{180}{ (1+x)^4 }
          + \frac{360}{ (1+x)^3 }
          - \frac{231}{ (1+x)^2 }
          + \frac{51}{ (1+x)}
          \Biggr) H(-1;x) \nn\\
& &  \hspace{12mm}
       -  \Biggl(
            \frac{9}{ (1-x)} \! 
          +  \! \frac{192}{ (1+x)^3 } \! 
          - \frac{288}{ (1+x)^2 } \! 
          +  \! \frac{87}{ (1+x)}
          \Biggr) H(-1,0;x)  \nn\\
& &  \hspace{12mm}
       +   \Biggl(
            \frac{1}{(1-x)^2 } \! 
          -  \! \frac{1}{(1-x)} \! 
          +  \! \frac{252}{ (1+x)^4 } \! 
          -  \! \frac{504}{ (1+x)^3 } \! 
          +  \! \frac{257}{ (1+x)^2 }  \nn\\
& &  \hspace{12mm}
          -  \frac{5}{ (1+x)}
          \Biggr) H(-1,0,0;x) \! 
       +  \! \zeta(2)   \Biggl(
          - \frac{19}{2 (1 \! - \! x)^2 } \! 
          +  \! \frac{81}{4 (1 \! - \! x)}  \nn\\
& &  \hspace{12mm}
          + \frac{60}{ (1+x)^5 }
          - \frac{156}{ (1+x)^4 }
          + \frac{131}{ (1+x)^3 }
          - \frac{10}{ (1+x)^2 }  \nn\\
& &  \hspace{12mm}
          - \frac{143}{4 (1+x)}
          \Biggr) H(0;x)
       + \zeta(3)   \Biggl(
          - \frac{7}{2 (1-x)}
          - \frac{168}{ (1+x)^5 }  \nn\\
& &  \hspace{12mm}
          + \frac{420}{ (1+x)^4 }
          - \frac{350}{ (1+x)^3 }
          + \frac{105}{ (1+x)^2 }
          - \frac{7}{2 (1+x)}
          \Biggr) H(0;x)  \nn\\
& &  \hspace{12mm}
       +   \Biggl(
            \frac{31}{4 (1-x)}
          - \frac{31}{4 (1+x)}
          \Biggr) H(0;x)  \nn\\
& &  \hspace{12mm}
       + \zeta(2)   \Biggl(
            \frac{9}{ (1-x)^3 }
          - \frac{27}{2 (1-x)^2 }
          + \frac{21}{4 (1-x)}
          - \frac{180}{ (1+x)^5 }  \nn\\
& &  \hspace{12mm}
          +  \frac{450}{ (1 \! + \! x)^4 } \! 
          -  \! \frac{348}{ (1+x)^3 } \! 
          +  \! \frac{72}{ (1+x)^2 } \! 
          +  \! \frac{21}{4 (1+x)}
          \Biggr) H(0,-1;x)  \nn\\
& &  \hspace{12mm}
       -   \Biggl(
            \frac{2}{ (1-x)^2 } \! 
          - \frac{2}{ (1-x)} \! 
          +  \! \frac{384}{ (1+x)^4 } \! 
          - \frac{768}{ (1+x)^3 } \! 
          +  \! \frac{394}{ (1+x)^2 }  \nn\\
& &  \hspace{12mm}
          - \frac{10}{ (1+x)}
          \Biggr) H(0,-1,0;x)
       +   \Biggl(
            \frac{1}{(1-x)^3 }
          - \frac{3}{2 (1-x)^2 }  \nn\\
& &  \hspace{12mm}
          -  \frac{7}{4 (1-x)} \! 
          +  \! \frac{252}{ (1+x)^5 } \! 
          -  \! \frac{630}{ (1+x)^4 } \! 
          +  \! \frac{496}{ (1+x)^3 } \! 
          -  \! \frac{114}{ (1+x)^2 }  \nn\\
& &  \hspace{12mm}
          - \frac{7}{4 (1+x)}
          \Biggr) H(0,-1,0,0;x)
       +  \zeta(2)   \Biggl(
          - \frac{7}{2 (1-x)^3 }  \nn\\
& &  \hspace{12mm}
          +  \frac{21}{4 (1-x)^2 } \! 
          -  \! \frac{1}{8 (1-x)} \! 
          +  \! \frac{66}{ (1+x)^5 } \! 
          -  \! \frac{165}{ (1 \! + \! x)^4 } \! 
          +  \! \frac{135}{ (1 \! + \! x)^3 }  \nn\\
& &  \hspace{12mm}
          -  \frac{75}{2 (1 \! + \! x)^2 } \! 
          -  \! \frac{1}{8 (1 \! + \! x)} \! 
          \Biggr) H(0,0;x) \! 
       +  \!   \Biggl(
            \frac{2}{ (1-x)^2 } \! 
          +  \! \frac{1}{2 (1-x)}  \nn\\
& &  \hspace{12mm}
          -  \frac{192}{ (1 \! + \! x)^4 } \! 
          +  \! \frac{504}{ (1 \! + \! x)^3 } \! 
          -  \! \frac{442}{ (1 \! + \! x)^2 } \! 
          +  \! \frac{255}{2 (1 \! + \! x)}
          \Biggr) H(0,0;x)  \nn\\
& &  \hspace{12mm}
       +   \Biggl(
            \frac{2}{ (1-x)^3 } \! 
          -  \! \frac{3}{ (1-x)^2 } \! 
          +  \! \frac{14 }{(1-x)} \! 
          -  \! \frac{384}{ (1+x)^5 } \! 
          +  \! \frac{960}{ (1 \! + \! x)^4 }  \nn\\
& &  \hspace{12mm}
          - \frac{714}{ (1+x)^3 }
          + \frac{111}{ (1+x)^2 }
          + \frac{14}{ (1+x)}
          \Biggr) H(0,0,-1,0;x)  \nn\\
& &  \hspace{12mm}
       -   \Biggl(
            \frac{7}{2 (1-x)^2 } \! 
          -  \! \frac{57}{4 (1-x)} \! 
          -  \! \frac{60}{ (1+x)^5 } \! 
          +  \! \frac{216}{ (1+x)^4 } \! 
          -  \! \frac{251}{ (1+x)^3 }  \nn\\
& &  \hspace{12mm}
          + \frac{88}{ (1+x)^2 }
          + \frac{71}{4 (1+x)}
          \Biggr) H(0,0,0;x)
       +   \Biggl(
          - \frac{3}{2 (1-x)^3 }  \nn\\
& &  \hspace{12mm}
          +  \frac{9}{4 (1-x)^2 } \! 
          -  \! \frac{11}{8 (1-x)} \! 
          +  \! \frac{6}{ (1 \! + \! x)^5 } \! 
          -  \! \frac{15}{ (1 \! + \! x)^4 } \! 
          +  \! \frac{8}{ (1 \! + \! x)^3 }  \nn\\
& &  \hspace{12mm}
          + \frac{3}{ (1+x)^2 }
          - \frac{11}{8 (1+x)}
          \Biggr) H(0,0,0,0;x)
       +   \Biggl(
          - \frac{6}{ (1-x)}  \nn\\
& &  \hspace{12mm}
          + \frac{96}{ (1+x)^5 }
          - \frac{240}{ (1+x)^4 }
          + \frac{168}{ (1+x)^3 }
          - \frac{12}{ (1+x)^2 }  \nn\\
& &  \hspace{12mm}
          - \frac{6}{ (1+x)}
          \Biggr) H(0,0,1,0;x)
       +   \Biggl(
          - \frac{4}{ (1-x)^2 }
          + \frac{4}{ (1-x)}  \nn\\
& &  \hspace{12mm}
          +  \frac{96}{ (1 \! + \! x)^4 } \! 
          -  \! \frac{192}{ (1 \! + \! x)^3 } \! 
          +  \! \frac{100}{ (1 \! + \! x)^2 } \! 
          -  \! \frac{4}{ (1 \! + \! x)}
          \Biggr) H(0,1,0;x)  \nn\\
& &  \hspace{12mm}
       +   \Biggl(
            \frac{25}{2 (1-x)} \! 
          -  \! \frac{360}{ (1+x)^5 } \! 
          +  \! \frac{900}{ (1+x)^4 } \! 
          -  \! \frac{670}{ (1+x)^3 } \! 
          +  \! \frac{105}{ (1+x)^2 }  \nn\\
& &  \hspace{12mm}
          + \frac{25}{2 (1+x)}
          \Biggr) H(0,1,0,0;x)
       + \zeta(2)   \Biggl(
            \frac{3}{ (1-x)}
          + \frac{144}{ (1+x)^5 }  \nn\\
& &  \hspace{12mm}
          - \frac{360}{ (1+x)^4 }
          + \frac{300}{ (1+x)^3 }
          - \frac{90}{ (1+x)^2 }
          + \frac{3}{ (1+x)}
          \Biggr) H(1,0;x)  \nn\\
& &  \hspace{12mm}
       -   \Biggl(
                \frac{4}{ (1-x)} \! 
          -  \! \frac{48}{ (1+x)^3 } \! 
          +  \! \frac{72}{ (1+x)^2 } \! 
          -  \! \frac{28}{ (1+x)} \! 
          \Biggr) H(1,0;x)  \nn\\
& &  \hspace{12mm}
       -   \Biggl(
            \frac{4}{ (1-x)^2 } \! 
          - \frac{4}{ (1-x)} \! 
          +  \! \frac{360}{ (1+x)^4 } \! 
          - \frac{720}{ (1+x)^3 } \! 
          +  \! \frac{356}{ (1+x)^2 }  \nn\\
& &  \hspace{12mm}
          +  \frac{4}{ (1+x)}
          \Biggr) H(1,0,0;x) \! 
       +  \!  \Biggl(
            \frac{3}{ (1-x)} \! 
          +  \! \frac{144}{ (1+x)^5 } \! 
          -  \! \frac{360}{ (1+x)^4 }  \nn\\
& &  \hspace{12mm}
          +  \frac{300}{ (1+x)^3 }
          - \frac{90}{ (1+x)^2 } \! 
          +  \! \frac{3}{ (1+x)}
          \Biggr) H(1,0,0,0;x) 
       \Biggr] \, .
\label{FF2fr}
\eea

\boldmath
\subsection{Form Factors for $\mu \neq m$ \label{munotm}}
\unboldmath

In this Section we report the expressions for the renormalized form factors
in the case we keep $\mu \neq m$.

At the one-loop level 
we do not have an explicit dependence on the logarithm of the ratio of the 
renormalization scale and the mass of the heavy quark, 
because an overall factor
$(\mu^2/m^2)^\epsilon$ can be taken out:
\be
F_{i,R}^{(1l)} \Bigl( \epsilon,s, \frac{\mu^2}{m^2} \Bigr) = 
C(\epsilon) \, \left( \frac{\mu^2}{m^2} \right)^{\epsilon} 
{\mathcal F}_{i,R}^{(1l)}(\epsilon,s) \, ,
\label{FF1loopmunotm}
\ee
where the functions ${\mathcal F}_{i,R}^{(1l)}(\epsilon,s)$ are given in 
Eqs.~(\ref{1lrenF1},\ref{1lrenF2}).

At the two-loop level, such a dependence results from the coupling constant 
renormalization, first appearing at this level.
Factoring an overall $(\mu^2/m^2)^{2 \epsilon}$, we have:
\bea
F_{i,R}^{(2l)} \Bigl( \epsilon,s, \frac{\mu^2}{m^2} \Bigr) &=& 
C^2(\epsilon) \, \left( \frac{\mu^2}{m^2} \right)^{2 \epsilon} 
\Biggl\{
{\mathcal F}_{i,R}^{(2l)}(\epsilon,s) 
+ {\mathcal M}_{i}^{(2l)}(\epsilon,s) \log{\left( \frac{\mu^2}{m^2} \right)}
\nn\\
& & \hspace*{30mm}
+ {\mathcal N}_{i}^{(2l)}(s) \log^2{\left( \frac{\mu^2}{m^2} \right)}
\Biggr\} 
\, ,
\label{FF2loopmunotm}
\eea
where the functions ${\mathcal F}_{i,R}^{(2l)}(\epsilon,s)$ are given in 
Eqs.~(\ref{FF1fr},\ref{FF2fr}) and the functions 
${\mathcal M}_{i}^{(2l)}(\epsilon,s)$ and ${\mathcal N}_{i}^{(2l)}(s)$ 
can be derived from the renormalization group equation.

Introducing 
$\beta_0$, the first coefficient of the QCD $\beta$-function:
\be
\beta_0 = \frac{11 C_A - 4 T_R (N_f+1)}{6} \, ,
\ee
we can write 
\bea
\hspace{-5mm}
F_{i,R}^{(2l)} \Bigl( \epsilon,s, \frac{\mu^2}{m^2} \Bigr) \! \! &=& 
\left( \frac{\alpha_S}{2 \pi} \right)^2 \,
C^2(\epsilon) \, \left( \frac{\mu^2}{m^2} \right)^{2 \epsilon}   \nn\\
& & \hspace*{15mm} 
\times \Biggl\{
{\mathcal F}_{i,R}^{(2l)}(\epsilon,s)
- \frac{\beta_0}{\epsilon} \left[ \left( \frac{\mu^2}{m^2} 
\right)^{- \epsilon} -1 \right]
{\mathcal F}_{i,R}^{(1l)}(\epsilon,s) \Biggr\}  
\label{eq:reno1} \\
\hspace{-5mm}
& = & \left( \frac{\alpha_S}{2 \pi} \right)^2 \,
C^2(\epsilon) \, \left( \frac{\mu^2}{m^2} \right)^{2 \epsilon}  \Biggl\{
{\mathcal F}_{i,R}^{(2l)}(\epsilon,s)
+ \beta_0 \,
{\mathcal F}_{1,R}^{(1l)}(\epsilon,s) \log{\left( \frac{\mu^2}{m^2} \right)}
\nn\\
& & \hspace*{45mm}
-  \epsilon \, \frac{\beta_0}{2}
{\mathcal F}_{1,R}^{(1l)}(\epsilon,s) 
\log^2{\left( \frac{\mu^2}{m^2} \right)}
\Biggr\} 
\, .\label{eq:reno2}
\eea
From these, the coefficients appearing in Eq.~(\ref{FF2loopmunotm}) can be 
readily read off:
\bea
{\mathcal M}_{1}^{(2l)}(\epsilon,s) & = & 
        \frac{1}{\epsilon} \Biggl\{
	- \, C_{F} C_{A} \, \frac{11}{6}  \Biggl[
            1
       - \Biggl(
            1
          - \frac{1}{ (1-x)}
          - \frac{1}{ (1+x)}
          \Biggr) H(0;x)  \nn\\
& & \hspace*{8mm}
       + \, C_{F} T_{R} ( N_f +1 ) \, \frac{2}{3}  \Biggl[
             1
       - \Biggl(
            1
          - \frac{1}{(1-x)}
          - \frac{1}{(1+x)}
          \Biggr) H(0;x) \Biggr] 
\Biggr\} \nn\\
& & 
       + C_{A} C_{F} \frac{11}{6} \Biggl[ 
          - 2
          + \frac{\zeta(2)}{ (1-x)}
          + \frac{\zeta(2)}{ (1+x)}
          - \zeta(2)
       -  \Biggl(
            2
          - \frac{2}{ (1-x)} \nn\\
& & \hspace*{12mm}
          - \frac{2}{ (1+x)}
          \Biggr) H(-1,0,x)
       +  \Biggl(
            \frac{3}{2}
          - \frac{1}{2 (1-x)}
          - \frac{1}{(1+x)}
          \Biggr) H(0,x) \nn\\
& & \hspace*{12mm}
       +   \Biggl(
            1
          - \frac{1}{ (1-x)}
          - \frac{1}{ (1+x)}
           \Biggr) H(0,0,x) \Biggl] \nn\\
& & 
       + C_{F} T_{R} ( N_{f} + 1 ) \, \frac{2}{3}   \Biggl[
            2
          - \frac{\zeta(2)}{ (1-x)}
          - \frac{\zeta(2)}{ (1+x)}
          + \zeta(2)
       +  \Biggl(
            2
          - \frac{2}{ (1-x)} \nn\\
& & \hspace*{12mm}
          - \frac{2}{ (1+x)}
          \Biggr) H(-1,0,x)
       -  \Biggl(
            \frac{3}{2}
          - \frac{2}{ (1-x)}
          - \frac{1}{ (1+x)}
          \Biggr) H(0,x) \nn\\
& & \hspace*{12mm}
       -  \Biggl(
            1
          - \frac{1}{ (1-x)}
          - \frac{1}{ (1+x)}
          \Biggr) H(0,0,x) \Biggl] \, , 
\label{UN} \\
{\mathcal N}_{1}^{(2l)}(s) & = & 
         C_{A} C_{F}  \frac{11}{12} \Biggl[
            1
       - \Biggl(
            1
          - \frac{1}{(1-x)}
          - \frac{1}{(1+x)}
          \Biggr) H(0;x) \Biggr] \nn\\
& & 
       - \, C_{F} T_{R} ( N_{f} + 1 )  \frac{1}{3} \Biggl[
            1
       - \Biggl(
            1
          - \frac{1}{(1-x)}
          - \frac{1}{(1+x)}
          \Biggr) H(0;x)\Biggr]  \, ,
\\
{\mathcal M}_{2}^{(2l)}(\epsilon,s) & = & 
       - C_{A} C_{F}  \frac{11}{6} \Biggl[ \Biggl(
            \frac{1}{(1-x)}
          - \frac{1}{(1+x)}
          \Biggr) H(0;x) \Biggr] \nn\\
& & 
       + \, C_{F} T_{R} (N_{f} + 1) \frac{2}{3}  \Biggl[ \Biggl(
            \frac{1}{(1-x)}
          - \frac{1}{(1+x)}
          \Biggr) H(0;x) \Biggr] \, ,
\\
{\mathcal N}_{2}^{(2l)}(s) & = & 0 \, . 
\label{UL} 
\eea

\section{Analytical Continuation above Threshold \label{analytical}}

Eqs.~(\ref{1lrenF1},\ref{1lrenF2},\ref{FF1fr},\ref{FF2fr}) are written in terms
of the ``space-like'' variable $x$, defined in Eq.~(\ref{xvar}), for negative
c.m. energy squared $S<0$. In particular, for $S<0$, $x$ is real and positive
and varies from $x=1$, when $S=0$, to $x=0$ when $S=- \infty$.

The physical form factors, defined in the time-like region $S=Q^2>0$ (in 
particular above the physical threshold $S>4m^2$, where an imaginary part
appears) can be recovered by analytical continuation with the usual 
$i \epsilon$-prescription, i.e. giving a small positive imaginary part to 
$S$: $S+i \epsilon$. 

In so doing, if $S>0$, but still $S<4m^2$, the variable $x$ becomes a phase
factor:
\be
x = r = \frac{\sqrt{4m^2-S}-\sqrt{-S-i \epsilon}}{\sqrt{4m^2-S}+\sqrt{-S-i
\epsilon}} = \frac{\sqrt{4m^2-S}+i\sqrt{S}}{\sqrt{4m^2-S}-i\sqrt{S}} 
= e^{i2 \phi} \, ,
\ee
where:
\be
\phi = {\mathrm{arctan}} \sqrt{\frac{S}{4m^2-S}} \, .
\ee

Above threshold, $S>4m^2$, we define:
\be
y = \frac{\sqrt{S} - \sqrt{S-4m^2} }{\sqrt{S} + \sqrt{S-4m^2} } \, ,
\ee
with $y=1$ at $S=4m^2$ and $y=0$ at $S= \infty$, and the continuation in $x$ is
performed by the replacement:
\be
x \rightarrow - y + i \epsilon \, .
\ee

The real and imaginary parts of the form factors are defined through the
relations:
\bea
F_{1,R}(\epsilon,s+i\epsilon) & = & \Re \, F_{1,R}(\epsilon,s) 
                         + i \pi \, \Im \, F_{1,R}(\epsilon,s) \, , \\ 
F_{2,R}(\epsilon,s+i\epsilon) & = & \Re \, F_{2,R}(\epsilon,s) 
                         + i \pi \, \Im \, F_{2,R}(\epsilon,s) \, ,
\eea
where $s=S/m^2$.

In the following two Sections we will give real and imaginary parts of
the one- and two-loop analytically continued form factors for $\mu=m$. 
The renormalization scale dependence follows from the pattern outlined 
in Section \ref{munotm}.

\subsection{One-Loop Form Factors above Threshold}

As in Section \ref{FFrenorm} we write:
\be
F_{i,R}^{(1l)}(\epsilon,s) = 
C(\epsilon) \, 
{\mathcal F}_{i,R}^{(1l)}(\epsilon,s) \qquad \mbox{with} \; \, i=1,2
\, .
\ee

We have:
\bea
\hspace{-5mm}
\Re \, {\mathcal F}^{(1l)}_{1,R}(\epsilon,s) & = & C_{F} \Biggl\{  
\frac{1}{\epsilon} \Biggl[ 
- 1 + \biggl( 1 - \frac{1}{1-y} - \frac{1}{1+y} \biggr) H(0;y)
\Biggr] \nn\\
\hspace{-5mm}
& & \hspace{6mm}
          - 2
          - \left( \frac{1}{2}
          - \frac{1}{1-y} \right) H(0;y) \nn\\
\hspace{-5mm}
& &  \hspace{6mm}
        - \biggl( 1 - \frac{1}{1-y} - \frac{1}{1+y} \biggr)  \bigl[
            4 \zeta(2)
          - 2 H(0;y) 
          - H(0,0;y) \nn\\
\hspace{-5mm}
& &  \hspace{11mm}
          - 2 H(1,0;y) \bigr]
 \nn\\
\hspace{-5mm}
& &  \hspace{6mm}
  - \, \epsilon \Biggl[ 
            4 \! 
       - \left( \frac{1}{2} 
          - \frac{1}{1-y} \right) \bigl[
            4 \zeta(2) \! 
          - H(0,0;y) \! 
          - 2 H(1,0;y)
             \bigr] \nn\\
\hspace{-5mm}
& &   \hspace{11mm}
       + \biggl( 1 - \frac{1}{1-y} - \frac{1}{1+y} \biggr)  \bigl[
            8 \zeta(2)
          + 2 \zeta(3) \nn\\
\hspace{-5mm}
& &   \hspace{16mm}
          - 4( 1- \zeta(2) ) H(0;y)
          + 8 \zeta(2) H(1;y) 
          - 2 H(0,0;y) \nn\\
\hspace{-5mm}
& &   \hspace{16mm}
          - 4 H(1,0;y)
          - H(0,0,0;y)
          - 2 H(0,1,0;y) \nn\\
\hspace{-5mm}
& &   \hspace{16mm}
          - 2 H(1,0,0;y)
          - 4 H(1,1,0;y)
           \bigr] 
      \Biggr\} 
+ \, {\mathcal O} \left( \epsilon^2 \right) \, , \\
\hspace{-5mm}
\hspace{-5mm}
\Im \, {\mathcal F}^{(1l)}_{1,R}(\epsilon,s) & = & C_{F} \Biggl\{  
\frac{1}{\epsilon} \Biggl[
1 - \frac{1}{1-y} - \frac{1}{1+y} 
\Biggr]  \nn\\
\hspace{-5mm}
& &  \hspace{6mm}
+ \frac{1}{2} - \frac{1}{(1+y)} + \biggl( 1 - 
\frac{1}{1-y} - \frac{1}{1+y} \biggr) \bigl[ 1 + H(0;y) \nn\\
\hspace{-5mm}
& & \hspace{10mm}
+ 2 H(1;y) \bigr] \nn\\
\hspace{-5mm}
& & \hspace{6mm}
+ \, \epsilon \, \Biggl[ \frac{1}{2} \biggl(
          1 
        - \frac{2}{1+y} \biggr) \bigl[
          H(0;y)
        + 2 H(1;y) \bigr] \nn\\
\hspace{-5mm}
& & \hspace*{13mm}
      + \biggl(
          1 \! 
        -  \! \frac{1}{1-y}  \! 
        -  \! \frac{1}{1+y} \biggr) \bigl[ 
          4 \! 
        -  \! 2 \zeta(2) \! 
        +  \! H(0;y) \nn\\
\hspace{-5mm}
& & \hspace*{18mm}
        + 2 H(1;y)
        + H(0,0;y)
        + 2 H(0,1;y) \nn\\
\hspace{-5mm}
& & \hspace*{18mm}
        + 2 H(1,0;y)
        + 4 H(1,1;y) \bigr] \Biggr] \Biggr\}  
+ \, {\mathcal O} \left( \epsilon^2 \right) \, , \\
\Re \, {\mathcal F}^{(1l)}_{2,R}(\epsilon,s) & = & C_{F} \, \Biggl\{
   \left( \frac{1}{1-y} -
          \frac{1}{1+y} \right) H(0;y) \nn\\
\hspace{-5mm}
&  &   \hspace*{6mm}
- \, \epsilon \Biggl[ \left( \frac{1}{1-y} -
\frac{1}{1+y} \right) \bigl[ 
             4 \zeta(2) 
           - 4 H(0;y)   \nn\\
\hspace{-5mm}
& & \hspace*{11mm}
           - H(0,0;y) 
           - 2 H(1,0;y)
\bigr] \Biggl] \Biggr\} 
+ \, {\mathcal O} \left( \epsilon^2 \right) \, , \\
\hspace{-5mm}
\Im \, {\mathcal F}^{(1l)}_{2,R}(\epsilon,s) & = & C_{F} \, \Biggl\{
        \frac{1}{1-y} - \frac{1}{1+y} 
+ \, \epsilon \Biggl[  \left( \frac{1}{1-y} -
\frac{1}{1+y} \right) \bigl[
          4
        + H(0;y) \nn\\
\hspace{-5mm}
& & \hspace*{8mm}
        + 2 H(1;y) \bigr]
\Biggr] \Biggr\} 
+ \, {\mathcal O} \left( \epsilon^2 \right) \, .
\eea

\subsection{Two-Loop Form Factors above Threshold}

We define:
\be
F_{i,R}^{(2l)}(\epsilon,s) = 
C^2(\epsilon) \, 
{\mathcal F}_{i,R}^{(2l)}(\epsilon,s) \qquad \mbox{with} \; \, i=1,2
\, .
\ee

We have:
\bea
\hspace{-10mm}
\Re \, {\mathcal F}^{(2l)}_{1,R}(\epsilon,s) & = &  \frac{1}{\epsilon^2}
    \Biggl\{
           \frac{11}{12} C_{F} C_{A}  \Biggl[
         1
       -  \Biggl(
            1
          - \frac{1}{ (1-y)}
          - \frac{1}{ (1+y)}
          \Biggr) H(0,y) 
          \Biggr] \nn\\
\hspace{-10mm} & &  \hspace{6mm}
       - \frac{1}{3} C_{F} T_{R} N_{f}   
          \Biggl[
       1
       -   \Biggl(
           1
          - \frac{1}{ (1-y)}
          - \frac{1}{ (1+y)}
          \Biggr) H(0,y)
           \Biggl]  \nn\\
\hspace{-10mm} & &  \hspace{6mm}
       + \, C_{F}^2   \Biggl[
         \frac{1}{2}
       - \zeta(2)   \Biggl(
            3
          + \frac{3}{ (1-y)^2}
          - \frac{3}{ (1-y)}
          + \frac{3}{ (1+y)^2}  \nn\\
\hspace{-10mm} & &  \hspace{12mm}
          - \frac{3}{ (1+y)}
          \Biggr)
       -   \Biggl(
            1
          - \frac{1}{(1-y)}
          - \frac{1}{(1+y)}
          \Biggr) H(0,y)  \nn\\
\hspace{-10mm} & &  \hspace{12mm}
       +   \Biggl(
            1
          + \frac{1}{(1-y)^2}
          - \frac{1}{(1-y)}
          + \frac{1}{(1+y)^2}  \nn\\
\hspace{-10mm} & &  \hspace{12mm}
          - \frac{1}{(1+y)}
          \Biggr) H(0,0,y)
          \Biggr]
     \Biggr\} \nn\\
\hspace{-10mm} & & + \frac{1}{\epsilon}  \Biggl\{ 
        C_{F} C_{A}  \Biggl[ 
       - \frac{49}{36}
       - \zeta(2)   \Biggl(
            \frac{9}{2}
          - \frac{5}{2 (1-y)}
          - \frac{5}{2 (1+y)}
          \Biggr)  \nn\\
\hspace{-10mm} & &  \hspace{12mm}
       - \zeta(3)   \Biggl(
            \frac{1}{2}
          + \frac{1}{2 (1-y)^2}
          - \frac{1}{2 (1-y)}
          + \frac{1}{2 (1+y)^2}  \nn\\
\hspace{-10mm} & &  \hspace{12mm}
          - \frac{1}{2 (1\! +\! y)}
          \Biggr)
       -  \!  \Biggl(
            1\! 
          - \! \frac{1}{(1-y)}\! 
          - \! \frac{1}{(1\! + \! y)}
          \Biggr) H(-1,0,y)  \nn\\
\hspace{-10mm} & &  \hspace{12mm}
       + \zeta(2)   \Biggl(
            \frac{9}{2}
          + \frac{5}{2 (1-y)^2}
          - \frac{9}{2 (1-y)}
          + \frac{5}{2 (1+y)^2}  \nn\\
\hspace{-10mm} & &  \hspace{12mm}
          - \frac{9}{2 (1 + y)}
          \Biggr) H(0,y) 
       +    \Biggl(
            \frac{67}{36} 
          -  \frac{67}{36 (1-y)}  \nn\\
\hspace{-10mm} & &  \hspace{12mm}
          -  \frac{67}{36 (1 + y)}
          \Biggr) H(0,y) 
       +   \Biggl(
            1
          + \frac{1}{(1-y)^2}
          - \frac{1}{(1-y)} \nn\\
\hspace{-10mm} & &  \hspace{12mm}
          +  \! \frac{1}{(1 \! + \! y)^2} 
          -  \! \frac{1}{(1 \! + \! y)}
          \Biggr) H(0,-1,0,y) \! 
       +   \!  \Biggl(
           2
          - \frac{1}{(1-y)} \nn\\
\hspace{-10mm} & &  \hspace{12mm}
          - \frac{1}{(1+y)}
          \Biggr) H(0,0,y)
       -   \Biggl(
            2
          + \frac{1}{(1-y)^2}
          - \frac{2}{ (1-y)} \nn\\
\hspace{-10mm} & &  \hspace{12mm}
          + \frac{1}{(1+y)^2}
          - \frac{2}{ (1+y)}
          \Biggr) H(0,0,0,y)
       -   \Biggl(
            1
          + \frac{1}{(1-y)^2} \nn\\
\hspace{-10mm} & &  \hspace{12mm}
          - \frac{1}{(1-y)}
          + \frac{1}{(1+y)^2}
          - \frac{1}{(1+y)}
          \Biggr) H(0,1,0,y) \nn\\
\hspace{-10mm} & &  \hspace{12mm}
       +   \Biggl(
            1
          - \frac{1}{(1-y)}
          - \frac{1}{(1+y)}
          \Biggr) H(1,0,y)
          \Biggr] \nn\\
\hspace{-10mm} & &  \hspace{6mm}
       + \, C_{F} T_{R} N_{f}   
          \Biggl[
         \frac{5}{9}
       -   \Biggl(
            \frac{5}{9}
          - \frac{5}{9 (1-y)}
          - \frac{5}{9 (1+y)}
          \Biggr) H(0,y)
           \Biggl] \nn\\
\hspace{-10mm} & &  \hspace{6mm}
       + \, C_{F}^2   \Biggl[ 
         2
       - \zeta(2)   \Biggl(
            5
          + \frac{6}{ (1-y)^2}
          - \frac{2}{ (1-y)}
          + \frac{12}{ (1+y)^2} \nn\\
\hspace{-10mm} & &  \hspace{12mm}
          - \frac{8}{ (1+y)}
          \Biggr)
       - \zeta(2)   \Biggl(
            10
          + \frac{10}{ (1-y)^2}
          - \frac{10}{ (1-y)} \nn\\
\hspace{-10mm} & &  \hspace{12mm}
          + \frac{10}{ (1+y)^2}
          - \frac{10}{ (1+y)}
          \Biggr) H(0,y)
       -   \Biggl(
            \frac{7}{2}
          - \frac{3}{ (1-y)} \nn\\
\hspace{-10mm} & &  \hspace{12mm}
          - \frac{4}{ (1+y)}
          \Biggr) H(0,y)
       +   \Biggl(
            2
          + \frac{2}{ (1-y)^2}
          - \frac{1}{(1-y)} \nn\\
\hspace{-10mm} & &  \hspace{12mm}
          + \frac{4}{ (1+y)^2}
          - \frac{3}{ (1+y)}
          \Biggr) H(0,0,y)
       +   \Biggl(
            3
          + \frac{3}{ (1-y)^2} \nn\\
\hspace{-10mm} & &  \hspace{12mm}
          - \frac{3}{ (1-y)}
          + \frac{3}{ (1+y)^2}
          - \frac{3}{ (1+y)}
          \Biggr) H(0,0,0,y) \nn\\
\hspace{-10mm} & &  \hspace{12mm}
       +   \Biggl(
            2
          + \frac{2}{ (1-y)^2}
          - \frac{2}{ (1-y)}
          + \frac{2}{ (1+y)^2} \nn\\
\hspace{-10mm} & &  \hspace{12mm}
          - \frac{2}{ (1+y)}
          \Biggr) H(0,1,0,y)
       - \zeta(2)   \Biggl(
            12
          + \frac{12}{ (1-y)^2} \nn\\
\hspace{-10mm} & &  \hspace{12mm}
          - \frac{12}{ (1-y)}
          + \frac{12}{ (1+y)^2}
          - \frac{12}{ (1+y)}
          \Biggr) H(1,y) \nn\\
\hspace{-10mm} & &  \hspace{12mm}
       -   \Biggl(
            2
          - \frac{2}{ (1-y)}
          - \frac{2}{ (1+y)}
          \Biggr) H(1,0,y) \nn\\
\hspace{-10mm} & &  \hspace{12mm}
       +  \Biggl(
            4
          + \frac{4}{ (1-y)^2}
          - \frac{4}{ (1-y)}
          + \frac{4}{ (1+y)^2} \nn\\
\hspace{-10mm} & &  \hspace{12mm}
          - \frac{4}{ (1+y)}
          \Biggr) H(1,0,0,y) 
          \Biggr]
\Biggr\} \nn\\
\hspace{-10mm} & & 
    + \, C_{F} C_{A}  \Biggl[ 
          - \frac{1595}{108} \! 
          +  \! \frac{3}{ (1 \! - \! y)^2} \! 
          -  \! \frac{3}{ (1 \! - \! y)} \! 
       +  \! \zeta(2)   \Biggl(
            \frac{76}{9} \! 
          +  \! \frac{36 \log{2}}{ (1 \! - \! y)^2} \nn\\
\hspace{-10mm} & &  \hspace{12mm}
          - \frac{36 \log{2}}{ (1-y)}
          + 6 \log{2}
          - \frac{240}{ (1-y)^4}
          + \frac{480}{ (1-y)^3}
          - \frac{276}{ (1-y)^2} \nn\\
\hspace{-10mm} & &  \hspace{12mm}
          + \frac{344}{9 (1-y)}
          + \frac{77}{9 (1+y)}
          \Biggr)
       - \zeta^2(2)   \Biggl(
            \frac{491}{20}
          - \frac{618}{5 (1-y)^5} \nn\\
\hspace{-10mm} & &  \hspace{12mm}
          + \frac{309}{ (1-y)^4}
          - \frac{1208}{5 (1-y)^3}
          + \frac{269}{4 (1-y)^2}
          - \frac{249}{10 (1-y)} \nn\\
\hspace{-10mm} & &  \hspace{12mm}
          + \frac{277}{20 (1+y)^2}
          - \frac{107}{5 (1+y)}
          \Biggr)
       + \zeta(3)   \Biggl(
            \frac{5}{6}
          + \frac{324}{ (1-y)^4} \nn\\
\hspace{-10mm} & &  \hspace{12mm}
          - \frac{648}{ (1-y)^3}
          + \frac{374}{ (1-y)^2}
          - \frac{269}{6 (1-y)}
          + \frac{31}{6 (1+y)}
          \Biggr) \nn\\
\hspace{-10mm} & &  \hspace{12mm}
       - \zeta(2)   \Biggl(
            11
          - \frac{72}{ (1-y)^4}
          + \frac{144}{ (1-y)^3}
          - \frac{114}{ (1-y)^2} \nn\\
\hspace{-10mm} & &  \hspace{12mm}
          + \frac{55}{ (1-y)}
          + \frac{13}{ (1+y)}
          \Biggr) H(-1,y)
       + \zeta(3)   \Biggl(
            2
          + \frac{2}{ (1-y)^2} \nn\\
\hspace{-10mm} & &  \hspace{12mm}
          - \frac{2}{ (1-y)}
          + \frac{2}{ (1+y)^2}
          - \frac{2}{ (1+y)}
          \Biggr) H(-1,y) \nn\\
\hspace{-10mm} & &  \hspace{12mm}
       +   \Biggl(
            2
          - \frac{2}{ (1-y)}
          - \frac{2}{ (1+y)}
          \Biggr) H(-1,-1,0,y) \nn\\
\hspace{-10mm} & &  \hspace{12mm}
       + \zeta(2)   \Biggl(
            2
          - \frac{672}{ (1-y)^5}
          + \frac{1680}{ (1-y)^4}
          - \frac{1376}{ (1-y)^3}
          + \frac{374}{ (1-y)^2} \nn\\
\hspace{-10mm} & &  \hspace{12mm}
          - \frac{30}{ (1-y)}
          - \frac{10}{ (1+y)^2}
          + \frac{10}{ (1+y)}
          \Biggr) H(-1,0,y) \nn\\
\hspace{-10mm} & &  \hspace{12mm}
       +   \Biggl(
            3
          - \frac{24}{ (1-y)^3}
          + \frac{36}{ (1-y)^2}
          - \frac{16}{ (1-y)} \nn\\
\hspace{-10mm} & &  \hspace{12mm}
          - \frac{2}{ (1+y)}
          \Biggr) H(-1,0,y)
       -   \Biggl(
            4
          + \frac{4}{ (1-y)^2}
          - \frac{4}{ (1-y)} \nn\\
\hspace{-10mm} & &  \hspace{12mm}
          + \frac{4}{ (1+y)^2}
          - \frac{4}{ (1+y)}
          \Biggr) H(-1,0,-1,0,y) \nn\\
\hspace{-10mm} & &  \hspace{12mm}
       +   \Biggl(
            4
          - \frac{24}{ (1-y)^4}
          + \frac{48}{ (1-y)^3}
          - \frac{38}{ (1-y)^2}
          + \frac{18}{ (1-y)} \nn\\
\hspace{-10mm} & &  \hspace{12mm}
          + \frac{4}{ (1+y)}
          \Biggr) H(-1,0,0,y)
       -   \Biggl(
            2
          - \frac{336}{ (1-y)^5}
          + \frac{840}{ (1-y)^4} \nn\\
\hspace{-10mm} & &  \hspace{12mm}
          - \frac{688}{ (1-y)^3}
          + \frac{188}{ (1-y)^2}
          - \frac{16}{ (1-y)}
          - \frac{4}{ (1+y)^2} \nn\\
\hspace{-10mm} & &  \hspace{12mm}
          + \frac{4}{ (1+y)}
          \Biggr) H(-1,0,0,0,y)
       +   \Biggl(
            4
          + \frac{4}{ (1-y)^2} \nn\\
\hspace{-10mm} & &  \hspace{12mm}
          - \frac{4}{ (1-y)}
          + \frac{4}{ (1+y)^2}
          - \frac{4}{ (1+y)}
          \Biggr) H(-1,0,1,0,y) \nn\\
\hspace{-10mm} & &  \hspace{12mm}
       -   \Biggl(
            6
          - \frac{6}{ (1-y)}
          - \frac{6}{ (1+y)}
          \Biggr) H(-1,1,0,y)
       - \zeta(2)   \Biggl(
            \frac{119}{3} \nn\\
\hspace{-10mm} & &  \hspace{12mm}
          - \frac{516}{ (1-y)^5}
          + \frac{1704}{ (1-y)^4}
          - \frac{1990}{ (1-y)^3}
          + \frac{951}{ (1-y)^2} \nn\\
\hspace{-10mm} & &  \hspace{12mm}
          - \frac{554}{3 (1-y)}
          - \frac{23}{3 (1+y)}
          \Biggr) H(0,y)
       - \zeta(3)   \Biggl(
            \frac{15}{2}
          - \frac{324}{ (1-y)^5} \nn\\
\hspace{-10mm} & &  \hspace{12mm}
          + \frac{810}{ (1-y)^4}
          - \frac{662}{ (1-y)^3} 
          + \frac{367}{2 (1-y)^2}
          - \frac{20}{ (1-y)} \nn\\
\hspace{-10mm} & &  \hspace{12mm}
          + \frac{1}{2 (1+y)^2}
          - \frac{2}{ (1+y)}
          \Biggr) H(0,y)
       +   \Biggl(
            \frac{2545}{216}
          + \frac{9}{ (1-y)^3} \nn\\
\hspace{-10mm} & &  \hspace{12mm}
          - \frac{27}{2 (1-y)^2}
          - \frac{599}{108 (1-y)}
          - \frac{365}{27 (1+y)}
          \Biggr) H(0,y) \nn\\
\hspace{-10mm} & &  \hspace{12mm}
       - \zeta(2)   \Biggl(
            7
          - \frac{72}{ (1-y)^5}
          + \frac{180}{ (1-y)^4}
          - \frac{216}{ (1-y)^3} \nn\\
\hspace{-10mm} & &  \hspace{12mm}
          + \frac{157}{ (1-y)^2}
          - \frac{34}{ (1-y)}
          + \frac{13}{ (1+y)^2}
          - \frac{16}{ (1+y)}
          \Biggr) H(0,-1,y) \nn\\
\hspace{-10mm} & &  \hspace{12mm}
       -   \Biggl(
            2
          + \frac{2}{ (1-y)^2}
          - \frac{2}{ (1-y)}
          + \frac{2}{ (1+y)^2} \nn\\
\hspace{-10mm} & &  \hspace{12mm}
          - \frac{2}{ (1+y)}
          \Biggr) H(0,-1,-1,0,y)
       -   \Biggl(
            8
          + \frac{48}{ (1-y)^4} \nn\\
\hspace{-10mm} & &  \hspace{12mm}
          - \frac{96}{ (1-y)^3}
          + \frac{56}{ (1-y)^2}
          - \frac{12}{ (1-y)} \nn\\
\hspace{-10mm} & &  \hspace{12mm}
          - \frac{4}{ (1+y)}
          \Biggr) H(0,-1,0,y)
       +   \Biggl(
            2
          - \frac{24}{ (1-y)^5}
          + \frac{60}{ (1-y)^4} \nn\\
\hspace{-10mm} & &  \hspace{12mm}
          - \frac{72}{ (1-y)^3}
          + \frac{52}{ (1-y)^2}
          - \frac{11}{ (1-y)}
          + \frac{4}{ (1+y)^2} \nn\\
\hspace{-10mm} & &  \hspace{12mm}
          - \frac{5}{ (1+y)}
          \Biggr) H(0,-1,0,0,y)
       +   \Biggl(
            6
          + \frac{6}{ (1-y)^2} \nn\\
\hspace{-10mm} & &  \hspace{12mm}
          - \frac{6}{ (1-y)}
          + \frac{6}{ (1+y)^2}
          - \frac{6}{ (1+y)}
          \Biggr) H(0,-1,1,0,y) \nn\\
\hspace{-10mm} & &  \hspace{12mm}
       + \zeta(2)   \Biggl(
            37
          - \frac{78}{ (1-y)^5}
          + \frac{195}{ (1-y)^4}
          - \frac{307}{2 (1-y)^3} \nn\\
\hspace{-10mm} & &  \hspace{12mm}
          + \frac{217}{4 (1-y)^2} \! 
          -  \! \frac{295}{8 (1-y)} \! 
          +  \! \frac{19}{ (1 \! + \! y)^2} \! 
          -  \! \frac{287}{8 (1 \! + \! y)}
          \Biggr) H(0,0,y) \nn\\
\hspace{-10mm} & &  \hspace{12mm}
       -   \Biggl(
            \frac{217}{36}
          - \frac{24}{ (1-y)^4}
          + \frac{39}{ (1-y)^3}
          - \frac{3}{2 (1-y)^2}
          - \frac{130}{9 (1-y)} \nn\\
\hspace{-10mm} & &  \hspace{12mm}
          + \frac{25}{18 (1+y)}
          \Biggr) H(0,0,y)
       +   \Biggl(
            14
          - \frac{48}{ (1-y)^5}
          + \frac{120}{ (1-y)^4} \nn\\
\hspace{-10mm} & &  \hspace{12mm}
          - \frac{88}{ (1-y)^3}
          + \frac{20}{ (1-y)^2}
          - \frac{12}{ (1-y)}
          + \frac{8}{ (1+y)^2} \nn\\
\hspace{-10mm} & &  \hspace{12mm}
          - \frac{12}{ (1+y)}
          \Biggr) H(0,0,-1,0,y)
       +   \Biggl(
            \frac{89}{6}
          - \frac{258}{ (1-y)^5} \nn\\
\hspace{-10mm} & &  \hspace{12mm}
          + \frac{816}{ (1-y)^4}
          - \frac{923}{ (1-y)^3}
          + \frac{436}{ (1-y)^2}
          - \frac{259}{3 (1-y)} \nn\\
\hspace{-10mm} & &  \hspace{12mm}
          - \frac{4}{3 (1+y)}
          \Biggr) H(0,0,0,y)
       -   \Biggl(
            12
          - \frac{3}{ (1-y)^5} \nn\\
\hspace{-10mm} & &  \hspace{12mm}
          + \frac{15}{2 (1-y)^4}
          - \frac{17}{4 (1-y)^3}
          + \frac{39}{8 (1-y)^2}
          - \frac{177}{16 (1-y)} \nn\\
\hspace{-10mm} & &  \hspace{12mm}
          + \frac{6}{ (1+y)^2}
          - \frac{193}{16 (1+y)}
          \Biggr) H(0,0,0,0,y) \nn\\
\hspace{-10mm} & &  \hspace{12mm}
       -   \Biggl(
            22
          + \frac{12}{ (1-y)^5}
          - \frac{30}{ (1-y)^4}
          + \frac{49}{ (1-y)^3}
          - \frac{63}{2 (1-y)^2} \nn\\
\hspace{-10mm} & &  \hspace{12mm}
          - \frac{51}{4 (1-y)}
          + \frac{12}{ (1+y)^2}
          - \frac{75}{4 (1+y)}
          \Biggr) H(0,0,1,0,y) \nn\\
\hspace{-10mm} & &  \hspace{12mm}
       + \zeta(2)   \Biggl(
            35
          - \frac{756}{ (1-y)^5}
          + \frac{1890}{ (1-y)^4}
          - \frac{1575}{ (1-y)^3} \nn\\
\hspace{-10mm} & &  \hspace{12mm}
          + \frac{991}{2 (1-y)^2}\! 
          - \! \frac{299}{4 (1-y)}\! 
          + \! \frac{23}{ (1+y)^2}\! 
          - \! \frac{107}{4 (1+y)}
          \Biggr) H(0,1,y) \nn\\
\hspace{-10mm} & &  \hspace{12mm}
       +   \Biggl(
            6
          + \frac{6}{ (1-y)^2}
          - \frac{6}{ (1-y)}
          + \frac{6}{ (1+y)^2} \nn\\
\hspace{-10mm} & &  \hspace{12mm}
          - \frac{6}{ (1+y)}
          \Biggr) H(0,1,-1,0,y)
       +   \Biggl(
            \frac{68}{3}
          - \frac{12}{ (1-y)^4} \nn\\
\hspace{-10mm} & &  \hspace{12mm}
          + \frac{24}{ (1\! -\! y)^3}\! 
          - \! \frac{14}{ (1\! -\! y)^2}\! 
          - \! \frac{29}{3 (1\! -\! y)}\! 
          - \! \frac{35}{3 (1\! +\! y)}
          \Biggr) H(0,1,0,y) \nn\\
\hspace{-10mm} & &  \hspace{12mm}
       -   \Biggl(
            14
          - \frac{282}{ (1-y)^5}
          + \frac{705}{ (1-y)^4}
          - \frac{1171}{2 (1-y)^3}
          + \frac{725}{4 (1-y)^2} \nn\\
\hspace{-10mm} & &  \hspace{12mm}
          - \frac{227}{8 (1-y)}
          + \frac{8}{ (1+y)^2}
          - \frac{83}{8 (1+y)}
          \Biggr) H(0,1,0,0,y) \nn\\
\hspace{-10mm} & &  \hspace{12mm}
       -   \Biggl(
            10
          + \frac{10}{ (1-y)^2}
          - \frac{10}{ (1-y)}
          + \frac{10}{ (1+y)^2} \nn\\
\hspace{-10mm} & &  \hspace{12mm}
          - \frac{10}{ (1+y)}
          \Biggr) H(0,1,1,0,y)
       - \zeta(2)   \Biggl(
            \frac{149}{3} \nn\\
\hspace{-10mm} & &  \hspace{12mm}
          + \frac{756}{ (1-y)^4}
          - \frac{1512}{ (1-y)^3}
          + \frac{882}{ (1-y)^2}
          - \frac{491}{3 (1-y)} \nn\\
\hspace{-10mm} & &  \hspace{12mm}
          - \frac{113}{3 (1+y)}
          \Biggr) H(1,y)
       - \zeta(3)   \Biggl(
            2
          + \frac{2}{ (1-y)^2} \nn\\
\hspace{-10mm} & &  \hspace{12mm}
          - \frac{2}{ (1-y)}
          + \frac{2}{ (1+y)^2}
          - \frac{2}{ (1+y)}
          \Biggr) H(1,y) \nn\\
\hspace{-10mm} & &  \hspace{12mm}
       -   \Biggl(
            6
          - \frac{6}{ (1-y)}
          - \frac{6}{ (1+y)}
          \Biggr) H(1,-1,0,y) \nn\\
\hspace{-10mm} & &  \hspace{12mm}
       + \zeta(2)   \Biggl(
            6
          + \frac{10}{ (1-y)^2}
          - \frac{6}{ (1-y)}
          + \frac{10}{ (1+y)^2} \nn\\
\hspace{-10mm} & &  \hspace{12mm}
          - \frac{6}{ (1+y)}
          \Biggr) H(1,0,y)
       +   \Biggl(
            \frac{31}{9}
          - \frac{6}{ (1-y)^3}
          + \frac{9}{ (1-y)^2} \nn\\
\hspace{-10mm} & &  \hspace{12mm}
          - \frac{46}{9 (1-y)}
          - \frac{43}{9 (1+y)}
          \Biggr) H(1,0,y)
       +   \Biggl(
            4
          + \frac{4}{ (1-y)^2} \nn\\
\hspace{-10mm} & &  \hspace{12mm}
          - \frac{4}{ (1-y)}
          + \frac{4}{ (1+y)^2}
          - \frac{4}{ (1+y)}
          \Biggr) H(1,0,-1,0,y) \nn\\
\hspace{-10mm} & &  \hspace{12mm}
       +   \Biggl(
            \frac{53}{3}
          + \frac{282}{ (1-y)^4}
          - \frac{564}{ (1-y)^3}
          + \frac{339}{ (1-y)^2}
          - \frac{206}{3 (1-y)} \nn\\
\hspace{-10mm} & &  \hspace{12mm}
          - \frac{35}{3 (1+y)}
          \Biggr) H(1,0,0,y)
       -   \Biggl(
            2
          + \frac{4}{ (1-y)^2} \nn\\
\hspace{-10mm} & &  \hspace{12mm}
          - \frac{2 }{(1-y)}
          + \frac{4}{ (1+y)^2}
          - \frac{2}{ (1+y)}
          \Biggr) H(1,0,0,0,y) \nn\\
\hspace{-10mm} & &  \hspace{12mm}
       -   \Biggl(
            4
          + \frac{4}{ (1-y)^2}
          - \frac{4}{ (1-y)}
          + \frac{4}{ (1+y)^2} \nn\\
\hspace{-10mm} & &  \hspace{12mm}
          - \frac{4}{ (1+y)}
          \Biggr) H(1,0,1,0,y)
       +   \Biggl(
            \frac{52}{3}
          - \frac{52}{3 (1-y)} \nn\\
\hspace{-10mm} & &  \hspace{12mm}
          - \frac{52}{3 (1+y)}
          \Biggr) H(1,1,0,y)
          \Biggr] \nn\\
\hspace{-10mm} & &  \hspace{6mm}
       + \, C_{F} T_{R} N_{f}   
          \Biggl[
         \frac{106}{27}
       + \zeta(2)   \Biggl(
            \frac{88}{9}
          - \frac{64}{9 (1-y)}
          - \frac{88}{9 (1+y)}
          \Biggr) \nn\\
\hspace{-10mm} & &  \hspace{12mm}
       + \zeta(3)   \Biggl(
            \frac{4}{3}
          - \frac{4}{3 (1-y)}
          - \frac{4}{3 (1+y)}
          \Biggr) \nn\\
\hspace{-10mm} & &  \hspace{12mm}
       + \zeta(2)   \Biggl(
            \frac{4}{3}
          - \frac{4}{3 (1-y)}
          - \frac{4}{3 (1+y)}
          \Biggr) H(0,y) \nn\\
\hspace{-10mm} & &  \hspace{12mm}
       -   \Biggl(
            \frac{209}{54}
          - \frac{103}{27 (1-y)}
          - \frac{106}{27 (1+y)}
          \Biggr) H(0,y) \nn\\
\hspace{-10mm} & &  \hspace{12mm}
       -   \Biggl(
            \frac{19}{9}
          - \frac{16}{9 (1-y)}
          - \frac{22}{9 (1+y)}
          \Biggr) H(0,0,y) \nn\\
\hspace{-10mm} & &  \hspace{12mm}
       -   \Biggl(
            \frac{2}{3}
          - \frac{2}{3 (1-y)}
          - \frac{2}{3 (1+y)}
          \Biggr) H(0,0,0,y) \nn\\
\hspace{-10mm} & &  \hspace{12mm}
       -   \Biggl(
            \frac{4}{3}
          - \frac{4}{3 (1-y)}
          - \frac{4}{3 (1+y)}
          \Biggr) H(0,1,0,y) \nn\\
\hspace{-10mm} & &  \hspace{12mm}
       + \zeta(2)   \Biggl(
            \frac{16}{3}
          - \frac{16}{3 (1-y)}
          - \frac{16}{3 (1+y)}
          \Biggr) H(1,y) \nn\\
\hspace{-10mm} & &  \hspace{12mm}
       -   \Biggl(
            \frac{38}{9}
          - \frac{32}{9 (1-y)}
          - \frac{44}{9 (1+y)}
          \Biggr) H(1,0,y) \nn\\
\hspace{-10mm} & &  \hspace{12mm}
       -   \Biggl(
            \frac{4}{3}
          - \frac{4}{3 (1-y)}
          - \frac{4}{3 (1+y)}
          \Biggr) H(1,0,0,y) \nn\\
\hspace{-10mm} & &  \hspace{12mm}
       -   \Biggl(
            \frac{8}{3}
          - \frac{8}{3 (1-y)}
          - \frac{8}{3 (1+y)}
          \Biggr) H(1,1,0,y)
          \Biggr] \nn\\
\hspace{-10mm} & &  \hspace{6mm}
       + \, C_{F} T_{R}   
          \Biggl[
            \frac{383}{27}
          + \frac{196}{9 (1-y)^2}
          - \frac{196}{9 (1-y)} \nn\\
\hspace{-10mm} & &  \hspace{12mm}
       - \zeta(2)   \Biggl(
            \frac{22}{3}
          - \frac{48}{ (1-y)^4}
          + \frac{96}{ (1-y)^3}
          - \frac{44}{ (1-y)^2} \nn\\
\hspace{-10mm} & &  \hspace{12mm}
          - \frac{4}{ (1-y)}
          \Biggr)
       + \zeta(2)   \Biggl(
            \frac{4}{3}
          + \frac{48}{ (1-y)^5}
          - \frac{120}{ (1-y)^4} \nn\\
\hspace{-10mm} & &  \hspace{12mm}
          + \frac{88}{ (1-y)^3}
          - \frac{12}{ (1-y)^2}
          - \frac{10}{3 (1-y)}
          - \frac{10}{3 (1+y)}
          \Biggr) H(0,y) \nn\\
\hspace{-10mm} & &  \hspace{12mm}
       -   \Biggl(
            \frac{265}{54}
          - \frac{356}{9 (1-y)^3}
          + \frac{178}{3 (1-y)^2}
          - \frac{563}{27 (1-y)} \nn\\
\hspace{-10mm} & &  \hspace{12mm}
          - \frac{236}{27 (1+y)}
          \Biggr) H(0,y)
       +   \Biggl(
            \frac{19}{9}
          + \frac{248}{9 (1-y)^4}
          - \frac{496}{9 (1-y)^3} \nn\\
\hspace{-10mm} & &  \hspace{12mm}
          + \frac{326}{9 (1-y)^2}
          - \frac{26}{3 (1-y)}
          \Biggr) H(0,0,y)
       -   \Biggl(
            \frac{2}{3}
          + \frac{24}{ (1-y)^5} \nn\\
\hspace{-10mm} & &  \hspace{12mm}
          - \frac{60}{ (1-y)^4}
          + \frac{44}{ (1-y)^3}
          - \frac{6}{ (1-y)^2}
          - \frac{5}{3 (1-y)} \nn\\
\hspace{-10mm} & &  \hspace{12mm}
          - \frac{5}{3 (1+y)}
          \Biggr) H(0,0,0,y)
           \Biggl] \nn\\
\hspace{-10mm} & &  \hspace{6mm}
       + \, C_{F}^2   \Biggl[ 
         \frac{23}{2}
       - \zeta(2)   \Biggl(
            158
          + \frac{72 \log{2}}{ (1-y)^2}
          - \frac{72 \log{2}}{ (1-y)}
          + 12 \log{2} \nn\\
\hspace{-10mm} & &  \hspace{12mm}
          + \frac{816}{ (1-y)^4}\! 
          - \! \frac{2040}{ (1-y)^3}\! 
          + \! \frac{1698}{ (1-y)^2}\! 
          - \! \frac{520}{ (1-y)}\! 
          + \! \frac{48}{ (1\! +\! y)^2} \nn\\
\hspace{-10mm} & &  \hspace{12mm}
          - \frac{128}{ (1+y)}
          \Biggr)
       + \zeta^2(2)  \Biggl(
            \frac{98}{5}
          + \frac{18}{ (1-y)^5}
          - \frac{45}{ (1-y)^4} \nn\\
\hspace{-10mm} & &  \hspace{12mm}
          + \frac{327}{5 (1-y)^3}
          - \frac{73}{2 (1-y)^2}
          - \frac{159}{20 (1-y)}
          + \frac{83}{5 (1+y)^2} \nn\\
\hspace{-10mm} & &  \hspace{12mm}
          - \frac{331}{20 (1+y)}
          \Biggr)
       - \zeta(3)   \Biggl(
            7
          - \frac{168}{ (1-y)^4}
          + \frac{336}{ (1-y)^3} \nn\\
\hspace{-10mm} & &  \hspace{12mm}
          - \frac{190}{ (1-y)^2}
          + \frac{24}{ (1-y)}
          + \frac{4}{ (1+y)^2}
          - \frac{2}{ (1+y)}
          \Biggr) \nn\\
\hspace{-10mm} & &  \hspace{12mm}
       + \zeta(2)   \Biggl(
            48
          + \frac{1080}{ (1-y)^4}
          - \frac{2160}{ (1-y)^3}
          + \frac{1182}{ (1-y)^2} \nn\\
\hspace{-10mm} & &  \hspace{12mm}
          - \frac{102}{ (1-y)}
          \Biggr) H(-1,y)
       - \zeta(3)   \Biggl(
            4
          + \frac{4}{ (1-y)^2}
          - \frac{4}{ (1-y)} \nn\\
\hspace{-10mm} & &  \hspace{12mm}
          + \frac{4}{ (1+y)^2}\! 
          - \! \frac{4}{ (1+y)}
          \Biggr) H(-1,y)\! 
       + \! \zeta(2)   \Biggl(
            20
          - \frac{288}{ (1-y)^5} \nn\\
\hspace{-10mm} & &  \hspace{12mm}
          + \frac{720}{ (1-y)^4}\! 
          - \! \frac{624}{ (1-y)^3}\! 
          + \! \frac{236}{ (1-y)^2}\! 
          - \! \frac{40}{ (1-y)}\! 
          + \! \frac{20}{ (1\! +\! y)^2} \nn\\
\hspace{-10mm} & &  \hspace{12mm}
          - \frac{24}{ (1+y)}
          \Biggr) H(-1,0,y)
       -   \Biggl(
            16
          - \frac{48}{ (1-y)^3}
          + \frac{72}{ (1-y)^2} \nn\\
\hspace{-10mm} & &  \hspace{12mm}
          - \frac{40}{ (1-y)}
          - \frac{16}{ (1+y)}
          \Biggr) H(-1,0,y)
       +   \Biggl(
            8
          + \frac{8}{ (1-y)^2} \nn\\
\hspace{-10mm} & &  \hspace{12mm}
          - \frac{8}{ (1-y)}
          + \frac{8}{ (1+y)^2}
          - \frac{8}{ (1+y)}
          \Biggr) H(-1,0,-1,0,y) \nn\\
\hspace{-10mm} & &  \hspace{12mm}
       -   \Biggl(
            16
          + \frac{360}{ (1-y)^4}
          - \frac{720}{ (1-y)^3}
          + \frac{394}{ (1-y)^2} \nn\\
\hspace{-10mm} & &  \hspace{12mm}
          - \frac{34}{ (1-y)}
          \Biggr) H(-1,0,0,y)
       -   \Biggl(
            8
          - \frac{144}{ (1-y)^5} \nn\\
\hspace{-10mm} & &  \hspace{12mm}
          + \frac{360}{ (1-y)^4}\! 
          - \! \frac{312}{ (1-y)^3}\! 
          + \! \frac{116}{ (1-y)^2}\! 
          - \! \frac{18}{ (1-y)}\! 
          + \! \frac{8}{ (1\! +\! y)^2} \nn\\
\hspace{-10mm} & &  \hspace{12mm}
          - \frac{10}{ (1+y)}
          \Biggr) H(-1,0,0,0,y)
       -   \Biggl(
            8
          + \frac{8}{ (1-y)^2} \nn\\
\hspace{-10mm} & &  \hspace{12mm}
          - \frac{8}{ (1-y)}
          + \frac{8}{ (1+y)^2}
          - \frac{8}{ (1+y)}
          \Biggr) H(-1,0,1,0,y) \nn\\
\hspace{-10mm} & &  \hspace{12mm}
       + \zeta(2)   \Biggl(
            13 \! 
          + \! \frac{120}{ (1-y)^5}\! 
          - \! \frac{492}{ (1-y)^4}\! 
          + \! \frac{632}{ (1-y)^3}\! 
          - \! \frac{261}{ (1-y)^2} \nn\\
\hspace{-10mm} & &  \hspace{12mm}
          - \frac{50}{ (1-y)}
          - \frac{20}{ (1+y)^2}
          + \frac{27}{ (1+y)}
          \Biggr) H(0,y) \nn\\
\hspace{-10mm} & &  \hspace{12mm}
       + \zeta(3)   \Biggl(
            8
          + \frac{168}{ (1-y)^5}
          - \frac{420}{ (1-y)^4}
          + \frac{364}{ (1-y)^3} \nn\\
\hspace{-10mm} & &  \hspace{12mm}
          - \frac{126}{ (1-y)^2}
          + \frac{5}{ (1-y)}
          - \frac{7}{ (1+y)}
          \Biggr) H(0,y) \nn\\
\hspace{-10mm} & &  \hspace{12mm}
       -   \Biggl(
            \frac{85}{8}
          - \frac{47}{4 (1-y)}
          - \frac{19}{2 (1+y)}
          \Biggr) H(0,y) \nn\\
\hspace{-10mm} & &  \hspace{12mm}
       - \zeta(2)   \Biggl(
            12 \! 
          - \! \frac{1080}{ (1-y)^5}\! 
          + \! \frac{2700}{ (1-y)^4}\! 
          - \! \frac{2100}{ (1-y)^3}\! 
          + \! \frac{450}{ (1-y)^2} \nn\\
\hspace{-10mm} & &  \hspace{12mm}
          - \frac{15}{ (1-y)}
          + \frac{21}{ (1+y)}
          \Biggr) H(0,-1,y)
       +   \Biggl(
            10
          + \frac{96}{ (1-y)^4} \nn\\
\hspace{-10mm} & &  \hspace{12mm}
          - \frac{192}{ (1-y)^3}
          + \frac{116}{ (1-y)^2}
          - \frac{20}{ (1-y)}
          + \frac{8}{ (1+y)^2} \nn\\
\hspace{-10mm} & &  \hspace{12mm}
          - \frac{8}{ (1+y)}
          \Biggr) H(0,-1,0,y)
       +   \Biggl(
            4
          - \frac{360}{ (1-y)^5} \nn\\
\hspace{-10mm} & &  \hspace{12mm}
          + \frac{900}{ (1-y)^4}
          - \frac{700}{ (1-y)^3}
          + \frac{150}{ (1-y)^2}
          - \frac{5}{ (1-y)} \nn\\
\hspace{-10mm} & &  \hspace{12mm}
          + \frac{7}{ (1+y)}
          \Biggr) H(0,-1,0,0,y)
       - \zeta(2)   \Biggl(
            68
          + \frac{48}{ (1-y)^5} \nn\\
\hspace{-10mm} & &  \hspace{12mm}
          - \frac{120}{ (1-y)^4}
          + \frac{115}{ (1-y)^3}
          - \frac{13}{2 (1-y)^2}
          - \frac{233}{4 (1-y)} \nn\\
\hspace{-10mm} & &  \hspace{12mm}
          + \frac{46}{ (1+y)^2}
          - \frac{273}{4 (1+y)}
          \Biggr) H(0,0,y)
       +   \Biggl(
            \frac{229}{4}
          + \frac{192}{ (1-y)^4} \nn\\
\hspace{-10mm} & &  \hspace{12mm}
          - \frac{504}{ (1-y)^3}
          + \frac{466}{ (1-y)^2}
          - \frac{343}{2 (1-y)}
          + \frac{16}{ (1+y)^2} \nn\\
\hspace{-10mm} & &  \hspace{12mm}
          - \frac{42}{ (1+y)}
          \Biggr) H(0,0,y)
       -   \Biggl(
            12
          - \frac{96}{ (1-y)^5}
          + \frac{240}{ (1-y)^4} \nn\\
\hspace{-10mm} & &  \hspace{12mm}
          - \frac{176}{ (1-y)^3}
          + \frac{28}{ (1-y)^2}
          - \frac{8}{ (1-y)}
          + \frac{4}{ (1+y)^2} \nn\\
\hspace{-10mm} & &  \hspace{12mm}
          - \frac{8}{ (1+y)}
          \Biggr) H(0,0,-1,0,y)
       -   \Biggl(
            10
          + \frac{60}{ (1-y)^5} \nn\\
\hspace{-10mm} & &  \hspace{12mm}
          - \frac{216}{ (1-y)^4}\! 
          + \! \frac{256}{ (1-y)^3}\! 
          - \! \frac{89}{ (1-y)^2}\! 
          - \! \frac{71}{2 (1-y)}\! 
          - \! \frac{4}{ (1\! +\! y)^2} \nn\\
\hspace{-10mm} & &  \hspace{12mm}
          + \frac{17}{2 (1+y)}
          \Biggr) H(0,0,0,y)\! 
       +  \!  \Biggl(
            27
          - \frac{6}{ (1-y)^5}\! 
          + \! \frac{15}{ (1-y)^4} \nn\\
\hspace{-10mm} & &  \hspace{12mm}
          - \frac{17}{2 (1-y)^3}
          + \frac{59}{4 (1-y)^2}
          - \frac{201}{8 (1-y)}
          + \frac{17}{ (1+y)^2} \nn\\
\hspace{-10mm} & &  \hspace{12mm}
          - \frac{217}{8 (1 \! + \! y)}
          \Biggr) H(0,0,0,0,y) \! 
       +   \!  \Biggl(
            22 \! 
          -  \! \frac{384}{ (1-y)^5} \! 
          + \!  \frac{960}{ (1-y)^4} \nn\\
\hspace{-10mm} & &  \hspace{12mm}
          - \frac{746}{ (1-y)^3}
          + \frac{169}{ (1-y)^2}
          - \frac{45}{2 (1-y)}
          + \frac{10}{ (1+y)^2} \nn\\
\hspace{-10mm} & &  \hspace{12mm}
          - \frac{21}{2 (1+y)}
          \Biggr) H(0,0,1,0,y)
       - \zeta(2)   \Biggl(
            32
          + \frac{936}{ (1-y)^5} \nn\\
\hspace{-10mm} & &  \hspace{12mm}
          -  \frac{2340}{ (1-y)^4} \! 
          +  \! \frac{1914}{ (1-y)^3} \! 
          -  \! \frac{499 }{(1-y)^2} \! 
          +  \! \frac{5}{2 (1-y)} \! 
          +  \! \frac{32}{ (1 \! + \! y)^2} \nn\\
\hspace{-10mm} & &  \hspace{12mm}
          - \frac{91}{2 (1+y)}
          \Biggr) H(0,1,y)
       -   \Biggl(
            16
          + \frac{384}{ (1-y)^4}
          - \frac{768}{ (1-y)^3} \nn\\
\hspace{-10mm} & &  \hspace{12mm}
          + \frac{442}{ (1-y)^2}
          - \frac{60}{ (1-y)}
          - \frac{2}{ (1+y)}
          \Biggr) H(0,1,0,y) \nn\\
\hspace{-10mm} & &  \hspace{12mm}
       +   \Biggl(
            14
          + \frac{252}{ (1-y)^5}
          - \frac{630}{ (1-y)^4}
          + \frac{517}{ (1-y)^3}
          - \frac{271}{2 (1-y)^2} \nn\\
\hspace{-10mm} & &  \hspace{12mm}
          - \frac{19}{4 (1-y)}
          + \frac{10}{ (1+y)^2}
          - \frac{67}{4 (1+y)}
          \Biggr) H(0,1,0,0,y) \nn\\
\hspace{-10mm} & &  \hspace{12mm}
       +   \Biggl(
            4
          + \frac{4}{ (1-y)^2}
          - \frac{4}{ (1-y)}
          + \frac{4}{ (1+y)^2} \nn\\
\hspace{-10mm} & &  \hspace{12mm}
          - \frac{4}{ (1+y)}
          \Biggr) H(0,1,1,0,y)
       - \zeta(2)   \Biggl(
            28
          + \frac{936}{ (1-y)^4} \nn\\
\hspace{-10mm} & &  \hspace{12mm}
          - \frac{1872}{ (1-y)^3}
          + \frac{1080}{ (1-y)^2}
          - \frac{136}{ (1-y)}
          + \frac{48}{ (1+y)^2} \nn\\
\hspace{-10mm} & &  \hspace{12mm}
          - \frac{40}{ (1+y)}
          \Biggr) H(1,y)
       - \zeta(2)   \Biggl(
            40
          + \frac{40}{ (1-y)^2} \nn\\
\hspace{-10mm} & &  \hspace{12mm}
          - \frac{40}{ (1-y)}
          + \frac{40}{ (1+y)^2}
          - \frac{40}{ (1+y)}
          \Biggr) H(1,0,y) \nn\\
\hspace{-10mm} & &  \hspace{12mm}
       +   \Biggl(
            \frac{55}{2}
          - \frac{192}{ (1-y)^3}
          + \frac{288}{ (1-y)^2}
          - \frac{115}{ (1-y)} \nn\\
\hspace{-10mm} & &  \hspace{12mm}
          - \frac{36}{ (1+y)}
          \Biggr) H(1,0,y)
       +   \Biggl(
            6
          + \frac{252}{ (1-y)^4}
          - \frac{504}{ (1-y)^3} \nn\\
\hspace{-10mm} & &  \hspace{12mm}
          +  \frac{270}{ (1-y)^2} \! 
          -  \! \frac{16}{ (1-y)} \! 
          +  \! \frac{16}{ (1 \! + \! y)^2} \! 
          -  \! \frac{14}{ (1 \! + \! y)}
          \Biggr) H(1,0,0,y) \nn\\
\hspace{-10mm} & &  \hspace{12mm}
       +   \Biggl(
            12
          + \frac{12}{ (1-y)^2}
          - \frac{12}{ (1-y)}
          + \frac{12}{ (1+y)^2} \nn\\
\hspace{-10mm} & &  \hspace{12mm}
          - \frac{12}{ (1+y)}
          \Biggr) H(1,0,0,0,y)
       +   \Biggl(
            8
          + \frac{8}{ (1-y)^2} \nn\\
\hspace{-10mm} & &  \hspace{12mm}
          - \frac{8}{ (1-y)}
          + \frac{8}{ (1+y)^2}
          - \frac{8}{ (1+y)}
          \Biggr) H(1,0,1,0,y) \nn\\
\hspace{-10mm} & &  \hspace{12mm}
       - \zeta(2)   \Biggl(
            48
          + \frac{48}{ (1-y)^2}
          - \frac{48}{ (1-y)}
          + \frac{48}{ (1+y)^2}  \nn\\
\hspace{-10mm} & &  \hspace{12mm}
          - \frac{48}{ (1+y)}
          \Biggr) H(1,1,y)
       -   \Biggl(
            4
          - \frac{4}{ (1-y)} \nn\\
\hspace{-10mm} & &  \hspace{12mm}
          - \frac{4}{ (1+y)}
          \Biggr) H(1,1,0,y)
       +   \Biggl(
            16
          + \frac{16}{ (1-y)^2} \nn\\
\hspace{-10mm} & &  \hspace{12mm}
          - \frac{16}{ (1-y)}
          + \frac{16}{ (1+y)^2}
          - \frac{16}{ (1+y)}
          \Biggr) H(1,1,0,0,y)
           \Biggl] \, , \\
%%%%%%%%%
%%%%%%%%%
%
\hspace{-10mm} 
\Im \, {\mathcal F}^{(2l)}_{1,R}(\epsilon,s) & = & \frac{1}{\epsilon^2}
    \Biggl\{
         - \frac{11}{12} C_{F} C_{A}  \Biggl( 
            1
          - \frac{1}{1-y}
          - \frac{1}{1+y} \Biggr) \nn\\
\hspace{-10mm} & &  \hspace{6mm}
       + \frac{1}{3} C_{F} T_{R} N_{f}   
          \Biggl(
            1
          - \frac{1}{1-y}
          - \frac{1}{1+y} \Biggl)  \nn\\
\hspace{-10mm} & &  \hspace{6mm}
       - \, C_{F}^2   \Biggl[
          \Biggl(
            1
          - \frac{1}{1-y}
          - \frac{1}{1+y} \Biggl)  \nn\\
\hspace{-10mm} & &  \hspace{12mm}
        - \Biggl(
            1 \! 
          - \! \frac{1}{1\! -\! y} \! 
          + \! \frac{1}{(1\! -\! y)^2}\! 
          - \! \frac{1}{1\! +\! y} \! 
          + \! \frac{1}{(1\! +\! y)^2} \Biggl) H(0,y)
          \Biggr]
     \Biggr\} \nn\\
\hspace{-10mm} & & + \frac{1}{\epsilon}  \Biggl\{  
         C_{F} C_{A} \Biggl[
          + \frac{67}{36}
          - \frac{67}{36 (1-y)}
          - \frac{67}{36 (1+y)}  \nn\\
\hspace{-10mm} & &  \hspace{12mm}
       + \zeta(2)   \Biggl(
            \frac{1}{2}
          + \frac{1}{2 (1-y)^2}
          - \frac{1}{2 (1-y)}
          + \frac{1}{2 (1+y)^2}  \nn\\
\hspace{-10mm} & &  \hspace{12mm}
          - \frac{1}{2 (1+y)}
          \Biggr)
       -   \Biggl(
            1
          - \frac{1}{(1-y)}
          - \frac{1}{(1+y)}
          \Biggr) H(-1;y)  \nn\\
\hspace{-10mm} & &  \hspace{12mm}
       +   \Biggl(
            2
          - \frac{1}{(1-y)}
          - \frac{1}{(1+y)}
          \Biggr) H(0;y)
       +   \Biggl(
            1
          + \frac{1}{(1-y)^2}  \nn\\
\hspace{-10mm} & &  \hspace{12mm}
          - \frac{1}{(1-y)}
          + \frac{1}{(1+y)^2}
          - \frac{1}{(1+y)}
          \Biggr) H(0,-1;y) \nn\\
\hspace{-10mm} & &  \hspace{12mm}
       -   \Biggl(
            2
          + \frac{1}{(1-y)^2}
          - \frac{2 }{(1-y)}
          + \frac{1}{(1+y)^2} \nn\\
\hspace{-10mm} & &  \hspace{12mm}
          - \frac{2}{ (1+y)}
          \Biggr) H(0,0;y)
       -   \Biggl(
            1
          + \frac{1}{(1-y)^2}
          - \frac{1}{(1-y)} \nn\\
\hspace{-10mm} & &  \hspace{12mm}
          + \frac{1}{(1+y)^2} 
          - \frac{1}{(1+y)}
          \Biggr) H(0,1;y)
       +   \Biggl(
            1
          - \frac{1}{(1-y)} \nn\\
\hspace{-10mm} & &  \hspace{12mm}
          - \frac{1}{(1+y)}
          \Biggr) H(1;y)
           \Biggr] \nn\\
\hspace{-10mm} & &  \hspace{6mm}
       - \frac{5}{9} C_{F} T_{R} N_{f}   \Biggl[
            1
          - \frac{1}{(1-y)}
          - \frac{1}{(1+y)}
          \Biggr] \nn\\
\hspace{-10mm} & &  \hspace{6mm}
       + C_{F}^2  \Biggl[
          - \frac{7}{2}
          + \frac{3}{ (1-y)}
          + \frac{4}{ (1+y)}
       - \zeta(2)   \Biggl(
            4
          + \frac{4}{ (1-y)^2} \nn\\
\hspace{-10mm} & &  \hspace{12mm}
          - \frac{4}{ (1-y)}
          + \frac{4}{ (1+y)^2}
          - \frac{4}{ (1+y)}
          \Biggr)
       +   \Biggl(
            2
          + \frac{2}{ (1-y)^2} \nn\\
\hspace{-10mm} & &  \hspace{12mm}
          - \frac{1}{(1-y)}
          + \frac{4}{ (1+y)^2}
          - \frac{3}{ (1+y)}
          \Biggr) H(0;y) \nn\\
\hspace{-10mm} & &  \hspace{12mm}
       +   \Biggl(
            3
          + \frac{3}{ (1-y)^2}
          - \frac{3}{ (1-y)}
          + \frac{3}{ (1+y)^2} \nn\\
\hspace{-10mm} & &  \hspace{12mm}
          - \frac{3}{ (1+y)}
          \Biggr) H(0,0;y)
       +   \Biggl(
            2
          + \frac{2}{ (1-y)^2}
          - \frac{2}{ (1-y)} \nn\\
\hspace{-10mm} & &  \hspace{12mm}
          + \frac{2 }{(1+y)^2}
          - \frac{2}{ (1+y)}
          \Biggr) H(0,1;y)
       -   \Biggl(
            2
          - \frac{2}{ (1-y)} \nn\\
\hspace{-10mm} & &  \hspace{12mm}
          - \frac{2}{ (1+y)}
          \Biggr) H(1;y)
       +   \Biggl(
            4
          + \frac{4}{ (1-y)^2}
          - \frac{4}{ (1-y)} \nn\\
\hspace{-10mm} & &  \hspace{12mm}
          + \frac{4}{ (1+y)^2}
          - \frac{4}{ (1+y)}
          \Biggr) H(1,0;y) 
          \Biggr] \Biggr\} \\ 
\hspace{-10mm} & &  \hspace{6mm}
        +  C_{F} C_{A} \Biggl[
            \frac{2545}{216}
          + \frac{9}{ (1-y)^3}
          - \frac{27}{2 (1-y)^2}
          - \frac{599}{108 (1-y)} \nn\\
\hspace{-10mm} & &  \hspace{12mm}
          - \frac{365}{27 (1+y)}
       - \zeta(2)   \Biggl(
            10
          + \frac{72}{ (1-y)^4}
          - \frac{144}{ (1-y)^3} \nn\\
\hspace{-10mm} & &  \hspace{12mm}
          + \frac{79}{ (1-y)^2}
          - \frac{12}{ (1-y)}
          - \frac{5}{ (1+y)}
          \Biggr)
       - \zeta(3)   \Biggl(
            \frac{15}{2} \nn\\
\hspace{-10mm} & &  \hspace{12mm}
          - \frac{324}{ (1-y)^5}
          + \frac{810}{ (1-y)^4}
          - \frac{662}{ (1-y)^3}
          + \frac{367}{2 (1-y)^2} \nn\\
\hspace{-10mm} & &  \hspace{12mm}
          - \frac{20}{ (1-y)}
          + \frac{1}{2 (1+y)^2}
          - \frac{2}{ (1+y)}
          \Biggr) \nn\\
\hspace{-10mm} & &  \hspace{12mm}
       - \zeta(2)   \Biggl(
            2
          + \frac{2}{ (1-y)^2}
          - \frac{2}{ (1-y)}
          + \frac{2}{ (1+y)^2} \nn\\
\hspace{-10mm} & &  \hspace{12mm}
          - \frac{2}{ (1+y)}
          \Biggr) H(-1;y)
       +   \Biggl(
            3
          - \frac{24}{ (1-y)^3}
          + \frac{36}{ (1-y)^2} \nn\\
\hspace{-10mm} & &  \hspace{12mm}
          - \frac{16}{ (1-y)}
          - \frac{2}{ (1+y)}
          \Biggr) H(-1;y)
       +   \Biggl(
            2
          - \frac{2}{ (1-y)}\nn\\
\hspace{-10mm} & &  \hspace{12mm}
          - \frac{2}{ (1+y)}
          \Biggr) H(-1,-1;y)
       +   \Biggl(
            4
          - \frac{24}{ (1-y)^4}
          + \frac{48}{ (1-y)^3}\nn\\
\hspace{-10mm} & &  \hspace{12mm}
          - \frac{38}{ (1-y)^2}
          + \frac{18}{ (1-y)}
          + \frac{4}{ (1+y)}
          \Biggr) H(-1,0;y) \nn\\
\hspace{-10mm} & &  \hspace{12mm}
       +   \Biggl(
          - 4
          - \frac{4}{ (1-y)^2}
          + \frac{4}{ (1-y)}
          - \frac{4}{ (1+y)^2} \nn\\
\hspace{-10mm} & &  \hspace{12mm}
          + \frac{4}{ (1+y)}
          \Biggr) H(-1,0,-1;y)
       +   \Biggl(
          - 2
          + \frac{336}{ (1-y)^5} \nn\\
\hspace{-10mm} & &  \hspace{12mm}
          - \frac{840}{ (1-y)^4}\! 
          + \frac{688}{ (1-y)^3}\! 
          - \! \frac{188}{ (1-y)^2}\! 
          + \! \frac{16}{ (1-y)}\! 
          + \! \frac{4}{ (1\! +\! y)^2} \nn\\
\hspace{-10mm} & &  \hspace{12mm}
          - \frac{4}{ (1+y)}
          \Biggr) H(-1,0,0;y)
       +   \Biggl(
            4
          + \frac{4}{ (1-y)^2}
          - \frac{4}{ (1-y)} \nn\\
\hspace{-10mm} & &  \hspace{12mm}
          + \frac{4}{ (1+y)^2}
          - \! \frac{4}{ (1+y)}
          \Biggr) H(-1,0,1;y)\! 
       +  \!  \Biggl(
          - \! 6\! 
          + \! \frac{6}{ (1-y)} \nn\\
\hspace{-10mm} & &  \hspace{12mm}
          + \frac{6}{ (1+y)}
          \Biggr) H(-1,1;y)
       + \zeta(2)   \Biggl(
            13
          - \frac{72}{ (1-y)^5} \nn\\
\hspace{-10mm} & &  \hspace{12mm}
          + \frac{180}{ (1-y)^4}
          - \frac{145}{ (1-y)^3}
          + \frac{89}{2 (1-y)^2}
          - \frac{59}{4 (1-y)} \nn\\
\hspace{-10mm} & &  \hspace{12mm}
          + \frac{7}{ (1+y)^2}
          - \frac{47}{4 (1+y)}
          \Biggr) H(0;y)
       +  \Biggl(
          - \frac{217}{36}
          + \frac{24}{ (1-y)^4} \nn\\
\hspace{-10mm} & &  \hspace{12mm}
          - \frac{39}{ (1-y)^3}\! 
          + \! \frac{3}{2 (1-y)^2}\! 
          + \! \frac{130}{9 (1-y)}\! 
          - \! \frac{25}{18 (1\! +\! y)}
          \Biggr) H(0;y)  \nn\\
\hspace{-10mm} & &  \hspace{12mm}
       +   \Biggl(
          - 8
          - \frac{48}{ (1-y)^4}
          + \frac{96}{ (1-y)^3}
          - \frac{56}{ (1-y)^2}
          + \frac{12}{ (1-y)} \nn\\
\hspace{-10mm} & &  \hspace{12mm}
          + \frac{4}{ (1+y)}
          \Biggr) H(0,-1;y)
       +   \Biggl(
          - 2
          - \frac{2}{ (1-y)^2}
          + \frac{2}{ (1-y)} \nn\\
\hspace{-10mm} & &  \hspace{12mm}
          - \frac{2}{ (1\! +\! y)^2}\! 
          + \! \frac{2 }{(1\! +\! y)}
          \Biggr) H(0,-1,-1;y)\! 
       +  \!  \Biggl(
            2\! 
          - \! \frac{24}{ (1-y)^5} \nn\\
\hspace{-10mm} & &  \hspace{12mm}
          + \frac{60}{ (1-y)^4}\! 
          - \! \frac{72}{ (1-y)^3}\! 
          + \! \frac{52}{ (1-y)^2}\! 
          - \! \frac{11}{ (1-y)}\! 
          + \! \frac{4}{ (1\! +\! y)^2} \nn\\
\hspace{-10mm} & &  \hspace{12mm}
          - \frac{5}{ (1+y)}
          \Biggr) H(0,-1,0;y)
       +   \Biggl(
            6
          + \frac{6}{ (1-y)^2}
          - \frac{6}{ (1-y)} \nn\\
\hspace{-10mm} & &  \hspace{12mm}
          + \frac{6}{ (1\! +\! y)^2}
          - \! \frac{6}{ (1\! +\! y)}
          \Biggr) H(0,-1,1;y)\! 
       +  \!  \Biggl(
            \frac{89}{6}\! 
          - \! \frac{258}{ (1-y)^5} \nn\\
\hspace{-10mm} & &  \hspace{12mm}
          + \frac{816}{ (1-y)^4}
          - \frac{923}{ (1-y)^3}
          + \frac{436}{ (1-y)^2}
          - \frac{259}{3 (1-y)} \nn\\
\hspace{-10mm} & &  \hspace{12mm}
          - \frac{4}{3 (1+y)}
          \Biggr) H(0,0;y)
       +   \Biggl(
            14
          - \frac{48}{ (1-y)^5}
          + \frac{120}{ (1-y)^4} \nn\\
\hspace{-10mm} & &  \hspace{12mm}
          - \frac{88}{ (1-y)^3}
          + \frac{20}{ (1-y)^2}
          - \frac{12}{ (1-y)}
          + \frac{8}{ (1+y)^2} \nn\\
\hspace{-10mm} & &  \hspace{12mm}
          - \frac{12}{ (1+y)}
          \Biggr) H(0,0,-1;y)
       +   \Biggl(
          - 12
          + \frac{3}{ (1-y)^5} \nn\\
\hspace{-10mm} & &  \hspace{12mm}
          - \frac{15}{2 (1-y)^4}
          + \frac{17}{4 (1-y)^3}
          - \frac{39}{8 (1-y)^2}
          + \frac{177}{16 (1-y)} \nn\\
\hspace{-10mm} & &  \hspace{12mm}
          - \frac{6}{ (1+y)^2}
          + \frac{193}{16 (1+y)}
          \Biggr) H(0,0,0;y) \nn\\
\hspace{-10mm} & &  \hspace{12mm}
       -   \Biggl(
            22
          + \frac{12}{ (1-y)^5}
          - \frac{30}{ (1-y)^4}
          + \frac{49}{ (1-y)^3}
          - \frac{63}{2 (1-y)^2} \nn\\
\hspace{-10mm} & &  \hspace{12mm}
          - \frac{51}{4 (1-y)}
          + \frac{12}{ (1+y)^2}
          - \frac{75}{4 (1+y)}
          \Biggr) H(0,0,1;y) \nn\\
\hspace{-10mm} & &  \hspace{12mm}
       +   \Biggl(
            \frac{68}{3}
          - \frac{12}{ (1-y)^4}
          + \frac{24}{ (1-y)^3}
          - \frac{14}{ (1-y)^2}
          - \frac{29}{3 (1-y)} \nn\\
\hspace{-10mm} & &  \hspace{12mm}
          - \frac{35}{3 (1+y)}
          \Biggr) H(0,1;y)
       +  \Biggl(
            6
          + \frac{6}{ (1-y)^2}
          - \frac{6}{ (1-y)} \nn\\
\hspace{-10mm} & &  \hspace{12mm}
          + \frac{6}{ (1+y)^2}
          - \frac{6}{ (1+y)}
          \Biggr) H(0,1,-1;y) 
       -   \Biggl(
            14
          - \frac{282}{ (1-y)^5} \nn\\
\hspace{-10mm} & &  \hspace{12mm}
          + \frac{705}{ (1\! -\! y)^4}\! 
          - \! \frac{1171}{2 (1-y)^3}\! 
          + \! \frac{725}{4 (1-y)^2}\! 
          - \! \frac{227}{8 (1\! -\! y)}\! 
          + \! \frac{8}{ (1\! +\! y)^2} \nn\\
\hspace{-10mm} & &  \hspace{12mm}
          - \frac{83}{8 (1+y)}
          \Biggr) H(0,1,0;y)
       -   \Biggl(
            10
          + \frac{10}{ (1-y)^2}
          - \frac{10}{ (1-y)} \nn\\
\hspace{-10mm} & &  \hspace{12mm}
          + \frac{10}{ (1\! +\! y)^2}
          - \frac{10}{ (1\! +\! y)}
          \Biggr) H(0,1,1;y)\! 
       + \! \zeta(2)   \Biggl(
            2\! 
          + \! \frac{2}{ (1-y)^2} \nn\\
\hspace{-10mm} & &  \hspace{12mm}
          - \frac{2}{ (1-y)}
          + \frac{2}{ (1+y)^2}
          - \frac{2}{ (1+y)}
          \Biggr) H(1;y) \nn\\
\hspace{-10mm} & &  \hspace{12mm}
       +  \Biggl(
            \frac{31}{9}\! 
          - \! \frac{6}{ (1\! -\! y)^3}\! 
          + \! \frac{9}{ (1\! -\! y)^2}\! 
          - \! \frac{46}{9 (1\! -\! y)}\! 
          - \! \frac{43}{9 (1\! +\! y)}
          \Biggr) H(1;y)  \nn\\
\hspace{-10mm} & &  \hspace{12mm}
       -   \Biggl(
            6
          - \frac{6}{ (1-y)}
          - \frac{6}{ (1+y)}
          \Biggr) H(1,-1;y) \nn\\
\hspace{-10mm} & &  \hspace{12mm}
       +  \Biggl(
            \frac{53}{3}\! 
          + \! \frac{282}{ (1-y)^4}
          - \frac{564}{ (1-y)^3}\! 
          + \! \frac{339}{ (1-y)^2}
          - \frac{206}{3 (1-y)}  \nn\\
\hspace{-10mm} & &  \hspace{12mm}
          - \frac{35}{3 (1+y)}
          \Biggr) H(1,0;y) 
       +   \Biggl(
            4
          + \frac{4}{ (1-y)^2}
          - \frac{4}{ (1-y)}  \nn\\
\hspace{-10mm} & &  \hspace{12mm}
          + \frac{4}{ (1\! +\! y)^2}\! 
          - \! \frac{4}{ (1\! +\! y)}
          \Biggr) H(1,0,-1;y)\! 
       +  \!  \Biggl(
          - 2
          - \frac{4}{ (1-y)^2}  \nn\\
\hspace{-10mm} & &  \hspace{12mm}
          + \frac{2}{ (1-y)}
          - \frac{4}{ (1+y)^2}
          + \frac{2}{ (1+y)}
          \Biggr) H(1,0,0;y)  \nn\\
\hspace{-10mm} & &  \hspace{12mm}
       -   \Biggl(
            4
          + \frac{4}{ (1-y)^2}
          - \frac{4}{ (1-y)}
          + \frac{4}{ (1+y)^2} \nn\\
\hspace{-10mm} & &  \hspace{12mm}
          - \frac{4}{ (1+y)}
          \Biggr) H(1,0,1;y)
       +   \Biggl(
            \frac{52}{3}
          - \frac{52}{3 (1-y)} \nn\\
\hspace{-10mm} & &  \hspace{12mm}
          - \frac{52}{3 (1+y)}
          \Biggr) H(1,1;y)
          \Biggr]  \nn\\ 
\hspace{-10mm} & &  \hspace{6mm}
        + C_{F} T_{R} N_{f} \Biggl[
          - \frac{209}{54}
          + \frac{103}{27 (1-y)}
          + \frac{106}{27 (1+y)} \nn\\
\hspace{-10mm} & &  \hspace{12mm}
       +   \Biggl(
          - \frac{19}{9}
          + \frac{16}{9 (1-y)}
          + \frac{22}{9 (1+y)}
          \Biggr) H(0;y) \nn\\
\hspace{-10mm} & &  \hspace{12mm}
       +   \Biggl(
          - \frac{2}{3}
          + \frac{2}{3 (1-y)}
          + \frac{2}{3 (1+y)}
          \Biggr) H(0,0;y) \nn\\
\hspace{-10mm} & &  \hspace{12mm}
       +   \Biggl(
          - \frac{4}{3}
          + \frac{4}{3 (1-y)}
          + \frac{4}{3 (1+y)}
          \Biggr) H(0,1;y) \nn\\
\hspace{-10mm} & &  \hspace{12mm}
       +   \Biggl(
          - \frac{38}{9}
          + \frac{32}{9 (1-y)}
          + \frac{44}{9 (1+y)}
          \Biggr) H(1;y) \nn\\
\hspace{-10mm} & &  \hspace{12mm}
       +   \Biggl(
          - \frac{4}{3}
          + \frac{4}{3 (1-y)}
          + \frac{4}{3 (1+y)}
          \Biggr) H(1,0;y) \nn\\
\hspace{-10mm} & &  \hspace{12mm}
       +   \Biggl(
          - \frac{8}{3}
          + \frac{8}{3 (1-y)}
          + \frac{8}{3 (1+y)}
          \Biggr) H(1,1;y)
          \Biggr]  \nn\\ 
\hspace{-10mm} & &  \hspace{6mm}
        + \, C_{F} T_{R} \Biggl[
          - \frac{265}{54}
          + \frac{356}{9 (1-y)^3}
          - \frac{178}{3 (1-y)^2}
          + \frac{563}{27 (1-y)} \nn\\
\hspace{-10mm} & &  \hspace{12mm}
          + \frac{236}{27 (1+y)}
       +   \Biggl(
            \frac{19}{9}
          + \frac{248}{9 (1-y)^4}
          - \frac{496}{9 (1-y)^3} \nn\\
\hspace{-10mm} & &  \hspace{12mm}
          + \frac{326}{9 (1-y)^2}
          - \frac{26}{3 (1-y)}
          \Biggr) H(0;y)
       -   \Biggl(
            \frac{2}{3}
          + \frac{24}{ (1-y)^5} \nn\\
\hspace{-10mm} & &  \hspace{12mm}
          - \frac{60}{ (1-y)^4}
          + \frac{44}{ (1-y)^3}
          - \frac{6}{ (1-y)^2}
          - \frac{5}{3 (1-y)} \nn\\
\hspace{-10mm} & &  \hspace{12mm}
          - \frac{5}{3 (1+y)}
          \Biggr) H(0,0;y)
          \Biggr]  \nn\\ 
\hspace{-10mm} & &  \hspace{6mm}
       +  \,C_{F}^2 \Biggl[
          - \frac{85}{8}
          + \frac{47}{4 (1-y)}
          + \frac{19}{2 (1+y)} \nn\\
\hspace{-10mm} & &  \hspace{12mm}
       + \zeta(2)   \Biggl(
          - 7
          - \frac{60}{ (1-y)^4}
          + \frac{120}{ (1-y)^3}
          - \frac{83}{ (1-y)^2} \nn\\
\hspace{-10mm} & &  \hspace{12mm}
          + \frac{21}{ (1-y)}
          - \frac{12}{ (1+y)^2}
          + \frac{10}{ (1+y)}
          \Biggr) \nn\\
\hspace{-10mm} & &  \hspace{12mm}
       + \zeta(3)   \Biggl(
            8
          + \frac{168}{ (1-y)^5}
          - \frac{420}{ (1-y)^4}
          + \frac{364}{ (1-y)^3} \nn\\
\hspace{-10mm} & &  \hspace{12mm}
          - \frac{126}{ (1-y)^2}\! 
          + \! \frac{5}{ (1-y)}\! 
          - \! \frac{7}{ (1+y)}
          \Biggr)\! 
       + \! \zeta(2)   \Biggl(
            4\! 
          + \! \frac{4}{ (1-y)^2} \nn\\
\hspace{-10mm} & &  \hspace{12mm}
          - \frac{4}{ (1-y)}
          + \frac{4}{ (1+y)^2}
          - \frac{4}{ (1+y)}
          \Biggr) H(-1;y) \nn\\
\hspace{-10mm} & &  \hspace{12mm}
       -   \Biggl(
            16
          - \frac{48}{ (1-y)^3}
          + \frac{72}{ (1-y)^2}
          - \frac{40}{ (1-y)} \nn\\
\hspace{-10mm} & &  \hspace{12mm}
          - \frac{16}{ (1+y)}
          \Biggr) H(-1;y)
       -   \Biggl(
            16
          + \frac{360}{ (1-y)^4}
          - \frac{720}{ (1-y)^3} \nn\\
\hspace{-10mm} & &  \hspace{12mm}
          + \frac{394}{ (1-y)^2}
          - \frac{34}{ (1-y)}
          \Biggr) H(-1,0;y)
       +   \Biggl(
            8
          + \frac{8}{ (1-y)^2} \nn\\
\hspace{-10mm} & &  \hspace{12mm}
          - \frac{8}{ (1-y)}
          + \frac{8}{ (1+y)^2}
          - \frac{8}{ (1+y)}
          \Biggr) H(-1,0,-1;y) \nn\\
\hspace{-10mm} & &  \hspace{12mm}
       -   \Biggl(
            8
          - \frac{144}{ (1-y)^5}
          + \frac{360}{ (1-y)^4}
          - \frac{312}{ (1-y)^3}
          + \frac{116}{ (1-y)^2} \nn\\
\hspace{-10mm} & &  \hspace{12mm}
          - \frac{18}{ (1-y)}
          + \frac{8}{ (1+y)^2}
          - \frac{10}{ (1+y)}
          \Biggr) H(-1,0,0;y) \nn\\
\hspace{-10mm} & &  \hspace{12mm}
       -   \Biggl(
            8
          + \frac{8}{ (1-y)^2}
          - \frac{8}{ (1-y)}
          + \frac{8}{ (1+y)^2} \nn\\
\hspace{-10mm} & &  \hspace{12mm}
          - \frac{8}{ (1+y)}
          \Biggr) H(-1,0,1;y)
       + \zeta(2)   \Biggl(
          - 14
          - \frac{60}{ (1-y)^5} \nn\\
\hspace{-10mm} & &  \hspace{12mm}
          +  \frac{150}{ (1-y)^4}\! 
          - \! \frac{132}{ (1-y)^3}\! 
          + \! \frac{36}{ (1-y)^2}\! 
          + \! \frac{8}{ (1-y)}\! 
          - \! \frac{12}{ (1+y)^2} \nn\\
\hspace{-10mm} & &  \hspace{12mm}
          + \frac{14}{ (1+y)}
          \Biggr) H(0;y)
       +   \Biggl(
            \frac{229}{4}
          + \frac{192}{ (1-y)^4}
          - \frac{504}{ (1-y)^3} \nn\\
\hspace{-10mm} & &  \hspace{12mm}
          + \frac{466}{ (1-y)^2}
          - \frac{343}{2 (1-y)}
          + \frac{16}{ (1+y)^2}
          - \frac{42}{ (1+y)}
          \Biggr) H(0;y) \nn\\
\hspace{-10mm} & &  \hspace{12mm}
       +   \Biggl(
            10
          + \frac{96}{ (1-y)^4}
          - \frac{192}{ (1-y)^3}
          + \frac{116}{ (1-y)^2}
          - \frac{20}{ (1-y)} \nn\\
\hspace{-10mm} & &  \hspace{12mm}
          + \frac{8}{ (1+y)^2}
          - \frac{8}{ (1+y)}
          \Biggr) H(0,-1;y)
       +   \Biggl(
            4
          - \frac{360}{ (1-y)^5} \nn\\
\hspace{-10mm} & &  \hspace{12mm}
          + \frac{900}{ (1-y)^4}
          - \frac{700}{ (1-y)^3}
          + \frac{150}{ (1-y)^2}
          - \frac{5}{ (1-y)} \nn\\
\hspace{-10mm} & &  \hspace{12mm}
          + \frac{7}{ (1\! +\! y)}
          \Biggr) H(0,-1,0;y)\! 
       +  \!  \Biggl(
          - 10\! 
          - \! \frac{60}{ (1-y)^5}\! 
          + \! \frac{216}{ (1-y)^4} \nn\\
\hspace{-10mm} & &  \hspace{12mm}
          - \frac{256}{ (1-y)^3}
          + \frac{89}{ (1-y)^2}
          + \frac{71}{2 (1-y)}
          + \frac{4}{ (1+y)^2} \nn\\
\hspace{-10mm} & &  \hspace{12mm}
          - \frac{17}{2 (1\! +\! y)}
          \Biggr) H(0,0;y)\! 
       +  \!  \Biggl(
          - 12\! 
          + \! \frac{96}{ (1-y)^5}\! 
          - \! \frac{240}{ (1-y)^4} \nn\\
\hspace{-10mm} & &  \hspace{12mm}
          + \frac{176}{ (1-y)^3}
          - \frac{28}{ (1-y)^2}
          + \frac{8}{ (1-y)}
          - \frac{4}{ (1+y)^2} \nn\\
\hspace{-10mm} & &  \hspace{12mm}
          + \frac{8}{ (1\! +\! y)}
          \Biggr) H(0,0,-1;y)\! 
       +  \!  \Biggl(
            27\! 
          - \! \frac{6}{ (1-y)^5}\! 
          + \! \frac{15}{ (1-y)^4} \nn\\
\hspace{-10mm} & &  \hspace{12mm}
          - \frac{17}{2 (1-y)^3}
          + \frac{59}{4 (1-y)^2}
          - \frac{201}{8 (1-y)}
          + \frac{17}{ (1+y)^2} \nn\\
\hspace{-10mm} & &  \hspace{12mm}
          - \frac{217}{8 (1+y)}
          \Biggr) H(0,0,0;y)
       +   \Biggl(
            22
          - \frac{384}{ (1-y)^5}
          + \frac{960}{ (1-y)^4} \nn\\
\hspace{-10mm} & &  \hspace{12mm}
          - \frac{746}{ (1-y)^3}
          + \frac{169}{ (1-y)^2}
          - \frac{45}{2 (1-y)}
          + \frac{10}{ (1+y)^2} \nn\\
\hspace{-10mm} & &  \hspace{12mm}
          - \frac{21}{2 (1\! +\! y)}
          \Biggr) H(0,0,1;y)\! 
       +  \!  \Biggl(
          - 16\! 
          - \! \frac{384}{ (1-y)^4}\! 
          + \! \frac{768}{ (1-y)^3} \nn\\
\hspace{-10mm} & &  \hspace{12mm}
          - \frac{442}{ (1-y)^2}
          + \frac{60}{ (1-y)}
          + \frac{2}{ (1+y)}
          \Biggr) H(0,1;y) \nn\\
\hspace{-10mm} & &  \hspace{12mm}
       +   \Biggl(
            14
          + \frac{252}{ (1-y)^5}
          - \frac{630}{ (1-y)^4}
          + \frac{517}{ (1-y)^3}
          - \frac{271}{2 (1-y)^2} \nn\\
\hspace{-10mm} & &  \hspace{12mm}
          - \frac{19}{4 (1-y)}
          + \frac{10}{ (1+y)^2}
          - \frac{67}{4 (1+y)}
          \Biggr) H(0,1,0;y) \nn\\
\hspace{-10mm} & &  \hspace{12mm}
       +   \Biggl(
            4
          + \frac{4}{ (1-y)^2}
          - \frac{4}{ (1-y)}
          + \frac{4}{ (1+y)^2} \nn\\
\hspace{-10mm} & &  \hspace{12mm}
          - \frac{4}{ (1+y)}
          \Biggr) H(0,1,1;y)
       + \zeta(2)   \Biggl(
          - 16
          - \frac{16}{ (1-y)^2} \nn\\
\hspace{-10mm} & &  \hspace{12mm}
          + \frac{16}{ (1-y)}
          - \frac{16}{ (1+y)^2}
          + \frac{16}{ (1+y)}
          \Biggr) H(1;y) \nn\\
\hspace{-10mm} & &  \hspace{12mm}
       +   \Biggl(
            \frac{55}{2}
          - \frac{192}{ (1-y)^3}
          + \frac{288}{ (1-y)^2}
          - \frac{115}{ (1-y)} \nn\\
\hspace{-10mm} & &  \hspace{12mm}
          - \frac{36}{ (1+y)}
          \Biggr) H(1;y)
       +   \Biggl(
            6
          + \frac{252}{ (1-y)^4}
          - \frac{504}{ (1-y)^3} \nn\\
\hspace{-10mm} & &  \hspace{12mm}
          + \frac{270}{ (1-y)^2}
          - \frac{16}{ (1-y)}
          + \frac{16}{ (1+y)^2}
          - \frac{14}{ (1+y)}
          \Biggr) H(1,0;y) \nn\\
\hspace{-10mm} & &  \hspace{12mm}
       +   \Biggl(
            12
          + \frac{12}{ (1-y)^2}
          - \frac{12}{ (1-y)}
          + \frac{12}{ (1+y)^2} \nn\\
\hspace{-10mm} & &  \hspace{12mm}
          - \frac{12}{ (1+y)}
          \Biggr) H(1,0,0;y)
       +   \Biggl(
            8
          + \frac{8}{ (1-y)^2}
          - \frac{8}{ (1-y)} \nn\\
\hspace{-10mm} & &  \hspace{12mm}
          + \frac{8}{ (1+y)^2}
          - \frac{8}{ (1+y)}
          \Biggr) H(1,0,1;y)
       -   \Biggl(
            4
          - \frac{4}{ (1-y)} \nn\\
\hspace{-10mm} & &  \hspace{12mm}
          - \frac{4}{ (1+y)}
          \Biggr) H(1,1;y)
       +  \Biggl(
            16
          + \frac{16}{ (1-y)^2}
          - \frac{16}{ (1-y)} \nn\\
\hspace{-10mm} & &  \hspace{12mm}
          + \frac{16}{ (1+y)^2}
          - \frac{16}{ (1+y)}
          \Biggr) H(1,1,0;y) 
          \Biggr] \, ,
\eea
%
%%%%%%
%%

and
\bea
\hspace{-10mm} 
\Re \, {\mathcal F}^{(2l)}_{2,R}(\epsilon,s) & = &
\frac{1}{\epsilon}  \Biggl\{
         C_{F}^2 \Biggl[
         \zeta(2)   \Biggl(
            \frac{6}{ (1-y)^2}
          - \frac{6}{ (1-y)}
          - \frac{6}{ (1+y)^2}
          + \frac{6}{ (1+y)}
          \Biggr) \nn\\
\hspace{-10mm} & &  \hspace{12mm}
       -   \Biggl(
            \frac{1}{(1 \! - \! y)} \! 
          -  \! \frac{1}{(1 \! + \! y)}
          \Biggr) H(0,y) \! 
       -    \! \Biggl(
            \frac{2}{ (1-y)^2} \! 
          -  \! \frac{2}{ (1-y)} \nn\\
\hspace{-10mm} & &  \hspace{12mm}
          - \frac{2}{ (1+y)^2}
          + \frac{2}{ (1+y)}
          \Biggr) H(0,0,y)
           \Biggl] \Biggr\}  \nn\\
\hspace{-10mm} & & 
    + \, C_{F} C_{A}  \Biggl[ 
          - \frac{3}{ (1-y)^2} \! 
          +  \! \frac{3}{ (1-y)}
       - \zeta(2)   \Biggl(
            \frac{24 \log{2}}{ (1 \! - \! y)^2} \! 
          -  \! \frac{24 \log{2}}{ (1 \! - \! y)} \nn\\
\hspace{-10mm} & &  \hspace{12mm}
          - \frac{240}{ (1-y)^4}
          + \frac{480}{ (1-y)^3}
          - \frac{256}{ (1-y)^2}
          + \frac{49}{3 (1-y)} \nn\\
\hspace{-10mm} & &  \hspace{12mm}
          +  \frac{6}{ (1 \! + \! y)^2} \! 
          -  \! \frac{19}{3 (1 \! + \! y)}
          \Biggr) \! 
       -  \! \zeta^2(2)   \Biggl(
            \frac{618}{5 (1-y)^5} \! 
          -  \! \frac{309}{ (1-y)^4} \nn\\
\hspace{-10mm} & &  \hspace{12mm}
          + \frac{2313}{10 (1-y)^3}
          - \frac{759}{20 (1-y)^2}
          - \frac{27}{8 (1-y)}
          + \frac{12}{5 (1+y)^3} \nn\\
\hspace{-10mm} & &  \hspace{12mm}
          - \frac{18}{5 (1 \! + \! y)^2} \! 
          -  \! \frac{27}{8 (1 \! + \! y)}
          \Biggr)
       - \zeta(3)   \Biggl(
            \frac{324}{ (1-y)^4} \! 
          -  \! \frac{648}{ (1-y)^3} \nn\\
\hspace{-10mm} & &  \hspace{12mm}
          +  \frac{344}{ (1-y)^2}
          - \frac{20}{ (1-y)}
          \Biggr)
       - \zeta(2)   \Biggl(
            \frac{72}{ (1-y)^4} \! 
          -  \! \frac{144}{ (1-y)^3} \nn\\
\hspace{-10mm} & &  \hspace{12mm}
          + \frac{120}{ (1-y)^2}
          - \frac{48}{ (1-y)}
          \Biggr) H(-1,y)
       + \zeta(2)   \Biggl(
            \frac{672}{ (1-y)^5} \nn\\
\hspace{-10mm} & &  \hspace{12mm}
          - \frac{1680}{ (1-y)^4}
          + \frac{1320}{ (1-y)^3}
          - \frac{300}{ (1-y)^2}
          - \frac{6}{ (1-y)} \nn\\
\hspace{-10mm} & &  \hspace{12mm}
          - \frac{6}{ (1+y)}
          \Biggr) H(-1,0,y)
       +   \Biggl(
            \frac{24}{ (1-y)^3}
          - \frac{36}{ (1-y)^2} \nn\\
\hspace{-10mm} & &  \hspace{12mm}
          + \frac{12}{ (1-y)}
          \Biggr) H(-1,0,y)
       +   \Biggl(
            \frac{24}{ (1-y)^4}
          - \frac{48}{ (1-y)^3} \nn\\
\hspace{-10mm} & &  \hspace{12mm}
          + \frac{40}{ (1-y)^2}
          - \frac{16}{ (1-y)}
          \Biggr) H(-1,0,0,y)
       -   \Biggl(
            \frac{336}{ (1-y)^5} \nn\\
\hspace{-10mm} & &  \hspace{12mm}
          - \frac{840}{ (1-y)^4}
          + \frac{660}{ (1-y)^3}
          - \frac{150}{ (1-y)^2}
          - \frac{3}{ (1-y)} \nn\\
\hspace{-10mm} & &  \hspace{12mm}
          - \frac{3}{ (1+y)}
          \Biggr) H(-1,0,0,0,y)
       - \zeta(2)   \Biggl(
            \frac{516}{ (1-y)^5} \nn\\
\hspace{-10mm} & &  \hspace{12mm}
          - \frac{1704}{ (1-y)^4}
          + \frac{1947}{ (1-y)^3}
          - \frac{840}{ (1-y)^2}
          + \frac{319}{4 (1-y)} \nn\\
\hspace{-10mm} & &  \hspace{12mm}
          + \frac{6}{ (1+y)^3}
          - \frac{17}{2 (1+y)^2}
          + \frac{15}{4 (1+y)}
          \Biggr) H(0,y) \nn\\
\hspace{-10mm} & &  \hspace{12mm}
       - \zeta(3)   \Biggl(
            \frac{324}{ (1-y)^5} \! 
          -  \! \frac{810}{ (1-y)^4} \! 
          +  \! \frac{635}{ (1-y)^3} \! 
          -  \! \frac{285}{2 (1-y)^2} \nn\\
\hspace{-10mm} & &  \hspace{12mm}
          -  \frac{13}{4 (1-y)} \! 
          -  \! \frac{13}{4 (1 \! + \! y)}
          \Biggr) H(0,y) \! 
       -   \!  \Biggl(
            \frac{9}{ (1 \! - \! y)^3} \! 
          -  \! \frac{27}{2 (1 \! - \! y)^2} \nn\\
\hspace{-10mm} & &  \hspace{12mm}
          - \frac{23}{9 (1-y)}
          + \frac{127}{18 (1+y)}
          \Biggr) H(0,y)
       - \zeta(2)   \Biggl(
            \frac{72}{ (1-y)^5} \nn\\
\hspace{-10mm} & &  \hspace{12mm}
          - \frac{180}{ (1-y)^4}
          + \frac{210}{ (1-y)^3}
          - \frac{135}{ (1-y)^2}
          + \frac{33}{2 (1-y)} \nn\\
\hspace{-10mm} & &  \hspace{12mm}
          + \frac{33}{2 (1+y)}
          \Biggr) H(0,-1,y)
       +   \Biggl(
            \frac{48}{ (1-y)^4}
          - \frac{96}{ (1-y)^3} \nn\\
\hspace{-10mm} & &  \hspace{12mm}
          + \frac{48}{ (1-y)^2}
          \Biggr) H(0,-1,0,y)
       +   \Biggl(
            \frac{24}{ (1-y)^5}
          - \frac{60}{ (1-y)^4} \nn\\
\hspace{-10mm} & &  \hspace{12mm}
          + \frac{70}{ (1-y)^3}
          - \frac{45}{ (1-y)^2}
          + \frac{11}{2 (1-y)} \nn\\
\hspace{-10mm} & &  \hspace{12mm}
          + \frac{11}{2 (1+y)}
          \Biggr) H(0,-1,0,0,y)
       + \zeta(2)   \Biggl(
            \frac{78}{ (1-y)^5} \nn\\
\hspace{-10mm} & &  \hspace{12mm}
          - \frac{195}{ (1-y)^4}
          + \frac{147}{ (1-y)^3}
          - \frac{51}{2 (1-y)^2}
          - \frac{19}{8 (1-y)} \nn\\
\hspace{-10mm} & &  \hspace{12mm}
          - \frac{1}{2 (1+y)^3}
          + \frac{3}{4 (1+y)^2}
          - \frac{19}{8 (1+y)}
          \Biggr) H(0,0,y) \nn\\
\hspace{-10mm} & &  \hspace{12mm}
       -   \Biggl(
            \frac{24}{ (1-y)^4}
          - \frac{39}{ (1-y)^3}
          + \frac{7}{2 (1-y)^2}
          + \frac{137}{12 (1-y)} \nn\\
\hspace{-10mm} & &  \hspace{12mm}
          - \frac{3}{ (1+y)^2}
          + \frac{37}{12 (1+y)}
          \Biggr) H(0,0,y)
       +   \Biggl(
            \frac{48}{ (1-y)^5} \nn\\
\hspace{-10mm} & &  \hspace{12mm}
          - \frac{120}{ (1-y)^4}
          + \frac{84}{ (1-y)^3}
          - \frac{6}{ (1-y)^2}
          - \frac{3}{ (1-y)} \nn\\
\hspace{-10mm} & &  \hspace{12mm}
          - \frac{3}{ (1+y)}
          \Biggr) H(0,0,-1,0,y)
       +  \Biggl(
            \frac{258}{ (1-y)^5}
          - \frac{816}{ (1-y)^4} \nn\\
\hspace{-10mm} & &  \hspace{12mm}
          + \frac{1803}{2 (1-y)^3}
          - \frac{767}{2 (1-y)^2}
          + \frac{315}{8 (1-y)}
          + \frac{3}{ (1+y)^3} \nn\\
\hspace{-10mm} & &  \hspace{12mm}
          - \frac{15}{4 (1+y)^2}
          + \frac{11}{8 (1+y)}
          \Biggr) H(0,0,0,y) 
       -   \Biggl(
            \frac{3}{ (1-y)^5} \nn\\
\hspace{-10mm} & &  \hspace{12mm}
          - \frac{15}{2 (1-y)^4}
          + \frac{4}{ (1-y)^3}
          + \frac{3}{2 (1-y)^2}
          - \frac{11}{16 (1-y)} \nn\\
\hspace{-10mm} & &  \hspace{12mm}
          - \frac{3}{4 (1 \! + \! y)^3} \! 
          +  \! \frac{9}{8 (1 \! + \! y)^2}  \! 
          -  \! \frac{11}{16 (1+y)}
          \Biggr) H(0,0,0,0,y) \nn\\
\hspace{-10mm} & &  \hspace{12mm}
       +   \Biggl(
            \frac{12}{ (1-y)^5}
          - \frac{30}{ (1-y)^4} 
          + \frac{48}{ (1-y)^3}
          - \frac{42}{ (1-y)^2} \nn\\
\hspace{-10mm} & &  \hspace{12mm}
          +  \frac{25}{4 (1-y)} 
          +  \frac{1}{(1 + y)^3} 
          -  \frac{3}{2 (1 + y)^2} \nn\\
\hspace{-10mm} & &  \hspace{12mm}
          +  \frac{25}{4 (1 \! + \! y)}
          \Biggr) H(0, \! 0, \! 1, \! 0,y) \! 
       +  \! \zeta(2)   \Biggl(
            \frac{756}{ (1-y)^5} \! 
          -  \! \frac{1890}{ (1-y)^4} \nn\\
\hspace{-10mm} & &  \hspace{12mm}
          + \frac{1512}{ (1-y)^3}
          - \frac{378}{ (1-y)^2}
          + \frac{3}{4 (1-y)}
          + \frac{3}{ (1+y)^3} \nn\\
\hspace{-10mm} & &  \hspace{12mm}
          - \frac{9}{2 (1+y)^2}
          + \frac{3}{4 (1+y)}
          \Biggr) H(0,1,y)
       +   \Biggl(
            \frac{12}{ (1-y)^4} \nn\\
\hspace{-10mm} & &  \hspace{12mm}
          - \frac{24}{ (1-y)^3}
          + \frac{21}{ (1-y)^2}
          - \frac{9}{ (1-y)}
          + \frac{1}{(1+y)^2} \nn\\
\hspace{-10mm} & &  \hspace{12mm}
          - \frac{1}{(1+y)}
          \Biggr) H(0,1,0,y)
       -   \Biggl(
            \frac{282}{ (1-y)^5}
          - \frac{705}{ (1-y)^4} \nn\\
\hspace{-10mm} & &  \hspace{12mm}
          + \frac{562}{ (1-y)^3}
          - \frac{138}{ (1-y)^2}
          - \frac{5}{8 (1-y)}
          - \frac{1}{2 (1+y)^3} \nn\\
\hspace{-10mm} & &  \hspace{12mm}
          + \frac{3}{4 (1+y)^2}
          - \frac{5}{8 (1+y)}
          \Biggr) H(0,1,0,0,y) \nn\\
\hspace{-10mm} & &  \hspace{12mm}
       + \zeta(2)   \Biggl(
            \frac{756}{ (1-y)^4}
          - \frac{1512}{ (1-y)^3}
          + \frac{819}{ (1-y)^2}
          - \frac{63}{ (1-y)} \nn\\
\hspace{-10mm} & &  \hspace{12mm}
          + \frac{3}{ (1+y)^2}
          - \frac{3}{ (1+y)}
          \Biggr) H(1,y)
       +   \Biggl(
            \frac{6}{ (1-y)^3} \nn\\
\hspace{-10mm} & &  \hspace{12mm}
          - \frac{9}{ (1-y)^2}
          + \frac{19}{6 (1-y)}
          - \frac{1}{6 (1+y)}
          \Biggr) H(1,0,y) \nn\\
\hspace{-10mm} & &  \hspace{12mm}
       -   \Biggl(
            \frac{282}{ (1-y)^4}
          - \frac{564}{ (1-y)^3}
          + \frac{623}{2 (1-y)^2}
          - \frac{59}{2 (1-y)} \nn\\
\hspace{-10mm} & &  \hspace{12mm}
          - \frac{1}{2 (1+y)^2}
          + \frac{1}{2 (1+y)}
          \Biggr) H(1,0,0,y)
          \Biggr] \nn\\
\hspace{-10mm} & &  \hspace{6mm}
       + \, C_{F} T_{R} N_{f}   
          \Biggl[
         \zeta(2)   \Biggl(
            \frac{8}{3 (1-y)}
          - \frac{8}{3 (1+y)}
          \Biggr)
       -   \Biggl(
            \frac{25}{9 (1-y)} \nn\\
\hspace{-10mm} & &  \hspace{12mm}
          - \frac{25}{9 (1 \! + \! y)}
          \Biggr) H(0,y) \! 
       -   \!  \Biggl(
            \frac{2}{3 (1 \! - \! y)} \! 
          -  \! \frac{2}{3 (1 \! + \! y)} \! 
          \Biggr) H(0,0,y) \nn\\
\hspace{-10mm} & &  \hspace{12mm}
       -   \Biggl(
            \frac{4}{3 (1-y)}
          - \frac{4}{3 (1+y)}
          \Biggr) H(1,0,y)
          \Biggr] \nn\\
\hspace{-10mm} & &  \hspace{6mm}
       + \, C_{F} T_{R}  
          \Biggl[
          - \frac{68}{3 (1-y)^2}
          + \frac{68}{3 (1-y)}
       - \zeta(2)   \Biggl(
            \frac{48}{ (1-y)^4} \nn\\
\hspace{-10mm} & &  \hspace{12mm}
          - \frac{96}{ (1-y)^3}
          + \frac{40}{ (1-y)^2}
          + \frac{8}{ (1-y)}
          \Biggr)
       - \zeta(2)   \Biggl(
            \frac{48}{ (1-y)^5} \nn\\
\hspace{-10mm} & &  \hspace{12mm}
          - \frac{120}{ (1-y)^4}
          + \frac{84}{ (1-y)^3}
          - \frac{6}{ (1-y)^2}
          - \frac{3}{ (1-y)} \nn\\
\hspace{-10mm} & &  \hspace{12mm}
          - \frac{3}{ (1+y)}
          \Biggr) H(0,y)
       -   \Biggl(
            \frac{124}{3 (1-y)^3}
          - \frac{62}{ (1-y)^2} \nn\\
\hspace{-10mm} & &  \hspace{12mm}
          + \frac{118}{9 (1-y)}
          + \frac{68}{9 (1+y)}
          \Biggr) H(0,y)
       -   \Biggl(
            \frac{88}{3 (1-y)^4} \nn\\
\hspace{-10mm} & &  \hspace{12mm}
          - \frac{176}{3 (1-y)^3}
          + \frac{92}{3 (1-y)^2}
          - \frac{4}{3 (1-y)}
          \Biggr) H(0,0,y) \nn\\
\hspace{-10mm} & &  \hspace{12mm}
       +   \Biggl(
            \frac{24}{ (1-y)^5}
          - \frac{60}{ (1-y)^4}
          + \frac{42}{ (1-y)^3}
          - \frac{3}{ (1-y)^2} \nn\\
\hspace{-10mm} & &  \hspace{12mm}
          - \frac{3}{2 (1-y)}
          - \frac{3}{2 (1+y)}
          \Biggr) H(0,0,0,y)
          \Biggr] \nn\\
\hspace{-10mm} & &  \hspace{6mm}
       + \, C_{F}^2   
          \Biggl[
         \zeta(2)   \Biggl(
            \frac{48 \log{2}}{ (1 \! - \! y)^2}
          - \frac{48 \log{2}}{ (1 \! - \! y)} \! 
          +  \! \frac{816}{ (1-y)^4}
          - \frac{2040}{ (1-y)^3} \nn\\
\hspace{-10mm} & &  \hspace{12mm}
          + \frac{1630}{ (1-y)^2}
          - \frac{390}{ (1-y)}
          - \frac{6}{ (1+y)^2}
          - \frac{10}{ (1+y)}
          \Biggr) \nn\\
\hspace{-10mm} & &  \hspace{12mm}
       - \zeta^2(2)   \Biggl(
            \frac{18}{ (1 \! - \! y)^5} \! 
          -  \! \frac{45}{ (1 \! - \! y)^4} \! 
          +  \! \frac{639}{10 (1 \! - \! y)^3} \! 
          -  \! \frac{1017}{20 (1 \! - \! y)^2} \nn\\
\hspace{-10mm} & &  \hspace{12mm}
          +  \frac{231}{40 (1-y)}
          - \frac{24}{5 (1+y)^3} \! 
          +  \! \frac{36}{5 (1 \! + \! y)^2} \! 
          +  \! \frac{231}{40 (1 \! + \! y)}
          \Biggr) \nn\\
\hspace{-10mm} & &  \hspace{12mm}
       - \zeta(3)   \Biggl(
            \frac{168}{ (1-y)^4}
          - \frac{336}{ (1-y)^3}
          + \frac{182}{ (1-y)^2}
          - \frac{14}{ (1-y)} \nn\\
\hspace{-10mm} & &  \hspace{12mm}
          + \frac{10}{ (1+y)^2}
          - \frac{10}{ (1+y)}
          \Biggr)
       + \zeta(2)   \Biggl(
          - \frac{1080}{ (1-y)^4} \nn\\
\hspace{-10mm} & &  \hspace{12mm}
          + \frac{2160}{ (1-y)^3}
          - \frac{1068}{ (1-y)^2}
          - \frac{12}{ (1-y)}
          - \frac{12}{ (1+y)^2} \nn\\
\hspace{-10mm} & &  \hspace{12mm}
          + \frac{12}{ (1+y)}
          \Biggr) H(-1,y)
       + \zeta(2)   \Biggl(
            \frac{288}{ (1-y)^5}
          - \frac{720}{ (1-y)^4} \nn\\
\hspace{-10mm} & &  \hspace{12mm}
          +  \frac{600}{ (1-y)^3} \! 
          -  \! \frac{180}{ (1-y)^2} \! 
          +  \! \frac{6}{ (1-y)} \! 
          +  \! \frac{6}{ (1+y)}
          \Biggr) H(-1,0,y) \nn\\
\hspace{-10mm} & &  \hspace{12mm}
       -   \Biggl(
            \frac{48}{ (1 \! - \! y)^3} \! 
          -  \! \frac{72 }{(1 \! - \! y)^2} \! 
          +  \! \frac{28 }{(1 \! - \! y)} \! 
          -  \! \frac{4}{ (1 \! + \! y)}
          \Biggr) H(-1,0,y) \nn\\
\hspace{-10mm} & &  \hspace{12mm}
       +    \Biggl(
            \frac{360}{ (1-y)^4}
          -  \! \frac{720}{ (1-y)^3} \! 
          +  \! \frac{356}{ (1-y)^2} \! 
          +  \! \frac{4}{ (1-y)} \! 
          +  \! \frac{4}{ (1 \! + \! y)^2} \nn\\
\hspace{-10mm} & &  \hspace{12mm}
          - \frac{4}{ (1 \! + \! y)}
          \Biggr) H(-1,0,0,y) \! 
       -   \!  \Biggl(
            \frac{144}{ (1 \! - \! y)^5} \! 
          -  \! \frac{360}{ (1 \! - \! y)^4} \! 
          +  \! \frac{300}{ (1 \! - \! y)^3} \nn\\
\hspace{-10mm} & &  \hspace{12mm}
          - \frac{90}{ (1-y)^2}
          + \frac{3}{ (1-y)}
          + \frac{3}{ (1+y)}
          \Biggr) H(-1,0,0,0,y) \nn\\
\hspace{-10mm} & &  \hspace{12mm}
       - \zeta(2)   \Biggl(
            \frac{120}{ (1-y)^5}
          - \frac{492}{ (1-y)^4}
          + \frac{622}{ (1-y)^3}
          - \frac{254}{ (1-y)^2} \nn\\
\hspace{-10mm} & &  \hspace{12mm}
          - \frac{35}{2 (1-y)}
          - \frac{1}{(1+y)^2}
          + \frac{45}{2 (1+y)}
          \Biggr) H(0,y) \nn\\
\hspace{-10mm} & &  \hspace{12mm}
       - \zeta(3)   \Biggl(
            \frac{168}{ (1-y)^5}
          - \frac{420}{ (1-y)^4}
          + \frac{350}{ (1-y)^3}
          - \frac{105}{ (1-y)^2} \nn\\
\hspace{-10mm} & &  \hspace{12mm}
          + \frac{7}{2 (1-y)}
          + \frac{7}{2 (1+y)}
          \Biggr) H(0,y)
       -   \Biggl(
            \frac{31}{4 (1-y)} \nn\\
\hspace{-10mm} & &  \hspace{12mm}
          - \frac{31}{4 (1+y)}
          \Biggr) H(0,y)
       - \zeta(2)   \Biggl(
            \frac{1080}{ (1-y)^5}
          - \frac{2700}{ (1-y)^4} \nn\\
\hspace{-10mm} & &  \hspace{12mm}
          +  \! \frac{2010}{ (1-y)^3} \! 
          -  \! \frac{315}{ (1-y)^2} \! 
          -  \! \frac{75}{2 (1-y)} \! 
          -  \! \frac{75}{2 (1 \! + \! y)}
          \Biggr) H(0,-1,y) \nn\\
\hspace{-10mm} & &  \hspace{12mm}
       -   \Biggl(
            \frac{96}{ (1-y)^4} \! 
          -  \! \frac{192}{ (1-y)^3} \! 
          +  \! \frac{100}{ (1-y)^2} \! 
          -  \! \frac{4}{ (1-y)} \! 
          - \!  \frac{4}{ (1 \! + \! y)^2} \nn\\
\hspace{-10mm} & &  \hspace{12mm}
          +  \frac{4}{ (1 \! + \! y)}
          \Biggr) H(0, \! -1, \! 0,y) \! 
       +  \!   \Biggl(
            \frac{360}{ (1 \! - \! y)^5} \! 
          -  \! \frac{900}{ (1 \! - \! y)^4} \! 
          +  \! \frac{670}{ (1 \! - \! y)^3} \nn\\
\hspace{-10mm} & &  \hspace{12mm}
          - \frac{105}{ (1-y)^2}
          - \frac{25}{2 (1-y)}
          - \frac{25}{2 (1+y)}
          \Biggr) H(0,-1,0,0,y) \nn\\
\hspace{-10mm} & &  \hspace{12mm}
       + \zeta(2)   \Biggl(
            \frac{48}{ (1-y)^5}
          - \frac{120}{ (1-y)^4}
          + \frac{111}{ (1-y)^3}
          - \frac{93}{2 (1-y)^2} \nn\\
\hspace{-10mm} & &  \hspace{12mm}
          +  \frac{4}{ (1-y)} \! 
          +  \! \frac{1}{(1+y)^3} \! 
          -  \! \frac{3}{2 (1+y)^2} \! 
          +  \! \frac{4}{ (1 \! + \! y)}
          \Biggr) H(0,0,y) \nn\\
\hspace{-10mm} & &  \hspace{12mm}
       -   \Biggl(
            \frac{192}{ (1 \! - \! y)^4} \! 
          -  \! \frac{504}{ (1 \! - \! y)^3} \! 
          +  \! \frac{442}{ (1 \! - \! y)^2} \! 
          -  \! \frac{255}{2 (1 \! - \! y)}
          -  \! \frac{2}{ (1 \! + \! y)^2} \nn\\
\hspace{-10mm} & &  \hspace{12mm}
          - \frac{1}{2 (1 \! + \! y)}
          \Biggr) H(0,0,y) \! 
       -   \!  \Biggl(
            \frac{96}{ (1 \! - \! y)^5} \! 
          -  \! \frac{240}{ (1 \! - \! y)^4}
          +  \! \frac{168}{ (1 \! - \! y)^3} \nn\\
\hspace{-10mm} & &  \hspace{12mm}
          - \frac{12}{ (1-y)^2}
          - \frac{6}{ (1-y)}
          - \frac{6}{ (1+y)}
          \Biggr) H(0,0,-1,0,y) \nn\\
\hspace{-10mm} & &  \hspace{12mm}
       +   \Biggl(
            \frac{60}{ (1-y)^5}
          - \frac{216}{ (1-y)^4}
          + \frac{251}{ (1-y)^3}
          - \frac{88}{ (1-y)^2} \nn\\
\hspace{-10mm} & &  \hspace{12mm}
          - \frac{71}{4 (1-y)}
          - \frac{7}{2 (1+y)^2}
          + \frac{57}{4 (1+y)}
          \Biggr) H(0,0,0,y) \nn\\
\hspace{-10mm} & &  \hspace{12mm}
       +    \Biggl(
            \frac{6}{ (1-y)^5} \! 
          -  \! \frac{15}{ (1-y)^4} \! 
          +  \! \frac{8}{ (1 \! - \! y)^3} \! 
          +  \! \frac{3}{ (1 \! - \! y)^2} \! 
          -  \! \frac{11}{8 (1 \! - \! y)} \nn\\
\hspace{-10mm} & &  \hspace{12mm}
          - \frac{3}{2 (1+y)^3}
          + \frac{9}{4 (1+y)^2}
          - \frac{11}{8 (1+y)}
          \Biggr) H(0,0,0,0,y) \nn\\
\hspace{-10mm} & &  \hspace{12mm}
       +    \Biggl(
            \frac{384}{ (1 \! - \! y)^5} \! 
          - \!  \frac{960}{ (1 \! - \! y)^4} \! 
          +  \! \frac{714}{ (1 \! - \! y)^3} \! 
          -  \! \frac{111}{ (1 \! - \! y)^2} \! 
          -  \! \frac{14}{ (1 \! - \! y)} \nn\\
\hspace{-10mm} & &  \hspace{12mm}
          - \frac{2}{ (1+y)^3}
          + \frac{3}{ (1+y)^2}
          - \frac{14}{ (1+y)}
          \Biggr) H(0,0,1,0,y) \nn\\
\hspace{-10mm} & &  \hspace{12mm}
       + \zeta(2)   \Biggl(
            \frac{936}{ (1-y)^5}
          - \frac{2340}{ (1-y)^4}
          + \frac{1836}{ (1-y)^3}
          - \frac{414}{ (1-y)^2} \nn\\
\hspace{-10mm} & &  \hspace{12mm}
          -  \frac{21}{2 (1-y)} \! 
          -  \! \frac{6}{ (1 \! + \! y)^3} \! 
          +  \! \frac{9}{ (1 \! + \! y)^2} \! 
          -  \! \frac{21}{2 (1 \! + \! y)}
          \Biggr) H(0,1,y) \nn\\
\hspace{-10mm} & &  \hspace{12mm}
       +   \Biggl(
            \frac{384}{ (1-y)^4} \! 
          -  \! \frac{768}{ (1-y)^3} \! 
          +  \! \frac{394}{ (1-y)^2} \! 
          - \!  \frac{10}{ (1-y)} \! 
          +  \! \frac{2}{ (1 \! + \! y)^2} \nn\\
\hspace{-10mm} & &  \hspace{12mm}
          - \frac{2}{ (1 \! + \! y)}
          \Biggr) H(0, \! 1, \! 0,y) \! 
       -   \!  \Biggl(
            \frac{252}{ (1 \! - \! y)^5} \! 
          -  \! \frac{630}{ (1 \! - \! y)^4} \! 
          +  \! \frac{496}{ (1 \! - \! y)^3} \nn\\
\hspace{-10mm} & &  \hspace{12mm}
          - \frac{114}{ (1-y)^2}
          - \frac{7}{4 (1-y)}
          + \frac{1}{(1+y)^3}
          - \frac{3}{2 (1+y)^2} \nn\\
\hspace{-10mm} & &  \hspace{12mm}
          - \frac{7}{4 (1+y)}
          \Biggr) H(0,1,0,0,y)
       + \zeta(2)   \Biggl(
            \frac{936}{ (1-y)^4} \nn\\
\hspace{-10mm} & &  \hspace{12mm}
          - \frac{1872}{ (1-y)^3}
          + \frac{1002}{ (1-y)^2}
          - \frac{66}{ (1-y)}
          - \frac{6}{ (1+y)^2} \nn\\
\hspace{-10mm} & &  \hspace{12mm}
          + \frac{6}{ (1+y)}
          \Biggr) H(1,y)
       +   \Biggl(
            \frac{192}{ (1-y)^3}
          - \frac{288}{ (1-y)^2} \nn\\
\hspace{-10mm} & &  \hspace{12mm}
          + \frac{87}{ (1-y)}
          + \frac{9}{ (1+y)}
          \Biggr) H(1,0,y)
       -   \Biggl(
            \frac{252}{ (1-y)^4} \nn\\
\hspace{-10mm} & &  \hspace{12mm}
          - \frac{504}{ (1-y)^3}
          + \frac{257}{ (1-y)^2}
          - \frac{5}{ (1-y)}
          + \frac{1}{(1+y)^2} \nn\\
\hspace{-10mm} & &  \hspace{12mm}
          - \frac{1}{(1+y)}
          \Biggr) H(1,0,0,y)
          \Biggr] \, , \\
%
%%%%%%
%%
\hspace{-10mm} 
\Im \, {\mathcal F}^{(2l)}_{2,R}(\epsilon,s) & = & 
\frac{1}{\epsilon}  \Biggl\{
         C_{F}^2 \Biggl[
        \biggl(
            \frac{1}{1+y}
          - \frac{1}{1-y} \Biggl)
        + 2 \biggl(
            \frac{1}{1-y} 
          - \frac{1}{(1-y)^2} \nn\\
\hspace{-10mm} & &  \hspace{12mm}
          - \frac{1}{1+y}
          + \frac{1}{(1+y)^2} \Biggl) H(0,y)
          \Biggr] \Biggr\} \nn\\
\hspace{-10mm} & &  \hspace{6mm} 
        + C_{F} C_{A} \Biggl[
          - \frac{9}{ (1-y)^3}
          + \frac{27}{2 (1-y)^2}
          + \frac{23}{9 (1-y)}
          - \frac{127}{18 (1+y)} \nn\\
\hspace{-10mm} & &  \hspace{12mm}
       + \zeta(2)   \Biggl(
            \frac{72}{ (1-y)^4}
          - \frac{144}{ (1-y)^3}
          + \frac{73}{ (1-y)^2}
          - \frac{1}{(1-y)} \nn\\
\hspace{-10mm} & &  \hspace{12mm}
          + \frac{1}{(1+y)^2}
          - \frac{1}{(1+y)}
          \Biggr)
       + \zeta(3)   \Biggl(
          - \frac{324}{ (1-y)^5}
          + \frac{810}{ (1-y)^4} \nn\\
\hspace{-10mm} & &  \hspace{12mm}
          - \frac{635}{ (1-y)^3}
          + \frac{285}{2 (1-y)^2}
          + \frac{13}{4 (1-y)}
          + \frac{13}{4 (1+y)}
          \Biggr) \nn\\
\hspace{-10mm} & &  \hspace{12mm}
       +   \Biggl(
            \frac{24}{ (1-y)^3}
          - \frac{36}{ (1-y)^2}
          + \frac{12}{ (1-y)}
          \Biggr) H(-1;y) \nn\\
\hspace{-10mm} & &  \hspace{12mm}
       +  \Biggl(
            \frac{24}{ (1\! -\! y)^4}\! 
          - \! \frac{48}{ (1\! -\! y)^3}\! 
          + \! \frac{40}{ (1\! -\! y)^2}\! 
          - \! \frac{16}{ (1\! -\! y)}
          \Biggr) H(-1,0;y) \nn\\
\hspace{-10mm} & &  \hspace{12mm}
       +   \Biggl(
          - \frac{336}{ (1-y)^5}
          + \frac{840}{ (1-y)^4}
          - \frac{660}{ (1-y)^3}
          + \frac{150}{ (1-y)^2} \nn\\
\hspace{-10mm} & &  \hspace{12mm}
          + \frac{3}{ (1-y)}
          + \frac{3}{ (1+y)}
          \Biggr) H(-1,0,0;y)
       + \zeta(2)   \Biggl(
            \frac{72}{ (1-y)^5} \nn\\
\hspace{-10mm} & &  \hspace{12mm}
          - \frac{180}{ (1-y)^4}\! 
          + \! \frac{139}{ (1-y)^3}\! 
          - \! \frac{57}{2 (1\! -\! y)^2}\! 
          - \! \frac{1}{(1\! -\! y)}\! 
          + \! \frac{1}{(1\! +\! y)^3} \nn\\
\hspace{-10mm} & &  \hspace{12mm}
          - \frac{3}{2 (1\! +\! y)^2}\! 
          - \! \frac{1}{(1\! +\! y)}
          \Biggr) H(0;y)\! 
       +  \!  \Biggl(
          - \frac{24}{ (1\! -\! y)^4}\! 
          + \! \frac{39}{ (1\! -\! y)^3} \nn\\
\hspace{-10mm} & &  \hspace{12mm}
          - \frac{7}{2 (1\! -\! y)^2}\! 
          - \! \frac{137}{12 (1\! -\! y)}\! 
          + \! \frac{3}{ (1\! +\! y)^2}\! 
          - \! \frac{37}{12 (1\! +\! y)}
          \Biggr) H(0;y) \nn\\
\hspace{-10mm} & &  \hspace{12mm}
       +  \Biggl(
            \frac{48}{ (1-y)^4}
          - \frac{96}{ (1-y)^3}
          + \frac{48}{ (1-y)^2}
          \Biggr) H(0,-1;y)  \nn\\
\hspace{-10mm} & &  \hspace{12mm}
       +  \Biggl(
            \frac{24}{ (1-y)^5}\! 
          - \! \frac{60}{ (1-y)^4}\! 
          + \! \frac{70}{ (1\! -\! y)^3}\! 
          - \! \frac{45}{ (1\! -\! y)^2}\! 
          + \! \frac{11}{2 (1\! -\! y)} \nn\\
\hspace{-10mm} & &  \hspace{12mm}
          + \frac{11}{2 (1+y)}
          \Biggr) H(0,-1,0;y)
       +   \Biggl(
            \frac{258}{ (1-y)^5}
          - \frac{816}{ (1-y)^4} \nn\\
\hspace{-10mm} & &  \hspace{12mm}
          + \frac{1803}{2 (1\! -\! y)^3}\! 
          - \! \frac{767}{2 (1\! -\! y)^2}\! 
          + \! \frac{315}{8 (1\! -\! y)}\! 
          + \! \frac{3}{ (1\! +\! y)^3}\! 
          - \! \frac{15}{4 (1\! +\! y)^2} \nn\\
\hspace{-10mm} & &  \hspace{12mm}
          + \frac{11}{8 (1+y)}
          \Biggr) H(0,0;y)
       +   \Biggl(
            \frac{48}{ (1-y)^5}
          - \frac{120}{ (1-y)^4} \nn\\
\hspace{-10mm} & &  \hspace{12mm}
          + \frac{84}{ (1-y)^3}\! 
          - \! \frac{6}{ (1-y)^2}\! 
          - \! \frac{3}{ (1-y)}\! 
          - \! \frac{3}{ (1\! +\! y)}
          \Biggr) H(0,0,\! -1;y) \nn\\
\hspace{-10mm} & &  \hspace{12mm}
       +   \Biggl(
          - \frac{3}{ (1-y)^5}
          + \frac{15}{2 (1-y)^4}
          - \frac{4}{ (1-y)^3}
          - \frac{3}{2 (1-y)^2} \nn\\
\hspace{-10mm} & &  \hspace{12mm}
          + \frac{11}{16 (1\! -\! y)}\! 
          + \! \frac{3}{4 (1\! +\! y)^3}\! 
          - \! \frac{9}{8 (1\! +\! y)^2}\! 
          + \! \frac{11}{16 (1\! +\! y)}
          \Biggr) H(0\! ,0\! ,0;y) \nn\\
\hspace{-10mm} & &  \hspace{12mm}
       +   \Biggl(
            \frac{12}{ (1-y)^5}
          - \frac{30}{ (1-y)^4}
          + \frac{48}{ (1-y)^3}
          - \frac{42}{ (1-y)^2} \nn\\
\hspace{-10mm} & &  \hspace{12mm}
          + \frac{25}{4 (1-y)}\! 
          + \! \frac{1}{(1\! +\! y)^3}\! 
          - \! \frac{3}{2 (1\! +\! y)^2}\! 
          + \! \frac{25}{4 (1\! +\! y)}
          \Biggr) H(0,0,1;y) \nn\\
\hspace{-10mm} & &  \hspace{12mm}
       +   \Biggl(
            \frac{12}{ (1-y)^4}
          - \frac{24}{ (1-y)^3}
          + \frac{21}{ (1-y)^2}
          - \frac{9}{ (1-y)} \nn\\
\hspace{-10mm} & &  \hspace{12mm}
          + \frac{1}{(1\! +\! y)^2}\! 
          - \! \frac{1}{(1\! +\! y)}
          \Biggr) H(0,1;y)\! 
       +  \!  \Biggl(
          - \frac{282}{ (1\! -\! y)^5}\! 
          + \! \frac{705}{ (1\! -\! y)^4} \nn\\
\hspace{-10mm} & &  \hspace{12mm}
          - \frac{562}{ (1-y)^3}
          + \frac{138}{ (1-y)^2}
          + \frac{5}{8 (1-y)}
          + \frac{1}{2 (1+y)^3} \nn\\
\hspace{-10mm} & &  \hspace{12mm}
          - \frac{3}{4 (1+y)^2}
          + \frac{5}{8 (1+y)}
          \Biggr) H(0,1,0;y)
       +   \Biggl(
            \frac{6}{ (1-y)^3} \nn\\
\hspace{-10mm} & &  \hspace{12mm}
          - \frac{9}{ (1-y)^2}
          + \frac{19}{6 (1-y)}
          - \frac{1}{6 (1+y)}
          \Biggr) H(1;y) \nn\\
\hspace{-10mm} & &  \hspace{12mm}
       +   \Biggl(
          - \frac{282}{ (1-y)^4}
          + \frac{564}{ (1-y)^3}
          - \frac{623}{2 (1-y)^2}
          + \frac{59}{2 (1-y)} \nn\\
\hspace{-10mm} & &  \hspace{12mm}
          + \frac{1}{2 (1+y)^2}
          - \frac{1}{2 (1+y)}
          \Biggr) H(1,0;y)
          \Biggr] \nn\\
\hspace{-10mm} & &  \hspace{6mm}
        + \, C_{F} T_{R} N_{f} \Biggl[
       - \frac{25}{9 (1-y)}
          + \frac{25}{9 (1+y)}
       +   \Biggl(
          - \frac{2}{3 (1-y)} \nn\\
\hspace{-10mm} & &  \hspace{12mm}
          + \frac{2}{3 (1+y)}
          \Biggr) H(0;y)
       +   \Biggl(
          - \frac{4}{3 (1-y)}
          + \frac{4}{3 (1+y)}
          \Biggr) H(1;y)
          \Biggr] \nn\\
\hspace{-10mm} & &  \hspace{6mm}
        + \, C_{F} T_{R} \Biggl[
          - \frac{124}{3 (1-y)^3}
          + \frac{62}{ (1-y)^2}
          - \frac{118}{9 (1-y)}
          - \frac{68}{9 (1+y)} \nn\\
\hspace{-10mm} & &  \hspace{12mm}
       -  \Biggl(
            \frac{88}{3 (1\! -\! y)^4}\! 
          - \! \frac{176}{3 (1\! -\! y)^3}\! 
          + \! \frac{92}{3 (1\! -\! y)^2}\! 
          - \! \frac{4}{3 (1\! -\! y)}
          \Biggr) H(0;y) \nn\\
\hspace{-10mm} & &  \hspace{12mm}
       +  \Biggl(
            \frac{24}{ (1\! -\! y)^5}\! 
          - \! \frac{60}{ (1\! -\! y)^4}\! 
          + \! \frac{42}{ (1\! -\! y)^3}\! 
          - \! \frac{3}{ (1\! -\! y)^2}\! 
          - \! \frac{3}{2 (1\! -\! y)} \nn\\
\hspace{-10mm} & &  \hspace{12mm}
          - \frac{3}{2 (1+y)}
          \Biggr) H(0,0;y)
          \Biggr] \nn\\
\hspace{-10mm} & &  \hspace{6mm}
        + \, C_{F}^2 \Biggl[
          - \frac{31}{4 (1-y)}
          + \frac{31}{4 (1+y)}
       + \zeta(2)   \Biggl(
            \frac{60}{ (1-y)^4}
          - \frac{120}{ (1-y)^3} \nn\\
\hspace{-10mm} & &  \hspace{12mm}
          + \frac{78}{ (1-y)^2}
          - \frac{18}{ (1-y)}
          - \frac{6}{ (1+y)^2}
          + \frac{6}{ (1+y)}
          \Biggr) \nn\\
\hspace{-10mm} & &  \hspace{12mm}
       + \zeta(3)   \Biggl(
          - \frac{168}{ (1-y)^5}
          + \frac{420}{ (1-y)^4}
          - \frac{350}{ (1-y)^3}
          + \frac{105}{ (1-y)^2} \nn\\
\hspace{-10mm} & &  \hspace{12mm}
          - \frac{7}{2 (1-y)}
          - \frac{7}{2 (1+y)}
          \Biggr)
       +   \Biggl(
          - \frac{48}{ (1-y)^3}
          + \frac{72}{ (1-y)^2} \nn\\
\hspace{-10mm} & &  \hspace{12mm}
          - \frac{28}{ (1-y)}
          + \frac{4}{ (1+y)}
          \Biggr) H(-1;y)
       +   \Biggl(
            \frac{360}{ (1-y)^4}
          - \frac{720}{ (1-y)^3} \nn\\
\hspace{-10mm} & &  \hspace{12mm}
          + \frac{356}{ (1-y)^2}\! 
          + \! \frac{4}{ (1-y)}\! 
          + \! \frac{4}{ (1\! +\! y)^2}\! 
          - \! \frac{4}{ (1\! +\! y)}
          \Biggr) H(-1,0;y) \nn\\
\hspace{-10mm} & &  \hspace{12mm}
       -  \Biggl(
            \frac{144}{ (1-y)^5}\! 
          - \! \frac{360}{ (1-y)^4}\! 
          + \! \frac{300}{ (1-y)^3}\! 
          - \! \frac{90}{ (1-y)^2}\! 
          + \! \frac{3}{ (1\! -\! y)} \nn\\
\hspace{-10mm} & &  \hspace{12mm}
          + \frac{3}{ (1+y)}
          \Biggr) H(-1,0,0;y) 
       + \zeta(2)   \Biggl(
            \frac{60}{ (1-y)^5}
          - \frac{150}{ (1-y)^4} \nn\\
\hspace{-10mm} & &  \hspace{12mm}
          + \frac{127}{ (1-y)^3}\! 
          - \! \frac{81}{2 (1-y)^2}\! 
          + \! \frac{5}{4 (1-y)}\! 
          - \! \frac{2}{ (1\! +\! y)^3}\! 
          + \! \frac{3}{ (1\! +y)^2} \nn\\
\hspace{-10mm} & &  \hspace{12mm}
          + \frac{5}{4 (1\! +\! y)}
          \Biggr) H(0;y)\! 
       +  \!  \Biggl(
          - \frac{192}{ (1\! -\! y)^4}\! 
          + \! \frac{504}{ (1\! -\! y)^3}\! 
          - \! \frac{442}{ (1\! -\! y)^2} \nn\\
\hspace{-10mm} & &  \hspace{12mm}
          + \frac{255}{2 (1\!-\!y)}\! 
          + \! \frac{2}{ (1\! +\! y)^2}\! 
          + \! \frac{1}{2 (1\! +\! y)}
          \Biggr) H(0;y)\! 
       +   \! \Biggl(
          - \frac{96}{ (1\! -\! y)^4} \nn\\
\hspace{-10mm} & &  \hspace{12mm}
          + \frac{192}{ (1-y)^3}
          - \frac{100}{ (1-y)^2}
          + \frac{4}{ (1-y)}
          + \frac{4}{ (1+y)^2} \nn\\
\hspace{-10mm} & &  \hspace{12mm}
          - \frac{4}{ (1\! +\! y)}
          \Biggr) H(0,-1;y)\! 
       +  \!  \Biggl(
            \frac{360}{ (1\! -\! y)^5}\! 
          - \! \frac{900}{ (1\! -\! y)^4}
          +\!  \frac{670}{ (1\! -\! y)^3} \nn\\
\hspace{-10mm} & &  \hspace{12mm}
          - \frac{105}{ (1-y)^2}
          - \frac{25}{2 (1-y)}
          - \frac{25}{2 (1+y)}
          \Biggr) H(0,-1,0;y) \nn\\
\hspace{-10mm} & &  \hspace{12mm}
       +  \Biggl(
            \frac{60}{ (1\! -\! y)^5}\! 
          - \! \frac{216}{ (1\! -\! y)^4}\! 
          + \! \frac{251}{ (1\! -\! y)^3}\! 
          - \! \frac{88}{ (1\! -\! y)^2}\! 
          - \! \frac{71}{4 (1\! -\! y)} \nn\\
\hspace{-10mm} & &  \hspace{12mm}
          - \frac{7}{2 (1+y)^2}
          + \frac{57}{4 (1+y)}
          \Biggr) H(0,0;y)
       +   \Biggl(
          - \frac{96}{ (1-y)^5} \nn\\
\hspace{-10mm} & &  \hspace{12mm}
          + \frac{240}{ (1-y)^4}
          - \frac{168}{ (1-y)^3}
          + \frac{12}{ (1-y)^2}
          + \frac{6}{ (1-y)} \nn\\
\hspace{-10mm} & &  \hspace{12mm}
          + \frac{6}{ (1+y)}
          \Biggr) H(0,0,-1;y)
       +   \Biggl(
            \frac{6}{ (1-y)^5}
          - \frac{15}{ (1-y)^4} \nn\\
\hspace{-10mm} & &  \hspace{12mm}
          + \frac{8}{ (1-y)^3}\! 
          + \! \frac{3}{ (1-y)^2}\! 
          - \! \frac{11}{8 (1-y)}\! 
          - \! \frac{3}{2 (1\! +\! y)^3}\! 
          + \! \frac{9}{4 (1\! +\! y)^2} \nn\\
\hspace{-10mm} & &  \hspace{12mm}
          - \frac{11}{8 (1+y)}
          \Biggr) H(0,0,0;y)
       +   \Biggl(
            \frac{384}{ (1-y)^5}
          - \frac{960}{ (1-y)^4} \nn\\
\hspace{-10mm} & &  \hspace{12mm}
          + \frac{714}{ (1-y)^3}\! 
          - \! \frac{111}{ (1-y)^2}\! 
          - \! \frac{14}{ (1-y)}\! 
          - \! \frac{2}{ (1\! +\! y)^3}\! 
          + \! \frac{3}{ (1\! +\! y)^2} \nn\\
\hspace{-10mm} & &  \hspace{12mm}
          - \frac{14}{ (1+y)}
          \Biggr) H(0,0,1;y)
       +   \Biggl(
            \frac{384}{ (1-y)^4}
          - \frac{768}{ (1-y)^3} \nn\\
\hspace{-10mm} & &  \hspace{12mm}
          + \frac{394}{ (1-y)^2}
          - \frac{10}{ (1-y)}
          + \frac{2}{ (1+y)^2}
          - \frac{2}{ (1+y)}
          \Biggr) H(0,1;y) \nn\\
\hspace{-10mm} & &  \hspace{12mm}
       +   \Biggl(
          - \frac{252}{ (1-y)^5}
          + \frac{630}{ (1-y)^4}
          - \frac{496}{ (1-y)^3}
          + \frac{114}{ (1-y)^2} \nn\\
\hspace{-10mm} & &  \hspace{12mm}
          + \frac{7}{4 (1-y)}\! 
          - \! \frac{1}{(1\! +\! y)^3}\! 
          + \! \frac{3}{2 (1\! +\! y)^2}\! 
          + \! \frac{7}{4 (1\! +\! y)}
          \Biggr) H(0,\! 1,\! 0;y) \nn\\
\hspace{-10mm} & &  \hspace{12mm}
       +   \Biggl(
            \frac{192}{ (1-y)^3}
          - \frac{288}{ (1-y)^2}
          + \frac{87}{ (1-y)}
          + \frac{9}{ (1+y)}
          \Biggr) H(1;y) \nn\\
\hspace{-10mm} & &  \hspace{12mm}
       -   \Biggl(
            \frac{252}{ (1-y)^4}\! 
          - \! \frac{504}{ (1-y)^3}\! 
          + \! \frac{257}{ (1-y)^2}\! 
          - \! \frac{5}{ (1-y)}\! 
          + \! \frac{1}{(1\! +\! y)^2} \nn\\
\hspace{-10mm} & &  \hspace{12mm}
          - \frac{1}{(1+y)}
          \Biggr) H(1,0;y)
          \Biggr] \, .
\eea

The two-loop contribution to the heavy quark vector form factors 
arising from closed massive and massless fermion loops were computed 
previously in~\cite{teubner}. In this work, the result for the 
massless fermion loops was obtained from a more general
(unintegrated) result with different masses for the external and 
virtual quarks, by expanding in the mass of the virtual quark.

Comparing to~\cite{teubner}, we fully agree on the form factors 
for the massive fermion loop, and on the leading logarithmic terms for the 
massless fermion loop. Owing to the different regularization procedures 
used in the latter case (small quark mass versus dimensional regularization),
the finite terms differ.

\subsection{Threshold Expansions}

In this Section we provide the expansions of our results in the threshold 
limit $S \sim 4m^2$ ($y \rightarrow 1$ in the transformed variable). 
We define:
\be
\beta = \sqrt{1-\frac{4m^2}{S}} \, ,
\ee
as the small parameter in which we expand. Keeping terms up to the zeroth order 
in $\beta$, we have:
\bea
\Re \, {\mathcal F}^{(1l)}_{1,R}(\epsilon,s) & = &  C_{F} \Biggl[ 
\frac{3 \zeta(2)}{\beta} - 3
 \Biggr] \, , 
\label{REF11l} \\
\Im \, {\mathcal F}^{(1l)}_{1,R}(\epsilon,s) & = & C_{F} \Biggl\{  
\frac{1}{\epsilon} \Biggl[ - \frac{1}{2 \beta} \Biggr] 
+ \frac{1}{\beta} \Biggl[ - \frac{1}{2} + \ln{2} + \ln{\beta} \Biggr]
\Biggr\}
 \, , 
\label{IMF11l}\\
% ok
%
\Re \, {\mathcal F}^{(1l)}_{2,R}(\epsilon,s) & = &  - C_{F} \, ,  
\label{REF21l}\\
% ok
%
\Im \, {\mathcal F}^{(1l)}_{2,R}(\epsilon,s) & = & 
\frac{C_{F}}{2 \beta} 
 \, ,  
\label{IMF21l}\\
% ok
%
%
%
%
\Re \, {\mathcal F}^{(2l)}_{1,R}(\epsilon,s) & = & 
\frac{1}{\epsilon^2} \Biggl\{ 
      \frac{1}{\beta^2} C_{F}^2 \Bigl( - \frac{3}{4} \zeta(2)
          \Bigr) 
       + C_{F}^2 \Bigl( - \frac{3}{2} \zeta(2)
          \Bigr) \Biggr\} \nn\\
& &     
+ \frac{1}{\epsilon} \Biggl\{  
     \frac{1}{\beta^2} \Biggl[
        C_{F}^2 \Bigl( 
          - \frac{3}{2} \zeta(2)
          + 3 \zeta(2)  \ln{\beta}
          + 3 \zeta(2) \ln{2}
          \Bigr) \nn\\
& &   \hspace{6mm}
      + C_{F}^2 \Bigl(  
          - \frac{3}{2} \zeta(2)
          + 6 \zeta(2) \ln{\beta}
          + 6 \zeta(2) \ln{2} \Bigr) \Biggr\} \nn\\
& &     
     \frac{1}{\beta^2} \Biggl[
          C_{F}^2 \Bigl( 
          - 7 \zeta(2)
          + \frac{9}{2} \zeta^{2}(2)
          + 6 \zeta(2) \ln{\beta}
          - 6 \zeta(2) \ln^{2}{\beta}
          + 6 \zeta(2) \ln{2} \nn\\
& &   \hspace{12mm}
          - 12 \zeta(2) \ln{2} \ln{\beta}
          - 6 \zeta(2) \ln^{2}{2}
          \Bigr) \Biggr] \nn\\
& &     
     \frac{1}{\beta} \Biggl[
          C_{F}^2 \Bigl(
          - 9 \zeta(2)
          \Bigr) \nn\\
& &  \hspace{6mm} 
        + C_{F} C_{A} \Bigl(
            \frac{73}{6} \zeta(2)
          - 11 \zeta(2) \ln{\beta}
          - 11 \zeta(2) \ln{2}
\Bigr) \nn\\
& &  \hspace{6mm} 
        - C_{F} T_{R} N_{f} \Bigl(
            \frac{16}{3} \zeta(2) \! 
          - 4 \zeta(2) \ln{\beta} \! 
          - 4 \zeta(2) \ln{2}
\Bigr)   \Biggr] \nn\\
& &     
   + C_{F}^2 \Bigl(
            \frac{421}{60} \! 
          +  \! 5 \zeta(2) \ln{2} \! 
          - \frac{963}{50} \zeta(2) \! 
          +  \! 9 \zeta^2(2) \! 
          - \frac{81}{20} \zeta(3) \! 
          +  \! \frac{14}{5} \zeta(2) \ln{\beta} \nn\\
& &  \hspace{6mm} 
          - 12 \zeta(2) \ln^{2}{\beta}
          + \frac{14}{5} \zeta(2) \ln{2}
          - 24 \zeta(2) \ln{2} \ln{\beta} 
          - 12 \zeta(2) \ln^{2}{2}
          \Bigr) \nn\\
& &  
        + C_{F} C_{A} \Bigl(
          - \frac{379}{60}
          - \frac{42}{5} \zeta(2) \ln{2}
          + \frac{2741}{150} \zeta(2)
          - \frac{83}{10} \zeta(3)
          - \frac{36}{5} \zeta(2) \ln{\beta} \nn\\
& &  \hspace{6mm} 
          - \frac{36}{5} \zeta(2) \ln{2}
\Bigr) \nn\\
& &  
        + C_{F} T_{R} N_{f} 
        + C_{F} T_{R} \Bigl(
            \frac{37}{3}
          - \frac{104}{15} \zeta(2)
\Bigr) \, , \\
% ok
%
%
%
%
\Im \, {\mathcal F}^{(2l)}_{1,R}(\epsilon,s) & = & 
\frac{1}{\epsilon^2} \Biggl\{ 
   \frac{1}{\beta} \Biggl[
          C_{F} C_{A} \Bigl(
            \frac{11}{24}
\Bigr) 
        + C_{F} T_{R} N_{f} \Bigl(
          - \frac{1}{6}
\Bigr) \Biggr] \Biggr\} \nn\\
& &     
       + \frac{1}{\epsilon} \Biggl\{ 
   \frac{1}{\beta^2} \Biggl[
          C_{F}^2 \Bigl(
            \frac{3}{2} \zeta(2)
          \Bigr) \Biggr]\nn\\
& &  \hspace{6mm}
   + \frac{1}{\beta} \Biggl[
          C_{F}^2 \Bigl(
            \frac{3}{2}
          \Bigr)
        - C_{F} C_{A} \Bigl(
            \frac{31}{72}
\Bigr)
        + C_{F} T_{R} N_{f} \Bigl(
            \frac{5}{18}
\Bigr)   \Biggr]
        + C_F^2 \Bigl( - 3 \zeta(2)
	  \Bigr)
\Biggr\} \nn\\
& &   
   + \frac{1}{\beta^2} \Biggl[
         C_{F}^2 \Bigl(
          - 3 \zeta(2)
          + 6 \zeta(2) \ln{\beta}
          + 6 \zeta(2) \ln{2} 
          \Bigr) \Biggr] \nn\\
& &   
   + \frac{1}{\beta} \Biggl[
         C_{F}^2 \Bigl(
          - 3 \ln{\beta}
          - 3 \ln{2}
          \Bigr) \nn\\
& &  \hspace{6mm} 
        + C_{F} C_{A} \Bigl(
          - \frac{197}{54}
          + \frac{73}{18} \ln{\beta}
          - \frac{11}{6} \ln^{2}{\beta}
          + \frac{73}{18} \ln{2}
          - \frac{11}{3} \ln{2} \ln{\beta} \nn\\
& &  \hspace{18mm} 
          - \frac{11}{6} \ln^{2}{2}
\Bigr) \nn\\
& &  \hspace{6mm} 
        + C_{F} T_{R} N_{f} \Bigl(
            \frac{103}{54}
          - \frac{16}{9} \ln{\beta}
          + \frac{2}{3} \ln^{2}{\beta}
          - \frac{16}{9} \ln{2}
          + \frac{4}{3} \ln{2} \ln{\beta} \nn\\
& &  \hspace{18mm} 
          + \frac{2}{3} \ln^{2}{2}
\Bigr)  \Biggr] \nn\\
& &  
       + C_{F}^2 \Bigl(
          - \frac{7}{5} \zeta(2)
          + 12 \zeta(2) \ln{\beta}
          + 12 \zeta(2) \ln{2}
          \Bigr) \nn\\
& &  
        + C_{F} C_{A} \Bigl(
            \frac{18}{5} \zeta(2)
\Bigr) \, , \\
% ok
%
%
\Re \, {\mathcal F}^{(2l)}_{2,R}(\epsilon,s) & = & 
\frac{1}{\epsilon} \Biggl\{ 
    \frac{1}{\beta^2}  \Biggl[
          C_{F}^2 \Bigl(
            \frac{3}{2} \zeta(2)
          \Bigr)  \Biggr] \Biggr\} \nn\\
& &  
    + \frac{1}{\beta^2}  \Biggl[
          C_{F}^2 \Bigl(
          + 7 \zeta(2)
          - 6 \zeta(2) \ln{\beta}
          - 6 \zeta(2) \ln{2}
          \Bigr)  \Biggr] \nn\\
& &  
    + \frac{1}{\beta}  \Biggl[
          C_{F}^2 \Bigl(
          - 3 \zeta(2)
          \Bigr)
        + C_{F} C_{A} \Bigl(
          - 7 \zeta(2)
\Bigr) 
        + C_{F} T_{R} N_{f} \Bigl(
            2 \zeta(2)
\Bigr)  \Biggr] \nn\\
& &  
        + C_{F}^2 \Bigl(
            \frac{269}{60}
          + 19 \zeta(2) \ln{2}
          - \frac{2461}{150} \zeta(2)
          + \frac{41}{20} \zeta(3)
          + \frac{6}{5} \zeta(2) \ln{\beta} \nn\\
& &  \hspace{6mm} 
          + \frac{6}{5} \zeta(2) \ln{2}
          \Bigr) \nn\\
& &  
        + C_{F} C_{A} \Bigl(
          - \frac{373}{180}
          - \frac{58}{5} \zeta(2) \ln{2}
          + \frac{289}{25} \zeta(2)
          - \frac{47}{10} \zeta(3)
          - \frac{24}{5} \zeta(2) \ln{\beta} \nn\\
& &  \hspace{6mm} 
          - \frac{24}{5} \zeta(2) \ln{2}
\Bigr) \nn\\
& &  
        + C_{F} T_{R} N_{f} \Bigl(
            \frac{13}{9}
\Bigr) 
        + C_{F} T_{R} \Bigl(
          - \frac{23}{9}
          + \frac{8}{5} \zeta(2)
\Bigr) \, , \\
% ok
%
%
%
\Im \, {\mathcal F}^{(2l)}_{2,R}(\epsilon,s) & = & 
\frac{1}{\epsilon} \Biggl\{ 
   \frac{1}{\beta} \Biggl[
          C_{F}^2 \Bigl( \frac{1}{2} \Bigr) \Biggr] \Biggr\} \nn\\
& &      
  + \frac{1}{\beta^2} \Biggl[
          C_{F}^2 \Bigl( 
	    3 \zeta(2) 
          \Bigr) \Biggr] \nn\\
& &   
  + \frac{1}{\beta} \Biggl[
          C_{F}^2 \Bigl( 
          - \ln{\beta}
          - \ln{2}
          \Bigr) \nn\\
& &  \hspace{6mm} 
        + C_{F} C_{A} \Bigl(
            \frac{25}{9}
          - \frac{7}{3} \ln{\beta}
          - \frac{7}{3} \ln{2}
\Bigr) \nn\\
& &  \hspace{6mm} 
        + C_{F} T_{R} N_{f} \Bigl(
          - \frac{25}{18}
          + \frac{2}{3} \ln{\beta}
          + \frac{2}{3} \ln{2}
\Bigr) \Biggr] \nn\\
& & 
+ C_{F}^2 \Bigl(
          - \frac{3}{5} \zeta(2)
          \Bigr)
        + C_{F} C_{A} \Bigl(
            \frac{12}{5} \zeta(2)
\Bigr) \, .
% ok
\eea

The above results are in perfect agreement with the results already known
in the literature. In particular, the threshold limits of the 1-loop form
factors are in agreement with Eqs.~(9,10) of \cite{Hoang1} (if we put
$C_F=1$ in our results), once the following replacement is carried out:
\be
\ln{\left( \frac{\lambda}{M} \right)} = \frac{1}{2 \epsilon} \, .
\label{translation}
\ee

Consequently, we found also agreement with the 1-loop correction to the 
cross-section of $e^{+}e^{-} \rightarrow f \bar{f}$ given in 
Eqs.~(6,11,12) of \cite{Hoang1}.

At 2-loop level, Eqs.~(26,28) of \cite{Hoang1}, concerning the
abelian contributions matching our $C_F^2$ terms, are written 
regularizing the IR divergences with a small mass for the photon. If we 
replace $\ln{( \lambda/M )}$ using Eq.~(\ref{translation}), we are able 
to match the poles in $1/\epsilon$ with our expressions calculated 
directly in dimensional regularization. Nevertheless, we can not find 
agreement for the finite parts because of the differences between the two
regularization schemes. Eqs.~(27,29) of \cite{Hoang1}, instead, are
in complete agreement with our $C_F T_R$ terms. 
Let us note, however, that the contributions at two loops to the cross
section $e^{+}e^{-} \rightarrow f \bar{f}$ are IR finite and they can not
depend on the regularization scheme. This is actually the case and we 
found complete agreement with Eqs.~(31,32) of \cite{Hoang1}.

The 2-loop corrections in the threshold limit to the cross-section
$e^{+}e^{-} \rightarrow Q \bar{Q}$ were calculated in \cite{mel2,beneke}:
\be
\sigma_{e^{+}e^{-} \to Q \bar{Q}} = \sigma^{(0)} \Biggl[ 1 + 
\left( \frac{\alpha_S}{2 \pi} \right) \tilde{\Delta}^{(1)} + 
\left( \frac{\alpha_S}{2 \pi} \right)^2 \tilde{\Delta}^{(2)} \Biggr] ,
\ee
where $\sigma^{(0)}$ is the tree-level cross-section and
$\tilde{\Delta}^{(1)}$ and $\tilde{\Delta}^{(2)}$ can be expressed, up
to ${\mathcal O}( \beta^2)$, in terms of the form factors as:
\bea
\tilde{\Delta}^{(1)} & = & 2 \Bigl( \Re {\mathcal F}^{(1l)}_{1,R}(\epsilon,s)
              + \Re {\mathcal F}^{(1l)}_{2,R}(\epsilon,s) \Bigr) \, , \\
\tilde{\Delta}^{(2)}  & = & 
                  \Bigl( \Re {\mathcal F}^{(1l)}_{1,R}(\epsilon,s) \Bigr)^2
              + 2 \Bigl( \Re {\mathcal F}^{(1l)}_{1,R}(\epsilon,s) \,
	                 \Re {\mathcal F}^{(1l)}_{2,R}(\epsilon,s) \Bigr)
	      + 2 \, \Re {\mathcal F}^{(2l)}_{1,R}(\epsilon,s) \nn\\
& & 	      + \pi^2 \Bigl( \Im {\mathcal F}^{(1l)}_{1,R}(\epsilon,s) \Bigr)^2
	      + 2 \pi^2 \Bigl( \Im {\mathcal F}^{(1l)}_{1,R}(\epsilon,s) \, 
	                 \Im {\mathcal F}^{(1l)}_{2,R}(\epsilon,s) \Bigr)
	      +   \Bigl( \Re {\mathcal F}^{(1l)}_{2,R}(\epsilon,s) \Bigr)^2 
	      \nn\\
& & 	      + 2 \, \Re {\mathcal F}^{(2l)}_{2,R}(\epsilon,s) 
	      +  \pi^2 \Bigl( \Im {\mathcal F}^{(1l)}_{2,R}(\epsilon,s) \Bigr)^2
	       \, .
\eea
We found complete agreement with the results presented in these papers.

\subsection{Asymptotic Expansions}

In this Section we provide the expansions of our results in the limit 
$S \gg m^2$ ($y \rightarrow 0$ in the transformed variable). Putting 
$L = \ln{(S/m^2)}$ and keeping terms up to the second order in $(m^2/S)$, 
we have:
\bea
\Re \, {\mathcal F}^{(1l)}_{1,R}(\epsilon,s) & = & C_{F} \Biggl\{  
\frac{1}{\epsilon} \Biggl[ 
            \left( \frac{m^2}{S} \right)^2 \Bigl( -3 + 2L \Bigr)
	   - 2 \left( \frac{m^2}{S} \right)
	   - 1 + L
\Biggr] \nn\\
& & \hspace{8mm}
+ \left( \frac{m^2}{S} \right)^2 \Bigl( -4 +8 \zeta(2) + 9 L - L^2 \Bigr)
- \left( \frac{m^2}{S} \right) \Bigl( 1 -3L \Bigr) \nn\\
& & \hspace{8mm}
-2 + 4 \zeta(2) + \frac{3}{2} L - \frac{1}{2} L^2 \Biggr\}
 \, , \\
\Im \, {\mathcal F}^{(1l)}_{1,R}(\epsilon,s) & = & C_{F} \Biggl\{  
\frac{1}{\epsilon} \Biggl[ 
   - 2 \left( \frac{m^2}{S} \right)^2 - 1   
\Biggr] \nn\\
& & \hspace{8mm}
+ \left( \frac{m^2}{S} \right)^2 \Bigl( - 9 + 2L \Bigr)
- 3 \left( \frac{m^2}{S} \right) - \frac{3}{2} + L \Biggr\}
 \, , \\
\Re \, {\mathcal F}^{(1l)}_{2,R}(\epsilon,s) & = & C_{F} \Biggl\{ 
            \left( \frac{m^2}{S} \right)^2 \Bigl( 4 - 4L \Bigr)
	  - \left( \frac{m^2}{S} \right) \Bigl( 2L \Bigr) \Biggr\}
 \, , \\
\Im \, {\mathcal F}^{(1l)}_{2,R}(\epsilon,s) & = & C_{F} \Biggl\{  
            4 \left( \frac{m^2}{S} \right)^2 
	  + 2 \left( \frac{m^2}{S} \right) 
\Biggr\}
 \, , \\
\Re \, {\mathcal F}^{(2l)}_{1,R}(\epsilon,s) & = & 
\frac{1}{\epsilon^2} \Biggl\{ 
   \left( \frac{m^2}{S} \right)^2 \Biggl[
         C_{F}^2 \Bigl(
            5
          - 5 L
          + 2 L^2
          - 12 \zeta(2)
          \Bigr)
        + C_{F} C_{A} \Bigl(
            \frac{11}{4}
          - \frac{11}{6} L
\Bigr) \nn\\
& &  \hspace{18mm} 
        + C_{F} T_{R} N_{f} \Bigl(
          - 1
          + \frac{2}{3} L
\Bigr) \Biggr]  \nn\\
& &  \hspace{6mm} 
 - \left( \frac{m^2}{S} \right) \Biggl[
         C_{F}^2 \Bigl(
          - 2
          + 2 L
          \Bigr)
        - C_{F} C_{A} \Bigl(
            \frac{11}{6}
\Bigr) 
        + C_{F} T_{R} N_{f} \Bigl(
            \frac{2}{3}
\Bigr) \Biggr]  \nn\\
& &  \hspace{6mm} 
        + C_{F}^2 \Bigl(
            \frac{1}{2}
          - L
          + \frac{1}{2} L^2
          - 3 \zeta(2)
          \Bigr)
        + C_{F} C_{A} \Bigl(
            \frac{11}{12}
          - \frac{11}{12} L
\Bigr) \nn\\
& &  \hspace{6mm} 
        + C_{F} T_{R} N_{f} \Bigl(
          - \frac{1}{3}
          + \frac{1}{3} L
\Bigr) 
\Biggr\} \nn\\
& & + \frac{1}{\epsilon} \Biggl\{ 
   \left( \frac{m^2}{S} \right)^2 \Biggl[
         C_{F}^2 \Bigl(
            12
          + 40 \zeta(2) L
          - \frac{55}{2} L
          + \frac{29}{2} L^2
          - 2 L^3
          - 92 \zeta(2)
          \Bigr) \nn\\
& &  \hspace{18mm} 
        - C_{F} C_{A} \Bigl(
            \frac{55}{12}
          + 6 \zeta(2) L
          - \frac{47}{9} L
          + L^2
          - \frac{1}{3} L^3
          - \frac{13}{2} \zeta(2) \nn\\
& &  \hspace{18mm} 
          + 2 \zeta(3)
\Bigr) 
        + C_{F} T_{R} N_{f} \Bigl(
            \frac{5}{3}
          - \frac{10}{9} L
\Bigr) \Biggr]  \nn\\
& &  \hspace{6mm} 
 - \left( \frac{m^2}{S} \right) \Biggl[
         C_{F}^2 \Bigl(
          - 5
          + 7 L
          - 4 L^2
          + 26 \zeta(2)
          \Bigr)
        + C_{F} C_{A} \Bigl(
            \frac{67}{18}
          - \zeta(2)
\Bigr) \nn\\
& &  \hspace{18mm} 
        - C_{F} T_{R} N_{f} \Bigl(
            \frac{10}{9}
\Bigr) \Biggr]  \nn\\
& &  \hspace{6mm} 
        + C_{F}^2 \Bigl(
            2
          + 10 \zeta(2) L
          - \frac{7}{2} L
          + 2 L^2
          - \frac{1}{2} L^3
          - 13 \zeta(2)
          \Bigr) \nn\\
& &  \hspace{6mm} 
        + C_{F} C_{A} \Bigl(
          - \frac{49}{36}
          - \frac{1}{2} \zeta(2) L
          + \frac{67}{36} L
          + \frac{1}{2} \zeta(2)
          - \frac{1}{2} \zeta(3)
\Bigr) \nn\\
& &  \hspace{6mm} 
        + C_{F} T_{R} N_{f} \Bigl(
            \frac{5}{9}
          - \frac{5}{9} L
\Bigr) 
 \Biggr\} \nn\\
& & 
  + \left( \frac{m^2}{S} \right)^2 \Biggl[
         C_{F}^2 \Bigl(
            \frac{909}{4}
          + 434 \zeta(2) L
          - 148 \zeta(3) L
          - \frac{1}{2} L
          - 111 \zeta(2) L^2 \nn\\
& &  \hspace{24mm} 
          + 108 L^2
          - \frac{101}{6} L^3
          + \frac{7}{3} L^4
          - 288 \zeta(2) \ln{2}
          - 905 \zeta(2) \nn\\
& &  \hspace{24mm} 
          + \frac{727}{5} \zeta^2(2)
          + 246 \zeta(3)
          \Bigr) \nn\\
& &  \hspace{18mm} 
        + C_{F} C_{A} \Bigl(
            \frac{3749}{36} 
          +  \frac{119}{3} \zeta(2) L 
          -  238 \zeta(3) L 
          +  \frac{12607}{108} L  \nn\\
& &  \hspace{24mm} 
          +  2 \zeta(2) L^2
          + \frac{419}{18} L^2
          + \frac{77}{18} L^3
          - \frac{2}{3} L^4
          + 144 \zeta(2) \ln{2} \nn\\
& &  \hspace{24mm} 
          - \frac{7879}{18} \zeta(2) 
          + \frac{118}{5} \zeta^2(2)
          + \frac{3527}{6} \zeta(3)
\Bigr) \nn\\
& &  \hspace{18mm} 
        + C_{F} T_{R} N_{f} \Bigl(
            \frac{199}{18}
          + \frac{8}{3} \zeta(2) L
          - \frac{731}{27} L
          + \frac{37}{9} L^2
          - \frac{2}{9} L^3 \nn\\
& &  \hspace{24mm} 
          - \frac{260}{9} \zeta(2)
          - \frac{8}{3 }\zeta(3)
\Bigr)   \nn\\
& &  \hspace{18mm} 
        + C_{F} T_{R} \Bigl(
            \frac{2345}{18}
          - \frac{16}{3} \zeta(2) L
          - \frac{3730}{27} L
          + \frac{289}{9} L^2
          + \frac{4}{9} L^3 \nn\\
& &  \hspace{24mm} 
          + 28 \zeta(2)
\Bigr) \Biggr]  \nn\\
& & 
 - \left( \frac{m^2}{S} \right) \Biggl[
         C_{F}^2 \Bigl(
          - \frac{107}{4}
          - 99 \zeta(2) L
          + 12 \zeta(3) L
          + \frac{97}{4} L
          + 5 \zeta(2) L^2 \nn\\
& &  \hspace{24mm} 
          - \frac{77}{4} L^2
          + \frac{14}{3} L^3
          - \frac{1}{12} L^4
          + 72 \zeta(2) \ln{2}
          + 110 \zeta(2) \nn\\
& &  \hspace{24mm} 
          - \frac{43}{5} \zeta^2(2)
          - 6 \zeta(3)
          \Bigr) \nn\\
& &  \hspace{18mm} 
        + C_{F} C_{A} \Bigl(
          - \frac{167}{108}
          + 3 \zeta(2) L
          + 18 \zeta(3) L
          - \frac{269}{12} L
          + \frac{1}{2} \zeta(2) L^2 \nn\\
& &  \hspace{24mm} 
          - \frac{1}{4} L^2
          - \frac{2}{3} L^3
          + \frac{1}{24} L^4
          - 36 \zeta(2) \ln{2}
          + \frac{139}{3} \zeta(2) \nn\\
& &  \hspace{24mm} 
          - \frac{7}{2} \zeta^2(2)
          - 63 \zeta(3)
\Bigr) \nn\\
& &  \hspace{18mm} 
        + C_{F} T_{R} N_{f} \Bigl(
          - \frac{59}{27}
          + \frac{25}{3} L
          - L^2
          + \frac{16}{3} \zeta(2)
\Bigr)  \nn\\
& &  \hspace{18mm} 
        + C_{F} T_{R} \Bigl(
          - \frac{853}{27}
          + \frac{49}{3} L
          - 5 L^2
          + \frac{20}{3} \zeta(2)
\Bigr) \Biggr]  \nn\\
& &  
        + C_{F}^2 \Bigl(
            \frac{23}{2}
          + 31 \zeta(2) L
          + 8 \zeta(3) L
          - \frac{85}{8} L
          - 12 \zeta(2) L^2
          + \frac{55}{8} L^2  \nn\\
& &  \hspace{18mm} 
          - \frac{5}{3} L^3 \! 
          + \! \frac{7}{24} L^4\! 
          - \! 12 \zeta(2) \ln{2}\! 
          - \! 32 \zeta(2)\! 
          + \! \frac{68}{5} \zeta^2(2)\! 
          - \! 11 \zeta(3)
          \Bigr) \nn\\
& &  
        + C_{F} C_{A} \Bigl(
          - \frac{1595}{108}
          - \frac{11}{3} \zeta(2) L
          - \frac{13}{2} \zeta(3) L
          + \frac{2545}{216} L
          + \frac{1}{2} \zeta(2) L^2  \nn\\
& &  \hspace{18mm} 
          - \frac{233}{72} L^2
          + \frac{11}{36} L^3
          + 6 \zeta(2) \ln{2}
          + \frac{173}{9} \zeta(2)
          - \frac{63}{20} \zeta^2(2)  \nn\\
& &  \hspace{18mm} 
          + \frac{67}{6 }\zeta(3)
\Bigr) \nn\\
& &  
        + C_{F} T_{R} N_{f} \Bigl(
            \frac{106}{27}
          + \frac{4}{3} \zeta(2) L
          - \frac{209}{54} L
          + \frac{19}{18} L^2
          - \frac{1}{9} L^3
          - \frac{64}{9} \zeta(2)  \nn\\
& &  \hspace{18mm} 
          - \frac{4}{3} \zeta(3)
\Bigr) \nn\\
& &  
        + C_{F} T_{R} \Bigl(
            \frac{383}{27}
          + \frac{4}{3} \zeta(2) L
          - \frac{265}{54} L
          + \frac{19}{18} L^2
          - \frac{1}{9} L^3
          - \frac{22}{3} \zeta(2)
\Bigr) , \\
\Im \, {\mathcal F}^{(2l)}_{1,R}(\epsilon,s) & = & 
\frac{1}{\epsilon^2} \Biggl\{ 
   \left( \frac{m^2}{S} \right)^2 \Biggl[
         C_{F}^2 \Bigl(
            5
          - 4 L
          \Bigr)
        + C_{F} C_{A} \Bigl(
            \frac{11}{6}
\Bigr) 
        - C_{F} T_{R} N_{f} \Bigl(
            \frac{2}{3}
\Bigr) \Biggr]  \nn\\
& &  \hspace{6mm} 
 + \left( \frac{m^2}{S} \right) \Biggl[
         C_{F}^2 \Bigl(
            2
          \Bigr)  \Biggr]  \nn\\
& &  \hspace{6mm} 
        + C_{F}^2 \Bigl(
            1
          - L
          \Bigr)
        + C_{F} C_{A} \Bigl(
            \frac{11}{12}
\Bigr)
        - C_{F} T_{R} N_{f} \Bigl(
            \frac{1}{3}
\Bigr) 
\Biggr\} \nn\\
& & + \frac{1}{\epsilon} \Biggl\{ 
   \left( \frac{m^2}{S} \right)^2 \Biggl[
         C_{F}^2 \Bigl(
            \frac{55}{2}
          - 29 L
          + 6 L^2
          - 16 \zeta(2)
          \Bigr) \nn\\
& &  \hspace{18mm} 
        + C_{F} C_{A} \Bigl(
          - \frac{47}{9}
          + 2 L
          - L^2
          + 2 \zeta(2)
\Bigr)
        + C_{F} T_{R} N_{f} \Bigl(
            \frac{10}{9}
\Bigr) \Biggr]  \nn\\
& &  \hspace{6mm} 
 + \left( \frac{m^2}{S} \right) \Biggl[
         C_{F}^2 \Bigl(
            7
          - 8 L
          \Bigr)  \Biggr]  \nn\\
& &  \hspace{6mm} 
        + C_{F}^2 \Bigl(
            \frac{7}{2}
          - 4 L
          + \frac{3}{2} L^2
          - 4 \zeta(2)
          \Bigr)
        + C_{F} C_{A} \Bigl(
          - \frac{67}{36}
          + \frac{1}{2} \zeta(2)
\Bigr)  \nn\\
& &  \hspace{6mm} 
        + C_{F} T_{R} N_{f} \Bigl(
            \frac{5}{9}
\Bigr) 
 \Biggr\} \nn\\
& & 
  + \left( \frac{m^2}{S} \right)^2 \Biggl[
         C_{F}^2 \Bigl(
            \frac{1}{2}
          + 110 \zeta(2) L
          - 216 L
          + \frac{101}{2} L^2
          - \frac{28}{3} L^3
          - 232 \zeta(2) \nn\\
& &  \hspace{24mm} 
          + 148 \zeta(3)
          \Bigr) \nn\\
& &  \hspace{18mm} 
        + C_{F} C_{A} \Bigl(
          - \frac{12607}{108}
          + 28 \zeta(2) L
          - \frac{419}{9} L
          - \frac{77}{6} L^2
          + \frac{8}{3} L^3 \nn\\
& &  \hspace{24mm} 
          - 91 \zeta(2)
          + 238 \zeta(3)
\Bigr) \nn\\
& &  \hspace{18mm} 
        + C_{F} T_{R} N_{f} \Bigl(
            \frac{731}{27}
          - \frac{74}{9} L
          + \frac{2}{3} L^2
\Bigr)   \nn\\
& &  \hspace{18mm} 
        + C_{F} T_{R} \Bigl(
            \frac{3730}{27}
          - \frac{578}{9} L
          - \frac{4}{3} L^2
\Bigr) \Biggr]  \nn\\
& & 
 - \left( \frac{m^2}{S} \right) \Biggl[
         C_{F}^2 \Bigl(
          -  \frac{97}{4} \! 
          -  \! 6 \zeta(2) L \! 
          +  \! \frac{77}{2} L \! 
          -  \! 14 L^2 \! 
          +  \! \frac{1}{3} L^3 \! 
          +  \! 43 \zeta(2) \! 
          -  \! 12 \zeta(3)
          \Bigr) \nn\\
& &  \hspace{18mm} 
        + C_{F} C_{A} \Bigl(
            \frac{269}{12} \! 
          -  \! 3 \zeta(2) L \! 
          +  \! \frac{1}{2} L \! 
          + \!  2 L^2 \! 
          -  \! \frac{1}{6} L^3 \! 
          +  \! 5 \zeta(2) \! 
          -  \! 18 \zeta(3)
\Bigr) \nn\\
& &  \hspace{18mm} 
        + C_{F} T_{R} N_{f} \Bigl(
          - \frac{25}{3}
          + 2 L
\Bigr)
        + C_{F} T_{R} \Bigl(
          - \frac{49}{3}
          + 10 L
\Bigr) \Biggr]  \nn\\
& &  
        + C_{F}^2 \Bigl(
            \frac{85}{8}
          + 10 \zeta(2) L
          - \frac{55}{4} L
          + 5 L^2
          - \frac{7}{6} L^3
          - 11 \zeta(2)
          - 8 \zeta(3)
          \Bigr) \nn\\
& &  
        + C_{F} C_{A} \Bigl(
          - \frac{2545}{216}
          - \zeta(2) L
          + \frac{233}{36} L
          - \frac{11}{12} L^2
          + \frac{13}{2} \zeta(3)
\Bigr) \nn\\
& &  
        + C_{F} T_{R} N_{f} \Bigl(
            \frac{209}{54}
          - \frac{19}{9} L
          + \frac{1}{3} L^2
\Bigr) \nn\\
& & 
        + C_{F} T_{R} \Bigl(
            \frac{265}{54}
          - \frac{19}{9} L
          + \frac{1}{3} L^2
\Bigr) \, , \\
\Re \, {\mathcal F}^{(2l)}_{2,R}(\epsilon,s) & = & 
\frac{1}{\epsilon} \Biggl\{ 
   \left( \frac{m^2}{S} \right)^2 \Biggl[
         C_{F}^2 \Bigl(
          - 4
          + 12 L
          - 4 L^2
          + 24 \zeta(2)
          \Bigr)  \Biggr]  \nn\\
& &  \hspace{6mm} 
 - \left( \frac{m^2}{S} \right) \Biggl[
         C_{F}^2 \Bigl(
          - 2 L
          + 2 L^2
          - 12 \zeta(2)
          \Bigr) \Biggr] 
\Biggr\} \nn\\
& & 
  + \left( \frac{m^2}{S} \right)^2 \Biggl[
         C_{F}^2 \Bigl(
          - 89
          - 244 \zeta(2) L
          + 112 \zeta(3) L
          + 7 L
          + 28 \zeta(2) L^2 \nn\\
& &  \hspace{24mm} 
          - 79 L^2
          + \frac{28}{3} L^3
          - \frac{1}{3} L^4
          + 192 \zeta(2) \ln{2}
          + 656 \zeta(2)\nn\\
& &  \hspace{24mm} 
          - \frac{276}{5} \zeta^2(2)
          - 224 \zeta(3)
          \Bigr) \nn\\
& &  \hspace{18mm} 
        + C_{F} C_{A} \Bigl(
          - \frac{343}{9}
          - 44 \zeta(2) L
          + 136 \zeta(3) L
          - \frac{341}{9} L \nn\\
& &  \hspace{24mm} 
          + 10 \zeta(2) L^2
          - \frac{55}{3} L^2
          - \frac{8}{3} L^3
          + \frac{1}{6} L^4
          - 96 \zeta(2) \ln{2} \nn\\
& &  \hspace{24mm} 
          + \frac{944}{3} \zeta(2)
          - \frac{174}{5} \zeta^2(2)
          - 404 \zeta(3)
\Bigr) \nn\\
& &  \hspace{18mm} 
        + C_{F} T_{R} N_{f} \Bigl(
          - \frac{76}{9}
          + \frac{148}{9} L
          - \frac{4}{3} L^2
          + \frac{32}{3} \zeta(2)
\Bigr)   \nn\\
& &  \hspace{18mm} 
        + C_{F} T_{R} \Bigl(
          - \frac{916}{9}
          + \frac{868}{9} L
          - \frac{52}{3} L^2
          - 16 \zeta(2)
\Bigr) \Biggr]  \nn\\
& &  
 - \left( \frac{m^2}{S} \right) \Biggl[
         C_{F}^2 \Bigl(
            48 \zeta(2) L
          - \frac{31}{2} L
          + \frac{17}{2} L^2
          - 2 L^3
          - 48 \zeta(2) \ln{2} \nn\\
& &  \hspace{24mm} 
          - 36 \zeta(2)
          + 4 \zeta(3)
          \Bigr) \nn\\
& &  \hspace{18mm} 
        + C_{F} C_{A} \Bigl(
            3
          + \frac{173}{18} L
          + \frac{1}{6} L^2
          + 24 \zeta(2) \ln{2}
          - \frac{64}{3} \zeta(2) \nn\\
& &  \hspace{24mm} 
          + 20 \zeta(3)
\Bigr) \nn\\
& &  \hspace{18mm} 
        + C_{F} T_{R} N_{f} \Bigl(
          - \frac{50}{9} L
          + \frac{2}{3} L^2
          - \frac{16}{3} \zeta(2)
\Bigr) \nn\\
& &  \hspace{18mm} 
        + C_{F} T_{R} \Bigl(
            \frac{68}{3}
          - \frac{50}{9} L
          + \frac{2}{3} L^2
          - 8 \zeta(2)
\Bigr) \Biggr] \, , \\
\Im \, {\mathcal F}^{(2l)}_{2,R}(\epsilon,s) & = & 
\frac{1}{\epsilon} \Biggl\{ 
   \left( \frac{m^2}{S} \right)^2 \Biggl[
         C_{F}^2 \Bigl(
          - 12
          + 8 L
          \Bigr)  \Biggr]  \nn\\
& &  \hspace{6mm} 
 - \left( \frac{m^2}{S} \right) \Biggl[
         C_{F}^2 \Bigl(
            2
          - 4 L
          \Bigr) \Biggr] 
\Biggr\} \nn\\
& & 
   + \left( \frac{m^2}{S} \right)^2 \Biggl[
         C_{F}^2 \Bigl(
          - 7
          - 40 \zeta(2) L
          + 158 L
          - 28 L^2
          + \frac{4}{3} L^3
          + 132 \zeta(2) \nn\\
& &  \hspace{24mm} 
          - 112 \zeta(3)
          \Bigr) \nn\\
& &  \hspace{18mm} 
        + C_{F} C_{A} \Bigl(
            \frac{341}{9}
          - 28 \zeta(2) L
          + \frac{110}{3} L
          + 8 L^2
          - \frac{2}{3} L^3\nn\\
& &  \hspace{24mm} 
          + 76 \zeta(2)
          - 136 \zeta(3)
\Bigr) \nn\\
& &  \hspace{18mm} 
        - C_{F} T_{R} N_{f} \Bigl(
            \frac{148}{9}
          - \frac{8}{3} L
\Bigr)  
        - C_{F} T_{R} \Bigl(
            \frac{868}{9}
          - \frac{104}{3} L
\Bigr) \Biggr]  \nn\\
& &  \hspace{6mm} 
 - \left( \frac{m^2}{S} \right) \Biggl[
         C_{F}^2 \Bigl(
            \frac{31}{2}
          - 17 L
          + 6 L^2
          - 24 \zeta(2)
          \Bigr) \nn\\
& &  \hspace{18mm} 
        - C_{F} C_{A} \Bigl(
            \frac{173}{18}
          + \frac{1}{3} L
\Bigr)
        + C_{F} T_{R} N_{f} \Bigl(
            \frac{50}{9}
          - \frac{4}{3} L
\Bigr)  \nn\\
& &  \hspace{18mm} 
        + C_{F} T_{R} \Bigl(
            \frac{50}{9}
          - \frac{4}{3} L
\Bigr) \Biggr] \, .
\eea

All the results of this Section can be obtained in an electronic form 
by downloading the source of this manuscript from http://www.arxiv.org.

\section{Summary and Outlook}

In this paper, we calculated the two-loop QCD corrections 
to the vector vertex form factors for heavy quarks. The result for the
electric and magnetic form factors was obtained keeping 
the full dependence of the form factors on the mass of the heavy quarks,
as well as on the momentum transfer, which is maintained arbitrary.

The extraction of the form factors from the Feynman diagrams involved in the
calculation was carried out by means of standard projector operators. Each form
factor is expressed, in this way, as a combination of several hundreds of 
scalar integrals, whose UV and IR divergences are regularized, within the 
Dimensional Regularization procedure, by the same parameter $D$, dimension of
the space-time. Using the Laporta algorithm, it was possible to reduce the 
problem of the calculation of all these integrals to the calculation of 17 
master integrals, already present in the literature. 

The renormalization of the UV divergences was carried out in a hybrid scheme 
in which the coupling constant and the gluon wave function are renormalized 
in the $\overline{\mathrm{MS}}$ scheme, while the mass and wave function of 
the heavy quark are renormalized in the on-shell scheme. 

The expressions of the unsubtracted as well as the UV-renormalized form 
factors are given in a closed analytic form as a Laurent expansion in
$\epsilon=(4-D)/2$. The coefficients of this expansion have a suitable
representation in terms of 1-dimensional harmonic polylogarithms.
The presence of poles in $\epsilon$ in our results is related to the fact that 
IR divergences are still present. These divergences have to be canceled against
the divergences arising from the real radiation, which in this paper was not 
taken into account.

Besides being part of the full NNLO corrections to the forward-backward
asymmetry of heavy quarks, the results presented in this paper can, on
their own, already be used in a number of applications.

An immediate point of interest is the behaviour of the inclusive heavy
quark production cross section above the threshold.  
In the continuum the $4 \pi$-integrated cross section was computed to order
$\alpha_s^2$ \cite{Chetyrkin:1997pn} and, in view of top quark pair 
production at a future linear collider, very detailed NNLO studies 
have been carried out in the threshold region (see~\cite{hoang} for 
a review), where the cross section is most sensitive on the top quark mass. 
The form factors derived here can be extrapolated at the two-loop level to 
their threshold values. We find complete agreement with all threshold 
results available in the literature~\cite{Hoang1,mel2,beneke}.

All available calculations of NNLO QCD corrections to inclusive
heavy quark production at asymptotically large energies were made using
the optical theorem (see~\cite{hs} for a review), which avoids the
explicit calculation of the two-loop form factors. Computing more
differential quantities such as rapidity distributions requires the
explicit knowledge of the two-loop form factors as well as of the
massive (or leading-mass) single and double real radiation corrections.
In the massless case, such calculations could be performed only very
recently~\cite{twoj}, and a first step towards a fully massive calculation
would certainly only consider the leading mass terms. In view of this
application, we also provided the leading mass expansions of our full
two-loop corrections to the form factors.

The singularity structure of the two-loop heavy quark form factors could
also provide insight into the generic singularity structure of two-loop
integrals involving massive quarks. The corresponding structure of
massless two-loop QCD amplitudes has been predicted from non-abelian
exponentiation in~\cite{catani,st}, and proven very valuable in the
calculation of massless two-loop four-point amplitudes~\cite{glover}. For
QCD amplitudes involving massive quarks, the singularity structure is only
understood at the one-loop level at present~\cite{cditt}. At the two-loop
level, we observe that the infrared divergent contributions to the form
factors which are proportional to the colour factor $C_F^2$ exponentiate
naively~\cite{yfs}, as already seen in the QED calculation
\cite{RoPieRem2}. The other colour factors contain explicit $1/\epsilon$
singularities which can not be explained from naive abelian
exponentiation and deserve further investigation.

Finally, this calculation demonstrates that it is possible to analytically
compute two-loop vertex functions with at least one internal mass scale.
These functions appear in a variety of applications, ranging from
electroweak corrections to heavy quark physics, where first genuine
two-loop results were obtained only very recently
\cite{ba,kniehl,seidel,mel3}, using the same methods (reduction to master
integrals, differential equations, and basis of harmonic polylogarithms)
as employed here. Many more results in this domain are yet outstanding,
and the methods here can be clearly instrumental for their derivation.

To compute the NNLO QCD corrections to the forward-backward asymmetry
of heavy quarks, one also needs the two-loop corrections to the
axial vector form factors. These corrections subdivide into two
independent classes: anomalous and non-anomalous diagrams. The
calculation of these is currently in progress, and results will be
reported in two future publications.

\section*{Acknowledgment}

We are grateful to J.~Vermaseren for his kind assistance in the use
of the algebra manipulating program {\tt FORM}~\cite{FORM}, by which
all our calculations were carried out.

T.G. would like to thank Nigel Glover and Martin Beneke for useful 
discussions.

T.G. and P.M. wish to thank the Kavli Institute for Theoretical Physics, 
Santa Barbara, where part of this work was carried out, for its kind 
hospitality.

This work was partially supported by the European Union under
contract HPRN-CT-2000-00149, by Deutsche Forschungsgemeinschaft (DFG), 
SFB/TR9, by DFG-Graduiertenkolleg RWTH Aachen, by the Swiss 
National Funds (SNF) under contract 200021-101874, by the National 
Science Foundation under Grant No.\ PHY99-07949, and by the USA DoE under 
the grant DE-FG03-91ER40662, Task J.

%%%%%%%%%%%%%%%%%%%%%%%%%%%%%%%%%%%%%%%%%%%%%%%%%%%%%%%%%%%%%%%%%%%%%

\end{fmffile}

\end{document}